\documentclass[pdflatex,sn-mathphys-num]{sn-jnl}

\usepackage{mathtools}

\usepackage{color,xcolor,ucs}
\usepackage{mathtools}   \usepackage{tikz} 
\usepackage{ amssymb }
\usepackage{extarrows} 
\usepackage{pgf,tikz}
\usepackage{float}
\usetikzlibrary{positioning}
\usetikzlibrary{shapes.geometric}
\usetikzlibrary{shapes.misc}
\usetikzlibrary{arrows}
\usepackage{caption}
\usepackage{mathrsfs}
\usetikzlibrary{arrows,shapes,automata,backgrounds,petri,positioning}
\usetikzlibrary{decorations.pathmorphing}
\usetikzlibrary{decorations.shapes}
\usetikzlibrary{decorations.text}
\usetikzlibrary{decorations.fractals}
\usetikzlibrary{decorations.footprints}
\usetikzlibrary{shadows}
\usetikzlibrary{calc}
\usetikzlibrary{spy}
\usepackage{amsmath}
\usepackage{array}
\usepackage{ amssymb }
\usepackage{braket}
\usepackage{qcircuit}
\usepackage{soul}
\usepackage{braket} 
\usepackage{relsize}

\usepackage{amsmath}
\usepackage{ amssymb }
\usepackage{braket}
\usepackage{qcircuit}
\usepackage{soul}
\usepackage{braket}

\usepackage{graphicx}%
\usepackage{multirow}%
\usepackage{amsmath,amssymb,amsfonts}%
\usepackage{amsthm}%
\usepackage{mathrsfs}%
\usepackage[title]{appendix}%
\usepackage{xcolor}%
\usepackage{textcomp}%
\usepackage{manyfoot}%
\usepackage{booktabs}%
\usepackage{algorithm}%
\usepackage{algorithmicx}%
\usepackage{algpseudocode}%
\usepackage{listings}%

\theoremstyle{thmstyleone}%
\theoremstyle{thmstyletwo}%

\theoremstyle{thmstylethree}%

\begin{document}

\title[Article Title]{Parallel repetition of expanded, and multiplayer, Quantum games: anchoring, optimal values, generalized error bounds, dependency-breaking as symmetry-breaking}


\author[1]{\fnm{Pete} \sur{Rigas}}\email{pbr43@cornell.edu}

\affil[1]{\city{Newport Beach}, \postcode{92625}, \state{CA}, \country{USA}}


\abstract{We demonstrate that parallel repetition of the multiplayer anchored optimal value, $\omega \big( G_{\bot} \big)^{\otimes n}$, decays exponentially. Central to our approach are several probabilistic computations, pertaining to: (1) the computation of expected values for quantifying how the winning probability of the game is likely to change under the anchoring transformation; (2) the computation of positive operator valued measurements,  which can be placed into direct correspondence with several probabilistically defined quantities; (3) the computation of Relative, and Relative-min entropies; (4) and lastly, the computation of generalized error bounds, which have previously been analyzed by the author in several multiplayer game-theoretic settings (arXiv: 2505.06322, and arXiv: 2507.03035). This work builds upon observations originally provided by Bavarian, Vidick, and Yuen (arXiv: 1509.07466).\footnote{\textbf{MSC Class}: 81P02; 81Q02} }

\keywords{Multiplayer games, Quantum games, error bounds, optimal values, anchoring}



\maketitle

\section{Introduction}

\subsection{Overview}

Quantum interpretations of Game Theory have exposed striking differences between Classical and Quantum information, ranging from: pseudo-telepathy, such as encountered in the Magic square game, which has quantum value, ie the winning probability, equaling exactly $1$ [7]; computational complexity of training variational quantum algorithms, and related, preprocessing, steps [9, 28, 29, 31]; various classes of nonlinear problems [8, 11, 12, 13, 14, 19, 20, 51]; non-locality, and other related fundamental, properties of Quantum information [3, 4, 8, 13, 14, 15, 23]; amongst several related problems [1, 5, 21, 22, 23, 24, 25, 27]. In previous work of the author, significant effort was devoted towards elaborating, upon the structure of correlations, along with the impact of entanglement, on various information processing tasks [50, 52]. Within such an approach, as an adaptation of argument originally provided in [42], two-player XOR, $\mathrm{XOR}^{*}$, and FFL, games were shown to satisfy different properties, particularly with regards to closed form representations for optimal values, whether classical or quantum.

In general, the optimal value of a game reflects upon basic aspects of classical and quantum sources of information. Relatedly, the operation of perform ordinary, or strong, parallel repetition of a game can drastically impact the maximum probability of winning. In various multiplayer settings, previous work of the author analyzed ordinary, and generalized, error bounds, in addition to optimality and duality gaps [52]. To further contribute to the entanglement structures, and barriers to achieving quantum advantage, it remains of interest to examine how parallel repetition of a value for some game is reflected through generalized error bounds. In various multiplayer settings, one of which is described in [18], the parallel repetition of a game, 

\begin{align*}
  G^{\otimes n}  \text{, }
\end{align*}

\noindent for some $n>0$, is encoded through the product distributions of questions, and answers, from all participants,

\begin{align*}
  Q_1 \times \cdots \times Q_i \times \cdots \times Q_N  \text{, } \\ A_1 \times \cdots \times  A_i \times  \cdots \times A_N \text{, }
\end{align*}

\noindent the referee's scoring function, or predicate, in the case of an arbitrary number of questions which can be distributed to each participant, [21],

\begin{align*}
  p \big( q_1, \cdots, q_i, \cdots, q_n \big)   \text{, }
\end{align*}

\noindent would take the form,

\begin{align*}
  V \big( a_1, \cdots, a_i, \cdots, a_n | q_1, \cdots, q_i, \cdots, q_n \big)   \text{. }
\end{align*}

\noindent As it will be discussed, under parallel repetition the tensor product of the set of questions for each player $i$, $\mathcal{Q}_i$, satisfies,

\begin{align*}
  \mathcal{Q}^{\otimes k}_i \equiv   \bigg\{ \big( \mathcal{Q}_1, \cdots, \mathcal{Q}_k \big) , \cdots , \big( \mathcal{Q}^{\prime\cdots\prime}_1, \cdots, \mathcal{Q}^{\prime\cdots\prime}_k \big) \bigg\}       \text{. }
\end{align*}

\noindent In previous work of the author, [52], several error bounds, and generalizations of error bounds, were analyzed for multiplayer games, and parallel repetition of multiplayer games. Under such circumstances, the XOR, $\mathrm{XOR}^{*}$, and FFL, games present striking differences. That is, albeit the fact that the XOR and FFL games differ in a probability of $\frac{1}{2}$, versus $\frac{1}{4}$, for the answers that a player provides to one question drawn from the referee's probability distribution, the XOR value satisfies,

\begin{align*}
  \omega_{\mathrm{XOR}  \wedge \mathrm{XOR}} \big( G_{\mathrm{XOR} \wedge  \mathrm{XOR}} \big)  \equiv \omega \big(  \mathrm{XOR} \wedge \mathrm{XOR} \big) \equiv \underset{1\leq j \leq 2}{\prod}   \omega \big( \mathrm{XOR}\big)^{j} \equiv \big( \omega \big( \mathrm{XOR} \big) \big)^2 \equiv \frac{1}{2} \\  \neq \omega \big( \mathrm{XOR}  \big)  \neq \frac{1}{\sqrt{2}} \text{,} \end{align*}

\noindent while for the $\mathrm{FFL}$ value satisfies,

\begin{align*}
  \omega_{\mathrm{FFL} \wedge  \mathrm{FFL}} \big( G_{\mathrm{FFL}} \big) \equiv \omega_{\mathrm{FFL}} \big( G_{\mathrm{FFL}} \wedge G_{\mathrm{FFL}} \big)   \equiv \omega \big(  \mathrm{FFL}  \wedge \mathrm{FFL} \big) \equiv  \omega \big( \mathrm{FFL}\big) \equiv \frac{2}{3} \neq \bigg( \frac{2}{3} \bigg)^2  \text{.} \end{align*}

\subsection{This paper's contributions}

\noindent In comparison to previous work of the author which formalized notions of error bounds, exact optimality, and approximate optimality, as a related topic of research it remains of interest to determine how rates of decay, whether polynomial or exponential, of optimal values for different games determine the range of parameters, and hence, strategies, that participants can assume for winning. While modest prospects for Quantum advantage have been realized for CHSH, XOR, XOR*, and FFL games, [42, 50, 52], determining how the operation of parallel repetition impacts error bounds for expanded games has not yet been thoroughly examined within the previously aforementioned framework. To determine whether error bounds would be capable of determining dynamic, evolving, strategies that players can assume for maximizing his or her respective utility, and as a result, winning probability, we: (1) provide statements of error bounds previously obtained by the author, under parallel repetition, for XOR, $\mathrm{XOR}^{*}$, and FFL, games; (2) discuss the parallel repetition operation on expanded games, particularly the impact of the repetition operation on the decay of the optimal value; (3) argue, along the lines of those obtained for error bounds of XOR, $\mathrm{XOR}^{*}$ and FFL, games, given optimal strategies $\ket{\psi^{\prime}} \equiv \ket{\psi_{\mathrm{Expanded}}}$ of the expanded game; (4) discuss the significance of dependency-breaking variables, which in the case of expanded games, unlike in previous arguments for XOR, $\mathrm{XOR}^{*}$ and FFL, games, can describe how correlations between Quantum states can be broken. Quantum correlations, and associated dependency-breaking, or symmetry-breaking, routines, can be of use for determining whether, with high probability, players can still make use of optimal strategies for satisfying winning conditions from the referee's scoring predicate. However, in comparison to classically defined correlations introduced for models of Statistical Physics, such as in the Ising, Potts, 4-vertex, 6-vertex, 20-vertex, and closely related, models, Quantum correlations dictates,

\begin{align*}
  \textbf{P} \big[ \Omega_{\mathrm{Expanded}} \big| \textbf{X}_i \equiv x, W_C \big] \approx \textbf{P} \big[ \Omega_{\mathrm{Expanded}} \big| \textbf{X}_i \equiv x, \textbf{Y}_i \equiv y,  W_C  \big] \approx \textbf{P} \big[ \Omega_{\mathrm{Expanded}} \\ \big|  \textbf{Y}_i \equiv y,  W_C \big]   \text{, }
\end{align*}

\noindent for the reference probability measure $\textbf{P} \big[ \cdot \big]$ over the set of all possible responses for the expanded game, set of dependency breaking variables $\Omega_{\mathrm{Expanded}}$ for the expanded game, and the \textit{winning} event $W$. Together, these objects are not only used for defining several useful properties for breaking correlations amongst possible strategies that players can use when playing expanded games, but also for other game-theoretic settings. In the conditioning of the probabilities above, the event $W$ denotes,

\begin{align*}
  W_C \equiv \big\{ \text{Players win the expanded game for all coordinates } C \big\}   \text{,}
\end{align*}

\noindent In the above approximation of probability measures over the set of possible questions for each player in the expanded game, approximately entails that,

\begin{align*}
 \big|   \textbf{P} \big[ \Omega_{\mathrm{Expanded}} \big| \textbf{X}_i \equiv x, W_C \big] - \textbf{P} \big[ \Omega_{\mathrm{Expanded}} \big| \textbf{X}_i \equiv x, \textbf{Y}_i \equiv y,  W_C  \big] \big| \approx 0 , \\ \big| \textbf{P} \big[ \Omega_{\mathrm{Expanded}} \big| \textbf{X}_i \equiv x, \textbf{Y}_i \equiv y,  W_C  \big] -  \textbf{P} \big[ \Omega_{\mathrm{Expanded}}  \big|  \textbf{Y}_i \equiv y,  W_C \big]   \big| \approx 0  \text{, }
\end{align*}

\noindent holds with respect to the standard $l$-1, absolute-value, norm $\big| \cdot \big|$. The two underlying assumptions guarantee that the two sets of conditional probabilities above, which are dependent upon,

\begin{align*}
 \big\{  \textbf{X}_i \equiv x, W_C        \big\}    \text{, } \\ \big\{    \textbf{X}_i \equiv x, \textbf{Y}_i \equiv y,  W_C   \big\} \text{, } \\ \big\{ \textbf{X}_i \equiv x, \textbf{Y}_i \equiv y,  W_C  \big\} \text{, } \\ \big\{ \textbf{Y}_i \equiv y,  W_C  \big\} \text{, }
\end{align*}

\noindent rely upon \textit{usefulness} and \textit{sampleability}. These two conditions not only ensure, for expanded games, that the measurements that any of the two players can make on the set of all possible dependency breaking states,

\begin{align*}
  \ket{\Psi_{x,y}} \equiv \underset{\psi \in \Psi}{\bigcup} \ket{\psi_{x,y}}  \text{, }
\end{align*}

\noindent can be made arbitrarily close to the conditional probability,

\begin{align*}
  \textbf{P} \big[ A_i B_i \big| X_i \equiv x, Y_i \equiv y, W_C \big]  \text{, }
\end{align*}

\noindent but also that,

\begin{align*}
\big|  \big( U_x \otimes V_y \big) \ket{\Psi}  - \ket{\Psi_{x,y}} \big| \approx 0   \text{, }
\end{align*}

\noindent for some strategy $\ket{\Psi}$, and unitary transformations $U_x$ and $V_y$. The two conditions described above, \textit{usefulness} and \textit{sampleability}, relate to several other topics investigated Quantum Information theory. For example, from previous work of the author on converse results on the bit transmission rate $r$, it was demonstrated that, [53],

\begin{align*}
\underset{P_{\textbf{X}}}{\mathrm{sup}}  \big\{ \mathrm{min} \big\{  I \big( \textbf{X} , \textbf{Y} \big)  , \underset{z}{\mathrm{min}}  \big\{ H_Q \big( \textbf{Y} \big| \textbf{Z} = z \big)   -  H_P \big( \textbf{Y} \big| \textbf{X} \big)  \big\}  \big\}    \big\} 
\end{align*}
\[ <  \left\{\!\begin{array}{ll@{}>{{}}l} 
     \mathrm{log} \mathrm{log} \bigg[   \frac{  \mathrm{log} \big| \textbf{Y}^{*} \big|   }{ \big| \textbf{X}^{*} \big|  }          \bigg]   +   \mathrm{log}  \bigg[   \frac{   \mathrm{log}\big|  \textbf{Z}  \big|  }{  \big|  \textbf{Y}^{*}  \big|   } \bigg]  \Longleftrightarrow \big| \textbf{X} \big| > \big| \textbf{Y}^{*} \big| ,  \big| \textbf{Y}^{*} \big| > \big| \textbf{Z} \big|    ,     \\ \mathrm{log} \mathrm{log} \bigg[     \frac{  \mathrm{log} \big| \textbf{X} \big|  }{ \big| \textbf{Y}^{*} \big| }          \bigg]   +   \mathrm{log}  \bigg[  \frac{ \mathrm{log} \big|  \textbf{Y}^{*} \big|   }{  \big| \textbf{X} \big|   } \bigg]   \Longleftrightarrow \big| \textbf{X} \big| <  \big| \textbf{Y}^{*} \big| ,  \big|  \textbf{Y}^{*} \big| < \big| \textbf{Z} \big|        ,  \\ \mathrm{log} \mathrm{log} \bigg[     \frac{  \mathrm{log} \big| \textbf{Y}^{*} \big|  }{ \big| \textbf{X}^{*} \big| }          \bigg]   +   \mathrm{log}  \bigg[  \frac{  \mathrm{log}\big|  \textbf{Y}^{*} \big|   }{ \big| \textbf{X} \big|   } \bigg]   \Longleftrightarrow \big| \textbf{X} \big| >   \big| \textbf{Y}^{*} \big| ,  \big|  \textbf{Y}^{*} \big| < \big| \textbf{Z} \big|        ,     \\ \mathrm{log} \mathrm{log} \bigg[     \frac{  \mathrm{log} \big| \textbf{X}^{*} \big|  }{\big| \textbf{X}^{*} \big| }          \bigg]   +   \mathrm{log}  \bigg[  \frac{  \mathrm{log}\big|  \textbf{Z}  \big|   }{ \big| \textbf{Y}^{*} \big|   } \bigg]   \Longleftrightarrow \big| \textbf{X} \big| <    \big| \textbf{Y}^{*} \big| ,  \big|  \textbf{Y}^{*} \big| >   \big| \textbf{Z} \big|        .                          
\end{array}\right. \equiv  r \text{, }
\]

\noindent holds, which allows Alice and Bob to achieve composable, unconditionally secure, message authentication without Quantum key distribution. The dependency-breaking result for expanded games, as does the result for being able to ensure that authentication, error correction, and decoding error, protocols can be implemented with high probability given the converse result on $r$ above, both inherently rely upon \textit{shared} aspects of Quantum information between multiple participants. That is, participants being able to share aspects of entangled Quantum information with each other enables them to potentially, should they decide to use protocols for generating long Quantum secret keys, from limited information that is initially shared. For the purposes of dependency-breaking, in expanded games, previous arguments due to [2, 3, 4, 56] demonstrated that, under parallel repetition, setting questions to \textit{anchored} ones implies,

\begin{align*}
  \textbf{P} \big[ A_i B_i \big| X_i \equiv x, Y_i \equiv y, W_C \big] \approx  \textbf{P} \big[ A_i B_i \big| X_i \equiv \bot ,  Y_i \equiv y, W_C \big] \\ \approx \textbf{P} \big[ A_i B_i \big| X_i \equiv \bot ,  Y_i \equiv \bot , W_C \big] \text{, }
\end{align*}

\noindent for anchors $\bot$, iff,

\begin{align*}
  \big| \textbf{P} \big[ A_i B_i \big| X_i \equiv x, Y_i \equiv y, W_C \big] -  \textbf{P} \big[ A_i B_i \big| X_i \equiv \bot ,  Y_i \equiv y, W_C \big] \big| \approx 0 \text{, } \\  \big| \textbf{P} \big[ A_i B_i \big| X_i \equiv \bot ,  Y_i \equiv y, W_C \big]  -  \textbf{P} \big[ A_i B_i \big| X_i \equiv \bot ,  Y_i \equiv \bot , W_C \big] \big| \approx 0 \text{. }
\end{align*}

\noindent Hence, one would like to study properties of the dependency-breaking set,

\begin{align*}
   \Omega_{\mathrm{Multiplayer} } \equiv  \underset{\omega \in \Omega_{\mathrm{Multiplayer}}}{\bigcup} \big\{ \omega  : \omega       \text{ is a dependency breaking variable for} \\ \text{position j}       \big\} \text{. }
\end{align*}

\noindent To this end, determining the extent to which the two participants of other two-player games, rather than in expanded games, can share entangled Quantum information for breaking correlations,  through \textit{dependency-breaking}, remains of interest. In the forthcoming arguments, we identify the following connections:

\begin{itemize}
    \item[$\bullet$] \textit{Connections between Multiplayer, and Expanded, Quantum games}. We demonstrate how previous arguments of the author, [52], in light of exponential rates of decay for the optimal value under parallel repetition, can be employed to construct error bounds, and closely related objects related to generalizations of error bounds.

    \item[$\bullet$] \textit{Multiplayer anchored parallel repetition}. In tandem with the first item, we also demonstrate how connections with error bounds of expanded games can be employed to formulate arguments for multiplayer anchored parallel repetition, an analog of the two-player setting previously analyzed in [2, 3, 4].
\end{itemize}

\subsection{The approach}

To apply such arguments for games with more than two participants, we pursue the following programme:

\begin{itemize}
    \item[$\bullet$] \textit{Introduce dependency breaking variables for multiplayer settings}. In comparison to the set of all dependency-breaking variables that one would like to sample conditionally upon the responses from each player for the Expanded game, such collection of variables for other, potentially, multiplayer games, must depend on product strategies. To formulate product strategies from parallel repetition of games, recall, as discussed in the next section from primal feasible solutions of semidefinite programs, that,

\begin{align*}
  \underset{1 \leq j \leq n}{\bigwedge} \bigg[  \underset{\forall Z_{\mathrm{XOR}\wedge \cdots \wedge \mathrm{XOR}} \succcurlyeq 0 , 1 \leq i \leq m, F^{(j)}_{\wedge i} \cdot Z_{\mathrm{XOR} \wedge \cdots \wedge {\mathrm{XOR}}} \equiv C^{(j)}_i}{\mathrm{sup}}    G_{\mathrm{XOR}} Z_{\mathrm{XOR}}      \bigg]       \text{, }
\end{align*}

\noindent corresponds to the primal feasible solution under strong parallel repetition for multiplayer XOR games, with respect to the partial ordering,

\begin{align*}
 \succcurlyeq    \text{, }
\end{align*}

\noindent of the positive semidefinite cone. Under ordinary parallel repetition, namely the parallel repetition operation for which the decay of the optimal value of the game is dependent upon an exponential, the product strategy for $n$ parallel repetitions,

\begin{align*}
 \mathscr{S}^n \equiv \underset{\mathcal{S}^n \in \mathscr{S}^n}{\bigcup} \big\{ \text{strategies } \mathcal{S}^n :  \textbf{P} \big[ \text{Players can win using } \mathcal{S}^n \big] \cap \textbf{P} \big[ \mathcal{S}^n \text{ satisfies} \\ \text{ usefulness and sampleability simutlaneously} \big] > 0               \big\}   \text{, }
\end{align*}

\noindent admits the factorization,

\begin{align*}
 \mathscr{S}^n \equiv   \underset{1 \leq j \leq n}{\bigwedge} \mathscr{S}^j \equiv \underset{1 \leq j \leq n}{\prod} \mathscr{S}^j     \equiv \underset{1 \leq j \leq n}{\prod} \big\{ \text{strategies } \ket{\psi}^j : \text{Players participate} \\ \text{  using } \ket{\psi}^j      \big\}       \text{. }
\end{align*}

      \item[$\bullet$] \textit{Multiplayer product strategies}. As discussed in [2, 3, 4], considering the product strategies, as given in the form above, is essential for simultaneously being able to satisfy the two properties of dependency-breaking variables for several participants. While one can consider strategies that are not of the above product form for two-player settings, such game-theoretic settings are far less rich, primarily for the following reasons: (1) minimal systems in which entanglement can arise, through Bell states and EPR pairs, are dependent only upon the equalities,

\begin{align*}      
\bigg(  \textbf{I} \otimes \textbf{I} \bigg)  \frac{\ket{00} + \ket{11}}{\sqrt{2}} = \frac{\ket{00} + \ket{11}}{\sqrt{2}}   \text{ } \text{ , }     \bigg(     \sigma_x \otimes \textbf{I} \bigg) \frac{\ket{00} + \ket{11}}{\sqrt{2}}  = \frac{\ket{10} + \ket{01}}{\sqrt{2}}   \text{, }   \\   \bigg(   \sigma_z    \otimes \textbf{I} \bigg)  \frac{\ket{00} + \ket{11}}{\sqrt{2}}  = \frac{\ket{00} - \ket{11}}{\sqrt{2}}       \text{ } \text{ , }    \bigg( \sigma_x \sigma_z     \otimes   \textbf{I}  \bigg) \frac{\ket{00}+\ket{11}}{\sqrt{2}} = \frac{\ket{10}- \ket{01}}{\sqrt{2}}     \text{, } 
\end{align*}

      \noindent of Pauli, and identity, operators; (2) optimal values for two-player settings, namely the maximum winning probability,

\begin{align*}
 \omega_{\mathrm{XOR}} \big( G \big) \equiv \omega \big( \mathrm{XOR} \big)    \text{, }
\end{align*}

\noindent from the referree's scoring predicate,

\begin{align*}
    V \big(  s , t \big)   ab  \equiv 1 \Longleftrightarrow  \text{ Alice and Bob win,}    \\    V \big(  s , t \big)   ab \equiv -1 \Longleftrightarrow  \text{ Alice and Bob lose,}      
\end{align*}

\noindent given two questions $s,t \sim \pi$ from his probability distribution, with respective answers $a$, $b$ from Alice and Bob.

    \item[$\bullet$] \textit{Dependency-breaking variables for novel multiplayer game-theoretic settings}. For the set of multiplayer dependency-breaking variables, if one were to be able to conclude that,

\begin{align*}
  \textbf{P} \big[ \Omega_{\mathrm{Multiplayer}} \big| \textbf{X}_{1,2,\cdots,N} \equiv x, W_C \big] \approx \textbf{P} \big[ \Omega_{\mathrm{Multiplayer}} \big| \textbf{X}_{1,2,\cdots,N-1} \equiv x \backslash \big\{ N \big\} , \textbf{X}_N \equiv y,  W_C  \big] \\\approx \textbf{P} \big[ \Omega_{\mathrm{Multiplayer}} \big| \textbf{X}_{1,2,\cdots,N-2} \equiv x \backslash \big\{ N -1 , N \big\} , \textbf{X}_{N-1} \equiv y^{\prime } , \textbf{X}_N \equiv y,  W_C  \big].  \cdots \\  \approx \textbf{P} \big[ \Omega_{\mathrm{Multiplayer}} \big|  \textbf{X}_1 \equiv y^{\prime\cdots \prime}, \cdots, \textbf{X}_{N-1} \equiv y^{\prime} , \textbf{X}_N \equiv y,   W_C  \big]   \text{, }
\end{align*}

\noindent for,

\begin{align*}
    \textbf{X}_{1,2,\cdots, N} \equiv \underset{\# \text{ of players}}{\bigcup}  \big\{ \text{Responses from each player to questions drawn from} \\ \text{the referre's probability distribution } \pi \big\}           \text{,}
\\      \\ \textbf{X}_{1,2,\cdots, N-1} \equiv   \textbf{X}_{1,2,\cdots, N}  \backslash \textbf{X}_N  \equiv \bigg\{ \underset{\# \text{ of players}-1 }{\bigcup}  \big\{ \text{Responses from each player to questions} \\ \text{ drawn from the referre's probability distribution } \pi \big\} \bigg\}    \cap \bigg\{  \underset{N  \text{th player}}{\bigcup}  \big\{ \text{Responses from }  \end{align*}

\begin{align*} N \text{ th player to questions drawn from the referee's probability distribution } \pi \big\}    \bigg\}    \text{,} \\ \vdots \\   \textbf{X}_{1} \equiv   \textbf{X}_1 \backslash \textbf{X}_{1,2,\cdots, N-1}  \equiv \bigg\{ \big\{ \text{Responses from the First player to questions drawn} \\ \text{ from the referre's probability distribution } \pi \big\} \bigg\}    \cap \bigg\{  \underset{\# \text{ of players} - 1 }{\bigcup}  \big\{ \text{Responses from } \\  N-1 \text{ players to questions drawn from the referee's probability  distribution } \pi \big\}    \bigg\}    \text{,}    \\ \\ W_C \equiv       \big\{ \text{Players win the multiplayer game for all coordinates } C \big\} \text{, }   \end{align*}

    \noindent then the following sequences of computations can be performed:

    \begin{itemize}
    \item[$\bullet$] \underline{\textit{(1)}}. \textit{Approximating the probability of sampling an random variable from the set of all dependency-breaking variables by changing the response of one player, while holding the responses of all remaining } $(n-1)$ \textit{ players fixed}. One has that,

    \begin{align*} \textbf{P} \big[ \Omega_{\mathrm{Multiplayer}} \big| \textbf{X}_{1,2,\cdots,N} \equiv x, W_C \big] \approx \textbf{P} \big[ \Omega_{\mathrm{Multiplayer}} \big| \textbf{X}_{1,2,\cdots,N-1} \equiv x \backslash \big\{ N \big\} \\ , \textbf{X}_N \equiv y,  W_C  \big]
    \text{,} \end{align*}

    \noindent where,

\[ \mathcal{Q} \equiv \underset{1 \leq j \leq n}{\bigcup} \mathcal{Q}_j  \equiv   \left\{\!\begin{array}{ll@{}>{{}}l} 
y \equiv \text{Question distributed to Player } (n-1) \text{ from } \pi \text{, } \\ y^{\prime} \equiv \text{Question distributed to Player } (n-2) \text{ from } \pi \text{, } \\ \vdots \\   y^{\prime \overset{n-2}{\cdots} \prime}   \equiv \text{Question distributed to Player } 1 \text{ from } \pi          \text{, }
\end{array}\right. 
\]

\noindent denote a series of questions that Players $2, \cdots, n$ respond to, which are drawn from the referee's probability distribution $\pi$. After being able to initially perform a computation to demonstrate that the two  probabilities, from, the conditioning,

\begin{align*}
    \big\{ \textbf{X}_{1,2,\cdots,N} \equiv x, W_C \big\} \equiv  \big\{ \textbf{X}_{1,2,\cdots,N} \equiv x \big\} \cup \big\{ W_C \big\} \text{, }
\end{align*}

\noindent and,

\begin{align*}
   \big\{ \textbf{X}_{1,2,\cdots,N-1} \equiv x \backslash \big\{ N \big\}  , \textbf{X}_N \equiv y,  W_C\big\} \equiv  \big\{ \textbf{X}_{1,2,\cdots,N-1} \equiv x \backslash \big\{ N \big\}  , \textbf{X}_N \equiv y \big\}\\  \cup \big\{   W_C\big\}   \text{, }
\end{align*}

\noindent are statistically close, the next step asserts that there exists countably many more probabilities that are also statistically close, with respect to the same distance metric.

  \item[$\bullet$] \underline{\textit{(2)}}. \textit{Countably more conditional probabilities, such as the one defined in the previous step above, are 'statistically close'.} One has that,

\begin{align*}
  \textbf{P} \big[ \Omega_{\mathrm{Multiplayer}} \big| \textbf{X}_{1,2,\cdots,N} \equiv x, W_C \big] \approx \textbf{P} \big[ \Omega_{\mathrm{Multiplayer}} \big| \textbf{X}_{1,2,\cdots,N-1} \equiv x \backslash \big\{ N \big\} \\ , \textbf{X}_N \equiv y,  W_C  \big] \\\approx \textbf{P} \big[ \Omega_{\mathrm{Multiplayer}} \big| \textbf{X}_{1,2,\cdots,N-2} \equiv x \backslash \big\{ N -1 , N \big\} , \textbf{X}_{N-1} \equiv y^{\prime } , \textbf{X}_N \equiv y,  W_C  \big]  \cdots \\  \approx \textbf{P} \big[ \Omega_{\mathrm{Multiplayer}} \big|  \textbf{X}_1 \equiv y^{\prime\cdots \prime}, \cdots, \textbf{X}_{N-1} \equiv y^{\prime} , \textbf{X}_N \equiv y,   W_C  \big]   \text{, }
\end{align*}

\noindent for,

\begin{align*}
  y^{\prime}, \cdots, y^{\prime\cdots\prime} \in \mathcal{Q} \sim \pi   \text{. }
\end{align*}

    \item[$\bullet$] \underline{\textit{(3)}}. \textit{Extending the approximation of conditional probabilities, such as those defined in the previous two steps, to anchored questions from the referee's probability distribution}. Given $\mathcal{Q} \sim \pi$, a novelty of the current work lies in determining how \textit{anchored} questions shape optimality for multiplayer game, and more intricate game-theoretic settings, that have not yet been examined. Under such circumstances, \textit{anchored} questions serve the purpose of asserting that, instead of modifying the conditioning,

\begin{align*}
    \big\{ \textbf{X}_{1,2,\cdots,N-1} \equiv x \backslash \big\{ N \big\}  , \textbf{X}_N \equiv y,  W_C \big\}  \text{, }
\end{align*}

    \noindent to,

    \begin{align*}
    \big\{ \textbf{X}_{1,2,\cdots,N-1} \equiv x \backslash \big\{ N \big\}  , \textbf{X}_N \equiv \bot ,  W_C  \big\} \text{, }
\end{align*}

    \noindent with $y^{\prime} \in \mathcal{Q}$, one has that,

 \begin{align*} \textbf{P} \big[ \Omega_{\mathrm{Multiplayer}} \big| \textbf{X}_{1,2,\cdots,N} \equiv x, W_C \big] \approx \textbf{P} \big[ \Omega_{\mathrm{Multiplayer}} \big| \textbf{X}_{1,2,\cdots,N-1} \equiv x \backslash \big\{ N \big\} \\ , \textbf{X}_N \equiv y,  W_C  \big]
   \approx  \textbf{P} \big[ \Omega_{\mathrm{Multiplayer}} \big| \textbf{X}_{1,2,\cdots,N-1} \equiv x \backslash \big\{ N \big\} , \textbf{X}_N \equiv \bot ,  W_C  \big]     \text{,} \end{align*}

\noindent upon setting $X_N \equiv \bot$, where $\bot$ denotes the replacement, from the set of anchored responses, for the $N$ th player. In the next step of the procedure, we provide a statement of how one would replace, and subsequently, remove, the response from each player with an anchored response, $\bot$.

    \end{itemize}

    \item[$\bullet$] \textit{Constructing statistically close conditional probabilities with responses from the anchored set}. One has that,

\begin{align*}
     \textbf{P} \big[ \Omega_{\mathrm{Multiplayer}} \big| \textbf{X}_{1,2,\cdots,N-1} \equiv x \backslash \big\{ N \big\} , \textbf{X}_N \equiv \bot ,  W_C  \big]  \approx       \textbf{P} \big[ \Omega_{\mathrm{Multiplayer}} \\ \big| \textbf{X}_{1,2,\cdots,N-2} \equiv x \backslash \big\{ N -1 , N \big\} , \textbf{X}_{N-1} \equiv \bot  , \textbf{X}_N \equiv \bot ,  W_C  \big] \\ \vdots \\  \approx  \textbf{P} \big[ \Omega_{\mathrm{Multiplayer}}  \big| \textbf{X}_{1 } \equiv x \backslash \big\{ 2, \cdots , N -1 , N \big\} , \textbf{X}_2 \dots \equiv \textbf{X}_N \equiv \bot ,  W_C  \big]         \text{. }
\end{align*}

\noindent Iterating further along the lines of the observations above, one also has that,

\begin{align*}
        \textbf{P} \big[ \Omega_{\mathrm{Multiplayer}}  \big| \textbf{X}_{1 } \equiv x \backslash \big\{ 2, \cdots , N -1 , N \big\} , \textbf{X}_2 \dots \equiv \textbf{X}_N \equiv \bot ,  W_C  \big] \\   \approx    \textbf{P} \big[ \Omega_{\mathrm{Multiplayer}}  \big| \textbf{X}_{1 } \equiv x \backslash \big\{ 2, \cdots , N -1 , N \big\} , \textbf{X}_3 \dots \equiv \textbf{X}_N \equiv \bot ,  W_C  \big]   \\ \vdots \\ \approx   \textbf{P} \big[ \Omega_{\mathrm{Multiplayer}}  \big| \textbf{X}_{1 } \equiv x \backslash \big\{ 2, \cdots , N -1 , N \big\} , \textbf{X}_N \equiv \bot ,  W_C  \big]         \text{. }
\end{align*}

\item[$\bullet$] \textit{Putting it all together: Relating the conditional probability with anchored questions to each other}. Altogether, one has,

\begin{align*}
 \big\{  \textbf{P} \big[ \Omega_{\mathrm{Expanded}} \big| \textbf{X}_i \equiv x, W_C \big] \approx \textbf{P} \big[ \Omega_{\mathrm{Expanded}} \big| \textbf{X}_i \equiv x, \textbf{Y}_i \equiv y,  W_C  \big] \big\} \\  \Longrightarrow  \big\{   \textbf{P} \big[ \Omega_{\mathrm{Multiplayer}}  \big| \textbf{X}_{1 } \equiv x \backslash \big\{ 3, \cdots  , N -1 , N \big\} , \textbf{X}_{N-1} \equiv  \textbf{X}_N \equiv \bot ,  W_C  \big]   \\      \approx  \textbf{P} \big[ \Omega_{\mathrm{Multiplayer}}  \big| \textbf{X}_{1 } \equiv x \backslash \big\{ 2, \cdots , N -1 , N \big\} , \textbf{X}_N \equiv \bot ,  W_C  \big]  \big\}      \text{. }
\end{align*}

\end{itemize}

\noindent Through the sequence of steps described above, one can provide, straightforwardly, direct statements of how dependency-breaking variables would break Quantum correlations, under the assumption that players use product strategies. Determining the extent to which correlations persists in a system, whether Classical, or Quantum, remains of interest. In this context, characterizing the ways in which Quantum correlations can typically be broken would be useful for the following objectives:

\begin{itemize}
    \item[$\bullet$] \textit{Quantification of dependency-breaking from a fixed set of dependency-breaking variables}. Under objects that have been previously introduced for dependency-breaking, one would like to determine whether,

        \begin{align*}
         (*) \equiv   \frac{\textbf{P} \big[ \Omega_{\mathrm{Multiplayer}}  \big| \textbf{X}_{1 } \equiv x \backslash \big\{ 2, \cdots , N -1 , N \big\} , \textbf{X}_N \equiv \bot ,  W_C  \big] }{\textbf{P} \big[ \Omega  \big| \textbf{X}_{1 } \equiv x \backslash \big\{ 2, \cdots , N -1 , N \big\} , \textbf{X}_N \equiv \bot ,  W_C  \big] }       \text{,}
        \end{align*}

        \noindent for,

        \begin{align*}
          \Omega \equiv \big\{ \text{All possible answers in a multiplayer game} \big\}     \text{,}
        \end{align*}

\noindent are approximately comparable, namely, the ratio of the conditional probability of sampling the dependency-breaking variable set, with the probability, under identical conditioning, of sampling the entire set of variables.

\item[$\bullet$] \textit{Computing the threshold of the ratio of conditional probabilities in the previous item above}. Whether,

\begin{align*}
 (*) < 1   \text{, }
\end{align*}

\noindent or,

\begin{align*}
 (*) > 1   \text{, }
\end{align*}

\noindent hold depends upon the two conditions,

\begin{align*}
   \textbf{P} \bigg[  \Omega : \frac{\big| \Omega_{\mathrm{Multiplayer}} \big| }{\big| \Omega \big|} \approx 1           \bigg]  \text{, } \\    \textbf{P} \bigg[  \forall    \Omega , \exists c \in \textbf{N} : \frac{\big| \Omega_{\mathrm{Multiplayer}} \big| }{\big| \Omega \big|} <<  c               \bigg]   \text{, }
\end{align*}

\noindent being satisfied, namely occurring with positive probability, respectively. If the first condition provided above were to occur with positive probability, then, necessarily, one would have that nearly \textit{every} variable that the players can use to respond to questions drawn from $\pi$ are dependency-breaking variables. If this were the case, then a dependency-breaking variable could be used along \textit{any} position of player's response to decorrelate any long range order between the position of the clause at which the dependency-breaking variable is applied. More specifically, for,

\begin{align*}
 \mathcal{C} \equiv \underset{\text{coordinates }i}{\bigcup}   \mathcal{C}_i \equiv     \underset{\text{coordinates }i}{\bigcup}   \big\{ \text{Tuples } \mathcal{C} \text{ with finitely many coordinates} : \mathcal{C} \equiv \big( X_1, X_2,\\  \cdots  , X_i, \cdots, X_n \big)   \big\}      \text{, }
\end{align*}

\noindent \textit{quantum} long-range order would still occur whp, given the \textit{dependency-breaking} action $\mathscr{D}\mathscr{B}$,

\begin{align*}
    \mathscr{D}\mathscr{B}  \curvearrowright   \mathcal{C} \equiv   \mathscr{D}\mathscr{B}  \curvearrowright   \bigg[ \underset{\text{coordinates }i}{\bigcup} \mathcal{C}_i \bigg] \equiv        \mathscr{D}\mathscr{B}  \curvearrowright   \bigg[ \underset{\text{coordinates }i^{\prime}}{\bigcup} \mathcal{C}_{i^{\prime}} \bigg] \equiv \mathscr{D}\mathscr{B}  \curvearrowright   \mathcal{C}^{\prime}             \text{, }
\end{align*}

\noindent if, after Alice or Bob sample a dependency-breaking variable,

\begin{align*}
\textbf{P} \bigg[ \forall \mathscr{C} \text{sufficiently small}, i  > 0 , \exists j >i : \mathrm{Corr} \big( C_{i^{\prime}} , \ket{\psi_{\mathrm{Multiplayer}}} \big) > \mathscr{C}       \bigg] \text{, }
\end{align*}

\noindent for the quantum state corresponding to the optimal multiplayer strategy,

\begin{align*}
    \ket{\psi_{\mathrm{Multiplayer}}} \text{. }
\end{align*}

\noindent The length of each finite tuple would approach $+\infty$ as the number of parallel repetition operations itself approaches $+\infty$. If the above probability that the Quantum correlation function, $\mathrm{Corr} \big( \cdot , \cdot \big)$, were to exceed some strictly positive threshold $\mathscr{C}$, then it is, intuitively, natural to expect that the action of the multiplayer dependency-breaking variable could prevent the players from achieving optimality with $\ket{\psi_{\mathrm{Multiplayer}}}$. However, under the assumption,

\begin{align*}
\textbf{P} \bigg[  \forall    \Omega , \exists c \in \textbf{N} : \frac{\big| \Omega_{\mathrm{Multiplayer}} \big| }{\big| \Omega \big|} <<  c               \bigg]   \text{, }
\end{align*}

\noindent which is more likely to happen, as stochastic domination,

\begin{align*}
\textbf{P} \bigg[ \forall  \Omega, \exists c > 0  : \frac{\big| \Omega_{\mathrm{Multiplayer}}\big|}{\big| \Omega \big|} <<  c \bigg]  >   \textbf{P} \bigg[ \Omega : \frac{\big| \Omega_{\mathrm{Multiplayer}}\big|}{\big| \Omega \big|} \approx 1   \bigg]  \text{, }
\end{align*}

\noindent holds, dependency-breaking would necessarily require,

\begin{align*}
     \textbf{P} \bigg[ \forall  \omega \big( G \big) > \omega^{\prime} \big( G \big), \exists \ket{\psi^{\prime}_{\mathrm{Multiplayer}}}, \ket{\psi_{\mathrm{Multiplayer}}}  :  \omega^{\prime} \big( G \big)  \propto \bra{\psi^{\prime}_{\mathrm{Multiplayer}}} \\ \times \bigg[ \underset{\# \text{ players}}{\bigotimes} \text{Tensors of player observables} \bigg]   \ket{\psi^{\prime}_{\mathrm{Multiplayer}}}  \\   <  \bra{\psi_{\mathrm{Multiplayer}}}   \bigg[  \underset{\# \text{ players}}{\bigotimes} \text{Tensors of player observables} \bigg] \ket{\psi_{\mathrm{Multiplayer}}} \\ \propto w \big( G \big)     \bigg]         \text{. }
\end{align*}

\noindent To this end, we further manipulate several expectation values from multiplayer optimal strategies which have been previously introduced, and manipulated, by the author in [52]. The specific expectation values for quantifying exact, and approximate, optimality in several multiplayer settings include,

\begin{align*}
      \bra{\psi_{3\mathrm{XOR}}}   \bigg[   \underset{\# \text{ players}}{\bigotimes} \text{Tensors of player observables} \big( q_1, q_2, q_3 \big)        \bigg] \ket{\psi_{3\mathrm{XOR}}}   \text{, } \\ \\   \bra{\psi_{4\mathrm{XOR}}}   \bigg[   \underset{\# \text{ players}}{\bigotimes} \text{Tensors of player observables} \big( q_1, q_2, q_3, q_4  \big)        \bigg] \ket{\psi_{4\mathrm{XOR}}}  \text{, } \\ \\ \bra{\psi_{5\mathrm{XOR}}}   \bigg[   \underset{\# \text{ players}}{\bigotimes} \text{Tensors of player observables} \big( q_1, q_2, q_3, q_4, q_5   \big)        \bigg] \ket{\psi_{5\mathrm{XOR}}}   \text{, } \\ \vdots \\ \bra{\psi_{N\mathrm{XOR}}}   \bigg[   \underset{\# \text{ players}}{\bigotimes} \text{Tensors of player observables} \big( q_1, q_2, \cdots , q_N  \big)        \bigg] \ket{\psi_{N\mathrm{XOR}}}          \text{, } \\  \\ \bigg[ \bra{\psi_{2\mathrm{XOR}}} \wedge \bra{\psi_{2\mathrm{XOR}}} \bigg]   \bigg[   \underset{\# \text{ players}}{\bigotimes} \text{Tensors of player observables} \big( q_1, q_2  \big)  \\ \wedge  \text{Tensors of player observables} \big( q_1, q_2 \big)         \bigg]  \bigg[ \ket{\psi_{2\mathrm{XOR}}}  \wedge \ket{\psi_{2\mathrm{XOR}}}  \bigg]   \text{,} \\ \\  \bigg[ \bra{\psi_{3\mathrm{XOR}}} \wedge  \bra{\psi_{3\mathrm{XOR}}} \wedge \bra{\psi_{3\mathrm{XOR}}} \bigg]   \bigg[   \underset{\# \text{ players}}{\bigotimes} \text{Tensors of player observables} \big( q_1, q_2, q_3  \big)  \\ \wedge  \text{Tensors of player observables} \big( q_1, q_2, q_3  \big) \\  \wedge  \text{Tensors of player observables} \big( q_1, q_2, q_3  \big)         \bigg]  \bigg[ \ket{\psi_{N\mathrm{XOR}}}  \wedge  \ket{\psi_{N\mathrm{XOR}}}  \wedge \ket{\psi_{N\mathrm{XOR}}}  \bigg]  \text{,} \\ \\  \bigg[ \bra{\psi_{N\mathrm{XOR}}} \wedge \cdots \wedge \bra{\psi_{N\mathrm{XOR}}} \bigg]   \bigg[   \underset{\# \text{ players}}{\bigotimes} \text{Tensors of player observables} \big( q_1, q_2, \cdots , q_N  \big)  \\ \wedge \cdots \wedge \text{Tensors of player observables} \big( q_1, q_2, \cdots , q_N  \big)         \bigg]  \bigg[ \ket{\psi_{N\mathrm{XOR}}}  \wedge \cdots \wedge \ket{\psi_{N\mathrm{XOR}}}  \bigg]   \text{,} 
\end{align*}

\noindent for the set of responses,

\begin{align*}
  \underline{Q_{3\mathrm{XOR}}} \equiv   Q_{3\mathrm{XOR},1}  \cup  Q_{3\mathrm{XOR},2}  \cup  Q_{3\mathrm{XOR},3} \\  \equiv    \underset{q_1 \in Q_1, q_2 \in Q_2, q_3 \in Q_3}{\bigcup} \big\{ Q_1 \cup Q_2 \cup Q_3 \big\}  \text{,}  \\ \\  \underline{Q_{4\mathrm{XOR}}} \equiv   Q_{4\mathrm{XOR},1}  \cup  Q_{4\mathrm{XOR},2}  \cup  Q_{4\mathrm{XOR},3}  \cup Q_{4\mathrm{XOR},4}  \\ \equiv   \underset{q_1 \in Q_1, q_2 \in Q_2, q_3 \in Q_3, q_4 \in Q_4}{\bigcup} \big\{ Q_1 \cup Q_2 \cup Q_3 \cup Q_4 \big\}      \text{, } \\  \vdots \\  \underline{Q_{N\mathrm{XOR}}} \equiv   Q_{N\mathrm{XOR},1}  \cup  Q_{N\mathrm{XOR},2}  \cup  \cdots \cup  Q_{N\mathrm{XOR},N} \\  \equiv   \underset{q_1 \in Q_1, \cdots, q_N \in Q_N}{\bigcup} \big\{ Q_1 \cup \cdots  \cup Q_N \big\}   \text{, } \\ \\ \text{, }        
\end{align*}

\noindent prepared by each player in the multiplayer XOR, and strong parallel repetition of XOR and FFL games. Several closely related game-theoretic objects corresponding to multiplayer strong parallel repetition can be defined from the objects introduced above.

    \item[$\bullet$] \textit{Anchored parallel repetition}. One can make use of the steps provided in previous items above for anchoring each parallel repetition of a multiplayer game. That is, by replacing the answers from players with anchored responses, analytically the decay of the optimal value can be related to parallel repetition of the optimal value.
    
\end{itemize}

\subsection{Paper organization}

\noindent In the next section, we further describe objects that are required for anchored parallel repetition of multiplayer optimal values. That is, we demonstrate that a desired, explicit, representation for the optimal value under anchored, multiplayer parallel repetition holds. To obtain the desired representation for the multiplayer optimal value, we make use of several discrete probabilistic notions, through the formalism of: (1) probability distributions, inherited from the referee's probability distribution, for each player; (2) conditional probability distributions, inherited from the referee's probability distribution, for each player; (3) conditional probability distributions, inherited from tensor products of the referee
's probability distribution, for pairwise combinations of players; (4) products of ordinary, and conditional, probability distributions; (5) products of conditional probability distributions; (6) products of ordinary probability distributions with expectation values; (7) products of conditional probability distributions with conditional expectation values. Together, such objects are indispensable for obtaining estimates, or equalities, for multiplayer optimal values under anchored parallel repetition. In various game-theoretic multiplayer settings, the impact of anchoring on parallel repetition of the optimal value has not yet been rigorously characterized. To characterize possible impacts of anchoring on parallel repetition of the optimal value, following the introduction of previous results of the author, [50, 52, 53], which can alternatively be described in algorithmic terms from variational quantum algorithms, [51], we state the main results in \textit{1.5}. In this section, we first state Main results concerning parallel repetition of expanded games, which from previous work of the author, [50, 51, 53], exhibit connections between optimal, and approximate optimality, in several game-theoretic settings. Besides Main results on Expanded games, we also state several results for game-theoretic settings in games previously studied by the author in [52].

\subsection{Game-theoretic objects}

\noindent We introduce the following objects from Quantum Information theory.

\subsubsection{Norms, and related distance metrics, from Quantum Information Theory}

\noindent Denote:

\begin{itemize}
\item[$\bullet$] \textit{l-1 distance between two quantum states}. For the quantum states $\rho$ and $\sigma$, the l-1 distance equals,

\begin{align*}
   \big| \big| \rho - \sigma \big| \big|_1 = \mathrm{Tr} \big[ \big( \rho - \sigma \big)^{\dagger} \big( \rho - \sigma \big)   \big] \equiv \mathrm{Tr} \big| \rho - \sigma \big| .
\end{align*}

\item[$\bullet$] \textit{Fidelity between two quantum states}. With the same states introduced in the previous item, the fidelity is denoted with,

\begin{align*}
   F \big( \rho , \sigma \big) = \big| \big| \sqrt{\rho} \sqrt{\sigma} \big| \big|_1 .
\end{align*}

\item[$\bullet$] \textit{Fucsh-van de Graaf inequality}. One has that,

\begin{align*}
 2 \big( 1 - F \big( \rho , \sigma \big) \big) \leq \big| \big| \rho - \sigma \big| \big|_1 \leq 2 \sqrt{1 - F \big( \rho , \sigma \big)}           .
\end{align*}

\item[$\bullet$] \textit{Data processing inequality}. For a Quantum channel $\mathcal{E}$ and the two quantum states introduced in the first time, one has that,

\begin{align*}
 \big| \big| \mathcal{E} \big( \rho \big) - \mathcal{E} \big( \sigma \big) \big| \big| \leq \big| \big| \rho - \sigma \big| \big|_1     ,
\end{align*}

\noindent as well as,

\begin{align*}
  F \big( \mathcal{E} \big( \rho \big) , \mathcal{E} \big( \sigma \big) \big) \geq F \big( \rho , \sigma \big)   .
\end{align*}

\end{itemize}

\noindent The above inequality for data processing can straightforwardly be indexed with respect to the number of parallel repetitions, $n$, with,

\begin{align*}
\underset{1 \leq i \leq n}{\bigwedge} \big| \big| \mathcal{E} \big( \rho_{i} \big) - \mathcal{E} \big( \sigma_{i } \big) \big| \big|  \equiv    \big| \big| \mathcal{E} \big( \rho_{1\cdots n} \big) - \mathcal{E} \big( \sigma_{1\cdots n } \big) \big| \big| \leq \big| \big| \rho_{1 \cdots n } - \sigma_{1 \cdots n} \big| \big|_1 \\ \equiv \underset{1 \leq i \leq n}{\bigwedge} \big| \big| \rho_{i} - \sigma_{i} \big| \big|   .
\end{align*}

\noindent The above expression for the fidelity is used to conclude that the purification of certain Quantum states can be related to the set of dependency-breaking variables. The following objects below are introduced from both Probability theory, and Quantum Information theory.

\subsubsection{Discrete probabilistic objects}

\noindent We introduce a few objects from Probability theory which are indispensable for the following arguments in Quantum Information theory. To this end, denote:

\begin{align*}
    \underline{\text{Question superset}} \equiv  \mathcal{Q} \equiv \underset{\# \text{ players}}{\bigcup} \mathcal{Q} \sim \underset{\# \text{ players}}{\bigotimes}  \pi \equiv \pi^{\underset{\# \text{ players}}{\bigotimes} }\text{, } \\       \\ \underline{\text{Probability measures over the first player's responses}} \equiv         \textbf{P}_{\mathcal{Q}_1} \big[ \cdot \big]      \text{, } \\ \\   \underline{\text{Conditional Probability measures over the last player's responses}} \equiv         \textbf{P}_{\mathcal{Q}_1} \big[ \cdot \big| \cdot \big]              \text{, } \\ \vdots \\         \underline{\text{Probability measures over the last player's responses}} \equiv         \textbf{P}_{\mathcal{Q}_N} \big[ \cdot \big]      \text{, } \\ \\ \underline{\text{Conditional Probability measures over the first player's responses}} \equiv         \textbf{P}_{\mathcal{Q}_N} \big[ \cdot \big| \cdot \big]              \text{, }      \\ \\ \underline{\text{Joint probability measure over all possible responses from the first and second}} \\ \underline{\text{players} }    \equiv \textbf{P}_{\mathcal{Q}_{12} } \big[ \cdot \big]  \equiv         \textbf{P}_{\mathcal{Q}_{1} \mathcal{Q}_2 } \big[ \cdot \big]     \equiv \textbf{P}_{\mathcal{Q}\big|_{12} } \big[ \cdot \big]  \text{, } 
    \\ \\ \underline{\text{Conditional joint probability measure over all possible responses from the first}} \\   \underline{\text{and second players} } \equiv \textbf{P}_{\mathcal{Q}_{12} } \big[ \cdot \big| \cdot \big]  \equiv         \textbf{P}_{\mathcal{Q}_{1} \mathcal{Q}_2 } \big[ \cdot \big| \cdot  \big]     \equiv \textbf{P}_{\mathcal{Q}\big|_{12} } \big[ \cdot \big| \cdot  \big]   \text{, } \\  \vdots \\ \underline{\text{Joint probability measure over all possible responses from the first and second}} \\ \underline{\text{players} }    \equiv \textbf{P}_{\mathcal{Q}_{12\cdots N} } \big[ \cdot \big]  \equiv         \textbf{P}_{\mathcal{Q}_{1} \mathcal{Q}_2 \cdots \mathcal{Q}_N} \big[ \cdot \big]     \equiv \textbf{P}_{\mathcal{Q}\big|_{12\cdots N} } \big[ \cdot \big]  \text{, } 
    \\ \\ \underline{\text{Conditional joint probability measure over all possible responses from the first}} \\ \underline{\text{and second players} } \equiv \textbf{P}_{\mathcal{Q}_{12\cdots N} } \big[ \cdot \big| \cdot \big]  \equiv         \textbf{P}_{\mathcal{Q}_{1} \mathcal{Q}_2 \cdots \mathcal{Q}_N} \big[ \cdot \big| \cdot  \big]     \equiv \textbf{P}_{\mathcal{Q}\big|_{12\cdots N} } \big[ \cdot \big| \cdot  \big]   \text{. } 
\end{align*}

\noindent Furthermore, one can straightforwardly formulate joint, and conditional joint, probability measures over any pairwise combination of $\mathcal{Q}_1, \mathcal{Q}_2, \cdots, \mathcal{Q}_N$, by making use of an identical construction of the probability measures above. With such discrete probabilistic objects, one can introduce the \textit{total variation distance}, between any two probability distributions $\mathcal{P}_1$ and $\mathcal{P}_2$, with,

\begin{align*}
    \mathscr{T}\mathscr{V}\mathscr{D} \big( \mathcal{P}_1 , \mathcal{P}_2 \big) \equiv \underset{x \in X }{\sum} \big| \big| \mathcal{P}_1 \big( x \big) - \mathcal{P}_2 \big( x \big) \big| \big| \equiv \underset{q \in Q }{\sum} \big| \big| \mathcal{P}_1 \big( q \big) - \mathcal{P}_2 \big( q \big) \big| \big| \textbf{}\text{.}
\end{align*}

\noindent A few expressions for the distance between any two probability distributions above are provided in \textit{Table 1}.

\bigskip

\noindent Crucially, expressions for the decay of parallel repetition of the multiplayer anchored optimal value are obtained with computations with respect to expected values. Straightforwardly, given the fact that expected values are computed with respect to some probability measure, we list the set of possible expectations below:

\begin{align*}
  \underline{\text{Expectation with respect to the probability measure on the first player's responses}} \\ \equiv \textbf{E}_{\textbf{P}_{\mathcal{Q}_1} \big[ \cdot \big]} \equiv  \textbf{E}_{\textbf{P}_{\mathcal{Q}_1}} \big[ \cdot \big]  \equiv  \textbf{E}_{\textbf{P}_{1}}  \big[ \cdot \big]      \text{, } \\ \\   \underline{\text{Expectation with respect to the probability measure on the second player's responses}} \\ \equiv \textbf{E}_{\textbf{P}_{\mathcal{Q}_2} \big[ \cdot \big]} \equiv  \textbf{E}_{\textbf{P}_{\mathcal{Q}_2}} \big[ \cdot \big]  \equiv  \textbf{E}_{\textbf{P}_{2}}  \big[ \cdot \big]    \text{, }  \\ \\ \vdots \\ \underline{\text{Expectation with respect to the probability measure on the } N \text{th player's responses}} \\ \equiv \textbf{E}_{\textbf{P}_{\mathcal{Q}_N} \big[ \cdot \big]} \equiv  \textbf{E}_{\textbf{P}_{\mathcal{Q}_N}} \big[ \cdot \big]  \equiv  \textbf{E}_{\textbf{P}_{N}}  \big[ \cdot \big]   \text{, }   \end{align*}

    \begin{align*}  \underline{\text{Conditional expectation with respect to the probability measure on the first player's}} \\ \underline{\text{  responses,  and the second player's responses}}  \equiv    \textbf{E}_{\textbf{P}_{\mathcal{Q}_1 | \mathcal{Q}_2 } \big[ \cdot \big]} \equiv  \textbf{E}_{\textbf{P}_{\mathcal{Q}_1 | \mathcal{Q}_2}} \big[ \cdot  \big| \cdot \big] \\ \equiv  \textbf{E}_{\textbf{P}_{1,2}}  \big[ \cdot  \big| \cdot  \big]          \text{, }  \\ \\    \underline{\text{Conditional expectation with respect to the probability measure on the first player's}}  \end{align*}

\begin{tabular}{|l|l|}
\hline\parbox[t]{0.25\textwidth}{
\begin{itemize}
\item $\underline{\mathcal{P}_1, \mathcal{P}_2}$
\item $\mathcal{P}_{\mathcal{Q}_1} \big[ q \big]$, $ \mathcal{P}_{\mathcal{Q}_2} \big[ q \big]$
\item $\mathcal{P}_{\mathcal{Q}_1} \big[ q \big]$, $ \mathcal{P}_{\mathcal{Q}_3} \big[ q \big]$
\item $\mathcal{P}_{\mathcal{Q}_2} \big[ q \big]$, $ \mathcal{P}_{\mathcal{Q}_3} \big[ q \big]$
\item $\mathcal{P}_{\mathcal{Q}_N} \big[ q \big]$, $ \mathcal{P}_{\mathcal{Q}_1} \big[ q \big]$
\item $\mathcal{P}_{\mathcal{Q}_1} \big[ q \big| \cdot  \big]$, $ \mathcal{P}_{\mathcal{Q}_2} \big[ q \big| \cdot  \big]$
\item $\mathcal{P}_{\mathcal{Q}_N} \big[ q \big| \cdot  \big]$, $ \mathcal{P}_{\mathcal{Q}_2} \big[ q \big| \cdot  \big]$
\item $\mathcal{P}_{\mathcal{Q}_1} \big[ q \big| \cdot  \big]$, $ \mathcal{P}_{\mathcal{Q}_2} \big[ q \big]$
\end{itemize}}& 
\parbox[t]{0.73\textwidth}{
\begin{itemize}
\item \underline{\text{Total Variation Distance}}
\item $\underset{q \in Q }{\sum} \big| \big|  \mathcal{P}_{\mathcal{Q}_1} \big[ q \big] - \mathcal{P}_{\mathcal{Q}_2} \big[ q \big] \big| \big| $
\item $\underset{q \in Q }{\sum} \big| \big|  \mathcal{P}_{\mathcal{Q}_1} \big[ q \big] - \mathcal{P}_{\mathcal{Q}_3} \big[ q \big] \big| \big| $
\item $\underset{q \in Q }{\sum} \big| \big|  \mathcal{P}_{\mathcal{Q}_2} \big[ q \big] - \mathcal{P}_{\mathcal{Q}_3} \big[ q \big] \big| \big| $
\item $\underset{q \in Q }{\sum} \big| \big|  \mathcal{P}_{\mathcal{Q}_N} \big[ q \big] - \mathcal{P}_{\mathcal{Q}_1} \big[ q \big] \big| \big| $
\item $\underset{q \in Q }{\sum} \big| \big|  \mathcal{P}_{\mathcal{Q}_N} \big[ q \big] - \mathcal{P}_{\mathcal{Q}_1} \big[ q \big] \big| \big| $
\item $\underset{q \in Q }{\sum} \big| \big|  \mathcal{P}_{\mathcal{Q}_1} \big[ q \big| \cdot \big] - \mathcal{P}_{\mathcal{Q}_2} \big[ q \big| \cdot \big] \big| \big| $
\item $\underset{q \in Q }{\sum} \big| \big|  \mathcal{P}_{\mathcal{Q}_N} \big[ q \big| \cdot  \big] - \mathcal{P}_{\mathcal{Q}_2} \big[ q \big| \cdot  \big] \big| \big| $
\item $\underset{q \in Q }{\sum} \big| \big|  \mathcal{P}_{\mathcal{Q}_1} \big[ q  \big| \cdot  \big] - \mathcal{P}_{\mathcal{Q}_2} \big[ q \big] \big| \big| $
\end{itemize}}\\
\hline
\end{tabular}
\noindent \textit{Table 1}. The total variation distance from probability measures introduced at the beginning of subsection \textit{1.3.1}.

\bigskip

  \begin{align*}  \underline{\text{  responses,  and the third player's responses}}  \equiv    \textbf{E}_{\textbf{P}_{\mathcal{Q}_1 | \mathcal{Q}_3 } \big[ \cdot \big]} \equiv  \textbf{E}_{\textbf{P}_{\mathcal{Q}_1 | \mathcal{Q}_3}} \big[ \cdot  \big| \cdot \big] \\   \equiv  \textbf{E}_{\textbf{P}_{1,3}}  \big[ \cdot   \big| \cdot \big]        \text{, } \\   \vdots \\  \underline{\text{Conditional expectation with respect to the probability measure on the first player's}} \\   \underline{\text{  responses,  and the third player's responses}}  \equiv    \textbf{E}_{\textbf{P}_{\mathcal{Q}_1 | \mathcal{Q}_N } \big[ \cdot \big]} \equiv  \textbf{E}_{\textbf{P}_{\mathcal{Q}_1 | \mathcal{Q}_N}} \big[ \cdot  \big| \cdot \big]  \end{align*}

    \begin{align*}  \equiv  \textbf{E}_{\textbf{P}_{1,N}}  \big[ \cdot  \big| \cdot  \big]        \text{, }  \\\vdots  \\ \underline{\text{Conditional expectation with respect to the probability measure on the } N\text{ th player's}} \\ \underline{\text{  responses,  and the third player's responses}}  \equiv    \textbf{E}_{\textbf{P}_{\mathcal{Q}_N | \mathcal{Q}_{N-1} } \big[ \cdot \big]} \equiv  \textbf{E}_{\textbf{P}_{\mathcal{Q}_N | \mathcal{Q}_{N-1}}} \big[ \cdot  \big| \cdot \big] \\  \equiv  \textbf{E}_{\textbf{P}_{N,N-1}}  \big[ \cdot  \big| \cdot \big]        \text{, }  \end{align*}

  \begin{align*}  \vdots \\  \underline{\text{Conditional expectation with respect to the probability measure on the first player's}} \\ \underline{\text{  responses,  and the third player's responses}}  \equiv    \textbf{E}_{\textbf{P}_{\mathcal{Q}_1 \mathcal{Q}_2 \cdots \mathcal{Q}_N | \mathcal{Q}_{N-1} } \big[ \cdot \big]} \\  \equiv  \textbf{E}_{\textbf{P}_{\mathcal{Q}_1 \mathcal{Q}_2 \cdots \mathcal{Q}_N  | \mathcal{Q}_{N-1}}} \big[ \cdot  \big| \cdot  \big]   \equiv  \textbf{E}_{\textbf{P}_{12\cdots N,N-1}}  \big[ \cdot \big| \cdot \big]        \text{.}   
\end{align*}

\subsubsection{Probabilistic facts}

\noindent Related to discussions provided in [2, 3, 4, 56], to further generalize arguments regarding anchored, multiplayer, parallel repetition, we provide the following list of probabilistic results:

\begin{itemize}
\item[$\bullet$] \textit{The probability, under product supports of questions, can be upper bounded with the distribution of questions over all players}. One has that,

\begin{align*}
  \bigg| \bigg|           \textbf{P}_{\mathcal{Q}_1} - \textbf{P}_{\mathcal{Q}_2}          \bigg| \bigg| \leq     \bigg| \bigg|         \textbf{P}_{\mathcal{Q}_1} \textbf{P}_{\mathcal{Q}^{\prime}_1} - \textbf{P}_{\mathcal{Q}_2}   \textbf{P}_{\mathcal{Q}^{\prime}_2}          \bigg| \bigg|  \leq     \bigg| \bigg|      \textbf{P}_{\mathcal{Q}_1} \textbf{P}_{\mathcal{Q}^{\prime}_1} \textbf{P}_{\mathcal{Q}^{\prime\prime}_1} - \textbf{P}_{\mathcal{Q}_2}   \textbf{P}_{\mathcal{Q}^{\prime}_2}   \textbf{P}_{\mathcal{Q}^{\prime\prime}_2}                        \bigg| \bigg|  \\ \leq     \bigg| \bigg|    \textbf{P}_{\mathcal{Q}_1} \textbf{P}_{\mathcal{Q}^{\prime}_1} \textbf{P}_{\mathcal{Q}^{\prime\prime}_1} \textbf{P}_{\mathcal{Q}^{\prime\prime\prime}_1}  - \textbf{P}_{\mathcal{Q}_2}   \textbf{P}_{\mathcal{Q}^{\prime}_2}   \textbf{P}_{\mathcal{Q}^{\prime\prime}_2} \textbf{P}_{\mathcal{Q}^{\prime\prime\prime}_2}                          \bigg| \bigg|  \leq        \cdots \leq     \bigg| \bigg|             \textbf{P}_{\mathcal{Q}_1} \textbf{P}_{\mathcal{Q}^{\prime}_1} \textbf{P}_{\mathcal{Q}^{\prime\prime}_1} \textbf{P}_{\mathcal{Q}^{\prime\prime\prime}_1}  \\ \times \cdots \times \textbf{P}_{\mathcal{Q}^{\prime\cdots\prime}_1} - \textbf{P}_{\mathcal{Q}_2}   \textbf{P}_{\mathcal{Q}^{\prime}_2}   \textbf{P}_{\mathcal{Q}^{\prime\prime}_2} \textbf{P}_{\mathcal{Q}^{\prime\prime\prime}_2}    \times \cdots \times \textbf{P}_{\mathcal{Q}^{\prime\cdots\prime}_2}                                \bigg| \bigg|     \text{.}
\end{align*}

\item[$\bullet$] \textit{The probability in the previous item above can be upper bounded with $\frac{1}{2}$ of the l-$1$ norm}. One has that,

\[   \left\{\!\begin{array}{ll@{}>{{}}l} 
 \bigg|            \textbf{P}_{\mathcal{Q}_1} - \textbf{P}_{\mathcal{Q}_2}          \bigg|  \leq \frac{1}{2}  \bigg| \bigg|    \textbf{P}_{\mathcal{Q}_1} - \textbf{P}_{\mathcal{Q}_2}       \bigg| \bigg|_1  \text{, } \\     \bigg|         \textbf{P}_{\mathcal{Q}_1} \textbf{P}_{\mathcal{Q}^{\prime}_1} - \textbf{P}_{\mathcal{Q}_2}   \textbf{P}_{\mathcal{Q}^{\prime}_2}          \bigg| \leq \frac{1}{2}  \bigg|  \bigg|         \textbf{P}_{\mathcal{Q}_1} \textbf{P}_{\mathcal{Q}^{\prime}_1} - \textbf{P}_{\mathcal{Q}_2}   \textbf{P}_{\mathcal{Q}^{\prime}_2}          \bigg| \bigg|_1 \text{, } \\   \bigg|      \textbf{P}_{\mathcal{Q}_1} \textbf{P}_{\mathcal{Q}^{\prime}_1} \textbf{P}_{\mathcal{Q}^{\prime\prime}_1} - \textbf{P}_{\mathcal{Q}_2}   \textbf{P}_{\mathcal{Q}^{\prime}_2}   \textbf{P}_{\mathcal{Q}^{\prime\prime}_2}                         \bigg| \leq \frac{1}{2} \bigg|  \bigg|    \textbf{P}_{\mathcal{Q}_1} \textbf{P}_{\mathcal{Q}^{\prime}_1} \textbf{P}_{\mathcal{Q}^{\prime\prime}_1} - \textbf{P}_{\mathcal{Q}_2}   \textbf{P}_{\mathcal{Q}^{\prime}_2}   \textbf{P}_{\mathcal{Q}^{\prime\prime}_2}                         \bigg| \bigg|_1    \\            \vdots \\ \bigg|              \textbf{P}_{\mathcal{Q}_1} \textbf{P}_{\mathcal{Q}^{\prime}_1} \textbf{P}_{\mathcal{Q}^{\prime\prime}_1} \textbf{P}_{\mathcal{Q}^{\prime\prime\prime}_1}   \times \cdots \times \textbf{P}_{\mathcal{Q}^{\prime\cdots\prime}_1} - \textbf{P}_{\mathcal{Q}_2}   \textbf{P}_{\mathcal{Q}^{\prime}_2}   \textbf{P}_{\mathcal{Q}^{\prime\prime}_2} \textbf{P}_{\mathcal{Q}^{\prime\prime\prime}_2}    \times \cdots \times \textbf{P}_{\mathcal{Q}^{\prime\cdots\prime}_2}                                \bigg| \\ \leq \frac{1}{2} \bigg|  \bigg|    \textbf{P}_{\mathcal{Q}_1} \textbf{P}_{\mathcal{Q}^{\prime}_1} \textbf{P}_{\mathcal{Q}^{\prime\prime}_1} \textbf{P}_{\mathcal{Q}^{\prime\prime\prime}_1}   \times \cdots \times \textbf{P}_{\mathcal{Q}^{\prime\cdots\prime}_1}  - \textbf{P}_{\mathcal{Q}_2}   \textbf{P}_{\mathcal{Q}^{\prime}_2}   \textbf{P}_{\mathcal{Q}^{\prime\prime}_2} \textbf{P}_{\mathcal{Q}^{\prime\prime\prime}_2}    \times \cdots \times \textbf{P}_{\mathcal{Q}^{\prime\cdots\prime}_2}                                        \bigg| \bigg|_1    \text{. } 
\end{array}\right. 
\]

\item[$\bullet$] \textit{The probability in the previous item above can be upper bounded with the probability that none of the questions are equal}. One has that,

\[   \left\{\!\begin{array}{ll@{}>{{}}l} 
  \bigg| \bigg|    \textbf{P}_{\mathcal{Q}_1} - \textbf{P}_{\mathcal{Q}_2}       \bigg| \bigg|_1 \leq 2 \textbf{P}_{\mathcal{Q}_1 \mathcal{Q}_2}     \big[ \mathcal{Q}_1 \neq \mathcal{Q}_2   \big]         \text{, } \\    \bigg|  \bigg|         \textbf{P}_{\mathcal{Q}_1} \textbf{P}_{\mathcal{Q}^{\prime}_1} - \textbf{P}_{\mathcal{Q}_2}   \textbf{P}_{\mathcal{Q}^{\prime}_2}          \bigg| \bigg|_1  \leq 2 \textbf{P}_{\mathcal{Q}_1 \mathcal{Q}^{\prime}_1 \mathcal{Q}_2 \mathcal{Q}^{\prime}_2}  \big[ \mathcal{Q}_1 \neq \mathcal{Q}^{\prime}_1 \neq \mathcal{Q}_2  \neq \mathcal{Q}^{\prime}_2  \big]   \text{, } \\  \bigg|  \bigg|    \textbf{P}_{\mathcal{Q}_1} \textbf{P}_{\mathcal{Q}^{\prime}_1} \textbf{P}_{\mathcal{Q}^{\prime\prime}_1} - \textbf{P}_{\mathcal{Q}_2}   \textbf{P}_{\mathcal{Q}^{\prime}_2}   \textbf{P}_{\mathcal{Q}^{\prime\prime}_2}                         \bigg| \bigg|_1  \leq 2 \textbf{P}_{\mathcal{Q}_1 \mathcal{Q}^{\prime}_1\mathcal{Q}^{\prime\prime}_1 \mathcal{Q}_2 \mathcal{Q}^{\prime}_2 \mathcal{Q}^{\prime\prime}_2}   \big[ \mathcal{Q}_1  \neq \mathcal{Q}^{\prime}_1 \neq \mathcal{Q}^{\prime\prime}_1  \\ \neq \mathcal{Q}_2 \neq \mathcal{Q}^{\prime}_2 \neq \mathcal{Q}^{\prime\prime}_2 \big]    \\            \vdots \\ \bigg|  \bigg|    \textbf{P}_{\mathcal{Q}_1} \textbf{P}_{\mathcal{Q}^{\prime}_1} \textbf{P}_{\mathcal{Q}^{\prime\prime}_1} \textbf{P}_{\mathcal{Q}^{\prime\prime\prime}_1}   \times \cdots \times \textbf{P}_{\mathcal{Q}^{\prime\cdots\prime}_1}  - \textbf{P}_{\mathcal{Q}_2}   \textbf{P}_{\mathcal{Q}^{\prime}_2}   \textbf{P}_{\mathcal{Q}^{\prime\prime}_2} \textbf{P}_{\mathcal{Q}^{\prime\prime\prime}_2}    \times \cdots \times \textbf{P}_{\mathcal{Q}^{\prime\cdots\prime}_2}                                        \bigg| \bigg|_1  \\ \leq 2 \textbf{P}_{\mathcal{Q}_1 \mathcal{Q}^{\prime}_1  \mathcal{Q}^{\prime\prime}_1 \times \cdots \times \mathcal{Q}^{\prime\cdots\prime}_1  \mathcal{Q}_2 \mathcal{Q}^{\prime}_2 \mathcal{Q}^{\prime\prime}_2 \times \cdots \times \mathcal{Q}^{\prime\cdots\prime}_2}  \big[ \mathcal{Q}_1  \neq  \mathcal{Q}^{\prime}_1  \neq \mathcal{Q}^{\prime\prime}_1 \neq  \cdots \neq \mathcal{Q}^{\prime\cdots\prime}_1 \\ \neq  \mathcal{Q}_2 \neq \mathcal{Q}^{\prime}_2 \neq \mathcal{Q}^{\prime\prime}_2  \neq \cdots   \neq  \mathcal{Q}^{\prime\cdots\prime}_2 \big]      \text{. } 
\end{array}\right. 
\]

\item[$\bullet$] \textit{Upper bounding conditional probabilities with products of probabilities}. One has that,

\[   \left\{\!\begin{array}{ll@{}>{{}}l}                \bigg| \bigg|  \textbf{P}_{\mathcal{Q}_1 \mathcal{Q}_2 \mathcal{Q}_3}  - \textbf{P}_{\mathcal{Q}_1 | \mathcal{Q}_2 , q^{\prime}_3 } \bigg| \bigg|    \leq \frac{\big( N + 2 \big)}{\alpha} \bigg| \bigg|  \textbf{P}_{\mathcal{Q}_1 \mathcal{Q}_2 \mathcal{Q}_3}  - \textbf{P}_{\mathcal{Q}_1} \textbf{P}_{ \mathcal{Q}_2 , q^{\prime}_3 } \bigg| \bigg|         \text{, } \\  \bigg| \bigg|  \textbf{P}_{\mathcal{Q}_1 \mathcal{Q}_2 \mathcal{Q}_3 \mathcal{Q}_4}  - \textbf{P}_{\mathcal{Q}_1  | \mathcal{Q}_2, q^{\prime}_3 , q^{\prime}_4}   \bigg| \bigg|    \leq  \frac{\big( N + 2 \big)}{\alpha}  \bigg| \bigg|  \textbf{P}_{\mathcal{Q}_1 \mathcal{Q}_2 \mathcal{Q}_3 \mathcal{Q}_4}  - \textbf{P}_{\mathcal{Q}_1 } \textbf{P}_{\mathcal{Q}_2, q^{\prime}_3 , q^{\prime}_4}   \bigg| \bigg|          \text{, } \\   
\bigg| \bigg|  \textbf{P}_{\mathcal{Q}_1 \mathcal{Q}_2 \mathcal{Q}_3 \mathcal{Q}_4 \mathcal{Q}_5 }  - \textbf{P}_{\mathcal{Q}_1  | \mathcal{Q}_2, q^{\prime}_3 , q^{\prime}_4 q^{\prime}_5}   \bigg| \bigg|    \leq \frac{\big( N + 2 \big)}{\alpha}  \bigg| \bigg|  \textbf{P}_{\mathcal{Q}_1 \mathcal{Q}_2 \mathcal{Q}_3 \mathcal{Q}_4 \mathcal{Q}_5 }  - \textbf{P}_{\mathcal{Q}_1} \textbf{P}_{\mathcal{Q}_2, q^{\prime}_3 , q^{\prime}_4 q^{\prime}_5}   \bigg| \bigg|       \text{, } \\ \vdots \\  \bigg| \bigg|  \textbf{P}_{\mathcal{Q}_1 \mathcal{Q}_2 \mathcal{Q}_3 \mathcal{Q}_4 \mathcal{Q}_5 \times \cdots \times \mathcal{Q}_N}  - \textbf{P}_{\mathcal{Q}_1  | \mathcal{Q}_2, q^{\prime}_3 , q^{\prime}_4 q^{\prime}_5 , \cdots, q^{\prime}_N } \bigg| \bigg|    \leq  \frac{\big( N + 2 \big)}{\alpha}  \bigg| \bigg|  \textbf{P}_{\mathcal{Q}_1 \mathcal{Q}_2 \mathcal{Q}_3 \mathcal{Q}_4 \mathcal{Q}_5 \times \cdots \times \mathcal{Q}_N}  - \textbf{P}_{\mathcal{Q}_1 } \\   \times \textbf{P}_{\mathcal{Q}_2, q^{\prime}_3 , q^{\prime}_4 q^{\prime}_5 , \cdots, q^{\prime}_N } \bigg| \bigg|   \text{, } \\  \bigg| \bigg|  \textbf{P}_{\mathcal{Q}_1 \mathcal{Q}_2 \mathcal{Q}_3 \mathcal{Q}_4 \mathcal{Q}_5 \times \cdots \times \mathcal{Q}_N}  - \textbf{P}_{\mathcal{Q}_1 \mathcal{Q}_2 | \mathcal{Q}^{\prime}_3 , q^{\prime}_4 q^{\prime}_5 , \cdots, q^{\prime}_N } \bigg| \bigg|    \leq  \frac{\big( N + 2 \big)}{\alpha}   \bigg| \bigg|  \textbf{P}_{\mathcal{Q}_1 \mathcal{Q}_2 \mathcal{Q}_3 \mathcal{Q}_4 \mathcal{Q}_5 \times \cdots \times \mathcal{Q}_N}  - \textbf{P}_{\mathcal{Q}_1 \mathcal{Q}_2} \\ \times \textbf{P}_{ \mathcal{Q}^{\prime}_3 , q^{\prime}_4 q^{\prime}_5 , \cdots, q^{\prime}_N } \bigg| \bigg|  \text{,} \\  \bigg| \bigg|  \textbf{P}_{\mathcal{Q}_1 \mathcal{Q}_2 \mathcal{Q}_3 \mathcal{Q}_4 \mathcal{Q}_5 \times \cdots \times \mathcal{Q}_N}  - \textbf{P}_{\mathcal{Q}_1 \mathcal{Q}_2 \mathcal{Q}_3  | \mathcal{Q}_4 q^{\prime}_5 , \cdots, q^{\prime}_N } \bigg| \bigg|    \leq \frac{\big( N + 2 \big)}{\alpha}   \bigg| \bigg|  \textbf{P}_{\mathcal{Q}_1 \mathcal{Q}_2 \mathcal{Q}_3 \mathcal{Q}_4 \mathcal{Q}_5 \times \cdots \times \mathcal{Q}_N}  - \textbf{P}_{\mathcal{Q}_1 \mathcal{Q}_2 \mathcal{Q}_3 } \\  \times \textbf{P}_{ \mathcal{Q}_4 q^{\prime}_5 , \cdots, q^{\prime}_N } \bigg| \bigg|   \\ \vdots \\  \bigg| \bigg|  \textbf{P}_{\mathcal{Q}_1 \mathcal{Q}_2 \mathcal{Q}_3 \mathcal{Q}_4 \mathcal{Q}_5 \times \cdots \times \mathcal{Q}_N}  - \textbf{P}_{\mathcal{Q}_1 \mathcal{Q}_2 \mathcal{Q}_3  \times \cdots \times \mathcal{Q}_{N-2} | \mathcal{Q}_{N-1} q^{\prime}_N } \bigg| \bigg|    \leq \frac{\big( N + 2 \big)}{\alpha}  \bigg| \bigg|  \textbf{P}_{\mathcal{Q}_1 \mathcal{Q}_2 \mathcal{Q}_3 \mathcal{Q}_4 \mathcal{Q}_5 \times \cdots \times \mathcal{Q}_N}  \\ - \textbf{P}_{\mathcal{Q}_1 \mathcal{Q}_2 \mathcal{Q}_3  \times \cdots \times \mathcal{Q}_{N-2} } \textbf{P}_{ \mathcal{Q}_{N-1} q^{\prime}_N } \bigg| \bigg|   \text{, } \end{array}\right. 
\]

\[   \left\{\!\begin{array}{ll@{}>{{}}l}  \bigg| \bigg|  \textbf{P}_{\mathcal{Q}_1 \mathcal{Q}_2 \mathcal{Q}_3 \mathcal{Q}_4 \mathcal{Q}_5 \times \cdots \times \mathcal{Q}_{N-1}}  - \textbf{P}_{\mathcal{Q}_1|  \mathcal{Q}_2 q^{\prime}_3  \times \cdots \times q^{\prime}_{N-2}  q^{\prime}_{N-1} q^{\prime}_N } \bigg| \bigg|    \leq \frac{\big( N + 2 \big)}{\alpha}  \bigg| \bigg|  \textbf{P}_{\mathcal{Q}_1 \mathcal{Q}_2 \mathcal{Q}_3 \mathcal{Q}_4 \mathcal{Q}_5 \times \cdots \times \mathcal{Q}_{N-1}}  - \textbf{P}_{\mathcal{Q}_1}
\\  \times  \textbf{P}_{  \mathcal{Q}_2 q^{\prime}_3  \times \cdots \times q^{\prime}_{N-2}  q^{\prime}_{N-1} q^{\prime}_N } \bigg| \bigg|  \text{, } \\ \\ \vdots \\ \bigg| \bigg|  \textbf{P}_{\mathcal{Q}_1 \mathcal{Q}_2 \mathcal{Q}_3 \mathcal{Q}_4 \mathcal{Q}_5 \times \cdots \times \mathcal{Q}_{N-1}}  - \textbf{P}_{\mathcal{Q}_1 \mathcal{Q}_2 \mathcal{Q}_3  \times \cdots \times \mathcal{Q}_{N-2} | \mathcal{Q}_{N-1} q^{\prime}_N } \bigg| \bigg|    \leq \frac{\big( N + 2 \big)}{\alpha}   \bigg| \bigg|  \textbf{P}_{\mathcal{Q}_1 \mathcal{Q}_2 \mathcal{Q}_3 \mathcal{Q}_4 \mathcal{Q}_5 \times \cdots \times \mathcal{Q}_{N-1}}  \\ - \textbf{P}_{\mathcal{Q}_1 \mathcal{Q}_2 \mathcal{Q}_3  \times \cdots \times \mathcal{Q}_{N-2}} \textbf{P}_{ \mathcal{Q}_{N-1} q^{\prime}_N } \bigg| \bigg|  \text{, } \\ \vdots \\ \bigg| \bigg|  \textbf{P}_{\mathcal{Q}_1 \mathcal{Q}_2 \mathcal{Q}_3 }  - \textbf{P}_{\mathcal{Q}_2 | \mathcal{Q}_{1} , q^{\prime}_3 } \bigg| \bigg|    \leq  \frac{\big( N + 2 \big)}{\alpha}  \bigg| \bigg|  \textbf{P}_{\mathcal{Q}_1 \mathcal{Q}_2 \mathcal{Q}_3 }  - \textbf{P}_{\mathcal{Q}_2} \textbf{P}_{ \mathcal{Q}_{1} , q^{\prime}_3 } \bigg| \bigg|         \text{, }
\end{array}\right. 
\]

\noindent where, each occurrence of $\alpha$ in the upper bound for the inequalities provided above equals,

\begin{align*}
  \textbf{P}_{\mathcal{Q}_i} \big[ q_1 , q_2 \big]       =  \alpha      \textbf{P}_{\mathcal{Q}_i} \big[ q_1 \big]  , 
\end{align*}

\noindent and so on.

\bigskip

\item[$\bullet$] \textit{Upper bounding conditional probabilities minus probabilities with conditioning removed}. One has that,

\[   \left\{\!\begin{array}{ll@{}>{{}}l} 
\bigg| \bigg| \textbf{P}_{\mathcal{Q}_1 \mathcal{Q}_2 | \mathcal{Q}_3 \times \cdots \times \mathcal{Q}_{N-1} q^{*}_{N}} - \textbf{P}_{\mathcal{Q}_1 \mathcal{Q}_2}  \bigg| \bigg| \leq  \frac{\big( N + 10 \big)}{\alpha} \bigg| \bigg| \textbf{P}_{\mathcal{Q}_1 \times \cdots \times \mathcal{Q}_N} - \textbf{P}_{\mathcal{Q}_1 \mathcal{Q}_3 \times \cdots \times \mathcal{Q}_{N} } \textbf{P}_{\mathcal{Q}^{\prime}_2 | \mathcal{Q}^{\prime}_1} \bigg| \bigg|  \text{, } \\  \bigg| \bigg| \textbf{P}_{\mathcal{Q}_1 \mathcal{Q}_2  \mathcal{Q}_3 | \mathcal{Q}_4 \times \cdots \times \mathcal{Q}_{N-1} q^{*}_{N}} - \textbf{P}_{\mathcal{Q}_1 \mathcal{Q}_2 \mathcal{Q}_3 }  \bigg| \bigg| \leq \frac{\big( N + 10 \big)}{\alpha}  \bigg| \bigg|     \textbf{P}_{\mathcal{Q}_1 \times \cdots \times \mathcal{Q}_N}       -          \textbf{P}_{\mathcal{Q}_1 \mathcal{Q}_4 \times \cdots \times \mathcal{Q}_N} \textbf{P}_{\mathcal{Q}^{\prime}_2 \mathcal{Q}^{\prime}_3 | \mathcal{Q}^{\prime}_1}     \bigg| \bigg|   \text{, } \\ \vdots \\   \bigg| \bigg| \textbf{P}_{\mathcal{Q}_1 \mathcal{Q}_2 \mathcal{Q}_3 \times \cdots \times  \mathcal{Q}_{N-2} | \mathcal{Q}_{N-1} q^{*}_{N}} - \textbf{P}_{\mathcal{Q}_1 \mathcal{Q}_2 \mathcal{Q}_3 \times \cdots \times  \mathcal{Q}_{N-2}} \bigg| \bigg| \leq  \frac{\big( N + 10 \big)}{\alpha}  \bigg| \bigg|     \textbf{P}_{\mathcal{Q}_1 \times \cdots \times \mathcal{Q}_N}       -          \textbf{P}_{\mathcal{Q}_1 \mathcal{Q}_{N-1} \mathcal{Q}_N} \\  \times \textbf{P}_{\mathcal{Q}^{\prime}_2 \times \cdots \times \mathcal{Q}^{\prime}_{N-2} | \mathcal{Q}^{\prime}_1}               \bigg| \bigg|   \text{, }  \\ \bigg| \bigg| \textbf{P}_{\mathcal{Q}_1 | \mathcal{Q}_2 \times \cdots \times q^{*}_{N-1}q^{*}_{N}} - \textbf{P}_{\mathcal{Q}_1 }  \bigg| \bigg| \leq  \bigg| \bigg|     \textbf{P}_{\mathcal{Q}_1 \times \cdots \times \mathcal{Q}_N}       -           \textbf{P}_{\mathcal{Q}_1 \mathcal{Q}_3 \times \cdots \times \mathcal{Q}_{N} } \textbf{P}_{\mathcal{Q}^{\prime}_2 | \mathcal{Q}^{\prime}_1}      \bigg| \bigg|   \text{, } \\  \bigg| \bigg| \textbf{P}_{\mathcal{Q}_1 \mathcal{Q}_2 | \mathcal{Q}_3 \times \cdots \times \mathcal{Q}_{N-3} q^{*}_{N-2}  q^{*}_{N-1} q^{*}_{N}} - \textbf{P}_{\mathcal{Q}_1 \mathcal{Q}_2}  \bigg| \bigg| \leq   \frac{\big( N + 10 \big)}{\alpha}  \bigg| \bigg|     \textbf{P}_{\mathcal{Q}_1 \times \cdots \times \mathcal{Q}_N}       -            \textbf{P}_{\mathcal{Q}_1 \mathcal{Q}_4 \times \cdots \times \mathcal{Q}_N} \textbf{P}_{\mathcal{Q}^{\prime}_2 \mathcal{Q}^{\prime}_3 | \mathcal{Q}^{\prime}_1}                                                                                          \bigg| \bigg|   \text{, } \\ \vdots \end{array}\right. 
\]

\[   \left\{\!\begin{array}{ll@{}>{{}}l}       \bigg| \bigg| \textbf{P}_{\mathcal{Q}_1 \mathcal{Q}_2 \mathcal{Q}_3 \times \cdots \times  \mathcal{Q}_{N-3} | \mathcal{Q}_{N-2} q^{*}_{N-1} q^{*}_{N}} - \textbf{P}_{\mathcal{Q}_1 \mathcal{Q}_2 \mathcal{Q}_3 \times \cdots \times  \mathcal{Q}_{N-3} } \bigg| \bigg| \leq  \frac{\big( N + 10 \big)}{\alpha}  \bigg| \bigg|     \textbf{P}_{\mathcal{Q}_1 \times \cdots \times \mathcal{Q}_N}     \\       -        \textbf{P}_{\mathcal{Q}_1 \times \cdots \times \mathcal{Q}_{N-1}\mathcal{Q}_{N} } \textbf{P}_{\mathcal{Q}^{\prime}_2 \times \cdots \times  \mathcal{Q}^{\prime}_{N-2} | \mathcal{Q}^{\prime}_1}                                                                                             \bigg| \bigg|   \text{, } \\   \bigg| \bigg| \textbf{P}_{\mathcal{Q}_1| \mathcal{Q}_2 \times \cdots \times q^{*}_{N-2} q^{*}_{N-1} q^*_N}  - \textbf{P}_{\mathcal{Q}_1} \bigg| \bigg| \leq \frac{\big( N + 10 \big)}{\alpha}  \bigg| \bigg|     \textbf{P}_{\mathcal{Q}_1 \times \cdots \times \mathcal{Q}_N}         - \textbf{P}_{\mathcal{Q}_1 \mathcal{Q}_3 \times \cdots \times \mathcal{Q}_{N} } \textbf{P}_{\mathcal{Q}^{\prime}_2 | \mathcal{Q}^{\prime}_1} \bigg| \bigg| \text{,} \\     \bigg| \bigg| \textbf{P}_{\mathcal{Q}_1 \mathcal{Q}_2 | \mathcal{Q}_3 \times \cdots \times \mathcal{Q}_{N-4} q^{*}_{N-3} q^{*}_{N-2}  q^{*}_{N-1} q^{*}_{N}} - \textbf{P}_{\mathcal{Q}_1 \mathcal{Q}_2}  \bigg| \bigg| \leq   \frac{\big( N + 10 \big)}{\alpha}  \bigg| \bigg|     \textbf{P}_{\mathcal{Q}_1 \times \cdots \times \mathcal{Q}_N}       -           \textbf{P}_{\mathcal{Q}_1 \mathcal{Q}_4 \times \cdots \times \mathcal{Q}_N} \\ \times \textbf{P}_{\mathcal{Q}^{\prime}_2 \mathcal{Q}^{\prime}_3 | \mathcal{Q}^{\prime}_1}                              \bigg| \bigg|   \text{, } \\ \vdots \\   \bigg| \bigg| \textbf{P}_{\mathcal{Q}_1 \mathcal{Q}_2 \mathcal{Q}_3 \times \cdots \times  \mathcal{Q}_{N-2} | \mathcal{Q}_{N-3} q^{*}_{N-2} q^{*}_{N-1} q^{*}_{N}} - \textbf{P}_{\mathcal{Q}_1 \mathcal{Q}_2 \mathcal{Q}_3 \times \cdots \times  \mathcal{Q}_{N-3} } \bigg| \bigg| \leq  \frac{\big( N + 10 \big)}{\alpha}   \bigg| \bigg|     \textbf{P}_{\mathcal{Q}_1 \times \cdots \times \mathcal{Q}_N}    \\    -        \textbf{P}_{\mathcal{Q}_1 \times \cdots \times \mathcal{Q}_{N-1}\mathcal{Q}_{N} } \textbf{P}_{\mathcal{Q}^{\prime}_2 \times \cdots \times  \mathcal{Q}^{\prime}_{N-2} | \mathcal{Q}^{\prime}_1}                                                                                          \bigg| \bigg|   \text{, }  \\ \vdots \\ \bigg| \bigg| \textbf{P}_{\mathcal{Q}_1 \mathcal{Q}_2 \mathcal{Q}_4 \times \cdots \times \mathcal{Q}_{N-2} | \mathcal{Q}_3 q^{*}_{N-1} q^{*}_N } - \textbf{P}_{\mathcal{Q}_1 \mathcal{Q}_2 \mathcal{Q}_4 \times \cdots \times \mathcal{Q}_{N-2}} \bigg| \bigg| \leq \frac{\big( N + 10 \big)}{\alpha}  \bigg| \bigg| \textbf{P}_{\mathcal{Q}_1 \times \cdots \times \mathcal{Q}_N}    \\  -               \textbf{P}_{\mathcal{Q}_1 \mathcal{Q}_2 \mathcal{Q}_4 \times \cdots \times \mathcal{Q}_{N-1}} \textbf{P}_{\mathcal{Q}^{\prime}_2 \times \cdots \times \mathcal{Q}^{\prime}_N | \mathcal{Q}^{\prime}_1  } . 
\end{array}\right. 
\]

\noindent The marginal probability measures to ones introduced above take the form,

\begin{align*}
 \textbf{P}_{\mathcal{Q}_1 \times \cdots \times \mathcal{Q}_N} \bigg|_{\mathcal{Q}_1} \equiv \textbf{P}_{\mathcal{Q}_1 }   \text{, } \\  \textbf{P}_{\mathcal{Q}_1 \times \cdots \times \mathcal{Q}_N} \bigg|_{\mathcal{Q}_2} \equiv \textbf{P}_{\mathcal{Q}_2 }   \text{, } \\  \vdots  \\  \textbf{P}_{\mathcal{Q}_1 \times \cdots \times \mathcal{Q}_N} \bigg|_{\mathcal{Q}_N} \equiv \textbf{P}_{\mathcal{Q}_N }  \text{,  }
\end{align*}

\noindent in addition to any pairwise combinations of questions that are distributed to each participant from the referee's probability distribution. With respect to each marginal probability measure, one writes each corresponding expectation value as $\textbf{E}_{\textbf{P}_{\mathcal{Q}_1}} \equiv \textbf{E}_{\mathcal{Q}_1}, \dots, \textbf{E}_{\textbf{P}_{\mathcal{Q}_N}} \equiv \textbf{E}_{\mathcal{Q}_N}$. Besides defining probability measure supported over questions sets from countably many players simultaneously, we provide a statement of the following claim below, which straightforwardly generalizes the marginal probability distributions from the anchored two-player setting previously studied in [2, 3, 4, 56]. The first claim states:

\bigskip

\noindent \textbf{Claim} \textit{1} (\textit{factorization of random variables corresponding to the possible set of questions prepared by each player}). Conditionally on $\big( D, 
M \big)$, the random variables $\mathcal{Q}_1, \mathcal{Q}_2, \cdots, \mathcal{Q}_N$ are independent.

\bigskip

\noindent \textit{Proof of Claim 1}. With direct computation,

\begin{align*}
  \textbf{P}_{\mathcal{Q}_1 \mathcal{Q}_2 \times \cdots \times \mathcal{Q}_N | D \equiv d, M \equiv m} \big( q_1, q_2, \cdots , q_N \big) \equiv  \textbf{P}_{\mathcal{Q}_1  | D \equiv d, M \equiv m} \big( q_1 \big) \times \\ \cdots \times  \textbf{P}_{\mathcal{Q}_N  | D \equiv d, M \equiv m} \big( q_N \big)          \equiv \underset{1 \leq i \leq N}{\prod} \textbf{P}_{\mathcal{Q}_i  | D \equiv d, M \equiv m} \big( q_i \big)    ,
\end{align*}

\noindent from which we conclude the argument. \boxed{}

\bigskip

\noindent Besides the first claim above, we muat also make use of thue fact that marginal probability distributions over the set of questions for each player coincide with the marginal measure, $\mu$. To this end, the second claim below states:

\bigskip

\noindent \textbf{Claim 2} (\textit{the marginal probability distributions over the set of questions distributed to each player are equal}). For questions $q_1 \in \mathcal{Q}_1, q_2 \in \mathcal{Q}_2, \cdots , q_n \in \mathcal{Q}_N$, one has that,

\begin{align*}
\textbf{P}_{\mathcal{Q}_1 \mathcal{Q}_2 \cdots \mathcal{Q}_N | D \equiv A_1} \big( q_1, q_2, \cdots, q_N \big)  =    \textbf{P}_{\mathcal{Q}_1 \mathcal{Q}_2 \cdots \mathcal{Q}_N | D \equiv A_2} \big( q_1, q_2, \cdots, q_N \big)  = \cdots \\ =    \textbf{P}_{\mathcal{Q}_1 \mathcal{Q}_2 \cdots \mathcal{Q}_N | D \equiv A_N} \big( q_1, q_2, \cdots, q_N \big)  = \mu_{\mathcal{Q}_1 \mathcal{Q}_2 \cdots \mathcal{Q}_N}  \big( q_1, q_2, \cdots, q_N \big)   .
\end{align*}

\bigskip

\noindent \textit{Proof of Claim 2}. Directly apply the argument from \textbf{Claim} \textit{4.2} from [], from which we conclude the argument. \boxed{}

\item[$\bullet$]
 \textit{Upper bounding the following system of probabilities with upper bounds dependent upon the one-sided anchoring probability}. Denote some position $i \in [n] \backslash C$, where $C$ is some clause. For the following probabilistic statements below, denote,

 \begin{align*}
  \underline{(1)}: \textit{Dependency-breaking variables at }\text{ }  i \equiv \Omega_i \equiv \big( D_i , M_i \big)   \text{, } \\ 
       \underline{(2)}:  \textit{Question distributed to the first player} \equiv X_C      \text{, } \\     \underline{(3)}:    \textit{Question distributed to the first player} \equiv Y_C  
     \text{, } \\    \underline{(4)}:   \textit{First player's answer} \equiv A_C  
     \text{, }  \\   \underline{(5)}:   \textit{Second player's answer} \equiv B_C  
     \text{, }   \\   \underline{(6)}:    \textit{The set of all dependency-breaking variables} \equiv \Omega \equiv \big( \Omega_i \big)_{i \in [n] \backslash C} \\ \cup \big( X_C , Y_C \big) , \\  \underline{(7)}:   \textit{Dependency-breaking variables at a position away from } \text{ } i \equiv \Omega_{-i} \\ \equiv  \big( \Omega_j \big)_{j \in [n] \backslash ( C \cup \{ i \}} \cup \big( X_C , Y_C \big) , \forall i \in [n] \backslash C , 
     \\  \underline{(8)}:  \textit{Variables corresponding to} \text{ } \Omega \equiv R \equiv \big( \Omega , A_C , B_C \big), \\   \underline{(9)}: \textit{Variables corresponding to} \text{ } \Omega_{-i} \equiv R \equiv \big( \Omega_{-i} , A_C , B_C \big) \\ ,  \forall i \in [n] \backslash C, 
 \end{align*}

 \noindent and finally,

 \begin{align*}
   W_C \equiv \underset{i \in C}{\prod} W_i   \text{, }
 \end{align*}

 \noindent corresponding to the winning condition over all possible coordinates $i$.

 \noindent In the two-player game-theoretic setting, one has that, [2, 3, 4, 30, 56], for suitable unitaries $U$, such that,

\begin{align*}
 U  \curvearrowright  Y_{\bar{C}} \bar{Y}_{\bar{C}} B Z_{\bar{C}}  \equiv \textit{Purification applied to the question of Bob, the second player}  \text{, }
\end{align*}

 \noindent as well as projection operators $\Pi$, such that, for $c^{\prime}$ taken sufficiently small,

\begin{align*}
    \Pi \equiv \Pi_{x_i r_i}  \curvearrowright  \ket{\psi^{\prime}}_{r_i} \Longleftrightarrow \textbf{P} \big[  \Pi \equiv \Pi_{x_i r_i}  \text{ } \textit{succeeds}  \text{ } \ket{\psi^{\prime}}_{r_i} \big] = 2^{-c^{\prime}} > 0 
\end{align*}

 \noindent one has that,

\[   \left\{\!\begin{array}{ll@{}>{{}}l}  \xi \in \big[ 0 ,1 \big] :  \bigg| \bigg| \textbf{P}_{X_i Y_i R_i | \mathcal{E}}  - \textbf{P}_{X_i Y_i} \textbf{P}_{R_i | \mathcal{E}_i Y_i} \bigg| \bigg|_1 \leq \frac{7 \xi}{20}  \text{, } \\   \xi \in \big[ 0 ,1 \big] :  \bigg| \bigg|  \textbf{P}_{X_i Y_i R_i | \mathcal{E}}  - \textbf{P}_{X_i Y_i} \textbf{P}_{R_i | \mathcal{E}_i X_i}  \bigg| \bigg|_1 \leq \frac{7 \xi}{20}  \text{, } \\   \xi \in \big[ 0 ,1 \big] :  \underset{P_{X_i Y_i R_i | \mathcal{E}}}{\textbf{E}} \frac{1}{2^{-c^{\prime}}} \bigg| \bigg| \ket{\varphi}\bra{\varphi}   - \big( \textbf{I} \otimes U_{y_i, r_i} \big) \bra{\varphi} \ket{\varphi}_{x_i y^* r_i} \big( \textbf{I} \otimes U^{\dagger}_{y_i r_i} \big) \bigg| \bigg|_1     \leq 21 \xi     \text{. } 
\end{array}\right. 
\]

\item[$\bullet$] \textit{Upper bounding the expected values of quantum states with a constant proportional to the one-sided anchoring probability}. In the two-player game-theoretic setting, one has that, [30], for suitable unitaries introduced in the previous item, that,

\[   \left\{\!\begin{array}{ll@{}>{{}}l} 
\epsilon \equiv \epsilon \big( \xi \big) > 0 :  \underset{i \in \bar{C}}{\textbf{E}} \bigg| \bigg| \textbf{P}_{X_i Y_i R_i | \mathcal{E}}  - \textbf{P}_{X_i Y_i} \textbf{P}_{R_i | \mathcal{E}_i Y_i} \bigg| \bigg|_1 \leq \frac{7 \epsilon}{600} \text{, } \\   \epsilon \equiv \epsilon \big( \xi \big) > 0  :  \underset{i \in \bar{C}}{\textbf{E}} \bigg| \bigg|  \textbf{P}_{X_i Y_i R_i | \mathcal{E}}  - \textbf{P}_{X_i Y_i} \textbf{P}_{R_i | \mathcal{E}_i X_i}  \bigg| \bigg|_1 \leq \frac{7 \epsilon}{600}  \text{, } \\  \epsilon \equiv \epsilon \big( \xi \big) > 0  :  \underset{ i \in \bar{C}}{\textbf{E}} \underset{P_{X_i Y_i R_i | \mathcal{E}}}{\textbf{E}}     \bigg| \bigg| \ket{\varphi}\bra{\varphi}   - \big( \textbf{I} \otimes U_{y_i, r_i} \big) \bra{\varphi} \ket{\varphi}_{x_i y^* r_i} \big( \textbf{I} \otimes U^{\dagger}_{y_i r_i} \big) \bigg| \bigg|_1     \leq \frac{4 \epsilon}{5}         \text{. } 
\end{array}\right. 
\]

\item[$\bullet$] \textit{Computations with expectation values leading to generalized, closed form representations for the optimal value under anchored parallel repetition}. Obtaining the two previous items for the system of probabilities, and subsequently, for the system of expectations, can be decomposed into the following components, which we briefly recapitulate from [30], as a generalization of the rates of decay of the optimal value provided in [2, 3, 4, 56], below:

\begin{itemize}
    \item[$\bullet$] \textit{The success probability of the projection operator}. For $c^{\prime}$ taken to be sufficiently small, the success probability is given by,

\begin{align*}
    \textbf{P} \big[  \Pi \equiv \Pi_{x_i r_i}  \text{ } \textit{succeeds}  \text{ } \ket{\psi^{\prime}}_{r_i} \big] = 2^{-c^{\prime}} > 0 .
    \end{align*}

     \item[$\bullet$] \textit{Expected values over all possible positions i}. One upper bounds,

     \begin{align*}
       \underset{i \in C}{\textbf{E}} \bigg| \bigg| \textbf{P}_{X_i Y_i R_i | \mathcal{E}} - \textbf{P}_{ Y_i R_i | \mathcal{E}} \textbf{P}_{ X_i | Y_i } \bigg| \bigg|_1  \text{. }
     \end{align*}

      \item[$\bullet$] \textit{Upper bounding the expected value of the outer product of the quantum state corresponding to the success probability under the action of projection operators, in comparison the outer product of the }. One argues that,

    \begin{align*}
    \underset{i \in \bar{C}}{\textbf{E}} \underset{\textbf{P}_{X_i R_i | \mathcal{E}}}{\textbf{E}} \frac{1}{\alpha} \bigg| \bigg| \big( \Pi_{x_i r_i} \otimes \textbf{I} \big) \ket{\varphi^{\prime}} \bra{\varphi^{\prime}}_{y^* r_i} \big( \Pi_{x_i r_i} \otimes \textbf{I} \big)  -   \ket{\varphi} \bra{\varphi}_{x_i y^* r_i}   \bigg| \bigg|_1    \text{, }
    \end{align*}

      \noindent is upper bounded with a term dependent upon the one-side anchoring probability,

      \begin{align*}
       C \big( \xi \big) \equiv 3 \xi + 40 \frac{\sqrt{2 \delta_1}}{\xi^2}   \text{, }
      \end{align*}

      \noindent for some $\delta_1$ taken to be sufficiently small.

\end{itemize}

\item[$\bullet$] \textit{Anchoring probability distributions towards the anchored set of responses}. Accompanying the above objects, for the set of anchored questions, $\bot$, one defines the collection of probability measures,

\[   \left\{\!\begin{array}{ll@{}>{{}}l}  \textbf{P}_{M | D = \mathcal{A}_1} \big( q_1 \big) \equiv    \left\{\!\begin{array}{ll@{}>{{}}l} \frac{\textbf{P}_{\mathcal{Q}_1} [ q_1]}{1 - \eta } \Longleftrightarrow  q_1 \neq \bot  , \\  \frac{\alpha - \eta}{1 - \eta} \Longleftrightarrow  q_1  = \bot    \end{array}\right. 
    \text{, } \\  \textbf{P}_{M | D = \mathcal{A}_2} \big( q_2 \big) \equiv    \left\{\!\begin{array}{ll@{}>{{}}l}  \frac{\textbf{P}_{\mathcal{Q}_2} [ q_2]}{1 - \eta } \Longleftrightarrow q_2  \neq  \bot     ,   \\   \frac{\alpha - \eta}{1 - \eta} \Longleftrightarrow q_2 = \bot     ,        \end{array}\right. 
    \text{, }\\ \vdots \\     \textbf{P}_{M | D = \mathcal{A}_N} \big( q_N \big) \equiv    \left\{\!\begin{array}{ll@{}>{{}}l}  \frac{\textbf{P}_{\mathcal{Q}_N} [ q_N]}{1 - \eta } \Longleftrightarrow  q_N \neq  \bot      ,   \\   \frac{\alpha - \eta}{1 - \eta} \Longleftrightarrow       q_N  = \bot     ,   \end{array}\right. 
    \text{, }
\end{array}\right. 
\]

\noindent for the respective answers prepared by each player, $\mathcal{A}_1, \cdots, \mathcal{A}_N$, and questions $q_1, \cdots, q_N \sim \pi$, and $\alpha = 2 \eta$. The $\alpha$ parameter, which is taken under the condition that $0 < \alpha <1$, in comparison to ordinary anchored games, satisfy,

\begin{align*}
  \textbf{P}_{\mathcal{Q}_1} \big[ a_1 = \bot \big] \equiv \textbf{P}_{\mathcal{Q}_1} \big[ \textit{Player one responds with an answer from the anchored set} \big] \equiv  \alpha    \text{, } \\  \textbf{P}_{\mathcal{Q}_2} \big[ a_2 = \bot  \big] \equiv \textbf{P}_{\mathcal{Q}_1} \big[ \textit{Player two responds with an answer from the anchored set} \big]  \equiv \alpha   \text{,} \\ \vdots \\  \textbf{P}_{\mathcal{Q}_N} \big[  a_N = \bot \big]  \equiv \textbf{P}_{\mathcal{Q}_1} \big[ \textit{Player N responds with an answer from the anchored set} \big] \equiv \alpha    \text{. }
\end{align*}

\noindent Furthermore, the conditionings,

\begin{align*}
 \big\{ D = \mathcal{A}_1 \big\}    \text{, } \\ \vdots \\   \big\{ D = \mathcal{A}_N \big\}  \text{, }
\end{align*}

\noindent are enforced on each respective answer prepared by each participant.

\item[$\bullet$] \textit{Leveraging anchored probability distributions from the responses of each player for computing expectations of the relative min-entropy}. In expressions for the optimal value obtained in [2, 3, 4, 56], and in [30], alike, computations of the expected value of the Mutual Information, and Relative-Min, entropies are related to the expectation,

    \begin{align*}
    \underset{i \in \bar{C}}{\textbf{E}} \underset{\textbf{P}_{X_i R_i | \mathcal{E}}}{\textbf{E}} \frac{1}{\alpha} \bigg| \bigg| \big( \Pi_{x_i r_i} \otimes \textbf{I} \big) \ket{\varphi^{\prime}} \bra{\varphi^{\prime}}_{y^* r_i} \big( \Pi_{x_i r_i} \otimes \textbf{I} \big)  -   \ket{\varphi} \bra{\varphi}_{x_i y^* r_i}   \bigg| \bigg|_1    \text{. }
    \end{align*}

\noindent In generalizations of the arguments of [2, 3, 4, 56] provided in [30], one argues,

\begin{align*}
    \underset{\textbf{P}_{X_C Y_C Z_C D G | \mathcal{E}}}{\textbf{E}}   \big[ \textit{Relative-min entropy of State 1 with State 2} \big] \\ \equiv \underset{\textbf{P}_{X_C Y_C Z_C D G | \mathcal{E}}}{\textbf{E}}   S_{\infty} \big( \textit{State 1} || \textit{State 2} \big)  \text{,}
\end{align*}

\noindent for the relative-min entropy, given two quantum states $\rho$ and $\sigma$,

\begin{align*}
   S_{\infty} \big( \rho \big| \big| \sigma \big) \equiv \mathrm{min} \big\{   \lambda : \rho \leq 2^{\lambda} \sigma      \big\}  \text{, }
\end{align*}
\end{itemize}

\noindent along with states,

\begin{align*}
   \textit{State 1} \equiv \mathrm{Tr} \bigg[  \big( \textit{Alice's Unitary} \big)            \bigg( \theta   \otimes    \sigma   \bigg)_{dg}         \big( \textit{Alice's Unitary} \big)^{\dagger}  \bigg]     \text{, } \\ \\ \textit{State 2} \equiv  \mathrm{Tr}  \bigg[   \big( \textit{Bob's Unitary} \big)            \bigg( \theta^{\prime}   \otimes    \sigma^{\prime}   \bigg)_{dg}         \big( \textit{Bob's Unitary} \big)^{\dagger}   \bigg]      \text{, }
\end{align*}

\noindent for a dependency-breaking instance $dg \in \mathcal{D} \mathcal{G}$, namely the set of all dependency-breaking instances, and state $\sigma$ which satisfying,

\begin{align*}
  S_{\infty}  \bigg(       \bigg( \theta_{M}  \bigg)_{dg} \bigg| \bigg| \bigg( \theta \otimes \sigma \bigg)_{dg}        \bigg)  \leq 2 ck  \text{, }
\end{align*}

\noindent for strictly positive $c,k$, and also for which,

\begin{align*}
    S_{\infty}  \bigg(       \bigg( \theta_{M}  \bigg)_{dg} \bigg| \bigg| \big( \textit{Bob's Unitary} \big)   \bigg( \theta \otimes \sigma \bigg)_{dg}   \big( \textit{Bob's Unitary} \big)^{\dagger}      \bigg)  \leq 2 ck    \text{, }
\end{align*}

\noindent under the same choice of $c,k$. In reality, in the two expressions for the relative-min entropy above, we suppressed the questions, along with whether the question is purified or not, in the indices of $\theta$, and of $\sigma$. For example, the indices that one would imposed in the superscripts of $\theta$ and $\sigma$, for the purification of $X_C Y_C$, with $X_C \bar{X_C} Y_C \bar{Y_C}$, takes the form,

\begin{align*}
    \theta_{X_C \bar{X_C } Y_C \bar{Y_C} A M E^B}   =   \underset{xy}{\sum} P_{XY} \big( x y \big) \ket{xy} \bra{xy}  \bigotimes \ket{\theta} \bra{\theta}  \text{. }
\end{align*}

\noindent One can straightforwardly introduce similar expressions for $\theta^{\prime}$, and for $\sigma^{\prime}$, which for multiplayer dependency-breaking variables, takes the form,

\begin{align*}
    \theta_{1_C \bar{1_C } 2_C \bar{2_C} \times \cdots \times N_C \bar{N_C} A M E^B}   =   \underset{q_1 q_2 \cdots q_N}{\sum} P_{Q_1\cdots Q_N} \big(q_1 q_2 \cdots q_N \big) \ket{q_1 q_2 \cdots q_N} \bra{q_1 q_2 \cdots q_N} \\  \bigotimes \ket{\theta} \bra{\theta}  \text{. }
\end{align*}

\noindent In comparison to previous game-theoretic arguments formulated by the author, [50, 52, 53], purification has not yet been examined, and has intriguing connections not only with optimality of strategies that each participant can adopt, but also with impacts to the maximum probability of winning through the optimal value. Computations above, for upper bounding the relative-min entropy provided in [2, 3, 4, 56], and further generalized in [30], together stem from the observation that it is imperative to upper bound the expectation,

\begin{align*}
    \underset{\textbf{P}_{X_C Y_C Z_C D G | \mathcal{E}}}{\textbf{E}}   \big[ \textit{Relative entropy of State 1 with State 2} \big] \\ \equiv \underset{\textbf{P}_{X_C Y_C Z_C D G | \mathcal{E}}}{\textbf{E}}   S \big( \big( \textit{State 1} \big) ||  \big( \textit{State 2} \big) \big)  \text{,}
\end{align*}

\noindent corresponding to the expected value of the \textit{relative entropy}, $S$. Such an entropy is defined as,

\begin{align*}
 \textit{Relative entropy} \equiv  D \big( \rho \big| \big| \sigma \big) \equiv \mathrm{Tr} \big( \rho \big( \mathrm{log} \rho - \mathrm{log} \sigma \big)  \big)  \text{. }
\end{align*}

\noindent For the multiplayer setting, one argues that,

\begin{align*}
    \underset{\textbf{P}_{1_C 2_C 3_C \times \cdots \times N_C D G | \mathcal{E}}}{\textbf{E}}   \big[ \textit{Relative entropy of Multiplayer State 1 with  Multiplayer State 2} \big] \\ \equiv \underset{\textbf{P}_{1_C 2_C 3_C \times \cdots \times N_C  D G | \mathcal{E}}}{\textbf{E}}   S  \big( \big( \textit{Multiplayer State 1} \big) || \big( \textit{Multiplayer State 2} \big) \big)  \\ < \underset{\textbf{P}_{1_C 2_C 3_C \times \cdots \times N_C  D G | \mathcal{E}}}{\textbf{E}}   S_{\infty} \big( \big( \textit{Multiplayer State 1} \big)  || \big( \textit{Multiplayer State 2} \big)  \big) \text{,}
\end{align*}

\noindent for,

\begin{align*}
   \textit{State 1} \equiv \mathrm{Tr} \bigg[  \big( \textit{Player unitaries} \big)            \bigg( \big( \theta_1 \big)   \otimes    \big( \theta_2 \big) \otimes \cdots \otimes \big( \theta_N \big)   \bigg)_{dg}         \big[ \textit{Player unitaries} \big]^{\dagger}  \bigg]     \text{, } \\ \\ \textit{State 2} \equiv  \mathrm{Tr}  \bigg[   \big( \textit{Player unitaries} \big)            \bigg( \big( \theta_1 \big)^{\prime}   \otimes    \big( \theta_2 \big)^{\prime} \otimes \cdots \otimes \big( \theta_N \big)^{\prime}   \bigg)_{dg}         \big[ \textit{Player unitaries} \big]^{\dagger}   \bigg]      \text{, }
\end{align*}

\noindent where,

\begin{align*}
   \textit{Player unitaries} \equiv \underset{1 \leq i \leq N}{\prod} \textit{Player i th unitary} \text{,} \\    \bigg[ \textit{Player unitaries} \bigg]^{\dagger} \equiv \bigg[  \underset{1 \leq i \leq N}{\prod} \textit{Player i th unitary}\bigg]^{\dagger}   \equiv  \underset{1 \leq i \leq N}{\prod}  \big[  \textit{Player i th unitary}\big]^{\dagger}    \text{. }
\end{align*}

\noindent Straightforwardly, one can define,

\begin{align*}
  \textit{State 3}  \equiv \mathrm{Tr} \bigg[ \big( \textit{Third player unitary} \big) \bigg( \big( \theta_1 \big)^{\prime\prime\prime} \otimes  \big( \theta_2 \big)^{\prime \prime\prime} \otimes \cdots \otimes \big( \theta_N \big)^{\prime\prime\prime}  \bigg)_{dg}  \big[ \textit{Third player} \\ \textit{unitary} \big]^{\dagger} \bigg]   \text{, } \\ \vdots \\    \textit{State N}  \equiv \mathrm{Tr} \bigg[ \big( \textit{N th player unitary} \big) \bigg( \big( \theta_1 \big)^{\prime\overset{N-2}{\dots} \prime} \otimes  \big( \theta_2 \big)^{\prime\overset{N-2}{\dots} \prime} \otimes \cdots \otimes \big( \theta_N \big)^{\prime\overset{N-2}{\dots} \prime}  \bigg)_{dg}  \big[ \textit{N th player} \\ \textit{ unitary} \big]^{\dagger} \bigg]  \text{, }
\end{align*}

\noindent for the $\dagger$ of the unitary for each player, $\big[ \textit{Third player unitary} \big]^{\dagger}, \cdots, \big[ \textit{N th player unitary} \big]^{\dagger}$, respectively. Each one of these states corresponding to suitable unitary operations of each player, as defined above, is used in the computation of,

\begin{align*}
  \underset{\textbf{P}_{1_C 2_C 3_C \times \cdots \times N_C D G | \mathcal{E}}}{\textbf{E}}  \underset{1 \leq i \leq N}{\sum} S_{\infty} \bigg(  \bigg(  \textit{(i-1) th player state} \bigg)  \bigg | \bigg|\bigg(  \textit{i th player state} \bigg)   \bigg) \\ \equiv  \underset{1 \leq i \leq N}{\sum}  \underset{\textbf{P}_{1_C 2_C 3_C \times \cdots \times N_C D G | \mathcal{E}}}{\textbf{E}}    S_{\infty} \bigg(  \bigg(  \textit{(i-1) th player state} \bigg)  \bigg | \bigg| \bigg(  \textit{i th player state}  \bigg)  \bigg), 
\end{align*}

\noindent corresponding to the relative-min entrop of consecutive states of each participant.

\bigskip

\noindent Previous results [2, 3, 4, 56] which have leveraged anchored parallel repetition for expanded game have analytically characterized the decay of the maximum winning probability, namely the optimal value. As we will further describe in upcoming sections, how previous characterizations of exact, and approximate, optimality determine the \textit{total} number of optimal strategies, within the space of all possible strategies, continues to remain of great interest to explore. For such purposes, one can make use of previous arguments, and analyses, provided in [2, 3, 4, 30, 56], which draw the attention of the reader to possible research directions from complexity and quantum circuit architecture. Specifically, the following results have been established for the decay of the optimal value in expanded, and closely related, games. Generally speaking, for such estimates of the decay of the optimal value, one looks at functions of the form $f \big( x \big) \equiv \big( 1 - \big( 1 - x \big) \big)^N$, with $N$ being some function of the combinatorial space of possible answers, and hence of strategies, which participants can utilize for playing optimally, while the input $x$ is taken to be the corresponding winning probability. Such expressions for the optimal value include:

\begin{itemize}
\item[$\bullet$] \textit{Optimal values for expanded games under anchoring}, [2, 3, 4]. Fix $0 < \alpha < 1$. Then the optimal value, $\omega^*$, of an expanded game $G$ under parallel repetition,  in comparison to the original optimal value, $\omega$, of $G$ before performing parallel repetition, and subsequently, anchoring, satisfies,

\begin{align*}
  \omega^{*} \big( G_{\bot} \big) = 1 - \big( 1 - \alpha \big)^2 \big[ 1 - \omega^* \big( G \big) \big]     \text{, }
\end{align*}

\noindent for the $\alpha$-anchored game, $G_{\bot}$.

\item[$\bullet$] \textit{Parallel repetition of} $\alpha$-\textit{anchoring}. Fix the same choice of $\alpha$ in the previous result, as well as $\epsilon$ sufficiently small so that $\omega^{*} \big( G_{\bot} \big) < 1 - \epsilon$. After providing the estimate described in the previous item, the authors of [2, 3, 4, 30, 56] also describe how performing the parallel repetition operation of $\alpha$-anchoring impacts the estimate. In particular, parallel repetition of $\alpha$-anchored expanded games yields the following exponentially decaying estimate,

\begin{align*}
 \omega^{*} \big( G^n_{\bot} \big) \leq \frac{4}{\epsilon}         \mathrm{exp} \bigg[  - \frac{c \alpha^{48} \epsilon^{17} n}{s}   \bigg]    \text{, }
\end{align*}

\noindent for $s \equiv \mathrm{sup} \big\{ \mathrm{log} \big( \big| \mathcal{A} \mathcal{B} \big| , 1 \big\}$, ie the supremum of the natural logarithm of the product of the alphabets from Alice and Bob and $1$, as well as strictly positive $c$.

\item[$\bullet$] \textit{Generalizations of the previous item for expressions of optimal values under anchoring}. Following the authors of [2,3,4], the authors of [30] argued that a generalization of the anchored parallel repetition on the optimal value satisfies,

\begin{align*}
  \omega^* \big( {\widetilde{G}}^k \big) = \bigg[ 1 - \big[ 1 - \omega^* \big( \widetilde{G} \big) \big]^5 \bigg]^{\Omega \big( k \big( \mathrm{log} \big( \big| \mathcal{A} \big| \big| \mathcal{B} \big| \big) \big)^{-1} \big)}  \text{, }
\end{align*}

\noindent for the sets, and strictly positive parameter,

\begin{align*}
 \mathcal{A} \equiv \textit{Set of answers from first player}   \text{, } \\ \\ \mathcal{B} \equiv \textit{Set of answers from second player}  \text{, }\\ \\ k \equiv \textit{Number of parallel repetition operations} > 0  \text{, }
\end{align*}

\noindent as well as the function,

\begin{align*}
  \Omega \equiv \underset{a \in \mathcal{A}, b \in \mathcal{B} }{\bigcup} \big\{ \text{functions} \big( a , b \big) \big\}  \equiv \text{functions} \big( \mathcal{A} , \mathcal{B} \big)         \text{, }
\end{align*}

\item[$\bullet$] \textit{Anchoring and parallel repetition of the optimal value}. Simultaneously, one can combine each of the operations described in the previous two items above as follows. Namely, to obtain an optimal value for the maximum probability of satisfying a winning assignment from the referee's predicate under anchoring and parallel repetition, from arguments provided in [2,3,4, 56] to generalize the upper bound for $\omega^{*} \big( G^n_{\bot} \big)$ provided in [2, 3, 4] with the equality provided for $\omega^{*} \big( \widetilde{G}^k \big)$, one analyzes:

\begin{itemize}
    \item[$\bullet$]   \textit{The decay of the winning probability under parallel repetition with one-sided anchoring}. One-sided anchoring, along the lines of a previously mentioned condition,

\begin{align*}
  \textbf{P} \big[ \Omega_{\mathrm{Expanded}} \big| \textbf{X}_i \equiv x, W_C \big] \approx \textbf{P} \big[ \Omega_{\mathrm{Expanded}} \big| \textbf{X}_i \equiv x, \textbf{Y}_i \equiv y,  W_C  \big] \approx \textbf{P} \big[ \Omega_{\mathrm{Expanded}} \\ \big|  \textbf{Y}_i \equiv y,  W_C \big]   \text{, }
\end{align*}

    \noindent stipulates that there exists $y, y^{*}$, for which, $\textbf{P} \big[ y^{*} \big]$ is constant, as well as the conditional probability distribution,

\begin{align*}
    \textbf{P} \big[ y \big| x \big] \text{, }
\end{align*}

    \noindent equaling the probability distribution,

\begin{align*}
    \textbf{P} \big[ y  \big] \text{, }
\end{align*}

    \noindent or alternatively, that there exists $x, x^{*}$, for which, $\textbf{P} \big[ x^{*} \big]$ is constant, as well as the conditional probability distribution,

\begin{align*}
    \textbf{P} \big[ x \big| y \big] \text{, }
\end{align*}

    \noindent equaling the probability distribution,

\begin{align*}
    \textbf{P} \big[ x  \big] \text{. }
\end{align*}

\noindent The requirements for one-sided anchoring imply, for non-local two-player games, that it suffices to consider,

\begin{align*}
  \omega^* \big( G^k \big) = \bigg[ 1 - \big[ 1 - \omega^* \big( \widetilde{G} \big) \big]^5 \bigg]^{\Omega \big( \xi^2 k \big( \mathrm{log} \big( \big| \mathcal{A} \big| \big| \mathcal{B} \big| \big) \big)^{-1} \big)}  \text{, }
\end{align*}

\noindent for the one-sided anchoring probability $\xi$. In a similar work, [56], Yuen demonstrated that, under a different assumption on the decay of the optimal value,

\begin{align*}
    \mathrm{val} \big( G^n_{\mathrm{Entangled} } \big) \leq  \bigg[1 - \big[ 1 - \mathrm{val} \big( G_{\mathrm{Entangled}}\big)  \big]^3 \bigg]^{c_{G_{\mathrm{Entangled}}} n }  
\end{align*}

\noindent for a suitable constant $c$ depending upon the entangled game, corresponding to parallel repetyition in entangled games, one has that,

\begin{align*}
  \mathrm{val}^{*} \big( G_{\mathrm{Entangled}} \big) \leq c_G \frac{  \big|     \mathcal{A}      \big|     \big|           \big|           \mathrm{log} n \big| }{\epsilon^{17} n^{\frac{1}{4}}}    ,
\end{align*}

\noindent under the assumption that $\mathrm{val}^{*} \big( G_{\mathrm{Entangbled}} \big)= 1 - \epsilon$, for $\epsilon$ taken sufficiently small, and $\big| \mathcal{A} \big|  \equiv \underset{1 \leq i \leq \# \text{ of players}}{\sum} \big| \mathcal{A}_i \big|$, the length of the answers provided by all players.

    \item[$\bullet$] Several remarks regarding the previous item are in order. In particular, the rate of decay for the anchored, parallel repetition optimal value is clearly dependent upon:
        \begin{itemize}
        \item[$\bullet$] \underline{(1)}. The number of parallel repetition operations,
        \item[$\bullet$] \underline{(2)}. The size of the cardinality of $\mathcal{A}$, and of $\mathcal{B}$,
        \item[$\bullet$] \underline{(3)}. The ratios,

        \begin{align*}
        k \big[ \mathrm{log} \big[ \big|  \big| \mathcal{A} \big| \big| \mathcal{B} \big| \big| \big] \big]^{-1}    \text{,}
        \end{align*}

        \noindent and,

         \begin{align*}
          \xi^2 k \big[ \mathrm{log} \big[ \big| \big| \mathcal{A} \big| \big| \mathcal{B} \big| \big| \big] \big]^{-1}   \text{,}
        \end{align*}

        \noindent which are each respectively normalized in the summation of the logarithm of the size of each player's alphabet, which, unsurprisingly, from previous arguments of the author for upper bounding the bit transmission rate $r$, [53], are related to the Mutual Information entropy, $I \big( \cdot, \cdot \big)$. Straightforwardly, for the multiplayer settings one would make use of the ratios,
        
\begin{align*}
    k \bigg[ \mathrm{log} \bigg[ \bigg|   \underset{1 \leq i \leq N}{\prod} \big| \mathcal{A}_i  \big|     \bigg|  \bigg]     \bigg]^{-1}   ,
\end{align*}

\noindent and,

\begin{align*}
   \xi^N k \bigg[ \mathrm{log} \bigg[ \bigg|   \underset{1 \leq i \leq N}{\prod} \big| \mathcal{A}_i    \big|   \bigg|  \bigg]     \bigg]^{-1}  .
\end{align*}

        \bigskip
        
        \item[$\bullet$] \underline{(4)}. Parallel repetition of the worst-case scenario game. In comparison to typical arguments in Quantum Game theory which consider the optimal value from the supremum,

        \begin{align*}
         \omega \equiv \underset{\text{Strategies}}{\mathrm{sup}} \big\{ \text{winning probability} \big\}   \text{, }
        \end{align*}

        \noindent the worst-case game instead considers the behavior of the value,

             \begin{align*}
         \omega^{*}_{\mathrm{WC}} \equiv \underset{\text{Strategies}}{\mathrm{inf}} \big\{ \text{winning probability} \big\}   \text{, }
        \end{align*}

\noindent namely, the \textit{lowest} probability of satisfying the winning condition from the referee's predicate, $V$. Previous arguments in [] analyze the equalities provided for $\omega^{*} \big( \widetilde{G}^k \big)$, and for $\omega^{*} \big( G^k \big)$, through,

\begin{align*}
       \omega^{*}_{\mathrm{WC}} \big( G^k_{\mathrm{WC}} \big) \equiv \bigg[ 1 - \big[ 1 - \omega^{*}_{\mathrm{WC}} \big( G^k_{\mathrm{WC}} \big) \big]^7 \bigg]^{\Omega \big( k \big( \mathrm{log} \big( \big| \mathcal{A} \big| \big| \mathcal{B} \big| \big) \big)^{-1} \big)}       \text{. }
\end{align*}

        \item[$\bullet$] \underline{(5)}. Upper bounding $\omega^* \big(  G^k   \big)$. To obtain the desired upper bound on parallel repetition of the anchored optimal value, the authors of [2, 3] argue,

\begin{align*}
  \omega^{*} \big( G^k \big) \leq \big[ 1 - \epsilon \big]^{\frac{\xi^2 \epsilon^4 k }{\mathrm{log} ( | \mathcal{A}| | \mathcal{B} | )}}  \text{, }
\end{align*}

        \noindent for the same choice of $\epsilon$, and number of parallel repetition operations $k$. The estimate above, besides being very technical, is also being dependent upon the one-side anchoring probability, through the parameters,

        \begin{align*}
         \frac{\xi^2 \epsilon^4}{14440000}   \text{. }
        \end{align*}

        \noindent and,

 \begin{align*}
              \frac{\xi^2 \epsilon^4}{14440000 \mathrm{log} \big[ \big| \mathcal{A} \big| \big| \mathcal{B} \big| \big] }    \text{. }
        \end{align*}

     \noindent In generalizations to multiplayer game-theoretic settings, one expects to make use of the thresholds,

     \begin{align*}
       \frac{\xi^N \epsilon^{2N}}{20000000}  , 
     \end{align*}
        
     \noindent corresponding to the product of the one-side anchoring probability, raised to the total number of participants in the game, with the $\epsilon$ parameter raised to twice the number of participants, with a suitable normalization,as well as,

 \begin{align*}
              \frac{\xi^N \epsilon^{2N}}{20000000 \mathrm{log} \bigg[ \bigg| \underset{1 \leq i \leq N}{\prod} \big| \mathcal{A}_i \big| \bigg| \bigg]  }    \text{. }
        \end{align*}

        \noindent corresponding to the previous threshold provided above, additional subject to the normalization of the natural logarithm of the product,

        \begin{align*}
       \bigg|   \underset{1 \leq i \leq N}{\prod} \big| \mathcal{A}_i\big| \bigg|    ,
        \end{align*}

        \noindent of the set of possible responses from each player.

\end{itemize}

\end{itemize}

\noindent With the previous items discussed above, it remains of interest to determine how \textit{generalized} multiplayer game-theoretic settings would behave under anchored parallel repetition. As discussed in the concluding remarks of [4], such a research direction of interest is not only inextricably linked to communication complexity, and Quantum circuit, lower bounds, but also to connections between expanded, and multiplayer, games.

\item[$\bullet$] \textit{Contributions of this work: determining how parallel repetition with respect to anchoring behaves for multiplayer game-theoretic multiplayer settings}. Fix $q_1 \in \mathcal{Q}_1, \cdots, q_n \in \mathcal{Q}_N$. We argue that the desired expression for anchored parallel repetition of the multiplayer optimal value holds. Towards this purpose, denote,

\begin{align*}
  \ket{\Psi_{q_1,q_2, \cdots, q_n}} \equiv   \ket{\Psi_{1,2, \cdots, n}} \equiv \underset{\psi \in \Psi}{\bigcup} \ket{\psi_{q_1,q_2, \cdots, q_n}} \equiv \underset{\psi \in \Psi}{\bigcup} \ket{\psi_{1,2, \cdots, n}}  \text{, }
\end{align*}

\noindent corresponding to the multiplayer dependency-breaking state (by a slight abuse of notation, we take $\Psi$ as the same state in comparison to the space $\Psi$ used for the set of all two-player dependency breaking variables). Conveniently, from the set of all such dependency-breaking variables one has the identities,

\begin{align*}
 \ket{\Psi_{q_1, \cdots, q_N}}  \equiv   \bigg[  \underset{\# \text{ of players}}{\bigotimes}        \text{Tensors of player observables} \big( q_1, \cdots, q_N \big)    \bigg]    \ket{\psi^{\prime}} \text{, } \\ \\  \ket{\Psi_{q_1, \bot, q_2, \cdots, q_N}}  \equiv   \bigg[  \bigg[        \text{Tensors of player observables} \big( q_1, \cdots, q_N \big) \bigg] \otimes \bigg[   \text{Ten-} \\  \text{sors of player observables} \big( q_1, \bot , q_3,  \cdots, q_N \big)     \bigg] \\ \otimes \bigg[  \underset{3 \leq \# \text{ of players} \leq N-1}{\bigotimes}       \text{Tensors of player observables} \big( q_1, q_2 , q_3,  \cdots, q_N \big)     \bigg] \\  \\  \ket{\Psi_{q_1, \cdots, q_{N-1}, \bot}}  \equiv \bigg[ \bigg[  \underset{1 \leq \# \text{ of players} \leq N-1}{\bigotimes}       \text{Tensors of player observables} \big( q_1, \bot , q_3,  \cdots, q_N \big)     \bigg] \end{align*}

 \begin{align*} \otimes \bigg[  \text{Tensors of player observables} \big( q_1, \cdots, \bot  \big)  \bigg]  \bigg]     \ket{\psi^{\prime}} \text{,} \\ \\                   \ket{\Psi_{q_1, \bot, \bot, q_3,  \cdots, q_N}}  \equiv   \bigg[  \bigg[        \text{Tensors of player observables} \big( q_1, \cdots, q_N \big) \bigg] \otimes \bigg[   \text{Ten-}\\   \text{sors of player observables} \big( q_1, \bot , q_3,  \cdots, q_N \big)     \bigg]  \\ \otimes \bigg[        \text{Tensors of player observables} \big( q_1, \cdots, q_N \big) \bigg] \\  \otimes \bigg[   \text{Tensors of player observables} \big( q_1, q_2 , \bot ,  \cdots, q_N \big)     \bigg]   \bigg]  \ket{\psi^{\prime}}  \\  \vdots
\\ 
  \ket{\Psi_{\bot, \cdots \bot, q_N}} \equiv           \bigg[ \bigg[             \text{Tensors of player observables} \big( q , \big( \bot \big)^k \big)        \bigg]  \otimes   \bigg[ \text{Tensors of player observa-} \\ \text{bles} \big( \bot, \bot , \cdots, \bot, q_N \big)  \bigg] \bigg] \ket{\psi^{\prime}}             \text{, }
\end{align*}

\noindent where, in the last identity provided above, for,

\begin{align*}
   \text{Tensors of player observables} \big( q , \big( \bot \big)^k \big)  \equiv \text{Tensors of Player observables} \big( q_1 , \bot , q_3  , \cdots, q_N \big) + \cdots \\ + \text{Tensors of Player observables} \big( q_1 , \bot , \cdots, \bot ,  q_N \big)   \text{, }
\end{align*}

\noindent where,

\begin{align*}
 \text{Tensors of player observables} \big( q , \big( \bot \big)^k \big)   \equiv   \underset{1 \leq k \leq N-1}{\bigcup}  \big\{ \text{Tensors of player observables} \big( q , q^{\prime} , \big( \bot \big)^k  \big) \big) \big\}  \\ \Longleftrightarrow q^{\prime} \neq \bot  \text{, }
\end{align*}

\noindent for some $k> 0 $. Anchoring $q_i$ to $\bot$ individually, given the set of possible responses for each participant, implies that one can introduce,

\begin{align*}
 \big( \mathcal{Q}_1 \big)_{\omega_{-i}, q_1 \backslash \bot } \big( q_1 \big) \equiv  \eta \big( \mathcal{Q}_1 \big)_{\omega_{-i}, q_1} \big( q_1 \big)  + \big( 1 - \eta \big) \big( \mathcal{Q}_1 \big)_{\omega_{-i}, \bot } \big( q_1 \big)  \text{, } \\ \\   \big( \mathcal{Q}_2 \big)_{\omega_{-i}, q_2 \backslash \bot } \big( q_2 \big) \equiv  \eta \big( \mathcal{Q}_2 \big)_{\omega_{-i}, q_2 } \big( q_2 \big)  + \big( 1 - \eta \big) \big( \mathcal{Q}_2 \big)_{\omega_{-i}, \bot } \big( q_2 \big)   \text{, } \\ \vdots \\  \big( \mathcal{Q}_N \big)_{\omega_{-i}, q_N \backslash \bot } \big( q_N \big) \equiv  \eta \big( \mathcal{Q}_N \big)_{\omega_{-i}, q_N } \big( q_N \big)  + \big( 1 - \eta \big) \big( \mathcal{Q}_2 \big)_{\omega_{-i}, \bot } \big( q_N \big)   \text{, } \end{align*}

\noindent for the same choice of $\eta$ introduced for the anchored question in the probability distributions $\textbf{P}_{M | D = \mathcal{A}_1} \big( q_1 \big), \cdots, \textbf{P}_{M | D = \mathcal{A}_N} \big( q_N \big) $.

\noindent The forthcoming approach consists of the following steps:

\begin{itemize}
\item[$\bullet$] \textit{Computing the L-2 norm of the set of multiplayer dependency-breaking variables with respect to the intertwining of suitable unitary operations}. Define the collection of unitaries,

\[   \left\{\!\begin{array}{ll@{}>{{}}l} 
 \mathcal{U}_1 \equiv \text{\textit{First player unitary}}   \text{, } \\  \mathcal{U}_2 \equiv \text{\textit{Second player unitary}}  \text{, } \\ \vdots \\ \mathcal{U}_N \equiv \text{\textit{N th player unitary}}  \text{, } 
\end{array}\right. 
\]

\noindent for which the first component of the argument for demonstrating that anchored, parallel repetition of the multiplayer optimal value implies,

\[  \mathscr{E} \equiv  \left\{\!\begin{array}{ll@{}>{{}}l} 
   \underset{R_{-i} | W_C}{\textbf{E}} \underset{\mathcal{Q}_1}{\textbf{E}}       \big| \big|  \big[ \big( \mathcal{U}_1 \bigotimes \textbf{I}^{\otimes N-1} \big) - \textbf{I} \big] \ket{\psi}    \big| \big|\text{, } \\    \underset{R_{-i} | W_C}{\textbf{E}} \underset{\mathcal{Q}_2}{\textbf{E}}      \big| \big|  \big[ \big( \textbf{I} \bigotimes \mathcal{U}_2 \bigotimes \textbf{I}^{\otimes N-2} \big) - \textbf{I} \big] \ket{\psi}    \big| \big|\text{, } \\ \vdots \\    \underset{R_{-i} | W_C}{\textbf{E}} \underset{\mathcal{Q}_N}{\textbf{E}}     \big| \big|  \big[ \big( \textbf{I}^{\otimes N-1 } \bigotimes \mathcal{U}_N  \big) - \textbf{I} \big] \ket{\psi}   \big| \big|  \text{, } \\ \underset{R_{-i} | W_C}{\textbf{E}} \underset{\mathcal{Q}_1 \mathcal{Q}_2 }{\textbf{E}}     \big| \big|   \big[ \big( \textbf{I}^{\otimes N-1 } \bigotimes \mathcal{U}_N  \big) - \textbf{I} \big] \ket{\psi}    \big| \big| \text{, } \\ \vdots \\ \underset{R_{-i} | W_C}{\textbf{E}} \underset{\mathcal{Q}_1 \times \cdots \times \mathcal{Q}_N}{\textbf{E}}    \big| \big|  \big[ \big( \textbf{I}^{\otimes N-1 } \bigotimes \mathcal{U}_N  \big) - \textbf{I} \big] \ket{\psi}   \big| \big|  ,
\end{array}\right. 
\]

\noindent for unitarites $\mathcal{U}_1, \cdots, \mathcal{U}_N$, and multiplayer dependency-breaking variables $\ket{\psi} \equiv \ket{\psi_{1,\cdots, N}} \sim \ket{\Psi_{1,\cdots, N}}$. The set of all such variables, $\Psi$, can be decomposed as,

\begin{align*}
  \ket{\Psi} \equiv \underset{1 \leq k \leq N}{\underset{\textit{dependency-breaking among the first}\text{ }  k\text{ }  \textit{players}}{\bigcup}} \ket{\Psi_k}     \text{, }
\end{align*}

\noindent which implies that dependency-breaking states can have the following action,

\[   \left\{\!\begin{array}{ll@{}>{{}}l}  \textit{Dependency-breaking amongst the first player} \equiv \ket{\Psi_1} , \\ \vdots \\  \textit{Dependency-breaking amongst all players} \equiv \ket{\Psi_N} , 
\end{array}\right. 
\]

\noindent Equipped with the same unitaries $\mathcal{U}$ introduced above, one can also provide upper bounds for each expectation in the system,

\[  \widetilde{\mathscr{E}} \equiv  \left\{\!\begin{array}{ll@{}>{{}}l} 
   \underset{R_{-i} | W_C}{\textbf{E}} \underset{\mathcal{Q}_1}{\textbf{E}}       \big| \big|  \big[ \big( \mathcal{U}_1 \bigotimes \textbf{I}^{\otimes N-1} \big) - \textbf{I} \big] \widetilde{\ket{\psi} }   \big| \big|\text{, } \\    \underset{R_{-i} | W_C}{\textbf{E}} \underset{\mathcal{Q}_2}{\textbf{E}}      \big| \big|  \big[ \big( \textbf{I} \bigotimes \mathcal{U}_2 \bigotimes \textbf{I}^{\otimes N-2} \big) - \textbf{I} \big] \widetilde{\ket{\psi} }     \big| \big|\text{, } \\ \vdots \\    \underset{R_{-i} | W_C}{\textbf{E}} \underset{\mathcal{Q}_N}{\textbf{E}}     \big| \big|  \big[ \big( \textbf{I}^{\otimes N-1 } \bigotimes \mathcal{U}_N  \big) - \textbf{I} \big] \widetilde{\ket{\psi} }    \big| \big|  \text{, } \\ \underset{R_{-i} | W_C}{\textbf{E}} \underset{\mathcal{Q}_1 \mathcal{Q}_2 }{\textbf{E}}     \big| \big|  \big[  \big( \textbf{I}^{\otimes N-1 } \bigotimes \mathcal{U}_N  \big) - \textbf{I} \big] \widetilde{\ket{\psi} }   \big| \big| \text{, } \\ \vdots \\ \underset{R_{-i} | W_C}{\textbf{E}} \underset{\mathcal{Q}_1 \times \cdots \times \mathcal{Q}_N}{\textbf{E}}    \big| \big|  \big[  \big( \textbf{I}^{\otimes N-1 } \bigotimes \mathcal{U}_N  \big) - \textbf{I} \big] \widetilde{\ket{\psi} }  \big| \big|  ,
\end{array}\right. 
\]

\noindent for normalized strategies,

\begin{align*}
  \widetilde{\ket{\psi} }  \equiv \frac{\ket{\psi}}{\big| \big| \ket{\psi}\big| \big|} \propto \ket{\psi}  \text{. }
\end{align*}

\noindent Related to the system of expectation values provided in $\mathscr{E}$, and also in $\widetilde{\mathscr{E}}$, one can also study the following two system of expectation values. The first system, in comparison to the tensor product of operators with the respective unitary of each player subtracted from $\textbf{!}$, is instead replaced with a suitable normalization factor. This normalization factor, as described further later in this subsection, can take a variety of forms, which is not only dependent upon how many answers in the multiplayer game are anchored, but also upon the support of each expectation. With regards to the second system of expectation values below, straightforwardly one can formulate an accompanying system of expectation values with $\ket{\psi}$, instead of with $\widetilde{\ket{\psi}}$. We state the two systems below, the first of which is,

\[  \widetilde{\mathscr{E}} \equiv  \left\{\!\begin{array}{ll@{}>{{}}l} 
   \underset{R_{-i} | W_C}{\textbf{E}} \underset{\mathcal{Q}_1}{\textbf{E}}       \big| \big|  \big[ \big( \mathcal{U}_1 \bigotimes \textbf{I}^{\otimes N-1} \big) - \textit{Normalization} \big] \widetilde{\ket{\psi} }   \big| \big|\text{, } \\    \underset{R_{-i} | W_C}{\textbf{E}} \underset{\mathcal{Q}_2}{\textbf{E}}      \big| \big|  \big[ \big( \textbf{I} \bigotimes \mathcal{U}_2 \bigotimes \textbf{I}^{\otimes N-2} \big) - \textit{Normalization} \big] \widetilde{\ket{\psi} }     \big| \big|\text{, } \\ \vdots \\    \underset{R_{-i} | W_C}{\textbf{E}} \underset{\mathcal{Q}_N}{\textbf{E}}     \big| \big|  \big[ \big( \textbf{I}^{\otimes N-1 } \bigotimes \mathcal{U}_N  \big) - \textit{Normalization}  \big] \widetilde{\ket{\psi} }    \big| \big|  \text{, } \\ \underset{R_{-i} | W_C}{\textbf{E}} \underset{\mathcal{Q}_1 \mathcal{Q}_2 }{\textbf{E}}     \big| \big|  \big[  \big( \textbf{I}^{\otimes N-1 } \bigotimes \mathcal{U}_N  \big) - \textit{Normalization} \big] \widetilde{\ket{\psi} }   \big| \big| \text{, } \\ \vdots \\ \underset{R_{-i} | W_C}{\textbf{E}} \underset{\mathcal{Q}_1 \times \cdots \times \mathcal{Q}_N}{\textbf{E}}    \big| \big|  \big[  \big( \textbf{I}^{\otimes N-1 } \bigotimes \mathcal{U}_N  \big) - \textit{Normalization} \big] \widetilde{\ket{\psi} }  \big| \big|  ,
\end{array}\right. 
\] 

\noindent and the second of which is,

\[  \mathscr{E} \equiv  \left\{\!\begin{array}{ll@{}>{{}}l} 
   \underset{R_{-i} | W_C}{\textbf{E}} \underset{\mathcal{Q}_1}{\textbf{E}}       \big| \big|  \big[ \big( \mathcal{U}_1 \bigotimes \textbf{I}^{\otimes N-1} \big) - \textit{Normalization} \big] \ket{\psi}    \big| \big|\text{, } \\    \underset{R_{-i} | W_C}{\textbf{E}} \underset{\mathcal{Q}_2}{\textbf{E}}      \big| \big| \big( \textbf{I} \bigotimes \mathcal{U}_2 \bigotimes \textbf{I}^{\otimes N-2} \big) - \textit{Normalization} \big] \ket{\psi}    \big| \big|\text{, } \\ \vdots \\    \underset{R_{-i} | W_C}{\textbf{E}} \underset{\mathcal{Q}_N}{\textbf{E}}     \big| \big|  \big[ \big( \textbf{I}^{\otimes N-1 } \bigotimes \mathcal{U}_N  \big) - \textit{Normalization} \big] \ket{\psi}   \big| \big|  \text{, } \\ \underset{R_{-i} | W_C}{\textbf{E}} \underset{\mathcal{Q}_1 \mathcal{Q}_2 }{\textbf{E}}     \big| \big|   \big[ \big( \textbf{I}^{\otimes N-1 } \bigotimes \mathcal{U}_N  \big) - \textit{Normalization} \big] \ket{\psi}    \big| \big| \text{, } \\ \vdots \\ \underset{R_{-i} | W_C}{\textbf{E}} \underset{\mathcal{Q}_1 \times \cdots \times \mathcal{Q}_N}{\textbf{E}}    \big| \big|  \big[ \big( \textbf{I}^{\otimes N-1 } \bigotimes \mathcal{U}_N  \big) - \textit{Normalization}\big] \ket{\psi}   \big| \big|  .
\end{array}\right. 
\]

\noindent The existence of such a family of inequalities with respect to suitable unitaries, $\mathcal{U}$, is similar to upper bounds, with respect to the Frobenius norm, which have been previously obtained by the author in [52] for various multiplayer settings. Besides this result, the dependency-breaking variables, and corresponding dependency-breaking \textit{strategies}, can be expressed as,

\begin{align*}
  \ket{\Psi_{r_{-i}, q_1, q_2, \cdots, q_n}} \equiv    \bigg[    \big( \mathcal{Q}_1 \big)_{r_{-i} , q_1} \otimes \big(  \mathcal{Q}_2 \big)_{r_{-i} , q_2}  \otimes \cdots \otimes  \big( \mathcal{Q}_N \big)_{r_{-i} , q_N}     \bigg]  \ket{\psi} \text{, }
\end{align*}

\noindent namely, that the dependency breaking state which is determined by each instance of $r_{-i}$ can be factorised, with respect to the tensor product, amongst the possible responses from each player.

\bigskip

\noindent The first result claims:

\bigskip

\noindent \textbf{Lemma} \textit{1 -N $\mathrm{XOR}$} (\textit{computation of the Frobenius norm for the anticommutation rule of $T_{N\mathrm{XOR}}$ yields a desired up to constants $\sqrt{\epsilon}$ upper bound}). One has that,

\begin{align*}
      \underline{\text{Player $1$}:} \text{ } \bigg| \bigg|   \bigg(  A_i \bigotimes \bigg( \underset{1 \leq k \leq n-1}{\bigotimes} \textbf{I}_k  \bigg)        \bigg) \mathscr{T}  -   \mathscr{T} \bigg( \bigg( \underset{1 \leq k \leq n-1}{\bigotimes} \textbf{I}_k  \bigg)      \bigotimes \widetilde{A_i}      \bigg)      \bigg| \bigg|_F \\ < c_1 n^N \sqrt{\epsilon}  \text{,} \\     \vdots \\              \underline{\text{Player $N$}:} \text{ } \bigg| \bigg|   \bigg(  \bigg( \underset{1 \leq k \leq n-1}{\bigotimes} \textbf{I}_k  \bigg)   \bigotimes  A^{(n-1)}_{i_1, \cdots, i_{n-1}}     \bigg) \mathscr{T}  -   \mathscr{T} \bigg( \bigg(    \widetilde{A^{(n-1)}_{i_1, \cdots, i_{n-1}} }       \bigotimes \\  \bigg( \underset{1 \leq k \leq n-1}{\bigotimes} \textbf{I}_k  \bigg)     \bigg)      \bigg| \bigg|_F      < c_N n^N \sqrt{\epsilon}  \text{, }
\end{align*}

\noindent has the upper bound,

\begin{align*}
   \mathscr{C}_{N\mathrm{XOR}} \equiv    \mathscr{C} \equiv    \underset{1 \leq i \leq N}{\bigcup}  \big\{     C_i  \neq c_i\in \textbf{R}  :            C_i \equiv c_i \sqrt{\epsilon}     \big\}     \propto n^N \sqrt{\epsilon} \text{. }
\end{align*}

\noindent To prove the above result, one makes use of the identification,

\begin{align*}
 \bigg| \bigg|    \bigg[             \bigg( \bigg( \underset{1 \leq i \leq n}{\prod}  A^{j_i}_i \bigg) \bigotimes \bigg( \underset{1 \leq k \leq n-1}{\bigotimes} \textbf{I}_k \bigg) \bigg)   -  \bigg(  \omega_{N\mathrm{XOR}} \bigg(         \pm \mathrm{sign} \big( i_1, \cdots, i_n \big)            \\ \times     \bigg( \underset{1 \leq i \leq n}{\prod}  A^{j_i}_i \bigg)           \bigg)    \bigotimes \bigg( \underset{1 \leq k \leq n-1}{\bigotimes} \textbf{I}_k \bigg)\bigg)                       \bigg]   \ket{\psi_{N\mathrm{XOR}}}        \bigg| \bigg|_F  \\  <    \bigg( n_1 +       \big(    n_1 + 2    \big) \omega_{N\mathrm{XOR}}^{-1}                \bigg) n^N \sqrt{\epsilon}        \text{,}         \\ \\  \bigg| \bigg|    \bigg[ \bigg(  \textbf{I} \bigotimes            \bigg( \underset{1 \leq i \leq n}{\prod}  A^{1,j_i}_i \bigg) \bigotimes \bigg( \underset{1 \leq k \leq n-2}{\bigotimes} \textbf{I}_k \bigg) \bigg)   -  \bigg(  \textbf{I} \bigotimes  \bigg( \omega_{N\mathrm{XOR}} \\  \times \bigg(         \pm \mathrm{sign} \big( i_1, j_1, \cdots, i_n, \cdots, j_n \big)        \bigg( \underset{1 \leq i_2 \leq m}{\underset{1 \leq i_1 \leq n}{\prod}}  A^{1,j_{i_1,i_2}}_{i_1,i_2} \bigg)           \bigg)  \bigg) \\    \bigotimes \bigg( \underset{1 \leq k \leq n-2}{\bigotimes} \textbf{I}_k \bigg)\bigg)                       \bigg]  \ket{\psi_{N\mathrm{XOR}}}       \bigg| \bigg|_F   <   \bigg( n_2 +  \big( n_2 + 2 \big) \omega_{N\mathrm{XOR}}^{-1}            \bigg) n^N \sqrt{\epsilon} \\  \vdots \end{align*}

 \begin{align*} \bigg| \bigg|    \bigg[             \bigg(\bigg( \underset{1 \leq k \leq n-1}{\bigotimes} \textbf{I}_k \bigg) \bigotimes   \bigg( \underset{1 \leq i \leq n}{\prod}  A^{(n-1),j_{i_1,\cdots,i_n}}_{i_1,\cdots, i_n} \bigg)  \bigg)   -  \bigg( \bigg( \underset{1 \leq k \leq n-1}{\bigotimes} \textbf{I}_k \bigg) \\ \bigotimes \bigg(    \omega_{N\mathrm{XOR}}  \times  \bigg(         \pm \mathrm{sign} \big( i_1, \cdots, i_n, j_1  , \cdots, j_n \big)                         \bigg)   \bigg( \underset{1 \leq i \leq n}{\prod}  A^{(n-1),j_{i_1,\cdots,i_n}}_{i_1,\cdots, i_n} \bigg)  \bigg)                 \bigg)   \bigg] \\ \times \ket{\psi_{N\mathrm{XOR}}}    \bigg| \bigg|_F     <   \bigg( n_N        +   \big(  n_N + 2\big)     \omega_{N\mathrm{XOR}}^{-1}            \bigg) n^N \sqrt{\epsilon}  \text{, }
\end{align*}

\noindent for the quantum optimal strategy,

\begin{align*}
       \ket{\psi_{N\mathrm{XOR}}} \equiv 
 \ket{\psi_{N\mathrm{XOR}} \big( \mathcal{S} \big) } \equiv \underset{\mathcal{S}}{\mathrm{sup}}  \big\{ \text{payoff for all players with some quantum strategy } \mathcal{S}             \big\}                  \text{,}
\end{align*}

\noindent and suitable $n_i$, each of which are taken to be strictly positive.

\bigskip

\noindent The second result states:

\bigskip

\noindent \textbf{Lemma} \textit{1-$\mathrm{XOR}$ strong parallel repetition } (\textit{computation of the Frobenius norm for the anticommutation rule of $T_{\mathrm{XOR} \wedge \cdots \wedge \mathrm{XOR}}$ yields a desired up to constants $\sqrt{\epsilon^{\wedge}}$ upper bound}, \textbf{Theorem} \textit{2}, {[50]}). Fix constants as specified in \textbf{Lemma} \textit{1}. One has that,

\begin{align*}
      \underline{\text{Player $1$}:} \text{ } \bigg| \bigg|   \bigg(  A_i \bigotimes \bigg( \underset{1 \leq k \leq n-1}{\bigotimes} \textbf{I}_k  \bigg)        \bigg) \bigg(  \mathscr{T} \wedge \cdots \wedge \mathscr{T} \bigg)  -  \bigg(  \mathscr{T} \wedge \cdots \wedge \mathscr{T} \bigg) \\ \times \bigg( \bigg( \underset{1 \leq k \leq n-1}{\bigotimes} \textbf{I}_k  \bigg)      \bigotimes \widetilde{A_i}      \bigg)      \bigg| \bigg|_F  < \big( c_1 \big)^{\wedge}  \big( n^N \big)^{\wedge}   \sqrt{\epsilon^{\wedge}}  \text{,} \\     \vdots \\              \underline{\text{Player $N$}:} \text{ } \bigg| \bigg|   \bigg(  \bigg( \underset{1 \leq k \leq n-1}{\bigotimes} \textbf{I}_k  \bigg)   \bigotimes  A^{(n-1)}_{i_1, \cdots, i_{n-1}}     \bigg) \bigg(  \mathscr{T} \wedge \cdots \wedge \mathscr{T} \bigg)  -  \bigg(  \mathscr{T} \wedge \cdots \wedge \mathscr{T} \bigg) \\ \times \bigg( \bigg(    \widetilde{A^{(n-1)}_{i_1, \cdots, i_{n-1}} }       \bigotimes   \bigg( \underset{1 \leq k \leq n-1}{\bigotimes} \textbf{I}_k  \bigg)     \bigg)      \bigg| \bigg|_F     < \big(  c_N \big)^{\wedge} \big(  n^N \big)^{\wedge} \sqrt{\epsilon^{\wedge}}   \text{, }
\end{align*}

\noindent has the upper bound,

\begin{align*}
    \mathscr{C}_{\wedge} \equiv   \mathscr{C} \equiv    \underset{1 \leq i \leq N}{\bigcup}  \big\{    \big(  C_i \big)^{\wedge}  \neq  \big( c_i \big)^{\wedge} \in \textbf{R}  :           \big(  C_i \big)^{\wedge} \equiv  \big( c_i  \big)^{\wedge} \sqrt{\epsilon^{\wedge}}     \big\}     \propto \big( n^N \big)^{\wedge} \sqrt{\epsilon^{\wedge}}\\ \equiv \big( n^{\wedge} \big)^N \sqrt{\epsilon^{\wedge}} \text{. }
\end{align*}

\noindent For anchored parallel repetition in the multiplayer setting, one has that the optimal value can be expressed with,

\begin{align*}
 \underline{\text{Anchored 3-XOR value}} \equiv   \omega^{\bot}_{3\mathrm{XOR}} \big( G \big) \equiv  \omega^{\bot} \big( 3 \mathrm{XOR} \big) \propto \frac{1}{{n \choose 3}}   \bigg\{ \underset{\ket{\psi^{\bot}_{3\mathrm{XOR}}}}{\mathrm{sup}}  \bra{\psi^{\bot}_{3\mathrm{XOR}}}     \mathcal{P}^{\bot}_{3\mathrm{XOR}}   \\ \times  \ket{\psi^{\bot}_{3\mathrm{XOR}}}           \bigg\}     \text{, }    \\  \\ \underline{\text{Anchored 3-XOR bias}} \equiv  \beta^{\bot}_{3\mathrm{XOR}} \big( G \big) \equiv \beta^{\bot} \big( 3\mathrm{XOR} \big) \propto \frac{1}{{n \choose 3}}   \bigg\{  \underset{\ket{\psi^{\bot}_{3\mathrm{XOR}}}}{\mathrm{sup}}  G_{3\mathrm{XOR}} \bra{\psi^{\bot}_{3\mathrm{XOR}}}        \mathcal{P}^{\bot}_{3\mathrm{XOR}}  \\ \times            \ket{\psi^{\bot}_{3\mathrm{XOR}}}  \bigg\}             \text{, } \\ \\ \mathcal{P}^{\bot}_{3\mathrm{XOR}}     \equiv \mathscr{P}^{\bot}_{3\mathrm{XOR}} \big( - 1 \big)^{\textbf{I}_{\{\text{Win 3-XOR game} \} }}   \\   +   \mathscr{P}^{\bot}_{3\mathrm{XOR}} \big( - 1 \big)^{\textbf{I}_{\{\text{Lose 3-XOR game} \} }}      \text{, }  \end{align*}
 
 \noindent corresponding to the value, and bias, of the $3$-$\mathrm{XOR}$ game,
 
 \begin{align*} \underline{\text{Anchored 4-XOR value}}  \equiv  \omega^{\bot}_{4\mathrm{XOR}} \big( G \big) \equiv  \omega^{\bot} \big( 4 \mathrm{XOR} \big)  \propto   \frac{1}{{n \choose 4}}  \bigg\{ \underset{\ket{\psi^{\bot}_{4\mathrm{XOR}}}}{\mathrm{sup}} \bra{\psi^{\bot}_{4\mathrm{XOR}}}   \mathcal{P}^{\bot}_{4\mathrm{XOR}} \\ \times \ket{\psi^{\bot}_{4\mathrm{XOR}}} \bigg\}   \text{, } \\ \\  \underline{\text{Anchored 4-XOR bias}} \equiv  \beta^{\bot}_{4\mathrm{XOR}} \big( G \big) \equiv \beta^{\bot} \big( 4\mathrm{XOR} \big) \propto   \frac{1}{{n \choose 4}}  \bigg\{  \underset{\ket{\psi^{\bot}_{4\mathrm{XOR}}}}{\mathrm{sup}} G_{4\mathrm{XOR}} \bra{\psi^{\bot}_{4\mathrm{XOR}}} \mathcal{P}^{\bot}_{4\mathrm{XOR}}    \\ \times  \ket{\psi^{\bot}_{4\mathrm{XOR}}}   \bigg\}   \text{, } \\ \\ \mathcal{P}^{\bot}_{4\mathrm{XOR}} \equiv \mathscr{P}^{\bot}_{4\mathrm{XOR}} \big( - 1 \big)^{\textbf{I}_{\{\text{Win 4-XOR game} \} }}  \\   +   \mathscr{P}^{\bot}_{4\mathrm{XOR}} \big( - 1 \big)^{\textbf{I}_{\{\text{Lose 4-XOR game} \} }} \text{, } \end{align*}

 \noindent corresponding to the value, and bias, of the $4$-$\mathrm{XOR}$ game,

 \begin{align*} \underline{\text{Anchored 5-XOR value}}  \equiv  \omega^{\bot}_{5\mathrm{XOR}} \big( G \big) \equiv  \omega^{\bot} \big( 5 \mathrm{XOR} \big)  \propto  \frac{1}{{n \choose 5}}  \bigg\{ \underset{\ket{\psi^{\bot}_{5\mathrm{XOR}}}}{\mathrm{sup}}   \bra{\psi^{\bot}_{5\mathrm{XOR}}}    \mathcal{P}^{\bot}_{5\mathrm{XOR}}   \\ \times \ket{\psi^{\bot}_{5\mathrm{XOR}}} \bigg\}   \text{, } \\ \\  \underline{\text{Anchored 5-XOR bias}} \equiv \beta^{\bot}_{5\mathrm{XOR}} \big( G \big) \equiv \beta^{\bot} \big( 5\mathrm{XOR} \big) \propto  \frac{1}{{n \choose 5}}  \bigg\{ \underset{\ket{\psi^{\bot}_{5\mathrm{XOR}}}}{\mathrm{sup}}  G_{5\mathrm{XOR}} \bra{\psi^{\bot}_{5\mathrm{XOR}}}  \mathcal{P}^{\bot}_{5\mathrm{XOR}}  \\ \times    \ket{\psi^{\bot}_{5\mathrm{XOR}}}     \bigg\}    \text{, }\\    \\ \mathcal{P}_{5\mathrm{XOR}}^{\bot} \equiv \mathscr{P}^{\bot}_{5\mathrm{XOR}} \big( - 1 \big)^{\textbf{I}_{\{\text{Win 5-XOR game} \} }}  \\   +   \mathscr{P}^{\bot}_{5\mathrm{XOR}} \big( - 1 \big)^{\textbf{I}_{\{\text{Lose 5-XOR game} \} }}   ,        \end{align*}

  \noindent corresponding to the value, and bias, of the $5$-$\mathrm{XOR}$ game,

 \begin{align*} \underline{\text{N-XOR value}}  \equiv   \omega^{\bot}_{N\mathrm{XOR}} \big( G \big) \equiv  \omega^{\bot} \big( N \mathrm{XOR} \big)   \propto  \frac{1}{{n \choose N}} \bigg\{ \underset{\ket{\psi^{\bot}_{N\mathrm{XOR}}}}{\mathrm{sup}}  \bra{\psi^{\bot}_{N\mathrm{XOR}}}      \mathcal{P}^{\bot}_{N \mathrm{XOR}} \\ \times \ket{\psi^{\bot}_{N\mathrm{XOR}}}      \bigg\}    \text{, } \\ \end{align*}
 
 \begin{align*} \underline{\text{N-XOR bias}} \equiv  \beta^{\bot}_{N\mathrm{XOR}} \big( G \big) \equiv \beta^{\bot} \big( N\mathrm{XOR} \big)   \propto  \frac{1}{{n \choose N}} \bigg\{ \underset{\ket{\psi^{\bot}_{N\mathrm{XOR}}}}{\mathrm{sup}}  G_{N\mathrm{XOR}}\bra{\psi^{\bot}_{N\mathrm{XOR}}}            \mathcal{P}^{\bot}_{N \mathrm{XOR}} \\  \times \ket{\psi^{\bot}_{N\mathrm{XOR}}}       \bigg\}   \text{, }   \\ \mathcal{P}^{\bot}_{N \mathrm{XOR}} \equiv \mathscr{P}^{\bot}_{N\mathrm{XOR}}\big( - 1 \big)^{\textbf{I}_{\{\text{Win N-XOR game} \} }}   \\  +   \mathscr{P}^{\bot}_{N\mathrm{XOR}} \big( - 1 \big)^{\textbf{I}_{\{\text{Lose N-XOR game} \} }}  
\end{align*}

 \noindent corresponding to the value, and bias, of the $N$-$\mathrm{XOR}$ game. In each expression provided above for multiplayer optimal values, and biases, \textit{anchoring} of the intermediate terms between each bra and ket of the optimal strategy for winning, or losing, entails,

 \begin{align*}
     \mathcal{P}^{\bot}_{\mathrm{Game}} \equiv \mathscr{P}^{\bot}_{\mathrm{Game}}\big( - 1 \big)^{\textbf{I}_{\{\text{Win game} \} }}    +   \mathscr{P}^{\bot}_{\mathrm{Game}} \big( - 1 \big)^{\textbf{I}_{\{\text{Lose game} \} }}         \text{. }
 \end{align*}

 \noindent The fact that one can straightforwardly introduce constrained optimization problems for anchored, multiplayer XOR games, implies that one can also introduce, for each instance of $R_{-i}$,

\begin{align*}
      \bra{\psi^{\bot}_{3\mathrm{XOR}}}   \bigg[   \underset{\# \text{ players}}{\bigotimes} \text{Tensors of player observables} \big( a_{1,C}, a_{2,C} , a_{3,C} \big)        \bigg] \ket{\psi^{\bot}_{3\mathrm{XOR}}}   \text{, } \\ \\   \bra{\psi^{\bot}_{4\mathrm{XOR}}}   \bigg[   \underset{\# \text{ players}}{\bigotimes} \text{Tensors of player observables} \big( a_{1,C}, a_{2,C} , a_{3,C} , a_{4,C} \big)        \bigg] \ket{\psi^{\bot}_{4\mathrm{XOR}}}  \text{, } \\ \\ \bra{\psi^{\bot}_{5\mathrm{XOR}}}   \bigg[   \underset{\# \text{ players}}{\bigotimes} \text{Tensors of player observables} \big( a_{1,C}, a_{2,C} , a_{3,C} , a_{4,C} , a_{5,C}   \big)        \bigg] \ket{\psi^{\bot}_{5\mathrm{XOR}}}   \text{, } \\  \vdots \\ \bra{\psi^{\bot}_{N\mathrm{XOR}}}   \bigg[   \underset{\# \text{ players}}{\bigotimes} \text{Tensors of player observables} \big( a_{1,C}, a_{2,C} ,   \cdots , a_{N,C}  \big)        \bigg] \ket{\psi^{\bot}_{N\mathrm{XOR}}}          \text{, } \\  \bigg[ \bra{\psi^{\bot}_{2\mathrm{XOR}}} \wedge \bra{\psi^{\bot}_{2\mathrm{XOR}}} \bigg]   \bigg[   \underset{\# \text{ players}}{\bigotimes} \text{Tensors of player observables} \big( a_{1,C}, a_{2,C} \big)  \\  \wedge  \text{Tensors of player observables} \big( a_{1,C}, a_{2,C} \big)         \bigg]  \bigg[ \ket{\psi^{\bot}_{2\mathrm{XOR}}}  \wedge \ket{\psi^{\bot}_{2\mathrm{XOR}}}  \bigg]   \text{,} \\ \\     \bigg[ \bra{\psi^{\bot}_{3\mathrm{XOR}}} \wedge  \bra{\psi^{\bot}_{3\mathrm{XOR}}} \wedge \bra{\psi^{\bot}_{3\mathrm{XOR}}} \bigg]   \bigg[   \underset{\# \text{ players}}{\bigotimes} \text{Tensors of player observables} \big( a_{1,C} \end{align*}

      \begin{align*}  , a_{2,C} , a_{3,C } \big)   \wedge  \text{Tensors of player observables} \big( a_{1,C}, a_{2,C} , a_{3,C } \big) \\ \wedge  \text{Tensors of player observables} \big( a_{1,C}, a_{2,C} , a_{3,C } \big)         \bigg]  \bigg[ \ket{\psi^{\bot}_{N\mathrm{XOR}}}  \wedge  \ket{\psi^{\bot}_{N\mathrm{XOR}}}  \wedge \ket{\psi^{\bot}_{N\mathrm{XOR}}}  \bigg]  \text{,} \\ \\  \bigg[ \bra{\psi^{\bot}_{N\mathrm{XOR}}} \wedge \cdots \wedge \bra{\psi^{\bot}_{N\mathrm{XOR}}} \bigg]   \bigg[   \underset{\# \text{ players}}{\bigotimes} \text{Tensors of player observables} \big( a_{1,C} , a_{2,C} \cdots , a_{N,C} \big)  \\ \wedge \cdots \wedge \text{Tensors of player observables} \big( a_{1,C} , a_{2,C} \cdots , a_{N,C} \big)         \bigg]  \bigg[ \ket{\psi^{\bot}_{N\mathrm{XOR}}}  \wedge \cdots \wedge \ket{\psi^{\bot}_{N\mathrm{XOR}}}  \bigg]   \text{,} 
\end{align*}

\noindent for the collection of answers, 

\[  a_C \equiv \underset{1 \leq i \leq N}{\bigcup} a_{i,C} \equiv  \left\{\!\begin{array}{ll@{}>{{}}l} 
a_{1,C} , \\ a_{2,C} , \\ \vdots \\ a_{N,C},
\end{array}\right. 
\]

\noindent for which the game is won.

\item[$\bullet$] \textit{Imposing conditioning with respect to the action of multiplayer dependency-breaking variables}. For the properties,

\begin{itemize}
    \item[$\bullet$] \underline{(1)}. \textit{Sampleability}. There exists a state $\ket{\Theta}$, independent of $q_1, \cdots, q_N$, and $\ket{\theta_{q_1,\cdots, q_N}}$, dependent on $q_1, \cdots, q_N$,  so that,

    \begin{align*}
      \big| \big|   \big(  \mathcal{U}_1 \otimes \mathcal{U}_2 \otimes \cdots \otimes \mathcal{U}_N     \big) \ket{\Theta}     - \ket{\theta_{q_1,\cdots, q_N}}  \big| \big|_1 \approx 0  \text{. }
    \end{align*}

     \item[$\bullet$] \underline{(2)}. \textit{Usefulness}. Each participant in the multiplayer game can perform a measurement, respectively denoted with $M_1 , \cdots, M_N \equiv \mathcal{M}_1, \cdots, \mathcal{M}_N$, so that,

     \begin{align*}
          \big| \big|    \mathcal{M}_1 \ket{\theta_{q_1,\cdots, q_N}}    -   \textbf{P}_{Q_1\times \cdots \times Q_N | Q_1 \equiv q_1, \cdots, Q_N \equiv q_N, W_C}    \big| \big| \approx 0  ,  \\  \vdots \\     \big| \big|    \mathcal{M}_N \ket{\theta_{q_1,\cdots, q_N}}     -   \textbf{P}_{Q_1\times \cdots \times Q_N | Q_1 \equiv q_1, \cdots, Q_N \equiv q_N, W_C}       \big| \big| \approx 0        \text{. }
     \end{align*}

\end{itemize}

\noindent of multiplayer dependency-breaking variables, one has the following result, which captures a generalization of the system of expectations formulated in $\mathscr{E}$ above:

\bigskip

\textbf{Lemma} \textit{1} (\textit{computing expectation values with respect to suitable unitaries, as provided in the system of relations} $\mathscr{E}$). For all $\mathscr{r}_{-i}$, $q_1, \cdots, q_N$, there exists unitaries  $\mathcal{U}_1, \cdots, \mathcal{U}_N \equiv \mathcal{U}_{1,\mathscr{r}_{-i}}, \cdots, \mathcal{U}_{N,\mathscr{r}_{-i}}$, such that,

\[   \left\{\!\begin{array}{ll@{}>{{}}l} 
   \underset{R_{-i} | W_C}{\textbf{E}} \underset{\mathcal{Q}_1}{\textbf{E}}       \big| \big| \big( \mathcal{U}_{1,r_{-i}} \bigotimes \textbf{I}^{\otimes N-1} \big) \ket{\psi_{r_{-i},\bot , \cdots , \bot }}   - \ket{\psi_{r_{-i},q_1 , \bot \cdots , \bot}}    \big| \big| \text{, } \\    \underset{R_{-i} | W_C}{\textbf{E}} \underset{\mathcal{Q}_2}{\textbf{E}}      \big| \big|\big( \textbf{I} \bigotimes \mathcal{U}_{2,r_{-i}} \bigotimes \textbf{I}^{\otimes N-2} \big)\ket{\psi_{r_{-i},\bot , \cdots , \bot }}    -  \ket{\psi_{r_{-i}, \bot , q_2, \bot , \cdots , \bot}}    \big| \big|\text{, } \\ \vdots \\    \underset{R_{-i} | W_C}{\textbf{E}} \underset{\mathcal{Q}_N}{\textbf{E}}     \big| \big| \big( \textbf{I}^{\otimes N-1 } \bigotimes \mathcal{U}_{N,r_{-i}}  \big) \ket{\psi_{r_{-i},\bot , \cdots , \bot }}  - \ket{\psi_{r_{-i},\bot , \cdots , \bot,  q_N }}    \big| \big| \text{, } \\ \underset{R_{-i} | W_C}{\textbf{E}} \underset{\mathcal{Q}_1 \mathcal{Q}_2 }{\textbf{E}}    \big| \big| \big( \textbf{I}^{\otimes N-1 } \bigotimes \mathcal{U}_{N,r_{-i}}  \big) \ket{\psi_{r_{-i},\bot \backslash q_1 , q_2, \cdots, q_N}}  - \ket{\psi_{r_{-i},\bot \backslash q_1 ,\bot, q_3, \cdots, q_N}}   \big| \big|  \text{, } \\ \vdots \\ \underset{R_{-i} | W_C}{\textbf{E}} \underset{\mathcal{Q}_1 \times \cdots \times \mathcal{Q}_N}{\textbf{E}}    \big| \big|\big( \textbf{I}^{\otimes N-1 } \bigotimes \mathcal{U}_{N,r_{-i}}  \big) \ket{\psi_{r_{-i}, q_1 \backslash \bot, q_2 , \cdots ,  q_N} } \\  - \ket{\psi_{r_{-i}, q_1 \backslash \bot, q_2 \backslash \bot, \cdots , q_{N-1} \backslash \bot, q_N} }  \big| \big| ,
\end{array}\right. 
\]

\noindent for dependency-breaking unitaries,

\[   \left\{\!\begin{array}{ll@{}>{{}}l} 
 \mathcal{U}_{1,r_{-i}} \equiv \text{\textit{First player  dependency-breaking unitary}}   \text{, } \\  \mathcal{U}_{2,r_{-i}} \equiv \text{\textit{Second player dependency-breaking unitary}}  \text{, } \\ \vdots \\ \mathcal{U}_{N,r_{-i}} \equiv \text{\textit{N th player dependency-breaking unitary}}  \text{, } 
\end{array}\right. 
\]

\noindent where,

\begin{align*}
   \ket{\psi_{r_{-i},\bot , \cdots , \bot }}  , \ket{\psi_{r_{-i},q_1 , \bot \cdots , \bot}}  ,   \ket{\psi_{r_{-i}, \bot , q_2, \bot , \cdots , \bot}}  ,  \ket{\psi_{r_{-i},q_1 , \bot \cdots , \bot}} \\  , \ket{\psi_{r_{-i},\bot , \cdots , \bot,  q_N }}  ,   \ket{\psi_{r_{-i},\bot \backslash q_1 , q_2, \cdots, q_N}}  ,  \ket{\psi_{r_{-i},\bot \backslash q_1 ,\bot, q_3, \cdots, q_N}} ,    \cdots     \\ ,   \ket{\psi_{r_{-i}, q_1 \backslash \bot, q_2 , \cdots ,  q_N} }  ,   \ket{\psi_{r_{-i}, q_1 \backslash \bot, q_2 \backslash \bot, \cdots , q_{N-1} \backslash \bot, q_N}}             \in \ket{\Psi^{\prime}}  \text{, }
\end{align*}

\noindent given the collection of states,

\begin{align*}
 \ket{\Psi^{\prime}} \equiv \underset{1 \leq i \leq N}{\underset{\textit{dependency-breaking on answer from the i th player}}{\bigcup}} \ket{\Psi^{\prime}_i}  \text{. }
\end{align*}

\noindent One can generalize the system of expectation values above, either through the impact that each dependency-breaking variable can have on quantum correlations of player's strategies, or through normalizations of dependency-breaking strategies. Related to the system of expectation values defined above, one can introduce the system,

\[  \widetilde{\mathscr{E}_{\bot}} \equiv  \left\{\!\begin{array}{ll@{}>{{}}l} 
   \underset{R_{-i} | W_C}{\textbf{E}} \underset{\mathcal{Q}_1}{\textbf{E}}       \big| \big| \big( \mathcal{U}_{1,r_{-i}} \bigotimes \textbf{I}^{\otimes N-1} \big)  \widetilde{\ket{\psi_{r_{-i},\bot , \cdots , \bot }}}   - \widetilde{\ket{\psi_{r_{-i},q_1 , \bot \cdots , \bot}}}    \big| \big| \text{, } \\    \underset{R_{-i} | W_C}{\textbf{E}} \underset{\mathcal{Q}_2}{\textbf{E}}      \big| \big|\big( \textbf{I} \bigotimes \mathcal{U}_{2,r_{-i}} \bigotimes \textbf{I}^{\otimes N-2} \big)\widetilde{\ket{\psi_{r_{-i},\bot , \cdots , \bot }}}    -  \widetilde{\ket{\psi_{r_{-i}, \bot , q_2, \bot , \cdots , \bot}}}   \big| \big|\text{, } \\ \vdots \\    \underset{R_{-i} | W_C}{\textbf{E}} \underset{\mathcal{Q}_N}{\textbf{E}}     \big| \big| \big( \textbf{I}^{\otimes N-1 } \bigotimes \mathcal{U}_{N,r_{-i}}  \big) \widetilde{\ket{\psi_{r_{-i},\bot , \cdots , \bot }}}  - \widetilde{\ket{\psi_{r_{-i},\bot , \cdots , \bot,  q_N }}}    \big| \big| \text{, } \\ \underset{R_{-i} | W_C}{\textbf{E}} \underset{\mathcal{Q}_1 \mathcal{Q}_2 }{\textbf{E}}    \big| \big| \big( \textbf{I}^{\otimes N-1 } \bigotimes \mathcal{U}_{N,r_{-i}}  \big) \widetilde{\ket{\psi_{r_{-i},\bot \backslash q_1 , q_2, \cdots, q_N}}}  - \widetilde{\ket{\psi_{r_{-i},\bot \backslash q_1 ,\bot, q_3, \cdots, q_N}}}   \big| \big|  \text{, } \\ \vdots \\ \underset{R_{-i} | W_C}{\textbf{E}} \underset{\mathcal{Q}_1 \times \cdots \times \mathcal{Q}_N}{\textbf{E}}    \big| \big|\big( \textbf{I}^{\otimes N-1 } \bigotimes \mathcal{U}_{N,r_{-i}}  \big) \widetilde{\ket{\psi_{r_{-i}, q_1 \backslash \bot, q_2 , \cdots ,  q_N} }} \\  - \widetilde{\ket{\psi_{r_{-i}, q_1 \backslash \bot, q_2 \backslash \bot, \cdots , q_{N-1} \backslash \bot, q_N} }}  \big| \big| ,
\end{array}\right. 
\]

\noindent corresponding to the set of relations on normalized dependency-breaking states. The normalization introduced into $\widetilde{\ket{\psi}}$ for each $\ket{\psi}$, are of the form,

\begin{align*}
 \widetilde{\ket{\psi_{r_{-i}}}}  \equiv  \frac{\ket{\psi_{r_{-i}}} }{\big| \big| \ket{\psi_{r_{-i}}}\big| \big| } \propto \ket{\psi_{r_{-i}}} \text{, }
\end{align*}

\noindent corresponding to conjugation of operators of the form,

\begin{align*}
    \widetilde{\big( \mathcal{Q}_1 \big)_{q_1} \big( a_1 \big) } = \mathcal{U}_{1,r_{-i}}^{\dagger}   \bar{\big( \mathcal{Q}_1 \big)_{q_1} \big( a_1 \big) }\mathcal{U}_{1,r_{-i}}  \text{,} \\ \\    \widetilde{\big( \mathcal{Q}_2 \big)_{q_2} \big( a_2 \big) } = \mathcal{U}_{2,r_{-i}}^{\dagger} \bar{\big( \mathcal{Q}_2 \big)_{q_2} \big( a_2 \big) }\mathcal{U}_{2,r_{-i}}  \text{,}\\ \\ \widetilde{\big( \mathcal{Q}_3 \big)_{q_3} \big( a_3 \big) } = \mathcal{U}_{3,r_{-i}}^{\dagger} \bar{\big( \mathcal{Q}_3 \big)_{q_3} \big( a_3 \big) }\mathcal{U}_{3,r_{-i}}  \text{,} \\  \vdots \\     \widetilde{\big( \mathcal{Q}_N \big)_{q_N} \big( a_N \big) } = \mathcal{U}_{N,r_{-i}}^{\dagger} \bar{\big( \mathcal{Q}_N \big)_{q_N} \big( a_N \big) }\mathcal{U}_{N,r_{-i}}   \text{, }
\end{align*}

\noindent for dependency-breaking unitarites, $\mathcal{U}$, and,

\begin{align*}
 \bar{\big( \mathcal{Q}_1 \big)_{q_1} \big( a_1 \big) } \equiv   \underset{a | a_i, a_C}{\sum}   \big( \mathcal{Q}_1 \big)_{q_1} \big( a_C \big) \big)^{-\frac{1}{2}}  \big( \mathcal{Q}_1 \big)_{q_1} \big( a \big)   \big( \mathcal{Q}_1 \big)_{q_1} \big( a_C \big) \big)^{-\frac{1}{2}}  \text{, } \\ \\ \bar{\big( \mathcal{Q}_2 \big)_{q_2} \big( a_2 \big) } \equiv \underset{a | a_2, a_C}{\sum}      \big( \mathcal{Q}_2 \big)_{q_2} \big( a_C \big) \big)^{-\frac{1}{2}}  \big( \mathcal{Q}_2 \big)_{q_2} \big( a \big)   \big( \mathcal{Q}_2 \big)_{q_2} \big( a_C \big) \big)^{-\frac{1}{2}}          \text{, } \\ \\ \bar{\big( \mathcal{Q}_3 \big)_{q_3} \big( a_3 \big) } \equiv \underset{a | a_3, a_C}{\sum}      \big( \mathcal{Q}_3 \big)_{q_3} \big( a_C \big) \big)^{-\frac{1}{2}}  \big( \mathcal{Q}_3 \big)_{q_3} \big( a \big)   \big( \mathcal{Q}_3 \big)_{q_3} \big( a_C \big) \big)^{-\frac{1}{2}}          \text{, } \\  \vdots \\ \bar{\big( \mathcal{Q}_N \big)_{q_N} \big( a_N \big) } \equiv     \underset{a | a_N, a_C}{\sum}  \big( \mathcal{Q}_N \big)_{q_N} \big( a_C \big) \big)^{-\frac{1}{2}}  \big( \mathcal{Q}_N \big)_{q_N} \big( a \big)   \big( \mathcal{Q}_N \big)_{q_N} \big( a_C \big) \big)^{-\frac{1}{2}}              \text{, }
\end{align*}

\noindent corresponding transformations of tensor player observables. As one can expect, the normalization introduced above in $\widetilde{\ket{\psi_{r_{-i}}}}$ can be straightforwardly defined with,

\[    \left\{\!\begin{array}{ll@{}>{{}}l} 
\widetilde{\ket{\psi_{r_{-i}, \bot , q_2, q_3, \cdots, q_N}}} \equiv \frac{\widetilde{\ket{\psi_{r_{-i}, \bot , q_2, q_3, \cdots, q_N}}} }{\big| \big| \widetilde{\ket{\psi_{r_{-i}, \bot , q_2, q_3, \cdots, q_N}}} \big| \big| } ,  \\  \widetilde{\ket{\psi_{r_{-i}, \bot , \bot , q_3, \cdots, q_N}}} \equiv \frac{\widetilde{\ket{\psi_{r_{-i}, \bot , \bot , q_3, \cdots, q_N}}} }{\big| \big| \widetilde{\ket{\psi_{r_{-i}, \bot , \bot , q_3, \cdots, q_N}}} \big| \big| }  , \\ \widetilde{\ket{\psi_{r_{-i}, \bot , \bot , q_3, \cdots, q_N}}} \equiv \frac{\widetilde{\ket{\psi_{r_{-i}, \bot , \bot , \bot , q_4 , \cdots, q_N}}} }{\big| \big| \widetilde{\ket{\psi_{r_{-i}, \bot , \bot , \bot , \cdots, q_N}}} \big| \big| }   ,  \\  \vdots  \\  \widetilde{\ket{\psi_{r_{-i}, \bot , \bot , q_3, \cdots, q_N}}} \equiv \frac{\widetilde{\ket{\psi_{r_{-i}, \bot , \cdots , \bot , q_N}}} }{\big| \big| \widetilde{\ket{\psi_{r_{-i}, \bot , \cdots, \bot , q_N}}} \big| \big| }  .
\end{array}\right. 
\]

\noindent The system of expectation values defined above leads to several computations with a \textit{Normalization} factor described previously. With such a factor, recall from previous remarks that the system of expectation values with the \textit{Normalization} factor takes the form,

\[  \widetilde{\mathscr{E}} \equiv  \left\{\!\begin{array}{ll@{}>{{}}l} 
   \underset{R_{-i} | W_C}{\textbf{E}} \underset{\mathcal{Q}_1}{\textbf{E}}       \big| \big|  \big[ \big( \mathcal{U}_1 \bigotimes \textbf{I}^{\otimes N-1} \big) - \textit{Normalization} \big] \widetilde{\ket{\psi} }   \big| \big|\text{, } \\    \underset{R_{-i} | W_C}{\textbf{E}} \underset{\mathcal{Q}_2}{\textbf{E}}      \big| \big|  \big[ \big( \textbf{I} \bigotimes \mathcal{U}_2 \bigotimes \textbf{I}^{\otimes N-2} \big) - \textit{Normalization} \big] \widetilde{\ket{\psi} }     \big| \big|\text{, } \\ \vdots \\   \underset{R_{-i} | W_C}{\textbf{E}} \underset{\mathcal{Q}_N}{\textbf{E}}     \big| \big|  \big[  \big( \textbf{I}^{\otimes N-1 } \bigotimes \mathcal{U}_N  \big) - \textit{Normalization}  \big] \widetilde{\ket{\psi} }    \big| \big|  \text{, } \\ \underset{R_{-i} | W_C}{\textbf{E}} \underset{\mathcal{Q}_1 \mathcal{Q}_2 }{\textbf{E}}     \big| \big|  \big[  \big( \textbf{I}^{\otimes N-1 } \bigotimes \mathcal{U}_N  \big) - \textit{Normalization} \big] \widetilde{\ket{\psi} }   \big| \big| \text{, }  \end{array}\right. 
\]

\[   \left\{\!\begin{array}{ll@{}>{{}}l}      \vdots \\ \underset{R_{-i} | W_C}{\textbf{E}} \underset{\mathcal{Q}_1 \times \cdots \times \mathcal{Q}_N}{\textbf{E}}    \big| \big|  \big[ \big( \textbf{I}^{\otimes N-1 } \bigotimes \mathcal{U}_N  \big) - \textit{Normalization} \big] \widetilde{\ket{\psi} }  \big| \big|  .
\end{array}\right. 
\] 

\noindent As a result, given the system of expectation values with normalized dependency-breaking states $\widetilde{\ket{\Psi}}$, obtained the desired rate of decay for the optimal value under $\alpha$ anchored parallel repetition implies that one would like to upper bound terms of the system,

\[   \left\{\!\begin{array}{ll@{}>{{}}l} 
   \underset{R_{-i} | W_C}{\textbf{E}} \underset{\mathcal{Q}_1}{\textbf{E}}       \bigg| \bigg| \big( \mathcal{U}_{1,r_{-i}} \bigotimes \textbf{I}^{\otimes N-1} \big)  \widetilde{\ket{\psi_{r_{-i},\bot , \cdots , \bot }}}   - \bigg[       \frac{\big| \big|   \ket{\psi_{q_1,\cdots, q_N}}         \big| \big|}{\big| \big|   \ket{\psi_{\bot,\cdots,\bot}}       \big| \big| }     \bigg] \widetilde{\ket{\psi_{r_{-i},q_1 , \bot \cdots , \bot}}}    \bigg| \bigg| \text{, } \\    \underset{R_{-i} | W_C}{\textbf{E}} \underset{\mathcal{Q}_2}{\textbf{E}}      \bigg| \bigg|\big( \textbf{I} \bigotimes \mathcal{U}_{2,r_{-i}} \bigotimes \textbf{I}^{\otimes N-2} \big)\widetilde{\ket{\psi_{r_{-i},\bot , \cdots , \bot }}}    -  \bigg[   \frac{\big| \big|   \ket{\psi_{\bot, q_2,\cdots, q_N}}         \big| \big|}{\big| \big|   \ket{\psi_{\bot,\cdots,\bot}}       \big| \big| }   \bigg]  \widetilde{\ket{\psi_{r_{-i}, \bot , q_2, \bot , \cdots , \bot}}}   \bigg| \bigg|\text{, } \\ \vdots \\    \underset{R_{-i} | W_C}{\textbf{E}} \underset{\mathcal{Q}_N}{\textbf{E}}     \bigg| \bigg| \big( \textbf{I}^{\otimes N-1 } \bigotimes \mathcal{U}_{N,r_{-i}}  \big) \widetilde{\ket{\psi_{r_{-i},\bot , \cdots , \bot }}}  - \bigg[     \frac{\big| \big|   \ket{\psi_{\bot, \bot, \cdots, \bot,  q_N}}         \big| \big|}{\big| \big|   \ket{\psi_{\bot,\cdots,\bot}}       \big| \big| }   \bigg]  \widetilde{\ket{\psi_{r_{-i},\bot , \cdots , \bot,  q_N }}}    \bigg| \bigg| \text{, } \\ \underset{R_{-i} | W_C}{\textbf{E}} \underset{\mathcal{Q}_1 \mathcal{Q}_2 }{\textbf{E}}    \bigg| \bigg| \big( \textbf{I}^{\otimes N-1 } \bigotimes \mathcal{U}_{N,r_{-i}}  \big) \widetilde{\ket{\psi_{r_{-i},\bot \backslash q_1 , q_2, \cdots, q_N}}}  - \bigg[   \frac{\big| \big|   \ket{\psi_{\bot,  q_2,\cdots, q_N}}         \big| \big|}{\big| \big|   \ket{\psi_{\bot,\cdots,\bot}}       \big| \big| }  \bigg] \\ \times \widetilde{\ket{\psi_{r_{-i},\bot \backslash q_1 ,\bot, q_3, \cdots, q_N}}}   \bigg| \bigg|  \text{, } \end{array}\right. 
\]

\[   \left\{\!\begin{array}{ll@{}>{{}}l} \vdots \\ \underset{R_{-i} | W_C}{\textbf{E}} \underset{\mathcal{Q}_1 \times \cdots \times \mathcal{Q}_N}{\textbf{E}}    \bigg| \bigg|\big( \textbf{I}^{\otimes N-1 } \bigotimes \mathcal{U}_{N,r_{-i}}  \big) \widetilde{\ket{\psi_{r_{-i}, q_1 \backslash \bot, q_2 , \cdots ,  q_N} }}   - \bigg[              \frac{\big| \big|   \ket{\psi_{\bot, \cdots, \bot,  q_N}}         \big| \big|}{\big| \big|   \ket{\psi_{\bot,\cdots,\bot}}       \big| \big| }  \bigg]  \\ \times \widetilde{\ket{\psi_{r_{-i}, q_1 \backslash \bot, q_2 \backslash \bot, \cdots , q_{N-1} \backslash \bot, q_N} }}  \bigg| \bigg| ,
\end{array}\right. 
\]

\noindent of expectation values.

\bigskip

\noindent Probability measures over dependency-breaking variables, and the set of possible questions for each participant, which have previously been introduced can also be identified through \textit{positive operator value measurements}, POVMs, which for multiplayer game-theoretic settings are of the form,

\begin{align*}
 \mathrm{Tr} \bigg[    \bigg[  \bar{\big( \mathcal{Q}_1 \big)_{q_1} \big( a_1 \big) } \otimes    \bar{\big( \mathcal{Q}_2 \big)_{q_2} \big( a_2 \big)  \bigg] \widetilde{\Psi_{r_{-i}, q_1,q_2}}}  \bigg]  \equiv \textbf{P}_{\mathcal{Q}_1\mathcal{Q}_2 \mathcal{Q}_3| r_{-i}, q_1 q_2 q_3 } \big( q_1, q_2 , q_3 \big) \text{, } \\ \\  \mathrm{Tr} \bigg[ \bigg[  \bar{\big( \mathcal{Q}_1 \big)_{q_1} \big( a_1 \big) } \otimes    \bar{\big( \mathcal{Q}_2 \big)_{q_2} \big( a_2 \big)}  \otimes    \bar{\big( \mathcal{Q}_3 \big)_{q_3} \big( a_3 \big)}  \bigg]   \widetilde{\Psi_{r_{-i}, q_1,q_2}}    \bigg]   \equiv \textbf{P}_{\mathcal{Q}_1\mathcal{Q}_2 \mathcal{Q}_3 \mathcal{Q}_4| r_{-i}, q_1 q_2 q_3 q_4 } \big( q_1, q_2 \\ , q_3, q_4  \big)  \text{,} \end{align*}

 \begin{align*} \vdots \\  \mathrm{Tr} \bigg[  \bigg[  \bar{\big( \mathcal{Q}_1 \big)_{q_1} \big( a_1 \big) } \otimes    \bar{\big( \mathcal{Q}_2 \big)_{q_2} \big( a_2 \big)}  \otimes    \bar{\big( \mathcal{Q}_3 \big)_{q_3} \big( a_3 \big)} \otimes \cdots \otimes   \bar{\big( \mathcal{Q}_N \big)_{q_N} \big( a_N \big)} \bigg]  \widetilde{\Psi_{r_{-i}, q_1,q_2, \cdots, q_N}} \bigg] \\  \equiv \textbf{P}_{\mathcal{Q}_1\mathcal{Q}_2 \mathcal{Q}_3 \times \cdots \times \mathcal{Q}_N| r_{-i}, q_1 q_2 q_3 \times \cdots \times q_N} \big( q_1, q_2 , q_3, q_4, \cdots, q_N  \big)  \text{.}
\end{align*}

\noindent POVMs are of great significance in determining whether each expectation value, taken with respect to some $\mathcal{Q}_i$, or pairs of countable many $\mathcal{Q}_i \subsetneq \mathcal{Q}$ as provided above, are upper bounded by some suitable $\mathrm{O} \big( \cdot \big)$, or by a suitable constant depending upon the one-side anchoring probability. In the presence of a sequence of normalizations,

\begin{align*}
 \Gamma_i \equiv   \bigg\{            \frac{ \big| \big|                                   \textbf{1}_{\ \{         {\textit{the first i th answers are anchored}\} }}       \ket{\psi_{r_{-i}, \bot , \cdots , \bot, q_{i+1} , \cdots, q_N}}         \big| \big|}{\big| \big|  \textbf{1}_{\{          \textit{all answers are anchored}\} }         \ket{\psi_{r_{-i}, \bot , \cdots , \bot }}       \big| \big| }    \bigg\}_
   {1 \leq i \leq N}\text{, }
\end{align*}

\noindent one can quantify the impact of anchoring finitely many answers, through formulating upper bounds for the system,

\[  \widetilde{\mathscr{E}_{\Gamma_i}} \equiv  \left\{\!\begin{array}{ll@{}>{{}}l} 
   \underset{R_{-i} | W_C}{\textbf{E}} \underset{\mathcal{Q}_1}{\textbf{E}}       \big| \big|  \big[ \big( \mathcal{U}_1 \bigotimes \textbf{I}^{\otimes N-1} \big) - \Gamma_1  \big] \widetilde{\ket{\psi} }   \big| \big|\text{, } \\    \underset{R_{-i} | W_C}{\textbf{E}} \underset{\mathcal{Q}_2}{\textbf{E}}      \big| \big|  \big[ \big( \textbf{I} \bigotimes \mathcal{U}_2 \bigotimes \textbf{I}^{\otimes N-2} \big) - \Gamma_2 \big] \widetilde{\ket{\psi} }     \big| \big|\text{, } \\ \vdots \\    \underset{R_{-i} | W_C}{\textbf{E}} \underset{\mathcal{Q}_N}{\textbf{E}}     \big| \big|  \big[ \big( \textbf{I}^{\otimes N-1 } \bigotimes \mathcal{U}_N  \big) - \Gamma_N  \big] \widetilde{\ket{\psi} }    \big| \big|  \text{, } \\ \underset{R_{-i} | W_C}{\textbf{E}} \underset{\mathcal{Q}_1 \mathcal{Q}_2 }{\textbf{E}}     \big| \big|  \big[ \big( \textbf{I}^{\otimes N-1 } \bigotimes \mathcal{U}_N  \big) - \Gamma_1 \Gamma_2  \big] \widetilde{\ket{\psi} }   \big| \big| \text{, } \\ \vdots \\ \underset{R_{-i} | W_C}{\textbf{E}} \underset{\mathcal{Q}_1 \times \cdots \times \mathcal{Q}_N}{\textbf{E}}    \bigg| \bigg|  \bigg[ \big( \textbf{I}^{\otimes N-1 } \bigotimes \mathcal{U}_N  \big) - \bigg[ \underset{1 \leq i \leq N}{\prod} \Gamma_i \bigg] \bigg] \widetilde{\ket{\psi} }  \bigg| \bigg|  .
\end{array}\right. 
\]

\noindent Besides the fact that the collection of all dependency-breaking variables can be expressed with the above union, over all realizations $\omega_{-i}$, which is hence over all realizations of $R_{-i}$, one has that,

\begin{align*}
  \big( \mathcal{Q}_1 \big)_{\omega_{-i} , \bot}     \equiv              \underset{ ( \mathcal{Q}_1 )^n | \Omega_{-i}}  {\textbf{E}}   \big( \mathcal{Q}_1 \big)_{( \mathcal{Q}_1 )^n }   \Longleftrightarrow    ( \mathcal{Q}_1 )^n | \Omega_{-i} \in \omega_{-i} \bigwedge \big( \mathcal{Q}_1 \big)_i \in \bot   \text{, } \\ \vdots \\       \big( \mathcal{Q}_N \big)_{\omega_{-i} , \bot}     \equiv              \underset{ ( \mathcal{Q}_N )^n | \Omega_{-i}}  {\textbf{E}}   \big( \mathcal{Q}_N \big)_{( \mathcal{Q}_1 )^n }   \Longleftrightarrow    ( \mathcal{Q}_N )^n | \Omega_{-i} \in \omega_{-i} \bigwedge \big( \mathcal{Q}_1 \big)_i \in \bot     \text{, }
\end{align*}

\noindent corresponding to the expected value of the set of all possible responses from each participant. Relatedly, for any answer of a player satisfying the referee's scoring predicate over all coordinates,

\begin{align*}
  \big( \mathcal{Q}_1 \big)^{q_C}_{q_1} \equiv \underset{q^n | q^C}{\sum} \big( \mathcal{Q}_1 \big)^{q_n}_{q_1}   \text{, } \\ \vdots \\  \big( \mathcal{Q}_N \big)^{q_C}_{q_1} \equiv \underset{q^n | q^C}{\sum} \big( \mathcal{Q}_N \big)^{q_n}_{q_1} \text{,}
\end{align*}

\noindent corresponding to the tensor of possible responses from all players, $\mathcal{Q}_{1,2,\cdots, N}$.

\bigskip

\noindent The above system of relations is itself a generalization of the following result, first provided in [3]:

\bigskip

\noindent \textbf{Lemma} (\textbf{Lemma} \textit{5.12}, [2, 3]). For all $r_{-i}$, $x$, and $y$, there exists unitaries $U_{r_{-i}.x}$ acting on $E_A$, along with unitaries $V_{r_{-i},y}, V_{r_{-i},x,y}$ acting on $E_B$ such that with probability at least $1 - \mathrm{O
} \big( \delta^{\frac{1}{16}} \big)$, for $\delta$ taken to be sufficiently small, over the choice of uniformaly random $i \in [n] \backslash C$,

\begin{align*}
  \underset{R_{-i}| W_C}{\textbf{E}} \underset{X}{\textbf{E}} \big| \big|  \big( U_{r_{-i}, x } \bigotimes \textbf{I} \big) \ket{\Psi_{r_{-i}, \bot, \bot}} - \ket{\Psi_{r_{-i}, x, \bot}}     \big| \big|   = \mathrm{O} \bigg(  \frac{\delta^{\frac{1}{16}}}{\alpha^{\frac{5}{4}}} \bigg)     \text{,} \\ \\  \underset{R_{-i}| W_C}{\textbf{E}} \underset{Y}{\textbf{E}} \big| \big|  \big(\textbf{I} \bigotimes    V_{r_{-i}, x }  \big) \ket{\Psi_{r_{-i}, \bot, \bot}} - \ket{\Psi_{r_{-i}, \bot, y}}     \big| \big|   = \mathrm{O} \bigg(  \frac{\delta^{\frac{1}{16}}}{\alpha^{\frac{5}{4}}} \bigg)    \text{, } \\ \\  \underset{R_{-i}| W_C}{\textbf{E}} \underset{XY}{\textbf{E}} \big| \big|  \big(\textbf{I} \bigotimes    V_{r_{-i}, x,y }  \big) \ket{\Psi_{r_{-i}, \bot \backslash x, y}} - \ket{\Psi_{r_{-i}, \bot, \bot}}     \big| \big|        = \mathrm{O} \bigg(  \frac{\delta^{\frac{1}{16}}}{\alpha^{\frac{5}{4}}} \bigg)     \text{. }
\end{align*}

\bigskip

\noindent In a 2017 STOC conference proceeding, the same group of authors obtained a similar upper bound, as the one provided above for the two-player setting. Determining how the largest, most dominant order of expectation values, such as the ones above, is intimately related to the rate of decay of the optimal value under anchored parallel repetition. In a sense, to determine whether similar expressions, such as the one above for the optimal value that is dependent upon $\big( 1 - \big( 1 - x\big) \big)^N$, in multiplayer settings one must take account for anchoring amongst multiple players simultaneously. Furthermore, under higher-dimensional generalizations of anchoring, accompanying dependency-breaking correlations correspond to the probability of continuing to maintain a high optimal value for the game after breaking correlations amongst, possibly, finitely many players within a game with an arbitrary number of players. The result from the conference proceeding takes the form:

\bigskip

\noindent  \textbf{Lemma} (\textbf{Lemma} \textit{5.5}, [2, 3, 4], \textit{big} $\mathrm{O}$ \textit{values of the expected value for games with three participants under $\alpha$-anchoring, a generalization of ordinary anchoring}). For all $\big( x,y,z \big) \in \mathcal{X} \times \mathcal{Y} \times \mathcal{Z}$, namely the set of all possible questions for all three players, and the number of parallel repetition operations $k$, there exists unitaries $U_x, V_y, W_z$ acting on $E_A, E_B, E_C$, respectively, such that,

\begin{align*}
   \underset{XYZ}{\textbf{E}} \big| \big| \big( U_x \otimes V_y \otimes W_z \big) \ket{\Psi_{\bot,\bot,\bot}} - \ket{\Psi_{x,y,z}}\big| \big|  = \mathrm{O} \bigg( \frac{\delta^{\frac{1}{4}}}{\alpha^{2k}} \bigg)        \text{. }
\end{align*}

\bigskip

\noindent Determining how anchoring across the set of responses for multiple participants simultaneously not only demonstrates how higher-dimensional dependency-breaking destroys quantum correlations, but also how optimality of strategies is impacted by anchoring.

\bigskip

\noindent The system of expectation values above determines a previously cited rate of exponential decay of the optimal value,

\begin{align*}
 \omega^{*} \big( G^n_{\bot} \big) \leq \frac{4}{\epsilon}         \mathrm{exp} \bigg[  - \frac{c \alpha^{48} \epsilon^{17} n}{s}   \bigg]    \text{, }
\end{align*}

\noindent which also raised implications for the rate of exponential decay of the multiplayer optimal value provided in this work, under anchored parallel repetition. In comparison to anchored parallel repetition for games $G^n_{\bot}$ discussed above, a polynomial decay for the optimal value of the three player setting, which has upper bound of order $\mathrm{O} \bigg( \frac{\delta^{\frac{1}{4}}}{\alpha^{2k}} \bigg)$ provided above, the following result demonstrates one possible rate of decay that one would hope to obtain in the multiplayer setting:

\bigskip

\noindent \textbf{Theorem} (\textbf{Theorem} \textit{5.6}, [3], \textit{polynomial function for the rate of decay of the optimal value for the $\alpha$-anchored $k$-player game}). For an $\alpha$-anchored game $G$ such that the optimal value of the game is $\leq 1 - \epsilon$ for $\epsilon$ taken to be sufficiently small, for a game with at least $n > 1$ players,

\begin{align*}
    \bigg[ 1 - \frac{\gamma^9}{2} \bigg]^{c \alpha^{8k} \frac{n}{s}}     \text{,}
\end{align*}

\noindent corresponds to the probability of winning more than $\big( 1 - \epsilon + \gamma \big) n$ games, for $s$ being the length of all responses from each participant, and for a universal constant, $c$.

\item[$\bullet$] \textit{Computing the power of the exponential in the rate of decay of the optimal value under anchored parallel repetition}. To obtain the desired exponential bound, one performs computations associated with the quantity,

\begin{align*}
  \delta_{\mathrm{Multiplayer}} \equiv \frac{1}{n} \bigg[ \big| C \big| \mathrm{log} \bigg[ \bigg|  \underset{1 \leq i \leq N}{\prod} \big| \mathcal{Q}_i \big|  \bigg| \bigg] + \mathrm{log} \bigg[ \bigg| \frac{1}{\textbf{P} \big[ W_C \big] } \bigg| \bigg] \bigg]  \text{. }
\end{align*}

\noindent In comparison to a similar parameter, such as the one provided above for multiplayer game-theoretic settings, one takes the set of all coordinates as a prefactor for the first natural logarithm above as,

\begin{align*}
  \big| C \big| \equiv \underset{1 \leq i \leq N}{\bigcup } \big| C_i \big|   \text{, }
\end{align*}

\noindent corresponding to a union over the coordinates of each respective player.

\bigskip

\item[$\bullet$] \textit{Relating upper bounds of the relative-min entropy to the optimal value}. Depending upon the natural logarithm of the possible answers from each player,

\begin{align*}
  \mathrm{log}\big[  \big| \mathcal{Q}_1 \big| \cdot \big| \mathcal{Q}_2  \big|  \cdots \cdot \big|  \mathcal{Q}_N \big| \big]   \text{, }
\end{align*}

\noindent as well as upon the function of the optimal value,

\begin{align*}
   1 - \big[ 1 - \omega_{\mathrm{Multiplayer}} \big( G^{\bot} \big) \big]^{\alpha}       \text{, }
\end{align*}

\noindent for,

\begin{align*}
  \alpha \equiv   \alpha \big( G, k, N, \bot \big)         \text{, }
\end{align*}

\noindent one argues that the power of the exponential in the rate of decay of the anchored optimal value under parallel repetition implies that,

\begin{align*}
     \omega_{\mathrm{Multiplayer}} \big( \big( G_{\bot} \big)    \big)^{\otimes n}   \precsim  \mathrm{exp} \bigg[  -  \frac{c_{\mathrm{Multiplayer}} \alpha^{20N + 1}_{\mathrm{Multiplayer}}        \epsilon^{6N}_{\mathrm{Multiplayer}}  n }{s_{\mathrm{Multiplayer}}  }\bigg]          \text{, }
\end{align*}

\noindent for the parameters $c_{\mathrm{Multiplayer}}, \alpha_{\mathrm{Multiplayer}}, \epsilon_{\mathrm{Multiplayer}}$, and $s_{\mathrm{Multiplayer}}$ defined in the next item below.

\end{itemize}

\item[$\bullet$] \textit{Sharpening the up to constants upper bound with an appropriate prefactor}. Fix $\epsilon_{\mathrm{Multiplayer}}$ strictly positive, $s_{\mathrm{Multiplayer}} \equiv \mathrm{max} \bigg\{ \underset{1 \leq i \leq N}{\prod} \mathrm{log} \big| \mathcal{Q}_i \big| ,1\bigg\}$, $0 < \alpha_{\mathrm{Multiplayer}} \leq 1$, a multiplayer $\alpha$-anchored game $G$, $\epsilon_{\mathrm{Multiplayer}}$ strictly positive, and universal constant $c_{\mathrm{Multiplayer}}$, where,

\begin{align*}
  0 < c_{\mathrm{Multiplayer}} < \frac{1}{N^{2N} \mathrm{log} \big( e \big)}  .
\end{align*}

\noindent One has that,

\begin{align*}
   \omega_{\mathrm{Multiplayer}} \big( \big( G_{\bot} \big)    \big)^{\otimes n} \leq  \frac{10}{\epsilon_{\mathrm{Multiplayer}}} \mathrm{exp} \bigg[  -  \frac{c_{\mathrm{Multiplayer}} \alpha^{20N + 1}_{\mathrm{Multiplayer}}        \epsilon^{6N}_{\mathrm{Multiplayer}}  n }{s_{\mathrm{Multiplayer}}  }\bigg]          \text{, }
\end{align*}

\noindent under the assumption that,

\begin{align*}
    \omega_{\mathrm{Multiplayer}} \big( \big( G_{\bot} \big)    \big)^{\otimes n}   \precsim  \mathrm{exp} \bigg[  -  \frac{c_{\mathrm{Multiplayer}} \alpha^{20N + 1}_{\mathrm{Multiplayer}}        \epsilon^{6N}_{\mathrm{Multiplayer}}  n }{s_{\mathrm{Multiplayer}}  }\bigg]         \text{. }
\end{align*}

\item[$\bullet$] \textit{Relating computations for expected values of probabilities to expectations of quantum states through Pinkser's inequality}. The result on Pinkser's inequality takes the form below:

\bigskip

\noindent (\textit{Pinsker's inequality - upper bounding the expected value of the l-1 norm with the relative entropy}). Denote $\rho$ and $\sigma$ as two quantum states. One has that,

\begin{align*}
    \frac{1}{2 \mathrm{ln} 2 } \big| \big| \rho - \sigma \big| \big|_1 \leq D \big( \rho \big| \big| \sigma \big) . 
\end{align*}

\noindent The statement of Pinkser's inequality above was originally used in the two-player setting for obtaining desired upper bounds of the relative entropy, which directly relates to upper bounding the Mutual Information entropy. While the upper bound for expectations of probabilities, in the multiplayer setting, exhibits a different ratio of constants, which is directly proportional to $\delta_{\mathrm{Multiplayer}}$, and inversely proportional to the multiplayer anchoring probability, $\alpha_{\mathrm{Multiplayer}}$, generalizing the upper bound obtained from Pinkser's inequality to the multiplayer setting, from the two-player setting, is straightforward.

\begin{figure}
\begin{align*}
\includegraphics[width=0.8\columnwidth]{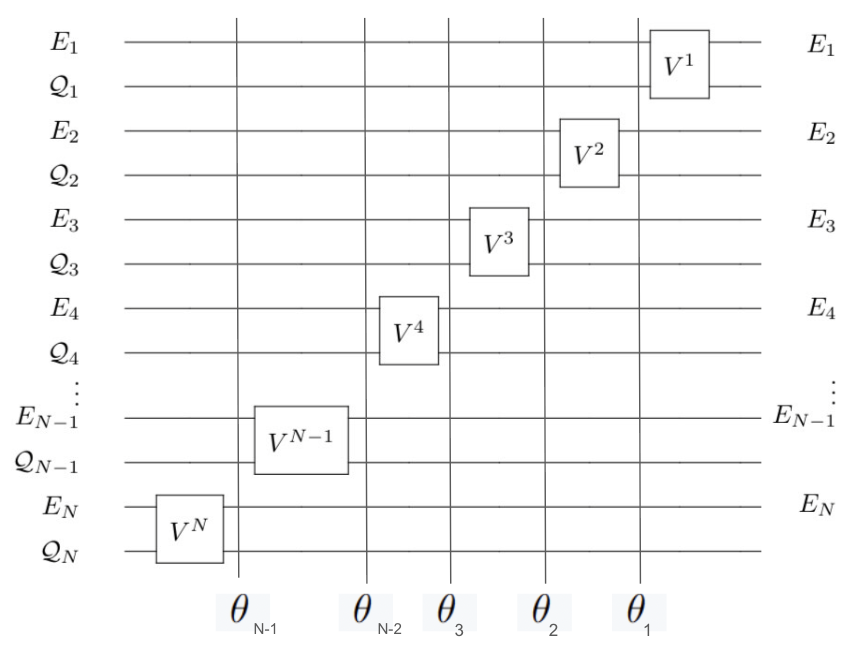}
\end{align*}
\caption{\textit{Quantum circuit representation of the registers for countably many players. $\theta_i$ are applied between entangling gates $V^i$.} }
\end{figure}

In the two-player setting, the system of expectation values which one upper bounds with Pinsker's inequality takes the form, [3, 4],

\[   \left\{\!\begin{array}{ll@{}>{{}}l} 
   \underset{I}{\textbf{E}} \underset{R_{-i} | W_C}{\textbf{E}} \underset{XY}{\textbf{E}}  \big| \big| \xi^{E_A}_{r_{-i}, x,y}  -  \xi^{E_A}_{r_{-i}, x,\bot} \big| \big|^2_1 = \mathrm{O} \big( \frac{\sqrt{\delta}}{\alpha^4} \big) , \\     \underset{I}{\textbf{E}} \underset{R_{-i} | W_C}{\textbf{E}} \underset{XY}{\textbf{E}}  \big| \big|          \lambda^{E_B}_{r_{-i}, x,y} - \lambda^{E_B}_{r_{-i}, \bot ,y}         \big| \big|^2_1 = \mathrm{O} \big( \frac{\sqrt{\delta}}{\alpha^4} \big) , 
\end{array}\right. 
\]

\noindent while for the multiplayer setting, straightforwardly, the system of expectation values which one upper bounds with Pinsker's inequality takes the form,

\[ \mathcal{P} \mathcal{I} \equiv   \left\{\!\begin{array}{ll@{}>{{}}l} 
\underset{I}{\textbf{E}} \underset{R_{-i} | W_C}{\textbf{E}} \underset{\mathcal{Q}_1 \mathcal{Q}_2 \times \cdots \times \mathcal{Q}_N}{\textbf{E}}  \big| \big| \textbf{I}^{E_1}_{r_{-i}, q_1,q_2, \cdots, q_N}  -  \textbf{I}^{E_1}_{r_{-i}, q_1,\bot,q_3, \cdots, q_N} \big| \big|^2_1   , \\ \underset{I}{\textbf{E}} \underset{R_{-i} | W_C}{\textbf{E}} \underset{\mathcal{Q}_1 \mathcal{Q}_2 \times \cdots \times \mathcal{Q}_N}{\textbf{E}}  \big| \big| \textbf{2}^{E_2}_{r_{-i}, q_1,q_2, \cdots, q_N}  -  \textbf{2}^{E_2}_{r_{-i}, \bot, q_2, q_3, \cdots, q_N} \big| \big|^2_1  ,  \\ \vdots \\ \underset{I}{\textbf{E}} \underset{R_{-i} | W_C}{\textbf{E}} \underset{\mathcal{Q}_1 \mathcal{Q}_2 \times \cdots \times \mathcal{Q}_N}{\textbf{E}}  \big| \big| \textbf{N}^{E_N}_{r_{-i}, q_1,q_2, \cdots, q_N}  -  \textbf{N}^{E_N}_{r_{-i}, q_1,\cdots, q_{N-2}, \bot, q_N} \big| \big|^2_1  ,
\end{array}\right. 
\]

\noindent for the collection of Quantum states,

\[   \left\{\!\begin{array}{ll@{}>{{}}l} 
  \textbf{I} \equiv \textit{First player's quantum state} , \\ \textbf{2} \equiv \textit{Second player's quantum state} ,  \\ \vdots  \\  \textbf{N} \equiv \textit{N th player's quantum state} ,  
\end{array}\right. 
\] 

\noindent along with the registers,

 \[   \left\{\!\begin{array}{ll@{}>{{}}l} 
 E_1 \equiv \textit{First player's quantum register} , \\ E_2 \equiv \textit{Second player's quantum register} ,  \\ \vdots  \\ E_N \equiv \textit{N th player's quantum register} ,  
\end{array}\right. 
\]

\noindent which are related to the collection of states,

 \[   \left\{\!\begin{array}{ll@{}>{{}}l} \xi^{\mathcal{Q}_1, E_1}_r
\equiv \underset{\mathcal{Q}_1 | R \equiv r, W_C}{\textbf{E}}  \big[ q_1 \big]^{\mathcal{Q}_1}  \bigotimes  \xi^{E_1}_{r,q_1} , \\ \xi^{\mathcal{Q}_2, E_2}_r
\equiv \underset{\mathcal{Q}_2 | R \equiv r, W_C}{\textbf{E}}  \big[ q_2 \big]^{\mathcal{Q}_2}  \bigotimes  \xi^{E_2}_{r,q_2} ,  \\ \vdots  \\ \xi^{\mathcal{Q}_n, E_1}_r
\equiv \underset{\mathcal{Q}_n | R \equiv r, W_C}{\textbf{E}}  \big[ q_n \big]^{\mathcal{Q}_n}  \bigotimes  \xi^{E_n}_{r,q_n}  .  
\end{array}\right. 
\]

\noindent Above each Quantum state in the system of expected values for the two-player, and multiplayer, systems alike, $E_A, E_B, \cdots$ denote Quantum registers associated with the strategies of each player for the two-player setting. $E_1, E_2, \cdots, E_N$ denote the registers for the multiplayer setting. Diagramatically, such registers are provided in  \textit{Figure 1}. Such Quantum circuit diagrams relate to previous work of the author, [51], currently under peer review, for performing time-evolution of Zalka-Grover-Rudolph (ZGR) Quantum Fourier Transform ansatzae for approximating solutions to various PDEs, provided in \textit{Figure 2}. Specifically, $18$ input ports (IPs) are depicted, which are fed into a collection of Adder gates. The resulting $18$ output ports (OPs) are manipulated for approximating solutions to a PDE of interest through measurement operations with the $\sigma_z$ operator.

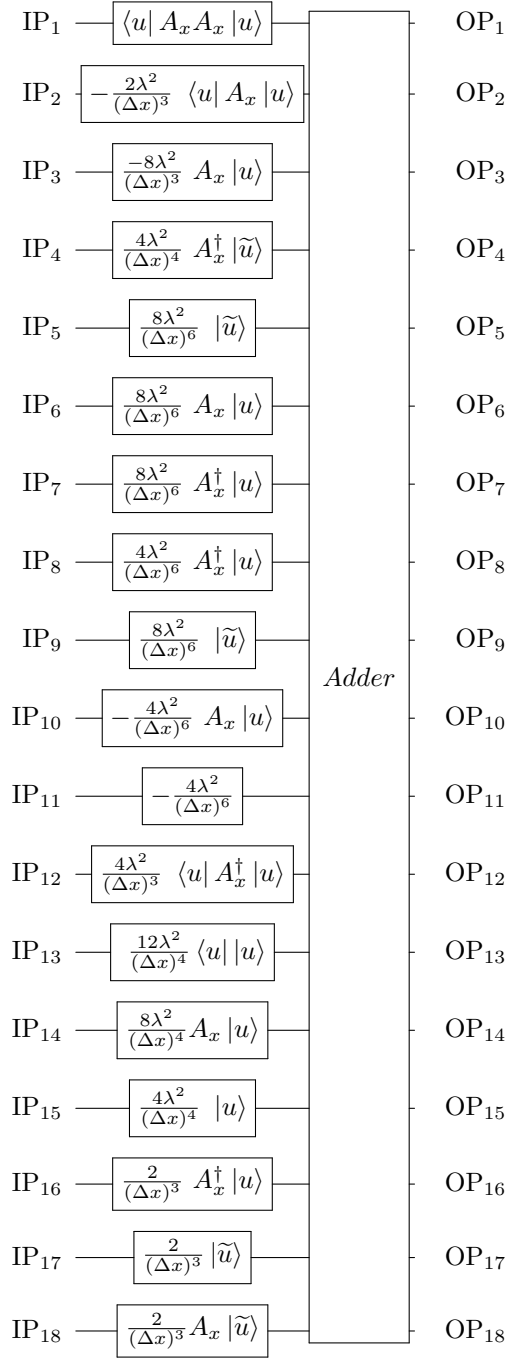
\begin{figure}
\begin{align*}
\Qcircuit @C=0.2em @R0.8em {
\lstick{ \mathrm{IP}_1       } & \gate{ \bra{u} A_x A_x \ket{u}  } &   \multigate{17}{Adder}  &    \ghost{Adder} & \lstick{\mathrm{OP}_1}     \\
\lstick{ \mathrm{IP}_2}  &   \gate{ - \frac{2 \lambda^2}{(\Delta x)^3} \text{ } \bra{u} A_x \ket{u}       } &   \ghost{Adder} & \ghost{Adder} & \lstick{\mathrm{OP}_2}        \\
\lstick{     \mathrm{IP}_3    } &   \gate{    \frac{-8 \lambda^2}{(\Delta x)^3} \text{ } A_x \ket{u}    }  & \ghost{Adder} & \ghost{Adder} & \lstick{\mathrm{OP}_3}      \\
\lstick{   \mathrm{IP}_4     } &  \gate{  \frac{4 \lambda^2}{(\Delta x)^4} \text{ } A^{\dagger}_x \ket{\widetilde{u}}  } &  \ghost{Adder} & \ghost{Adder}   & \lstick{\mathrm{OP}_4}    \\
\lstick{   \mathrm{IP}_5        } & \gate{             \frac{8 \lambda^2}{(\Delta x)^6} \text{ } \ket{\widetilde{u}}   }    &  \ghost{Adder} & \ghost{Adder}  & \lstick{\mathrm{OP}_5}     \\
\lstick{      \mathrm{IP}_6      } &   \gate{      \frac{8 \lambda^2}{(\Delta x)^6} \text{ } A_x \ket{u}          }   &  \ghost{Adder} & \ghost{Adder} & \lstick{\mathrm{OP}_6}      \\ \lstick{\mathrm{IP}_7} & \gate{  \frac{8 \lambda^2}{(\Delta x)^6} \text{ } A^{\dagger}_x \ket{u}     } &  \ghost{Adder} & \ghost{Adder} & \lstick{\mathrm{OP}_7}    \\ \lstick{\mathrm{IP}_8} & \gate{               \frac{4 \lambda^2}{(\Delta x)^6} \text{ } A^{\dagger}_x \ket{u}     } &  \ghost{Adder} & \ghost{Adder} & \lstick{\mathrm{OP}_8}    \\ \lstick{\mathrm{IP}_9} & \gate{         \frac{8 \lambda^2}{(\Delta x)^6} \text{ } \ket{\widetilde{u}}      } &  \ghost{Adder} & \ghost{Adder} & \lstick{\mathrm{OP}_9}   \\ \lstick{\mathrm{IP}_{10}} & \gate{  - \frac{4 \lambda^2}{(\Delta x)^6} \text{ } A_x \ket{u}     }&  \ghost{Adder} & \ghost{Adder} & \lstick{\mathrm{OP}_{10}}     \\ \lstick{\mathrm{IP}_{11}} & \gate{        - \frac{4 \lambda^2}{(\Delta x)^6}   } &  \ghost{Adder} & \ghost{Adder} & \lstick{\mathrm{OP}_{11}}     \\ \lstick{\mathrm{IP}_{12}} & \gate{      \frac{4 \lambda^2}{(\Delta x)^3}\text{ }  \bra{u} A^{\dagger}_x \ket{u} } &  \ghost{Adder} & \ghost{Adder}  & \lstick{\mathrm{OP}_{12}}      \\   \lstick{ \mathrm{IP}_{13} }  & \gate{  \text{ } \frac{12 \lambda^2}{(\Delta x)^4} \bra{u} \ket{u}   } &  \ghost{Adder} & \ghost{Adder} & \lstick{\mathrm{OP}_{13}}     \\ \lstick{\mathrm{IP}_{14}} & \gate{  \frac{8 \lambda^2}{(\Delta x)^4} A_x \ket{u}   }&  \ghost{Adder} & \ghost{Adder} & \lstick{\mathrm{OP}_{14}}     \\ \lstick{\mathrm{IP}_{15}} & \gate{      \frac{4 \lambda^2}{(\Delta x)^4} \text{ } \ket{u}    }  &  \ghost{Adder} & \ghost{Adder} & \lstick{\mathrm{OP}_{15}}       \\      \lstick{\mathrm{IP}_{16}} &  \gate{     \frac{2}{(\Delta x)^3} \text{ } A^{\dagger}_x \ket{u
} }  & \ghost{Adder} & \ghost{Adder} & \lstick{\mathrm{OP}_{16}}       \\   \lstick{\mathrm{IP}_{17}} &  \gate{         \frac{2}{(\Delta x)^3} \ket{\widetilde{u}}       }  & \ghost{Adder} & \ghost{Adder}  & \lstick{\mathrm{OP}_{17}}      \\   \lstick{\mathrm{IP}_{18}} &  \gate{ \frac{2}{(\Delta x)^3} A_x \ket{\widetilde{u}}  }  & \ghost{Adder} & \ghost{Adder}  & \lstick{\mathrm{OP}_{18}}      \\ }  \end{align*}
\caption{\textit{Camassa-Holm Quantum Nonlinear Processing Unit}. $A_x$, and $A^{\dagger}_x$, represent operators that are used for finite-differences, $\Delta x$ denotes the resolution of the Cartesian plane, and $\ket{u}, \ket{\widetilde{u}}, \bra{u}, \bra{\widetilde{u}}$ are Quantum states representing solutions of the C-H PDE [55].}
\end{figure}

\item[$\bullet$] \textit{Obtaining upper bounds through positive operator valued measurements, POVMs}. POVMs, as previously defined through the quantities,

\begin{align*}
\mathscr{P}\mathscr{O}\mathscr{V}\mathscr{M}_1 \equiv  \mathrm{Tr} \bigg[    \bigg[  \bar{\big( \mathcal{Q}_1 \big)_{q_1} \big( a_1 \big) } \otimes    \bar{\big( \mathcal{Q}_2 \big)_{q_2} \big( a_2 \big)  \bigg] \widetilde{\Psi_{r_{-i}, q_1,q_2}}}  \bigg]  \equiv \textbf{P}_{\mathcal{Q}_1\mathcal{Q}_2 \mathcal{Q}_3| r_{-i}, q_1 q_2 q_3 } \big( q_1, q_2 , q_3 \big) \text{, } \\ \\  \mathscr{P}\mathscr{O}\mathscr{V}\mathscr{M}_2 \equiv   \mathrm{Tr} \bigg[ \bigg[  \bar{\big( \mathcal{Q}_1 \big)_{q_1} \big( a_1 \big) } \otimes    \bar{\big( \mathcal{Q}_2 \big)_{q_2} \big( a_2 \big)}  \otimes    \bar{\big( \mathcal{Q}_3 \big)_{q_3} \big( a_3 \big)}  \bigg]   \widetilde{\Psi_{r_{-i}, q_1,q_2}}    \bigg]   \equiv \textbf{P}_{\mathcal{Q}_1\mathcal{Q}_2 \mathcal{Q}_3 \mathcal{Q}_4| r_{-i}, q_1 q_2 q_3 q_4 } \\ \big( q_1, q_2  , q_3, q_4  \big)  \text{,} \end{align*}

\begin{align*} \vdots \\  \mathscr{P}\mathscr{O}\mathscr{V}\mathscr{M}_N \equiv   \mathrm{Tr} \bigg[  \bigg[  \bar{\big( \mathcal{Q}_1 \big)_{q_1} \big( a_1 \big) } \otimes    \bar{\big( \mathcal{Q}_2 \big)_{q_2} \big( a_2 \big)}  \otimes    \bar{\big( \mathcal{Q}_3 \big)_{q_3} \big( a_3 \big)} \otimes \cdots \otimes   \bar{\big( \mathcal{Q}_N \big)_{q_N} \big( a_N \big)} \bigg]  \widetilde{\Psi_{r_{-i}, q_1,q_2, \cdots, q_N}} \bigg] \\  \equiv \textbf{P}_{\mathcal{Q}_1\mathcal{Q}_2 \mathcal{Q}_3 \times \cdots \times \mathcal{Q}_N| r_{-i}, q_1 q_2 q_3 \times \cdots \times q_N} \big( q_1, q_2 , q_3, q_4, \cdots, q_N  \big)  \text{,}
\end{align*}

\noindent continue to appear in forthcoming arguments, particularly for determining the structure of upper bounds in multiplayer game-theoretic settings. That is, in the same way that upper bounds, were obtained in,

\[   \left\{\!\begin{array}{ll@{}>{{}}l} 
\epsilon \equiv \epsilon \big( \xi \big) > 0 :  \underset{i \in \bar{C}}{\textbf{E}} \bigg| \bigg| \textbf{P}_{X_i Y_i R_i | \mathcal{E}}  - \textbf{P}_{X_i Y_i} \textbf{P}_{R_i | \mathcal{E}_i Y_i} \bigg| \bigg|_1 \leq \frac{7 \epsilon}{600} \text{, } \\   \epsilon \equiv \epsilon \big( \xi \big) > 0  :  \underset{i \in \bar{C}}{\textbf{E}} \bigg| \bigg|  \textbf{P}_{X_i Y_i R_i | \mathcal{E}}  - \textbf{P}_{X_i Y_i} \textbf{P}_{R_i | \mathcal{E}_i X_i}  \bigg| \bigg|_1 \leq \frac{7 \epsilon}{600}  \text{, } \\  \epsilon \equiv \epsilon \big( \xi \big) > 0  :  \underset{ i \in \bar{C}}{\textbf{E}} \underset{P_{X_i Y_i R_i | \mathcal{E}}}{\textbf{E}}     \bigg| \bigg| \ket{\varphi}\bra{\varphi}   - \big( \textbf{I} \otimes U_{y_i, r_i} \big) \bra{\varphi} \ket{\varphi}_{x_i y^* r_i} \big( \textbf{I} \otimes U^{\dagger}_{y_i r_i} \big) \bigg| \bigg|_1     \leq \frac{4 \epsilon}{5}         \text{, } 
\end{array}\right. 
\] 

\noindent which are dependent upon the anchoring probability, one can make use of POVMs, such as those reiterated for multiplayer games above, through the stochastic system,

\[  \mathscr{P}\mathscr{O}\mathscr{V} \mathscr{M} \mathscr{E} \equiv  \left\{\!\begin{array}{ll@{}>{{}}l} 
   \underset{I}{\textbf{E}} \bigg| \bigg|      \textbf{P}_{R_{-i}| W_C}   \textbf{P}_{\mathcal{Q}_1 \mathcal{Q}_2 \mathcal{Q}_3} \mathscr{P}\mathscr{O}\mathscr{V} \mathscr{M}_1 -   \textbf{P}_{\mathcal{Q}_1 \mathcal{Q}_2 \mathcal{Q}_3 R_{-i} \mathcal{A}_1 \mathcal{A}_2 \mathcal{A}_3 | W_C}           \bigg| \bigg| ,     \\     \underset{I}{\textbf{E}}  \bigg| \bigg|  \textbf{P}_{R_{-i}| W_C}   \textbf{P}_{\mathcal{Q}_1 \mathcal{Q}_2 \mathcal{Q}_3}  \mathscr{P}\mathscr{O}\mathscr{V} \mathscr{M}_2 - \textbf{P}_{\mathcal{Q}_1 \mathcal{Q}_2 \mathcal{Q}_3 \mathcal{Q}_4 R_{-i} \mathcal{A}_1 \mathcal{A}_2 \mathcal{A}_3 \mathcal{A}_4 | W_C}            \bigg| \bigg|  ,  \\ \vdots \\   \underset{I}{\textbf{E}}  \bigg| \bigg|  \textbf{P}_{R_{-i}| W_C}   \textbf{P}_{\mathcal{Q}_1 \mathcal{Q}_2 \mathcal{Q}_3}  \mathscr{P}\mathscr{O}\mathscr{V} \mathscr{M}_N - \textbf{P}_{\mathcal{Q}_1 \mathcal{Q}_2 \mathcal{Q}_3 \mathcal{Q}_4 \times \cdots \times \mathcal{Q}_N R_{-i} \mathcal{A}_1 \mathcal{A}_2 \mathcal{A}_3 \mathcal{A}_4 \times \cdots \times \mathcal{A}_N | W_C}            \bigg| \bigg|   . \end{array}\right. 
\]

\end{itemize}

\subsection{Connection with objects previously examined in multiplayer game-theoretic settings}

\subsubsection{Previously obtained error bounds for XOR, $\mathrm{XOR}^{*}$, and FFL, games}

\noindent The optimal value for the FFL game being \textit{invariant}, or, namely equal to itself under the operation of parallel repetition, allows for one to argue that error bounds of the following form should hold,

\begin{align*}
\underset{\mathcal{Q}_2 , \mathcal{Q}^{\prime}_2}{\underset{\mathcal{Q}_1 , \mathcal{Q}^{\prime}_2}{\sum}}      \bigg| \bigg|      \bigg[   \bigg(      \bigg( \bigg( \frac{A_i + A_j}{\sqrt{2}} \bigg) \wedge \bigg(  \frac{A_{i^{\prime}} + A_{j^{\prime}}}{\sqrt{2}} \bigg) \bigg) \otimes     \textbf{I}   \bigg)   -   \bigg(          \textbf{I} \otimes \bigg( B_{ij} \wedge B_{i^{\prime}j^{\prime}}      \bigg)                    \bigg)   \bigg] \ket{\psi_{\mathrm{FFL} \wedge \mathrm{FFL}}} \bigg| \bigg|  \text{, } \end{align*}

\noindent with the summation being taken over the total number of questions belonging to the sets $\mathcal{Q}_1$, and $\mathcal{Q}_2$, of the first and second players, respectively. In parallel repetition of multiplayer XOR games, the error bound of the form above should expect to generalize to the following form, for the tensor $C$ corresponding to the responses of the third player,

\begin{align*}
    \bigg| \bigg|  \bigg[  \bigg(   \big( A_i \wedge A_{i^{\prime}} \big) \bigotimes \textbf{I}      \bigotimes \bigg(    \bigg(   \underset{\mathcal{Q}_1, \mathcal{Q}_2 , \mathcal{Q}_3}{\prod}             C^{l_{ijk}}_{ijk}     \bigg) \wedge \bigg(    \underset{\mathcal{Q}^{\prime}_1, \mathcal{Q}^{\prime}_2 , \mathcal{Q}^{\prime}_3}{\prod}         C^{l^{\prime}_{i^{\prime}j^{\prime}k^{\prime}}}_{i^{\prime}j^{\prime}k^{\prime}}            \bigg)        \bigg) \bigotimes                \bigg(     \underset{1\leq k \leq n-3}{\bigotimes}      \textbf{I}_k       \bigg)      \bigg)     \\   -    \omega_{\mathrm{XOR} \wedge \cdots \wedge \mathrm{XOR}}    \bigg(       \pm \mathrm{sign} \big( i_1 , j_1 , k_1, i_{111}, \cdots , j_{111} , \cdots , j_{nm(n+m)} \big)    \mathrm{sign} \big(  i^{\prime}_1 , j^{\prime}_1 , k^{\prime}_1, i^{\prime}_{111}, \end{align*}
    
    \begin{align*}  \cdots , j^{\prime}_{111} , \cdots , j^{\prime}_{nm(n+m)}          \big)     \bigg) \bigg[                            \textbf{I} \bigotimes \textbf{I} \bigotimes \bigg(           \bigg(   \underset{\mathcal{Q}_3, \mathcal{Q}_3 \oplus 1 \equiv \mathcal{Q}_3 + 1}{\underset{\mathcal{Q}_2 , \mathcal{Q}_2 \oplus 1 \equiv \mathcal{Q}_2 + 1 }{\underset{\mathcal{Q}_1, \mathcal{Q}_1 \oplus 1 \equiv \mathcal{Q}_1 +1  }{\prod}}}             C^{l_{ijk}}_{ijk}     \bigg) \wedge \bigg(  \underset{\mathcal{Q}^{\prime}_3, \mathcal{Q}^{\prime}_3 \oplus 1 \equiv \mathcal{Q}^{\prime}_3 + 1}{\underset{\mathcal{Q}^{\prime}_2 , \mathcal{Q}^{\prime}_2 \oplus 1 \equiv \mathcal{Q}^{\prime}_2 + 1 }{\underset{\mathcal{Q}^{\prime}_1, \mathcal{Q}^{\prime}_1 \oplus 1 \equiv \mathcal{Q}^{\prime}_1 +1  }{\prod}}}       C^{l^{\prime}_{i^{\prime}j^{\prime}k^{\prime}}}_{i^{\prime}j^{\prime}k^{\prime}}            \bigg)         \bigg) \\  \bigotimes \bigg( \bigg(   \underset{1\leq k \leq n-3}{\prod}  \textbf{I}_k \bigg)      \bigg)             \bigg] \bigg) \bigg]          \ket{\psi_{\mathrm{XOR} \wedge \cdots \wedge \mathrm{XOR}}}   \bigg| \bigg|            \text{. }
\end{align*}

\noindent Primal feasible solutions $Z \equiv Z_{2\mathrm{XOR}}$ to semidefinite programs for the 2$\mathrm{XOR}$, and $\mathrm{FFL}$, games can also be formulated with the 3$\mathrm{XOR}$ game. In place of two-player game matrices, primal feasible solutions $Z^{\prime} \equiv Z_{3\mathrm{XOR}}$ three-player objects, which in the case of the 3-$\mathrm{XOR}$ game matrix $G$, satisfies,

\begin{align*}
    G_{3 \mathrm{XOR}} \approx \frac{1}{{n\choose 3}}                            \underset{1 \leq k \leq r-(n+m)}{\underset{1 \leq j \leq n+m}{\underset{1 \leq i \leq n}{\sum}}}                   \bigg[   \ket{i} \bigg(   \bra{ij} \bra{ijk}  \bigg)     +    \ket{j} \bigg( \bra{ji} \bra{ijk} \bigg) + \ket{k} \bigg(   \bra{ki} \bra{kij}      \bigg)   +     \ket{i} \\ \times  \bigg( \bra{ik} \bra{ijk} \bigg)   +  \ket{k} \bigg(   \bra{ik} \bra{ijk}     \bigg) + \ket{k} \bigg( \bra{ki} \bra{ijk} \bigg) +       \ket{k} \bigg( \bra{ij} \bra{ijk} \bigg) + \ket{k} \bigg( \bra{ji } \bra{ijk} \bigg)     \\  +             \bra{k} \bigg( \bra{jk} \bra{ijk} \bigg) + \ket{k} \bigg( \bra{kj} \bra{ijk} \bigg)       + \ket{j} \bigg( \bra{jk} \bra{ijk} \bigg) +    \ket{i} \bigg( \bra{ki} \bra{ijk} \bigg) \\ + \ket{i} \bigg( \bra{ji} \bra{ijk} \bigg)  +  \ket{i} \bigg( \bra{ij} \bra{jik} \bigg) + \ket{j} \bigg( \bra{ji} \bra{ijk} \bigg)  +        \ket{k} \bigg( \bra{ki} \bra{ikj} \bigg) \\ + \ket{i} \bigg( \bra{ik} \bra{kij} \bigg)  +  \ket{j} \bigg( \bra{jk} \bra{kij}        \bigg)    + \ket{k} \bigg( \bra{kj} \bra{jki} \bigg) + \ket{k} \bigg( \bra{ki} \bra{kij} \bigg) \\                  + \text{Higher order permutations}             \bigg]               
\end{align*}

\noindent as a generalization of the 2-$\mathrm{XOR}$ game matrix,

\begin{align*}
  \frac{1}{4 {n \choose 2}}   \underset{1\leq i \leq j \leq n}{\sum}   \bigg(    \ket{i}\bra{ij}  +   \ket{j} \bra{ij} + \ket{i} \bra{ji} - \ket{j} \bra{ji}             \bigg)   \text{,}
\end{align*}

\noindent are of the form,

\begin{align*}
      \underset{\exists c_i \in \textbf{R}:  F_i \cdot G \equiv c_i   }{\underset{\forall Z^{\prime} \succcurlyeq 0  \text{, } 1 \leq i \leq 3 , }{\mathrm{sup} }} \big[  G Z^{\prime} \big]   \text{.}
\end{align*}

\noindent Given the primal feasible solution from the constrained optimization procedure above over the number of players in the game, $i$, there exists another semidefinite program,

\begin{align*}
     \underset{\exists c_i \in \textbf{R}:  F_i \cdot G \equiv c_i   }{\underset{\forall Z^{\prime} \succcurlyeq 0  \text{, } 1 \leq i \leq 3 , }{\mathrm{sup} }} \bigg[  {\sum}   \big( y_i F_i - G \big) Z^{\prime}  \bigg]   \text{, }
\end{align*}

\noindent corresponding to the duality gap between $y_i F_i$, and $G$, with the primal feasible 3-$\mathrm{XOR}$ solution. Primal feasible solutions obtained by semidefinite programs, such as those above, have previously been characterized by the author, [44], in the simpler 2-player setting, for $\mathrm{XOR}^{*}$ and $\mathrm{FFL}$ games, by  extending the construction of error bounds and intertwining operations provided in [37]. Equipped with a restriction of feasible $F_i$ of $F$, under the assumption that there exists a primal feasible solution $Z^{\prime}$ that can be approximated with polynomial runtime, semidefinite programs for approximating the primal feasible solution, and duality gap, are well posed.

Such representations of semidefinite programs for 3-XOR primal feasible solutions determine the structure of primal feasible solutions under ordinary, and strong, parallel repetition. Under such an operation with an arbitrary number of repetitions, the primal feasible solution, $Z_{\mathrm{XOR}\wedge \cdots \wedge \mathrm{XOR}}$, for the semidefinite program,

\begin{align*}
  \underset{\forall Z_{\mathrm{XOR}\wedge \cdots \wedge \mathrm{XOR}} \succcurlyeq 0 , 1 \leq i \leq m, F^{(j)}_{\wedge i} \cdot Z_{\mathrm{XOR} \wedge \cdots \wedge {\mathrm{XOR}}} \equiv C^{(j)}_i}{\mathrm{sup}}  G_{\mathrm{XOR}\wedge \cdots \wedge \mathrm{XOR}} Z_{\mathrm{XOR}\wedge \cdots \wedge \mathrm{XOR}} \text{, }
\end{align*}

\noindent for some strictly positive $C^{(j)}_i$, which admits the decomposition,

\begin{align*}
  \underset{1 \leq j \leq n}{\bigwedge} \bigg[  \underset{\forall Z_{\mathrm{XOR}\wedge \cdots \wedge \mathrm{XOR}} \succcurlyeq 0 , 1 \leq i \leq m, F^{(j)}_{\wedge i} \cdot Z_{\mathrm{XOR} \wedge \cdots \wedge {\mathrm{XOR}}} \equiv C^{(j)}_i}{\mathrm{sup}}    G_{\mathrm{XOR}} Z_{\mathrm{XOR}}      \bigg]       \text{. }
\end{align*}

\noindent The primal feasible solution itself admits the decomposition,

\begin{align*}
      Z_{\mathrm{XOR} \wedge \cdots \wedge \mathrm{XOR}} \equiv \underset{1 \leq j \leq n}{\bigwedge}  Z^{(j)}_{\mathrm{XOR}}    \text{. }
\end{align*}

\noindent Under strong parallel repetition, the duality of the semidefinite program above is equivalent to the minimization,

\begin{align*}
\underset{1 \leq j \leq n}{\bigwedge} \bigg[ \underset{ \underset{1 \leq j \leq n}{\wedge}   \underset{\mathcal{Q}^{(1)}_N \wedge \cdots \wedge \mathcal{Q}^{(i)}_N \wedge  \cdots \wedge \mathcal{Q}^{(n)}_N}{\underset{\vdots}{{\underset{\mathcal{Q}^{(1)}_1 \wedge \cdots \wedge \mathcal{Q}^{(i)}_N \wedge \cdots  \wedge \mathcal{Q}^{(n)}_1}{\sum}}}}    y_{\mathcal{Q}^{(j)}_i} E^{(j)}_{ii}     \succcurlyeq  G^{(j)}_{\mathrm{Sym}}          }{\mathrm{inf}}  \bigg(    \underset{\mathcal{Q}^{(1)}_N \wedge \cdots \wedge \mathcal{Q}^{(i)}_N \wedge \cdots \wedge \mathcal{Q}^{(n)}_N}{\underset{\vdots}{\underset{\mathcal{Q}^{(1)}_1 \wedge \cdots \wedge \mathcal{Q}^{(i)}_N \wedge \cdots \wedge \mathcal{Q}^{(n)}_1}{\sum}}}       y_{\mathcal{Q}^{(j)}_i}   \bigg)         \bigg]  \text{, }
\end{align*}

\noindent where, under the constraints placed on the infimum above, are,

\begin{align*}
     G_{\mathrm{Sym}, \mathrm{XOR} \wedge \cdots \wedge \mathrm{XOR}} \equiv \underset{1 \leq j \leq n}{\bigwedge}  G^{(j)}_{\mathrm{Sym}, \mathrm{XOR}} \equiv  \underset{1 \leq j \leq n}{\bigwedge}          \begin{bmatrix}  0 & \big( G^{(j)}_{\mathrm{XOR}} \big)^{\textbf{T}} \\ G^{(j)}_{\mathrm{XOR}} & 0  \end{bmatrix}       \\ \equiv \begin{bmatrix} 0 &  \underset{1 \leq j \leq n}{\wedge}  \big( G^{(j)}_{\mathrm{XOR}} \big)^{\textbf{T}} \\ \underset{1 \leq j \leq n}{\wedge} G^{(j)}_{\mathrm{XOR}} & 0   \end{bmatrix} \text{. }
\end{align*}

\noindent We denote the dual feasible solution with $V_{\mathrm{Dual},\mathrm{XOR} \wedge \cdots \wedge \mathrm{XOR}}$.

Besides ordinary, and strong, parallel repetition of primal feasible solutions to semidefinite programs, the 2-$\mathrm{XOR}$ game, intertwining operations play a fundamental role in transforming representations of responses of one player to those of another player.

\subsubsection{Inequalities with suitable linear operators}

In three-player settings, the accompanying operation, $T^{\prime} \equiv T_{3\mathrm{XOR}}$, has the following collection of actions,

\begin{align*}
 \bigg| \bigg| \bigg[  \bigg( T^{\prime} \otimes B_{ij} \otimes C_{ijk} \bigg) - \frac{1}{\sqrt{2}}              \bigg( \bigg( \widetilde{B_{ij}} \otimes T^{\prime}  \otimes C_{ijk}  \bigg)+ \bigg( T^{\prime} \otimes  \widetilde{C_{ijk}} \otimes B_{ij} \bigg) \bigg)  \bigg] \ket{\psi_{3\mathrm{XOR}}}   \bigg| \bigg|   \text{, } \\ \\   \bigg| \bigg| \bigg[  \bigg( A_i \otimes  T^{\prime}  \otimes C_{ijk} \bigg) - \frac{1}{\sqrt{2}}              \bigg( \bigg( T^{\prime} \otimes \widetilde{A_i }  \otimes C_{ijk}  \bigg)+ \bigg( \widetilde{C_{ijk}} \otimes  T^{\prime}   \otimes A_i \bigg) \bigg)  \bigg] \ket{\psi_{3\mathrm{XOR}}}   \bigg| \bigg|            \text{, } \end{align*}

 \begin{align*} \bigg| \bigg| \bigg[  \bigg( A_i \otimes  B_{ij} \otimes T^{\prime} \bigg) - \frac{1}{\sqrt{2}}              \bigg( \bigg( T^{\prime} \otimes \widetilde{A_i }  \otimes B_{ij}  \bigg)+ \bigg( \widetilde{B_{ij}} \otimes  T^{\prime}   \otimes A_i  \bigg) \bigg)  \bigg] \ket{\psi_{3\mathrm{XOR}}}   \bigg| \bigg|           \text{, }
\end{align*}

\noindent where the linear operator is a mapping of the form,

\begin{align*}
      T^{3 \text{ } \mathrm{XOR}} :  \textbf{C}^{3 \lceil \frac{n}{3} \rceil } \otimes      \textbf{C}^{3 \lceil \frac{n}{3} \rceil }   \otimes  \textbf{C}^{3 \lceil \frac{n}{3} \rceil }      \longrightarrow        \textbf{C}^{d_A} \otimes \textbf{C}^{d_B} \otimes \textbf{C}^{d_C}    \text{,  }
\end{align*}

\noindent where $d_A, d_B$ and $d_C$ represent the dimensions of the Hilbert spaces for each of the three active players, in the same way that the suitable linear operator, $T$, introduced in [37] for upper bounding the Frobenius norm associated with the strategy of each player, is a mapping of the form,

\begin{align*}
    T^{\mathrm{XOR}} :  \textbf{C}^{2 \lceil \frac{n}{2} \rceil } \otimes      \textbf{C}^{2 \lceil \frac{n}{2} \rceil }   \longrightarrow        \textbf{C}^{d_A} \otimes \textbf{C}^{d_B}      \text{. }
\end{align*}

\noindent Associated mappings under ordinary, and strong, parallel repetition take the following forms,

\begin{align*}
 \bigotimes T^{\mathrm{XOR}} : \underset{N \text{ copies}}{\bigotimes} \bigg(  \textbf{C}^{2 \lceil \frac{n}{2} \rceil } \bigg)  \longrightarrow        \underset{1 \leq i \leq N}{\bigotimes} \bigg( \textbf{C}^{d_i} \bigg) \text{, }  \\  \bigwedge T^{\mathrm{XOR}} :  \textbf{C}^{2 \lceil \frac{n}{2} \rceil } \bigwedge      \textbf{C}^{2 \lceil \frac{n}{2} \rceil }  \bigwedge \cdots \bigwedge  \textbf{C}^{2 \lceil \frac{n}{2} \rceil }  \longrightarrow        \textbf{C}^{d_A} \bigwedge \textbf{C}^{d_B} \bigwedge \textbf{C}^{d^{(1)}_B} \bigwedge\cdots \bigwedge \textbf{C}^{d^{(n-2)}_B}  \text{, } \\     T^{\mathrm{XOR}\wedge \cdots \wedge \mathrm{XOR}} :     \textbf{C}^{2 \lceil \frac{n}{2} \rceil \wedge 2 \lceil \frac{n}{2} \rceil \wedge \cdots \wedge 2 \lceil \frac{n}{2} \rceil}  \longrightarrow        \textbf{C}^{d_A \wedge d_B \wedge d^{(1)}_B \wedge \cdots \wedge {d^{(n-2)}_B }}     \text{, } \\      T^{\mathrm{FFL}\wedge  \mathrm{FFL}} :     \textbf{C}^{2 \lceil \frac{n}{2} \rceil \wedge 2 \lceil \frac{n}{2} \rceil}  \longrightarrow        \textbf{C}^{d_A \wedge d_B }     \text{. }
\end{align*}

\noindent Determining precise rates of decay for the quantum value under strong parallel repetition remains of strong interest to examine. Under an arbitrary number of parallel repetition operations $n$, the value of expanding games, $G_{\mathrm{Exp}}$ the corresponding rate of decay of the optimal value has previously examined, [16], which takes the form,

\begin{align*}
    \mathrm{val} \big( G_{\mathrm{Exp}}^{\otimes n} \big) \leq \mathrm{exp} \bigg(  - \frac{c \epsilon^5 n}{ k^2\mathrm{log} \big| A \big|}       \bigg) \text{, }
\end{align*}

\noindent where $\big| A \big|$ denotes the alphabet size of the players of the expanding game, and strictly positive parameters $c \neq \epsilon \neq k$, and $n \equiv n \big( \epsilon \big)$. Depending upon the assumptions that are placed in the rate of exponential decay for the expanding game optimal value, the power of the exponential can be dependent upon several other contributions, some of which can depend upon whether the game is free, anchored to strictly positive parameter $\alpha$, or connected. In expanded games, the notion of connectedness is determined by constructing the vertices and edges of a hypergraph, that is obtained from randomly flipping bits, or individual responses, from each player while holding all of the other responses from the remaining players fixed.

Denote the Frobenius norm,

\begin{align*}
  \big|\big| A \big|\big|_F \equiv \sqrt{\overset{m}{\underset{i=1}{\sum}} \overset{n}{\underset{j=1}{\sum}} \big| a_{ij} \big|^2 } = \sqrt{\mathrm{Tr} \big[ A^{\dagger} A \big] }  \text{, } 
\end{align*}

\noindent of an $m \times n$ matrix $A$ with entries $a_{ij}$, there exists a \textit{linear bijection} $\mathcal{L}$ between the tensor product space, $\textbf{C}^{d_A} \otimes \textbf{C}^{d_B}$, and the space of $d_A \times d_B$ matrices with complex entries, $\mathrm{Mat}_{d_A , d_B} \big( \textbf{C} \big)$, satisfying (\textbf{Lemma} \textit{1}, [42]),

\begin{itemize}
\item[$\bullet$] \underline{\textit{Image of the tensor product of two quantum states under} $\mathcal{L}$}: $\forall \ket{u} \in \textbf{C}^{d_A}, \ket{w} \in \textbf{C}^{d_B}, \exists \ket{u^{*}} \in \textbf{C}^{d_B} : \mathcal{L} \big( \ket{u} \otimes \ket{w} \big) = \ket{u} \bra{u^{*}}  \text{, }$ 
\item[$\bullet$] \underline{\textit{Product of a matrix with the image of a quantum state under} $\mathcal{L}$}: $\forall \ket{u} \in \textbf{C}^{d_A}, \exists A \in \mathrm{Mat}_{d_A} \big( \textbf{C} \big) : A \mathcal{L} \big( \ket{u} \big) = \mathcal{L} \big( A \otimes I \ket{u} \big)\text{, }$
\item[$\bullet$] \underline{\textit{Product of the image of a quantum state under $\mathcal{L}$ with the transpose of a matrix}}:  $\forall \ket{w} \in \textbf{C}^{d_B}, \exists B \in \mathrm{Mat}_{d_B} \big( \textbf{C} \big) : \mathcal{L} \big( \ket{w} \big) B^T = \mathcal{L} \big( I \otimes B \ket{w} \big)  \text{, }$
\item[$\bullet$] \underline{\textit{Frobenius norm equality}}: $\forall \ket{w} \in \textbf{C}^{d_B} : \big|\big| \mathcal{L} \big(   \ket{w}     \big) 
 \big|\big|_F = \ket{w}  \text{. } $
\end{itemize}

\noindent As alluded to when introducing the error bounds, and operator $T^{\prime}$ associated with the 3-$\mathrm{XOR}$ game, there are a wide variety of error bounds that intimately depend upon the optimal value of a game under consideration. One such inequality includes,

\begin{align*}
 \bigg| \bigg|    \bigg[             \bigg( \bigg( \underset{1 \leq i \leq n}{\prod}  A^{j_i}_i \bigg) \bigotimes \bigg( \underset{1 \leq k \leq n-1}{\bigotimes} \textbf{I}_k \bigg) \bigg)   -  \bigg(  \omega_{N\mathrm{XOR}} \bigg(         \pm \mathrm{sign} \big( i_1, \cdots, i_n \big)              \\ \times   \bigg( \underset{1 \leq i \leq n}{\prod}  A^{j_i}_i \bigg)           \bigg)     \bigotimes \bigg( \underset{1 \leq k \leq n-1}{\bigotimes} \textbf{I}_k \bigg)\bigg)                       \bigg]  \ket{\psi_{N\mathrm{XOR}}}        \bigg| \bigg|_F    <    \bigg( n_1 +       \big(    n_1 + 2    \big) \omega_{N\mathrm{XOR}}^{-1}                \bigg) \\ \times  n^N \sqrt{\epsilon}        \text{,}         \\ \\  \bigg| \bigg|    \bigg[ \bigg(  \textbf{I} \bigotimes            \bigg( \underset{1 \leq i \leq n}{\prod}  A^{1,j_i}_i \bigg) \bigotimes \bigg( \underset{1 \leq k \leq n-2}{\bigotimes} \textbf{I}_k \bigg) \bigg)   -  \bigg(  \textbf{I} \bigotimes  \bigg( \omega_{N\mathrm{XOR}} \bigg(         \pm \mathrm{sign} \big( i_1, j_1, \\ \cdots, i_n, \cdots, j_n \big)             \bigg( \underset{1 \leq i_2 \leq m}{\underset{1 \leq i_1 \leq n}{\prod}}  A^{1,j_{i_1,i_2}}_{i_1,i_2} \bigg)           \bigg)  \bigg)    \bigotimes \bigg( \underset{1 \leq k \leq n-2}{\bigotimes} \textbf{I}_k \bigg)\bigg)                       \bigg]  \ket{\psi_{N\mathrm{XOR}}}       \bigg| \bigg|_F  <   \bigg( n_2 +  \big( n_2 + 2 \big) \\ \times  \omega_{N\mathrm{XOR}}^{-1}            \bigg)  n^N \sqrt{\epsilon}  \\  \vdots \\ \bigg| \bigg|    \bigg[             \bigg(\bigg( \underset{1 \leq k \leq n-1}{\bigotimes} \textbf{I}_k \bigg) \bigotimes   \bigg( \underset{1 \leq i \leq n}{\prod}  A^{(n-1),j_{i_1,\cdots,i_n}}_{i_1,\cdots, i_n} \bigg)  \bigg)   -  \bigg( \bigg( \underset{1 \leq k \leq n-1}{\bigotimes} \textbf{I}_k \bigg)  \bigotimes \bigg(    \omega_{N\mathrm{XOR}} \\ \times \bigg(         \pm \mathrm{sign} \big( i_1, \cdots, i_n,   j_1, \cdots, j_n \big)                         \bigg)    \bigg( \underset{1 \leq i \leq n}{\prod}  A^{(n-1),j_{i_1,\cdots,i_n}}_{i_1,\cdots, i_n} \bigg)  \bigg)                 \bigg)   \bigg] \\ \times \ket{\psi_{N\mathrm{XOR}}}    \bigg| \bigg|_F     <   \bigg( n_N        +   \big(  n_N + 2\big) \omega_{N\mathrm{XOR}}^{-1}            \bigg) n^N \sqrt{\epsilon}  \text{, } \tag{*}
\end{align*}

\noindent corresponding to a system of Frobenius norm inequalities, for strictly positive integers $n_1, n_2, \cdots, n_N$, and $\epsilon$ taken to be sufficiently small. In the 2-player setting, the system of Frobenius norm inequalities above is related to the system,

\begin{align*}
     \forall i,  \big|\big|  \big( A_i \otimes \textbf{I} \big) T^{\mathrm{FFL}}     -          T^{\mathrm{FFL}} \big( \widetilde{A_i} \otimes \textbf{I} \big)    \big|\big|_{\mathrm{F}} < 9  n^2 \sqrt{\epsilon_{\mathrm{FFL}} } \big| \big|     T^{\mathrm{FFL}}   \big|\big|_{\mathrm{F}}     \text{, } \\ \forall j \neq k,  \big|\big|    \big( \textbf{I} \otimes B_{jk} \big) T^{\mathrm{FFL}}   - T^{\mathrm{FFL}} \big( \textbf{I} \otimes \widetilde{B_{jk}}    \big) \big|\big|_{\mathrm{F}} <   \frac{44}{3} n^2 \sqrt{\epsilon_{\mathrm{FFL}} }  \big| \big|     T^{\mathrm{FFL}}   \big|\big|_{\mathrm{F}} \text{, }
\end{align*}

\noindent of inequalities with respect to the Frobenius norm, $\big| \big| \cdot \big| \big|_F$,

\begin{align*}
  \big|\big| A \big|\big|_F \equiv \sqrt{\overset{m}{\underset{i=1}{\sum}} \overset{n}{\underset{j=1}{\sum}} \big| a_{ij} \big|^2 } = \sqrt{\mathrm{Tr} \big[ A^{\dagger} A \big] }  \text{, } 
\end{align*}

\noindent of an $m \times n$ matrix $A$ with entries $a_{ij}$, for,

\begin{align*}
A \cdot B \equiv  \mathrm{Tr} \big( A B \big)  \equiv     \underset{\text{columns }i, \text{ rows } j}{\underset{ij}{\sum}}   A_{ij} B_{ij}      \text{, }
\end{align*}

\noindent given another matrix $B$.

\subsubsection{Optimal values under parallel repetition}

Each one of the inequalities for the FFL game above is upper bounded with a suitable linear operator $T^{\mathrm{FFL}}$ for the FFL game, with $\epsilon_{\mathrm{FFL}}$ taken to be sufficiently small. The transformation, $\widetilde{A_i}$, and $\widetilde{B_{jk}}$, of Alice's and Bob's tensor observables, respectively, is given by the action,

\begin{align*}
 \widetilde{\cdot} : M \otimes N \longrightarrow N \otimes M   \text{, }
\end{align*}

\noindent for $N \neq M > 0$.

Determining how rates of exponential decay for parallel repetition of the expanding game with optimality is of interest to explore. From previous work of the author, some connections between error bounds, of the form indicated on the previous page, with the optimal value under parallel repetition were obtained [46]. The arguments for obtaining such error bounds, and generalizations of error bounds, can not only depend upon the rate of exponential decay for multiplayer parallel repetition, but also upon the number of operations in tensor product representations. Such representations are obtained from incorporating all possible responses from each player, which in the $3$-$\mathrm{XOR}$ setting take the form,

     \begin{align*}
         \bigotimes \text{Player tensor observables} \equiv \bigg( \text{Alice's observables} \bigg) \bigotimes \bigg( \text{Bob's observables} \bigg) \bigotimes  \bigg(\text{Cleo's} \\ \text{observables} \bigg)   \text{, }
     \end{align*}

\noindent for players Alice, Bob and Cleo. Beyond three players, in more complicated multiplayer settings one must determine how the space of, potentially combinatorially large, space of optimal strategies is constrained by the rate of decay of the optimal value under ordinary, and strong, parallel repetition. In multiplayer XOR, and through the following duality notions:

\begin{itemize}
    \item[$\bullet$] \big(\underline{The $\epsilon$ bit $\mathrm{XOR}$ game, \textbf{Lemma} \textit{1}, [10]}\big) An $\mathrm{XOR}$ game for which $\mathrm{min} \big\{ \big| \mathcal{S} \big| ,    \big| \mathcal{T} \big|   \big\} \leq 4$        is an $\epsilon$-bit $\mathrm{XOR}$ game.          

\bigskip 
        \item[$\bullet$] \big(\underline{ Classical and quantum bounds for $\mathrm{XOR}$  and $\mathrm{{XOR}^{*}}$ games ,\textbf{Theorem} \textit{2}, [10]}\big). Denote $a$ and $b$ as the two possible measurements that Alice and Bob can observe from some $s \in \mathcal{S}$ and $t \in \mathcal{T}$. Furthermore, denote the single qubit measurements from each possible $s$ and $t$ with $A_{a | s}$ and $B_{b|t}$, and the probability, conditional upon each input, as, $P \big(    a , b \big| s , t        \big)$, which can be expressed with the trace of the inner product $\big(  A_{a|s}       \otimes    B_{b|t}     \big) \ket{\psi} \bra{\psi} $. The output of the XOR game, $m = a \oplus b$, is such that the classical and quantum bounds of the $\mathrm{XOR}$ and $\mathrm{XOR^{*}}$games are equal.
\end{itemize}

\noindent of the $\mathrm{XOR}^{*}$ game, one can establish, roughly, qualitative rates of decay for,

\begin{align*}
  \omega_{\mathrm{XOR}^{\otimes n}} \big( G \big) \equiv \omega \big( \mathrm{XOR}^{\otimes n}   \big)    \equiv \mathrm{val} \big( \mathrm{XOR}^{\otimes n} \big)   \text{,}
\end{align*}

\noindent and, hence, for,

\begin{align*}
    \omega_{(\mathrm{XOR}^{*} ) ^{\otimes n}} \big( G \big) \equiv \omega \big( (\mathrm{XOR}^{*} )^{\otimes n} \big)    \equiv \mathrm{val} \big( (\mathrm{XOR}^{*} )^{\otimes n}\big)   \text{.}
\end{align*}

\noindent In the next subsection, we provide a statement of results which incorporate the exponential rate of decay for parallel repetition of the optimal value. In comparison to several inequalities for error bounds, and generalized error bounds, in parallel repetition of multiplayer settings, [52], as extensions of error bounds from 2-player settings in the XOR and CHSH games, [42], and later further examined, [50], exponential rates of decay for parallel repetition of the optimal value restrict the range of possible optimal strategies that players can pursue.

\subsubsection{Error bounds under strong parallel repetition}

\noindent We state the collection of results from the 2-player setting which depend on the optimal value, and then discuss how exponential rates of decay for parallel repetition of the optimal value come into play. To determine the inequalities that one would like to bound from parallel repetition of multiplayer games, in the XOR setting one considers the following system,

\begin{align*}
 \bigg| \bigg|    \bigg[             \bigg( \bigg( \underset{1 \leq i \leq n}{\prod}  A^{j_i}_i \bigg) \bigotimes \bigg( \underset{1 \leq k \leq n-1}{\bigotimes} \textbf{I}_k \bigg) \bigg)   -  \bigg(  \omega^{\prime} \bigg(         \pm \mathrm{sign} \big( i_1, \cdots, i_n \big)                \bigg( \underset{1 \leq i \leq n}{\prod}  A^{j_i}_i \bigg)           \bigg)    \\ \bigotimes \bigg( \underset{1 \leq k \leq n-1}{\bigotimes} \textbf{I}_k \bigg)\bigg)                       \bigg] \ket{\psi^{\prime}}       \bigg| \bigg|_F   <    \bigg(  \big( n_1 \big)^{\wedge} +       \big(   \big(  n_1 \big)^{\wedge} + 2    \big) \big( \omega^{\prime} \big)^{-1}                \bigg)\big(  n^N \big)^{\wedge } \sqrt{\epsilon^{\wedge}}        \text{,}   \end{align*}

 \begin{align*}    \bigg| \bigg|    \bigg[ \bigg(  \textbf{I} \bigotimes            \bigg( \underset{1 \leq i \leq n}{\prod}  A^{1,j_i}_i \bigg) \bigotimes \bigg( \underset{1 \leq k \leq n-2}{\bigotimes} \textbf{I}_k \bigg) \bigg)   -  \bigg(  \textbf{I} \bigotimes   \bigg( \omega^{\prime}     \bigg(         \pm \mathrm{sign} \big( i_1, j_1, \cdots, i_n, \\ \cdots, j_n \big)          \bigg( \underset{1 \leq i_2 \leq m}{\underset{1 \leq i_1 \leq n}{\prod}}  A^{1,j_{i_1,i_2}}_{i_1,i_2} \bigg)           \bigg)  \bigg)   \bigotimes \bigg( \underset{1 \leq k \leq n-2}{\bigotimes} \textbf{I}_k \bigg)\bigg)                       \bigg]   \ket{\psi^{\prime}}    \bigg| \bigg|_F    \\  <   \bigg( \big( n_2  \big)^{\wedge} +  \big( \big( n_2 \big)^{\wedge} + 2 \big) \big( \omega^{\prime} \big)^{-1}               \bigg) \big( n^N  \big)^{\wedge} \sqrt{\epsilon^{\wedge}}  \\  \vdots \\  \bigg| \bigg|    \bigg[             \bigg(\bigg( \underset{1 \leq k \leq n-1}{\bigotimes} \textbf{I}_k \bigg) \bigotimes   \bigg( \underset{1 \leq i \leq n}{\prod}  A^{(n-1),j_{i_1,\cdots,i_n}}_{i_1,\cdots, i_n} \bigg)  \bigg)   -  \bigg( \bigg( \underset{1 \leq k \leq n-1}{\bigotimes} \textbf{I}_k \bigg) \bigotimes \bigg(    \omega^{\prime}  \\ \times  \bigg(         \pm \mathrm{sign} \big( i_1, \cdots, i_n, j_1, \cdots, j_n \big)                         \bigg)   \bigg( \underset{1 \leq i \leq n}{\prod}  A^{(n-1),j_{i_1,\cdots,i_n}}_{i_1,\cdots, i_n} \bigg)  \bigg)                 \bigg)   \bigg] \ket{\psi^{\prime}}     \bigg| \bigg|_F   <   \bigg(\big(  n_N  \big)^{\wedge }    \\    +   \big( \big(  n_N \big)^{\wedge} + 2\big)  \big( \omega^{\prime} \big)^{-1}              \bigg) \big(  n^N \big)^{\wedge } \sqrt{\epsilon^{\wedge}}  \text{, }
\end{align*}

\noindent of Frobenius norms, which serves as a counterpart to the system of norms provided in (*). Upper bounds for parallel repetition of operators in Frobenius norms above are used to obtain the following results in parallel repetition settings.

\bigskip

\noindent \textbf{Lemma} \textit{7} (\textit{second error bound}, \textit{6.6}, [42]). From previously defined quantities, one has,

\begin{align*}
 \bigg| \bigg|  \bigg( \bigg( \underset{1 \leq i \leq n}{\prod}   A^{j_i}_i \bigg)   \otimes B_{kl} \bigg)  \ket{\psi_{\mathrm{FFL}}}  - \frac{2}{3} \bigg[ \pm \bigg( \mathrm{sign} \big( i , j_1 , \cdots , j_n \big) \bigg[ \text{ } \bigg(     \underset{i = j_k + 1 , \text{ } \mathrm{set} \text{ } j_k + 1 \equiv j_k \oplus 1 }{\underset{1 \leq i \leq n}{\prod}}     A^{j_i}_i    \bigg) \\ + \bigg( \underset{i = j_l + 1 , \text{ } \mathrm{set} \text{ } j_l + 1 \equiv j_l \oplus 1 }{\underset{1 \leq i \leq n}{\prod}}     A^{j_i}_i   \bigg) \text{ } \bigg]   \otimes \textbf{I} \bigg)  \ket{\psi_{\mathrm{FFL}}} \bigg]           \bigg| \bigg|   <    \bigg(       \frac{8200 \sqrt{2} }{27}   \bigg)                   n^2 \sqrt{\epsilon }                   \text{. }
\end{align*}

\bigskip

\noindent \textbf{Lemma} $\textit{3}^{*}$ ($\epsilon$\textit{-optimality}, \textbf{Lemma} \textit{7}, [50]). For an $\epsilon$-optimal strategy $A_i$, $B_{jk}$ and $\ket{\psi_{\mathrm{FFL}}}$,

\begin{align*}
 \underset{1 \leq i < j \leq n}{\sum} \bigg| \bigg| \bigg( \bigg(         \frac{A_i A_j + A_j A_i}{2}        \bigg) \otimes \textbf{I} \bigg)  \ket{\psi_{\mathrm{FFL}}} \bigg| \bigg|^2 <   2 \big(   \frac{7}{3}    \big)^2 n \big( n - 1 \big) \epsilon \text{. } 
\end{align*}

\bigskip

\noindent \textbf{Lemma} \textit{4B} ($\sqrt{\epsilon}$- \textit{approximality}, \textbf{Lemma} \textit{8}, [50]). From the same quantities introduced in the previous result, one has,

\begin{align*}
   \bigg| \bigg|               \bigg(  A_k \otimes \textbf{I} \bigg) \ket{\psi_{\mathrm{FFL}}}    -  \bigg( \textbf{I} \otimes \bigg(     \frac{\pm B_{kl} + B_{lk}}{\big| \pm B_{kl} + B_{lk}  \big| }           \bigg) \bigg) \ket{\psi_{\mathrm{FFL}}}              \bigg| \bigg| < 17 \sqrt{n \epsilon}   \text{. } 
\end{align*}

\noindent In addition to the result above, the following result below is used to characterize the error bound resulting from permuting the indicates of the first player's tensor observable.

\bigskip

\noindent \textbf{Lemma} \textit{5} (\textit{error bound from permuting indices}, \textbf{Lemma} \textit{5}, [50]). One has,

\begin{align*}
 \bigg| \bigg|  \bigg(   \bigg( \underset{1 \leq i \leq n}{\prod} A^{j_i}_i  \bigg)   -    \bigg( \underset{\text{if } i \equiv j_1+1, \text{ } \mathrm{set} \text{ } j_1 + 1 \equiv j_1 \oplus 1}{ \underset{1 \leq i \leq n}{\prod} }A^{j_i}_i        \bigg) \otimes \textbf{I} \bigg) \ket{\psi_{\mathrm{FFL}}}   \bigg| \bigg|  \leq \frac{100}{9} n^2 \sqrt{\epsilon}  \text{. } 
\end{align*}

\bigskip

\noindent \textbf{Lemma} $\textit{5}^{*}$ (\textit{error bound from permuting indices in the N-player setting}, \textbf{Lemma} \textit{5}, [50]). One has the following error bound from permuting indices,

\begin{align*}
     \bigg| \bigg|  \bigg(    \bigg( \underset{1 \leq i \leq n}{\prod} A^{j_i}_i  \bigg)      \bigotimes \bigg( \underset{1 \leq z \leq N-1}{\bigotimes} \textbf{I}_z \bigg)       \bigg) \ket{\psi_{N\mathrm{XOR}}} - \bigg(     \bigg( \underset{\text{if } i \equiv j_1+1, \text{ } \mathrm{set} \text{ } j_1 + 1 \equiv j_1 \oplus 1}{ \underset{1 \leq i \leq n}{\prod} }A^{j_i}_i        \bigg)      \\  \bigotimes \bigg( \underset{1 \leq z \leq N-1}{\bigotimes} \textbf{I}_z \bigg)            \bigg) \ket{\psi_{N\mathrm{XOR}}}     \bigg| \bigg|           <  N n^{N+\epsilon_{N\mathrm{XOR}}} \omega^3_{N\mathrm{XOR}}  \text{. }
\end{align*}

\bigskip

\noindent \textbf{Lemma} $\textit{5}^{**}$ (\textit{error bound from permuting indices in the strong parallel repetition of the N-player setting}, \textbf{Lemma} \textit{5}, [50]). One has the following error bound from permuting indices,

\begin{align*}
              \bigg| \bigg|  \bigg(  \bigg(    \bigg( \underset{1 \leq i \leq n}{\prod} A^{j_i}_i  \bigg)   \wedge  \cdots \wedge \bigg( \underset{1 \leq i^{\prime\cdots\prime} \leq n^{\prime\cdots\prime}}{\prod} A^{j^{\prime\cdots\prime}_{i^{\prime\cdots\prime}}}_{i^{\prime\cdots\prime}}   \bigg) \bigg)       \bigotimes \bigg( \underset{1 \leq z \leq N-1}{\bigotimes}  \big( \textbf{I}_z \wedge \cdots \wedge \textbf{I}_z \big)  \bigg) 
 \bigg) \\ \times \ket{\psi_{N\mathrm{XOR}\wedge \cdots \wedge N\mathrm{XOR}}}  \\ - \bigg(   \bigg(   \bigg( \underset{\text{if } i \equiv j_1+1, \text{ } \mathrm{set} \text{ } j_1 + 1 \equiv j_1 \oplus 1}{ \underset{1 \leq i \leq n}{\prod} }A^{j_i}_i        \bigg)  \wedge \cdots \wedge \bigg(     \underset{\text{if } i^{\prime\cdots\prime} \equiv j^{\prime\cdots\prime}_1+1, \text{ } \mathrm{set} \text{ } j^{\prime\cdots\prime}_1 + 1 \equiv j^{\prime\cdots\prime}_1 \oplus 1}{ \underset{1 \leq i^{\prime\cdots\prime} \leq n^{\prime\cdots\prime}}{\prod} }A^{j^{\prime\cdots\prime}_{i^{\prime\cdots\prime}}}_{i^{\prime\cdots\prime}}           \bigg)    \bigg)   \bigotimes \\  \bigg( \underset{1 \leq z \leq N-1}{\bigotimes}\big(  \textbf{I}_z  \wedge \cdots \wedge \textbf{I}_z \big)   \bigg)    \bigg)    \ket{\psi_{N\mathrm{XOR} \wedge \cdots \wedge N\mathrm{XOR}}}     \bigg| \bigg|   <  n^{N+\epsilon}_{\wedge } +  \bigg( \frac{50 n^{N+\epsilon}_{\wedge}}{\sqrt{n^{N-1}}}   \bigg) \\ \times \omega_{N\mathrm{XOR} \wedge \cdots \wedge N \mathrm{XOR}}          \text{. }
\end{align*}

\bigskip

\noindent \textbf{Lemma} $\textit{5}^{***}$ (\textit{error bound from permuting indices in the strong parallel repetition of the $\mathrm{FFL}$ 2-player setting}, \textbf{Lemma} \textit{5}, [50]). One has the following error bound from permuting indices,

\begin{align*}
             \bigg| \bigg|  \bigg(     \bigg( \bigg( \underset{1 \leq i \leq n}{\prod} A^{j_i}_i  \bigg)  \wedge \bigg( \underset{1 \leq i^{\prime}\leq n^{\prime}}{\prod} A^{j^{\prime}_{i^{\prime}}}_{i^{\prime}}  \bigg) \bigg)     \bigotimes \bigg( \underset{1 \leq z \leq N-1}{\bigotimes}  \big( \textbf{I}_z \wedge \textbf{I}_z \big)  \bigg)  \bigg) \ket{\psi_{\mathrm{FFL}\wedge \mathrm{FFL}}} \\ - \bigg(   \bigg(    \bigg( \underset{\text{if } i \equiv j_1+1, \text{ } \mathrm{set} \text{ } j_1 + 1 \equiv j_1 \oplus 1}{ \underset{1 \leq i \leq n}{\prod} }A^{j_i}_i        \bigg)     \wedge \bigg( \underset{\text{if } i^{\prime} \equiv j^{\prime}_1+1, \text{ } \mathrm{set} \text{ } j^{\prime}_1 + 1 \equiv j^{\prime}_1 \oplus 1}{ \underset{1 \leq i^{\prime} \leq n^{\prime}}{\prod} }A^{j^{\prime}_{i^{\prime}}}_{i^{\prime}}     \bigg) \bigg)    \bigotimes \bigg( \underset{1 \leq z \leq N-1}{\bigotimes}  \big( \textbf{I}_z \\ \wedge \textbf{I}_z \big)  \bigg) \bigg)  \ket{\psi_{\mathrm{FFL}\wedge \mathrm{FFL}}}    \bigg| \bigg|          <    n^{N+\epsilon}_{\wedge }  +  \bigg( \frac{50 n^{N+\epsilon}_{\wedge}}{\sqrt{n}}  \bigg) \omega_{\mathrm{FFL} \wedge  \mathrm{FFL}}      \text{. }
\end{align*}

\noindent Underlying the results for parallel repetition stated at the beginning of the subsection are several semidefinite programs associated with primal feasible solutions are dependent upon the constraints, from the partial ordering $ \succcurlyeq$ induced by the positive semidefinite cone,

\begin{align*}
 \underset{1 \leq i \leq n^3}{\sum}      y_{3\mathrm{XOR} \wedge \cdots \wedge 3 \mathrm{XOR},i} E_{3\mathrm{XOR} \wedge \cdots \wedge 3 \mathrm{XOR} ,ii}     \succcurlyeq           G_{3\mathrm{XOR} \wedge \cdots \wedge 3 \mathrm{XOR}, \mathrm{Sym}}        \text{, }   \tag{$3\mathrm{XOR} \wedge \cdots \wedge 3 \mathrm{XOR}$, Sym}  \\  \underset{1 \leq i \leq n^4}{\sum}   y_{4\mathrm{XOR} \wedge \cdots \wedge 4 \mathrm{XOR},i} E_{4\mathrm{XOR}\wedge \cdots \wedge 4 \mathrm{XOR},ii}    \succcurlyeq G_{4\mathrm{XOR} \wedge \cdots \wedge 4 \mathrm{XOR},\mathrm{Sym}} \tag{$4\mathrm{XOR} \wedge \cdots \wedge 4 \mathrm{XOR}$, Sym}  \text{, }  \\  \underset{1 \leq i \leq n^5}{\sum}  y_{5\mathrm{XOR} \wedge \cdots \wedge 5 \mathrm{XOR},i} E_{5\mathrm{XOR} \wedge \cdots \wedge 5 \mathrm{XOR},ii}    \succcurlyeq  G_{5\mathrm{XOR}\wedge \cdots \wedge 5 \mathrm{XOR},\mathrm{Sym}} \tag{$5\mathrm{XOR} \wedge \cdots \wedge 5 \mathrm{XOR}$, Sym}  \text{, }\\  \underset{1 \leq i \leq n^N}{\sum}    y_{N\mathrm{XOR} \wedge \cdots \wedge N \mathrm{XOR},i} E_{N\mathrm{XOR} \wedge \cdots \wedge N \mathrm{XOR},ii}       \succcurlyeq G_{N\mathrm{XOR}\wedge \cdots \wedge N \mathrm{XOR},\mathrm{Sym}}  \tag{$N\mathrm{XOR} \wedge \cdots \wedge N \mathrm{XOR}$, Sym}  \text{, }  \\  \underset{1 \leq i \leq n^2}{\sum}    y_{\mathrm{FFL} \wedge \mathrm{FFL},i} E_{\mathrm{FFL} \wedge \mathrm{FFL},ii}      \succcurlyeq G_{\mathrm{FFL} \wedge \mathrm{FFL},\mathrm{Sym}} \tag{$\mathrm{FFL} \wedge\mathrm{FFL}$, Sym}  \text{. }
\end{align*}

\noindent The symmetrized game tensor under parallel repetition takes the form,

\begin{align*}
     G_{\mathrm{Sym}, \mathrm{XOR} \wedge \cdots \wedge \mathrm{XOR}} \equiv \underset{1 \leq j \leq n}{\bigwedge}  G^{(j)}_{\mathrm{Sym}, \mathrm{XOR}} \equiv  \underset{1 \leq j \leq n}{\bigwedge}          \begin{bmatrix}  0 & \big( G^{(j)}_{\mathrm{XOR}} \big)^{\textbf{T}} \\ G^{(j)}_{\mathrm{XOR}} & 0  \end{bmatrix}       \\ \equiv \begin{bmatrix} 0 &  \underset{1 \leq j \leq n}{\wedge}  \big( G^{(j)}_{\mathrm{XOR}} \big)^{\textbf{T}} \\ \underset{1 \leq j \leq n}{\wedge} G^{(j)}_{\mathrm{XOR}} & 0   \end{bmatrix} \text{. }
\end{align*}

\noindent Inequalities corresponding to error bounds, and generalized error bounds, provided above determine how \textit{approximate optimality} for games can be formulated. Such conditions, from exact optimality, are obtained by fixing sufficiently small parameters $\epsilon$. In comparison to previous results on parallel repetition of optimal values that do not incorporate exponential rates of decay that have been obtained for expanded games, deviations from exactly optimal strategies can be leveraged for approximate optimality. We collect results that have been shown to hold from previous work of the author in [52], below.

\bigskip

\noindent \textbf{Lemma} \textit{5B}, [52] (\textit{strong parallel repetition of }$\sqrt{\epsilon^{\wedge}}$- \textit{FFL approximality}, \textbf{Lemma} \textit{8}, [50]). From the same quantities introduced in the previous result, one has,

\begin{align*}
   \bigg| \bigg|               \bigg( \big(  A_k  \wedge A_{k^{\prime}} \big) \otimes \textbf{I} \bigg) \ket{\psi_{\mathrm{FFL} \wedge \mathrm{FFL}}}    -  \bigg( \textbf{I} \otimes \bigg(     \frac{\pm \big(  B_{kl} \wedge B_{k^{\prime} l^{\prime}} \big)  + \big( B_{lk} \wedge B_{k^{\prime} l^{\prime}} \big) }{\big| \pm \big(  B_{kl} \wedge B_{k^{\prime} l^{\prime}} \big)  + \big( B_{lk} \wedge B_{k^{\prime} l^{\prime}} \big) \big| }           \bigg) \bigg) \\ \times \ket{\psi_{\mathrm{FFL} \wedge \mathrm{FFL}}}              \bigg| \bigg|    < 20 \sqrt{N \epsilon^{\wedge}}     \text{. } 
\end{align*}

\bigskip

\noindent \textbf{Lemma} $\textit{5}^{*}\textit{B}$, [52] (\textit{an arbitrary number of strong parallel repetition applications of }$\sqrt{\epsilon^{\wedge}_{2\mathrm{XOR}}}$- \textit{2 XOR approximality}, \textbf{Lemma} \textit{8}, [50]). From the same quantities introduced in the previous result, one has,

\begin{align*}
   \bigg| \bigg|               \bigg( \big(  A_k  \wedge A_{k^{\prime}} \wedge \cdots \wedge A_{k^{\prime\cdots \prime}} \big) \otimes \textbf{I} \bigg) \ket{\psi_{2\mathrm{XOR} \wedge \cdots \wedge 2\mathrm{XOR}}}    \\ -  \bigg( \textbf{I} \otimes \bigg(     \frac{\pm \big(  B_{kl} \wedge B_{k^{\prime} l^{\prime}} \wedge \cdots \wedge B_{k^{\prime\cdots\prime}l^{\prime\cdots \prime}}  \big)  + \big( B_{lk} \wedge B_{ l^{\prime} k^{\prime}} \wedge \cdots \wedge B_{l^{\prime\cdots \prime} k^{\prime\cdots\prime} }  \big) }{\big| \pm \big(  B_{kl} \wedge B_{k^{\prime} l^{\prime}} \wedge \cdots \wedge B_{k^{\prime\cdots\prime}l^{\prime\cdots \prime}}  \big)  + \big( B_{lk} \wedge B_{ l^{\prime} k^{\prime}} \wedge \cdots \wedge B_{l^{\prime\cdots \prime} k^{\prime\cdots\prime} }  \big) \big| }           \bigg) \bigg) \\ \times \ket{\psi_{2\mathrm{XOR} \wedge \cdots \wedge 2\mathrm{XOR}}}              \bigg| \bigg|   < 18 \sqrt{N \epsilon^{\wedge}_{2\mathrm{XOR}}}   \text{. } 
\end{align*}

\bigskip

\noindent \textbf{Lemma} $\textit{5}^{**}B$, [52] (\textit{an arbitrary number of strong parallel repetition applications of }$\sqrt{\epsilon^{\wedge}_{3\mathrm{XOR}}}$- \textit{3 XOR approximality}, \textbf{Lemma} \textit{8}, [50]). From the same quantities introduced in the previous result, one has,

\begin{align*}
   \bigg| \bigg|               \bigg( \big(  A_k  \wedge A_{k^{\prime}} \wedge \cdots \wedge A_{k^{\prime\cdots \prime}} \big) \otimes \textbf{I} \otimes \textbf{I} \bigg) \ket{\psi_{3\mathrm{XOR} \wedge \cdots \wedge 2\mathrm{XOR}}}   \\   -  \bigg( \textbf{I} \otimes \textbf{I}  \otimes \bigg(   \frac{  \pm \underset{\sigma_1 \in S_3}{\sum} \big(  \wedge  C_{\sigma(ijk)} \big)  +  \underset{\sigma_2, \sigma_3, \sigma_4, \sigma_5, \sigma_6  \in S_3}{\sum} \big( \wedge  C_{\sigma(ijk)} \big)  }{  \bigg| \pm \underset{\sigma_1 \in S_3}{\sum} \big( \wedge C_{\sigma(ijk)} \big)  +  \underset{\sigma_2, \sigma_3, \sigma_4, \sigma_5, \sigma_6  \in S_3}{\sum} \big( \wedge C_{\sigma(ijk)}\big)   \bigg|  } \bigg)   \bigg) \ket{\psi_{3\mathrm{XOR} \wedge \cdots \wedge 3\mathrm{XOR}}}              \bigg| \bigg|  \\   < 18 \sqrt{N \epsilon^{\wedge}_{3\mathrm{XOR}}}   \text{, } 
\end{align*}

\noindent where,

\begin{align*}
 \wedge C_{\sigma(ijk)} \equiv  C_{\sigma(ijk)}  \wedge \cdots \wedge C_{\sigma(ijk)}  \text{. }
\end{align*}

\bigskip

\noindent \textbf{Lemma} $\textit{5}\textit{B}^{**}$, [52]  (\textit{an arbitrary number of strong parallel repetition applications of }$\sqrt{\epsilon^{\wedge}_{{N\mathrm{XOR} \wedge \cdots \wedge N\mathrm{XOR}}}}$- \textit{N XOR approximality}, \textbf{Lemma} \textit{8}, [50]). From the same quantities introduced in the previous result, one has, for the $\mathrm{XOR}$ game under an arbitrary number of strong parallel repetitions, that the quantities,

\begin{align*}
  \mathcal{I}_1 \equiv  \bigg| \bigg|               \bigg( \big(  A_k  \wedge A_{k^{\prime}} \wedge \cdots \wedge A_{k^{\prime\cdots \prime}} \big) \bigotimes  \bigg( \underset{1 \leq z \leq N-1}{\bigotimes}\textbf{I}_z \bigg)  \bigg) \\ \times \ket{\psi_{N\mathrm{XOR} \wedge \cdots \wedge N\mathrm{XOR}}} \\   -  \bigg( \textbf{I}  \otimes \bigg(     \frac{\pm \big(  B_{kl} \wedge B_{k^{\prime} l^{\prime}} \wedge \cdots \wedge B_{k^{\prime\cdots\prime}l^{\prime\cdots \prime}}  \big)  + \big( B_{lk} \wedge B_{ l^{\prime} k^{\prime}} \wedge \cdots \wedge B_{l^{\prime\cdots \prime} k^{\prime\cdots\prime} }  \big) }{\big| \pm \big(  B_{kl} \wedge B_{k^{\prime} l^{\prime}} \wedge \cdots \wedge B_{k^{\prime\cdots\prime}l^{\prime\cdots \prime}}  \big)  + \big( B_{lk} \wedge B_{ l^{\prime} k^{\prime}} \wedge \cdots \wedge B_{l^{\prime\cdots \prime} k^{\prime\cdots\prime} }  \big) \big| }    \bigg) \\  \bigotimes  \bigg( \underset{1 \leq z \leq N-2}{\bigotimes}\textbf{I}_z \bigg)       \bigg)  \ket{\psi_{N\mathrm{XOR} \wedge \cdots \wedge N\mathrm{XOR}}}              \bigg| \bigg| \\ \vdots \end{align*}

  \begin{align*} \mathcal{I}_N \equiv   \bigg| \bigg|    \bigg(   \bigg( \underset{1 \leq z \leq N-2}{\bigotimes} \textbf{I}_z \bigg) \bigotimes   \frac{1}{\sqrt{\# \sigma^{\prime} }}  \bigg(    \underset{\text{Permutations } \sigma^{\prime}}{\sum}  \big(   B^{(N-1)}_{\sigma^{\prime} ( i_1, \cdots, i_{N-1})}         \wedge   B^{(N-1)}_{\sigma^{\prime} ( i^{\prime}_1, \cdots, i^{\prime}_{N-1})}  \wedge  \cdots \\  \wedge   B^{(N-1)}_{\sigma^{\prime} ( i^{\prime\cdots\prime}_1, \cdots, i^{\prime\cdots\prime}_{N-1})} \big)      \bigg)       \bigotimes \textbf{I}  \bigg)    \ket{\psi_{N\mathrm{XOR} \wedge \cdots \wedge N\mathrm{XOR}}}   -  \bigg( \bigg( \underset{1 \leq z \leq N-1}{\bigotimes} \textbf{I}_z \bigg) \\ \bigotimes \frac{1}{\sqrt{\# \sigma }}  \bigg(    \underset{\text{Permutations } \sigma}{\sum}  \big(   B^{(N-1)}_{\sigma ( i_1, \cdots, i_{N-1})}          \wedge   B^{(N-1)}_{\sigma ( i^{\prime}_1, \cdots, i^{\prime}_{N-1})}  \wedge  \cdots \\  \wedge   B^{(N-1)}_{\sigma ( i^{\prime\cdots\prime}_1, \cdots, i^{\prime\cdots\prime}_{N-1})} \big)      \bigg) \bigg) \ket{\psi_{N\mathrm{XOR} \wedge \cdots \wedge N\mathrm{XOR}}}           \bigg| \bigg|      \text{, } 
\end{align*}

\noindent have the strict upper bound,

\begin{align*}
\underset{1 \leq j \leq N}{\sum} \mathcal{I}_j < 20 N \sqrt{N \epsilon^{\wedge}_{{N\mathrm{XOR} \wedge \cdots \wedge N\mathrm{XOR}}}} \text{,}
 \end{align*}

 \noindent where the tensors beyond that of the second player, $B$, are indexed as,

 \begin{align*}
     \textbf{I} \bigotimes B_{\sigma ( i_1,i_2)} \bigotimes \bigg( \underset{1 \leq z \leq N-2}{\bigotimes} \textbf{I}_z   \bigg) \equiv  \textbf{I} \bigotimes B^{(1)}_{\sigma ( i_1,i_2)} \bigotimes \bigg( \underset{1 \leq z \leq N-2}{\bigotimes} \textbf{I}_z   \bigg)   \text{, } \\ \vdots \\    \bigg( \underset{1 \leq z \leq N-1}{\bigotimes} \textbf{I}_z   \bigg) \bigotimes B_{\sigma ( i_1,i_2,\cdots, i_{n-2} )}  \equiv   \bigg( \underset{1 \leq z \leq N-1}{\bigotimes} \textbf{I}_z   \bigg) \bigotimes B^{(n-2)}_{\sigma ( i_1,i_2,\cdots, i_{n-2} )}  \text{. }
 \end{align*}

\noindent For expanding games, previous results in [14] which argue that exponential rates of decay exist for parallel repetition of the optimal value pose several implications from previous results of the author obtained in [50] for the 2-player setting, and in [52] for various multiplayer settings, including: determining how many optimal strategies exist for parallel repetition of multiplayer XOR, versus for expanding, games; qualitatively determining, again in comparison to $\epsilon$ parameters stated in previous results for parallel repetition of XOR, $\mathrm{XOR}^{*}$, and FFL, games,  whether similar parameters exist for for expanding games; distinguishing between barriers to quantum advantage in various information processing tasks.

To this end, in the next subsection we state exact, and approximate, optimality for expanding games. Given the fact that the exponential rate of decay of the optimal value of expanding games under parallel repetition holds, this presents a fundamental difference in error bounds, and generalizations of error bounds, for parallel repetition of FFL games. Drawing the attention of the reader to differences between ordinary, and strong, parallel repetition in various game-theoretic settings draws our attention to fundamental differences between Quantum and Classical information theory.

\subsection{Statement of main results}

\noindent We state the results below, which demonstrate how exponential rates of decay of the optimal value under parallel repetition can be related to approximately optimal conditions, with a sufficiently small parameter, $\epsilon_{\mathrm{Multiplayer}}$.

\subsubsection{Game-theoretic results which are dependent on the multiplayer set of dependency-breaking variables}

\noindent First, we state the Main Results pertaining to objects introduced in the previous subsection, followed by the next  subsection in which previous arguments of the author are applied given exponential rates of decay of the optimal value under parallel repetition.

\bigskip

\noindent \textbf{Theorem} \textit{1} (\textit{quantifying the largest order 90
of the system of expected values with normalization factors}). For the sequence of normalization factors $\Gamma_i$, $\widetilde{\mathscr{E}_{\Gamma_i}} \equiv    \mathrm{O} \bigg(  \frac{\delta^{\frac{N}{300}}_{\mathrm{Multiplayer}}}{\alpha^{N+1}_{\mathrm{Multiplayer}}} \bigg)    $.

\bigskip

\noindent \textbf{Theorem} \textit{2} (\textit{quantifying the largest order of the POVM system of expected values}). For the POVMs $\mathscr{P}\mathscr{O}\mathscr{V}\mathscr{M}_1, \cdots, \mathscr{P}\mathscr{O}\mathscr{V}\mathscr{M}_N$, $ \mathscr{P}\mathscr{O}\mathscr{V} \mathscr{M} \mathscr{E} \equiv \mathrm{O} \bigg( \frac{\sqrt{\sqrt{\delta^{\frac{N}{300}}_{\mathrm{Multiplayer}}}}}{\alpha^{N+5}_{\mathrm{Multiplayer}}} \bigg) $.

\bigskip

\noindent \textbf{Theorem} \textit{3} (\textit{quantifying the larges order of the system of expected values with Pinsker's inequality}). Given the quantum states $\textbf{I}, \cdots, \textbf{N}$, $\mathcal{P}\mathcal{I} \equiv    \mathrm{O} \bigg(  \frac{\sqrt{\delta^{\frac{N}{300}}_{\mathrm{Multiplayer}}}}{\alpha^{2(N+1)}_{\mathrm{Multiplayer}}} \bigg) $.

\bigskip

\noindent \textbf{Theorem} \textit{4} (\textit{rate of decay of the optimal value under anchored parallel repetition}). Fix $s_{\mathrm{Multiplayer}} \equiv \mathrm{max} \bigg\{ \underset{1 \leq i \leq N}{\prod} \mathrm{log} \big| \mathcal{Q}_i \big| ,1\bigg\}$, $0 < \alpha_{\mathrm{Multiplayer}} \leq 1$, a multiplayer $\alpha$-anchored game $G$, $\epsilon_{\mathrm{Multiplayer}}$ strictly positive, and universal constant $c_{\mathrm{Multiplayer}}$, where,

\begin{align*}
  0 < c_{\mathrm{Multiplayer}} < \frac{1}{N^{2N} \mathrm{log} \big( e \big)}  .
\end{align*}

\noindent The decay of the multiplayer optimal value, $\omega_{\mathrm{Multiplayer}} \big( G \big)$, under $\alpha$-anchored parallel repetition, $\omega_{\mathrm{Multiplayer}} \big( G_{\bot} \big)^{\otimes k} $, satisfies,

\begin{align*}
 \omega_{\mathrm{Multiplayer}} \big( G_{\bot} \big)^{\otimes k} \equiv   \omega_{\mathrm{Multiplayer}} \big( G^k_{\bot} \big) \leq \frac{10}{\epsilon_{\mathrm{Multiplayer}}} \mathrm{exp} \bigg[  -  \frac{c_{\mathrm{Multiplayer}} \alpha^{20N + 1}_{\mathrm{Multiplayer}}        \epsilon^{6N}_{\mathrm{Multiplayer}}  k }{s_{\mathrm{Multiplayer}}  }\bigg]  .
\end{align*}

\subsubsection{Game-theoretic results which are independent of the multiplayer set of dependency-breaking variables}

\noindent For the following results, introduce,

\begin{align*}
  \ket{\psi^{\prime}} \equiv \text{Optimal quantum strategy under parallel repetition of the expanding} \\ \text{game}  \text{,} \\ \\   \mathrm{Sign} \big( i_1, \cdots, i_n \big) \equiv   \mathrm{sign} \big( i_1, \cdots, i_n \big) \mathrm{sign} \big( i^{\prime}_1, \cdots, i^{\prime}_n \big) \times \cdots \\ \times    \mathrm{sign} \big( i^{\prime\cdots \prime}_1, \cdots, i^{\prime\cdots \prime}_n \big)   \text{, } \\    \\     \bigg( \underset{1 \leq i \leq n}{\prod}  A^{j_i}_i \bigg) \wedge \cdots \wedge   \bigg(     \underset{1 \leq i^{\prime\overset{n-2}{\cdots}\prime} \leq n}{\prod}  A^{j_{i^{\prime\overset{n-2}{\cdots}\prime} }}_{i^{\prime\overset{n-2}{\cdots}\prime} }    \bigg)  \equiv \text{Parallel repetition of the}  \\                 \text{  first player's tensor observable}       \text{, } \\          \bigg( \underset{1 \leq i \leq n}{\prod} A^{1,j_i}_i \bigg) \wedge \cdots \wedge \bigg( \underset{1 \leq i^{\prime \overset{n-2}{\cdots} \leq n}}{\prod}        A^{1,j_{i^{\prime\overset{n-2}{\cdots}}\prime}}_{i^{\prime\overset{n-2}{\cdots}}\prime}    \bigg) \equiv \text{Parallel repetition of the second player's} \end{align*}

  \begin{align*} \text{tensor observable}    \text{, } \\ \\  \bigg( \underset{1 \leq i \leq n}{\prod}  A^{(n-1),j_{i_1,\cdots,i_n}}_{i_1,\cdots, i_n} \bigg)  \wedge \cdots \wedge  \bigg( \underset{1 \leq i^{\prime \overset{n-2}{\cdots \prime}} \leq n}{\prod}  A^{(n-1),j_{i^{\prime \overset{n-2}{\cdots \prime}}_1,\cdots,i^{\prime \overset{n-2}{\cdots \prime}}_n}}_{i^{\prime \overset{n-2}{\cdots \prime}}_1,\cdots, i^{\prime \overset{n-2}{\cdots \prime}}_n} \bigg) \equiv \text{Parallel repetition of} \\ \text{the N th player's tensor observable} \text{. }
\end{align*}

\bigskip

\noindent \textbf{Lemma} \textit{1} (\textit{upper bounding the Frobenius norm under parallel repetition of the expanding game}). For the optimal strategy to the expanded game, $\ket{\psi^{\prime}} \equiv \ket{\psi_{\mathrm{Expanded}}},$ one has that,

\begin{align*}
 \bigg| \bigg|    \bigg[             \bigg( \bigg(  \bigg( \underset{1 \leq i \leq n}{\prod}  A^{j_i}_i \bigg) \wedge \cdots \wedge   \bigg(     \underset{1 \leq i^{\prime\overset{n-2}{\cdots}\prime} \leq n}{\prod}  A^{j_{i^{\prime\overset{n-2}{\cdots}\prime} }}_{i^{\prime\overset{n-2}{\cdots}\prime} }    \bigg)    \bigg) \bigotimes \bigg( \underset{1 \leq k \leq n-1}{\bigotimes} \textbf{I}_k \bigg) \bigg) \\   -  \bigg(   \mathrm{exp} \bigg(  - \frac{c \epsilon^5 n}{ k^2\mathrm{log} \big| A \big|}       \bigg) \bigg(         \pm \mathrm{Sign} \big( i_1, \cdots , i_n \big)             \bigg(  \bigg( \underset{1 \leq i \leq n}{\prod}  A^{j_i}_i \bigg) \wedge \cdots \wedge   \bigg(     \underset{1 \leq i^{\prime\overset{n-2}{\cdots}\prime} \leq n}{\prod}  A^{j_{i^{\prime\overset{n-2}{\cdots}\prime} }}_{i^{\prime\overset{n-2}{\cdots}\prime} }    \bigg)    \bigg)       \bigg)     \\ \bigotimes \bigg( \underset{1 \leq k \leq n-1}{\bigotimes} \textbf{I}_k \bigg)\bigg)                       \bigg]   \ket{\psi^{\prime}}       \bigg| \bigg|_F    <    \bigg(  \big( n_1 \big)^{\wedge} +       \big(   \big(  n_1 \big)^{\wedge} + 2    \big) \mathrm{exp} \bigg(  - \frac{k^2\mathrm{log} \big| A \big|}{c \epsilon^5 n }       \bigg)             \bigg)\big(  n^N \big)^{\wedge } \sqrt{\epsilon^{\wedge}}        \text{,} \end{align*}

 \begin{align*} \bigg| \bigg|    \bigg[ \bigg(  \textbf{I} \bigotimes  \bigg(           \bigg( \underset{1 \leq i \leq n}{\prod}  A^{1,j_i}_i \bigg) \wedge \cdots \wedge  \bigg( \underset{1 \leq i^{\prime\overset{n-2}{\cdots}\prime} \leq n}{\prod}  A^{1,j_{i^{\prime\overset{n-2}{\cdots}\prime}}}_{i^{\prime\overset{n-2}{\cdots}\prime}}       \bigg) \bigg)  \bigotimes \bigg( \underset{1 \leq k \leq n-2}{\bigotimes} \textbf{I}_k \bigg) \bigg)   \\ -  \bigg(  \textbf{I}   \bigotimes   \bigg(  \mathrm{exp} \bigg(  - \frac{c \epsilon^5 n}{ k^2\mathrm{log} \big| A \big|}       \bigg)   \bigg(         \pm \mathrm{Sign} \big( i_1, j_1, \cdots, i_n, \cdots, j_n \big)        \bigg(    \bigg( \underset{1 \leq i_2 \leq m}{\underset{1 \leq i_1 \leq n}{\prod}}  A^{1,j_{i_1,i_2}}_{i_1,i_2} \bigg)  \wedge \cdots \\  \wedge   \bigg( \underset{1 \leq i^{\prime\overset{n-2}{\cdots}\prime} \leq n}{\prod}  A^{1,j_{i^{\prime\overset{n-2}{\cdots}\prime}}}_{i^{\prime\overset{n-2}{\cdots}\prime}}       \bigg)    \bigg)    \bigg)  \bigg)   \bigotimes \bigg( \underset{1 \leq k \leq n-2}{\bigotimes} \textbf{I}_k \bigg)\bigg)                       \bigg]   \ket{\psi^{\prime}}    \bigg| \bigg|_F    <   \bigg( \big( n_2  \big)^{\wedge} +  \big( \big( n_2 \big)^{\wedge} + 2 \big) \\ \times  \mathrm{exp} \bigg(  - \frac{k^2\mathrm{log} \big| A \big|}{c \epsilon^5 n }       \bigg)                  \bigg) \big( n^N  \big)^{\wedge} \sqrt{\epsilon^{\wedge}}    \\  \vdots \\   \bigg| \bigg|    \bigg[             \bigg(\bigg( \underset{1 \leq k \leq n-1}{\bigotimes} \textbf{I}_k \bigg) \bigotimes  \bigg(  \bigg( \underset{1 \leq i \leq n}{\prod}  A^{(n-1),j_{i_1,\cdots,i_n}}_{i_1,\cdots, i_n} \bigg)  \wedge \cdots \wedge  \bigg( \underset{1 \leq i^{\prime \overset{n-2}{\cdots \prime}} \leq n}{\prod}  A^{(n-1),j_{i^{\prime \overset{n-2}{\cdots \prime}}_1,\cdots,i^{\prime \overset{n-2}{\cdots \prime}}_n}}_{i^{\prime \overset{n-2}{\cdots \prime}}_1,\cdots, i^{\prime \overset{n-2}{\cdots \prime}}_n} \bigg) \bigg) \bigg) \end{align*}
 
 \begin{align*}  -  \bigg( \bigg( \underset{1 \leq k \leq n-1}{\bigotimes} \textbf{I}_k \bigg)   \bigotimes \bigg(    \mathrm{exp} \bigg(  - \frac{c \epsilon^5 n}{ k^2\mathrm{log} \big| A \big|}       \bigg)  \bigg(         \pm \mathrm{Sign} \big( i_1, \cdots   , i_n, j_1, \cdots, j_n \big)                         \bigg) \\ \times   \bigg(  \bigg( \underset{1 \leq i \leq n}{\prod}  A^{(n-1),j_{i_1,\cdots,i_n}}_{i_1,\cdots, i_n} \bigg)  \wedge \cdots  \wedge  \bigg( \underset{1 \leq i^{\prime \overset{n-2}{\cdots \prime}} \leq n}{\prod}  A^{(n-1),j_{i^{\prime \overset{n-2}{\cdots \prime}}_1,\cdots,i^{\prime \overset{n-2}{\cdots \prime}}_n}}_{i^{\prime \overset{n-2}{\cdots \prime}}_1,\cdots, i^{\prime \overset{n-2}{\cdots \prime}}_n} \bigg) \bigg)  \bigg)                 \bigg)   \bigg] \ket{\psi^{\prime}}     \bigg| \bigg|_F   \\  <   \bigg(\big(  n_N  \big)^{\wedge }       +   \big( \big(  n_N \big)^{\wedge} + 2\big)   \mathrm{exp} \bigg(  - \frac{k^2\mathrm{log} \big| A \big|}{c \epsilon^5 n }       \bigg)                  \bigg) \big(  n^N \big)^{\wedge } \sqrt{\epsilon^{\wedge}}  \text{, }
\end{align*}

\bigskip

\noindent \textbf{Lemma} \textit{2} (\textit{parallel repetition of second order error bounds}). For the optimal strategy to the expanded game, $\ket{\psi^{\prime}} \equiv \ket{\psi_{\mathrm{Expanded}}},$ one has that,

\begin{align*}
 \bigg| \bigg|  \bigg( \bigg(  \bigg( \underset{1 \leq i \leq n}{\prod}  A^{j_i}_i \bigg) \wedge \cdots \wedge   \bigg(     \underset{1 \leq i^{\prime\overset{n-2}{\cdots}\prime} \leq n}{\prod}  A^{j_{i^{\prime\overset{n-2}{\cdots}\prime} }}_{i^{\prime\overset{n-2}{\cdots}\prime} }    \bigg)    \bigg)   \otimes \big( B_{kl} \wedge \cdots \wedge B_{k^{\prime\cdots\prime}l^{\prime\cdots\prime}} \big)  \bigg)  \ket{\psi^{\prime}} \\  -  \mathrm{exp} \bigg(  - \frac{k^2\mathrm{log} \big| A \big|}{c \epsilon^5 n }       \bigg)       \bigg[ \pm \bigg( \mathrm{Sign} \big( i , j_1 , \cdots , j_n \big)  \bigg[ \text{ } \bigg( \bigg(      \underset{i = j_k + 1 , \text{ } \mathrm{set} \text{ } j_k + 1 \equiv j_k \oplus 1 }{\underset{1 \leq i \leq n}{\prod}}     A^{j_i}_i \bigg) \wedge \cdots \\ \wedge \bigg(        \underset{i = j_k + 1 , \text{ } \mathrm{set} \text{ } j_k + 1 \equiv j_k \oplus 1 }{\underset{1 \leq i \leq n}{\prod}}     A^{j_i}_i            \bigg)    \bigg)   + \bigg( \bigg(  \underset{i = j_l + 1 , \text{ } \mathrm{set} \text{ } j_l + 1 \equiv j_l \oplus 1 }{\underset{1 \leq i \leq n}{\prod}}     A^{j_i}_i  \bigg) \wedge  \cdots  \\ \wedge \bigg(         \underset{i = j_k + 1 , \text{ } \mathrm{set} \text{ } j_k + 1 \equiv j_k \oplus 1 }{\underset{1 \leq i \leq n}{\prod}}     A^{j_i}_i              \bigg)  \bigg) \text{ } \bigg]   \otimes \textbf{I} \bigg)  \ket{\psi^{\prime}} \bigg]           \bigg| \bigg|   \lesssim    \bigg(      \mathrm{exp} \bigg(  - \frac{k^2\mathrm{log} \big| A \big|}{c \epsilon^5 n }       \bigg)               \bigg)                n_{\wedge} \sqrt{\epsilon_{\wedge} }                   \text{. }
\end{align*}

\bigskip

\noindent Besides the two results provided above for quantifying the impact of parallel repetition on expanded games, as an adaptation of the following result below, it is also important to quantify the manner in which the parallel repetition operation can be expressed from summations over Schmidt blocks.

\bigskip

\noindent \textbf{Lemma} \textit{6}, [50] (\textit{odd n product expansion}, \textit{6.1}, [42]). For odd $n$, one has an expansion for,

\begin{align*}
 \underset{1 \leq i \leq n}{\prod} \widetilde{A_i}   \text{, } 
\end{align*}

\noindent in terms of a signed block identity matrix,

\[
\big( - 1 \big)^n \begin{bmatrix}
\textbf{I} & 0 \\ 0 & - \textbf{I}
\end{bmatrix}  \text{. } 
\]

\noindent Hence, for,

\begin{align*}
 \ket{\widetilde{\psi_{\mathrm{FFL}}}} = \frac{1}{\sqrt{2 \times  2^{\lfloor \frac{n}{2} \rfloor }}} \bigg[   \text{ } \bigg(      \underset{1 \leq j \leq 2^{\lfloor \frac{n}{2}\rfloor}}{\sum}   + \underset{2^{\lfloor \frac{n}{2} \rfloor} + 1 \leq j \leq 2 \times 2^{\lfloor \frac{n}{2}\rfloor}}{\sum} \bigg) \bigg( \ket{j} \otimes \ket{j}  \bigg) \text{ }  \bigg]    \text{, } 
\end{align*}

\noindent one has,

\begin{align*}
 \bigg( \bigg(  \underset{1 \leq i \leq n}{\prod} \widetilde{A_i}  \bigg) \otimes \textbf{I} \bigg) \ket{\widetilde{\psi_{\mathrm{FFL}}}} = \frac{1}{\sqrt{2 \times  2^{\lfloor \frac{n}{2} \rfloor }}}  \bigg[    \text{ } \bigg(      \underset{1 \leq j \leq 2^{\lfloor \frac{n}{2}\rfloor}}{\sum}   + \underset{2^{\lfloor \frac{n}{2} \rfloor} + 1 \leq j \leq 2 \times 2^{\lfloor \frac{n}{2}\rfloor}}{\sum} \bigg) \\ \times \bigg( \bigg(       \bigg(  \underset{1 \leq i \leq n}{\prod} \widetilde{A_i}  \bigg) \ket{j}       \otimes \ket{j}      \bigg)  \otimes \bigg( \bigg(  \underset{1 \leq i \leq n}{\prod} \widetilde{A_i}  \bigg) \ket{j}    \otimes \ket{j} \bigg)   \bigg) \text{ }      \bigg]            \text{. } 
\end{align*}

\bigskip

\noindent Straightforwardly, the adaptation of the above result is given below.

\bigskip

\noindent \textbf{Lemma} \textit{3} (\textit{adaptation of Lemma 6 for parallel repetition of expanded games}). For odd n, one has an expansion for,

\begin{align*}
 \underset{1 \leq i \leq n}{\prod} \big( \widetilde{A_i} \wedge \widetilde{A_{i^{\prime}}} \big)    \text{, } 
\end{align*}

\noindent in terms of a signed block identity matrix,

\[
\big( - 1 \big)^n \begin{bmatrix}
\textbf{I} \wedge \textbf{I} & 0 \\ 0 & - \textbf{I} \wedge \textbf{I}
\end{bmatrix}  \text{. } 
\]

\noindent Hence, for,

\begin{align*}
 \ket{\psi^{\prime}} \wedge \ket{\psi^{\prime}} = \frac{1}{\sqrt{2 \times  2^{\lfloor \frac{n}{2} \rfloor }}} \bigg[   \text{ } \bigg(      \underset{1 \leq j \leq 2^{\lfloor \frac{n}{2}\rfloor}}{\sum}   + \underset{2^{\lfloor \frac{n}{2} \rfloor} + 1 \leq j \leq 2 \times 2^{\lfloor \frac{n}{2}\rfloor}}{\sum} \bigg) \bigg( \big( \ket{j} \wedge \ket{j} \big)  \otimes \big( \ket{j} \wedge \ket{j} \big)   \bigg) \text{ }  \bigg]    \text{, } 
\end{align*}

\noindent one has,

\begin{align*}
 \bigg( \bigg(  \underset{1 \leq i \leq n}{\prod} \widetilde{A_i}  \bigg) \otimes \textbf{I} \bigg) \bigg( \ket{\psi^{\prime}} \wedge \ket{\psi^{\prime}} \bigg)   = \frac{1}{\sqrt{2 \times  2^{\lfloor \frac{n}{2} \rfloor }}}  \bigg[    \text{ } \bigg(      \underset{1 \leq j \leq 2^{\lfloor \frac{n}{2}\rfloor}}{\sum}   + \underset{2^{\lfloor \frac{n}{2} \rfloor} + 1 \leq j \leq 2 \times 2^{\lfloor \frac{n}{2}\rfloor}}{\sum} \bigg) \\ \times \bigg( \bigg(       \bigg(  \underset{1 \leq i \leq n}{\prod} \widetilde{A_i}  \bigg)  \big( \ket{j} \wedge \ket{j} \big)     \otimes \big( \ket{j} \wedge \ket{j} \big)       \bigg) \otimes \bigg( \bigg(  \underset{1 \leq i \leq n}{\prod} \widetilde{A_i}  \bigg) \big( \ket{j} \wedge \ket{j} \big)   \otimes\big( \ket{j} \wedge \ket{j} \big)    \bigg)   \bigg) \text{ }      \bigg]            \text{. } 
\end{align*}

\noindent In a result described in the previous subsection, \textbf{Lemma} $\textit{5}^{**}$, an error bound is obtained for parallel repetition of the multiplayer XOR game.

\bigskip

\noindent \textbf{Lemma} $\textit{4}$ (\textit{adaptation of the error bound from permuting indices in the strong parallel repetition of the N-player setting}, \textbf{Lemma} \textit{5}, [50]). One has the following error bound from permuting indices,

\begin{align*}
              \bigg| \bigg|  \bigg(  \bigg(    \bigg( \underset{1 \leq i \leq n}{\prod} A^{j_i}_i  \bigg)   \wedge  \cdots \wedge \bigg( \underset{1 \leq i^{\prime\cdots\prime} \leq n^{\prime\cdots\prime}}{\prod} A^{j^{\prime\cdots\prime}_{i^{\prime\cdots\prime}}}_{i^{\prime\cdots\prime}}   \bigg) \bigg)       \bigotimes \bigg( \underset{1 \leq z \leq N-1}{\bigotimes}  \big( \textbf{I}_z \wedge \cdots \wedge \textbf{I}_z \big)  \bigg) 
 \bigg) \\ \times \ket{\psi^{\prime}}    - \bigg(   \bigg(   \bigg( \underset{\text{if } i \equiv j_1+1, \text{ } \mathrm{set} \text{ } j_1 + 1 \equiv j_1 \oplus 1}{ \underset{1 \leq i \leq n}{\prod} }A^{j_i}_i        \bigg)  \wedge \cdots \wedge \bigg(     \underset{\text{if } i^{\prime\cdots\prime} \equiv j^{\prime\cdots\prime}_1+1, \text{ } \mathrm{set} \text{ } j^{\prime\cdots\prime}_1 + 1 \equiv j^{\prime\cdots\prime}_1 \oplus 1}{ \underset{1 \leq i^{\prime\cdots\prime} \leq n^{\prime\cdots\prime}}{\prod} }A^{j^{\prime\cdots\prime}_{i^{\prime\cdots\prime}}}_{i^{\prime\cdots\prime}}           \bigg)    \bigg) \\   \bigotimes  \bigg( \underset{1 \leq z \leq N-1}{\bigotimes}\big(  \textbf{I}_z  \wedge \cdots \wedge \textbf{I}_z \big)   \bigg)    \bigg)    \ket{\psi^{\prime}}   \bigg| \bigg|  <  n^{N+\epsilon}_{\wedge } +  \bigg( 50 n^{N+\epsilon}_{\wedge}   \bigg)    \\ \times  \mathrm{exp} \bigg(  -  \bigg( \frac{c \epsilon^5 n }{k^2\mathrm{log} \big| A \big|}  + \mathrm{log} \big(    n^{N - 1}  \big)   \bigg)      \bigg)         \text{. }
\end{align*}

\bigskip

\noindent Altogether, variants to error bounds for the expander game, which ultimately depend upon the rate of decay of the optimal value under parallel repetition, can be related to upper bounds in error bounds for parallel repetition of FFL, XOR, and $\mathrm{XOR}^{*}$, games. Under ordinary, and strong, parallel repetition, in the 2-player setting one performs a computation involving,

\begin{align*}
  \bigg| \bigg|               \bigg( \big(  A_k  \wedge A_{k^{\prime}} \wedge \cdots \wedge A_{k^{\prime\cdots \prime}} \big) \bigotimes  \bigg( \underset{1 \leq z \leq N-1}{\bigotimes}\textbf{I}_z \bigg)  \bigg) \\ \times \ket{\psi_{N\mathrm{XOR} \wedge \cdots \wedge N\mathrm{XOR}}}    \\  -  \bigg( \textbf{I}  \bigotimes \bigg(     \frac{\pm \big(  B_{kl} \wedge B_{k^{\prime} l^{\prime}} \wedge \cdots \wedge B_{k^{\prime\cdots\prime}l^{\prime\cdots \prime}}  \big)  + \big( B_{lk} \wedge B_{ l^{\prime} k^{\prime}} \wedge \cdots \wedge B_{l^{\prime\cdots \prime} k^{\prime\cdots\prime} }  \big) }{\big| \pm \big(  B_{kl} \wedge B_{k^{\prime} l^{\prime}} \wedge \cdots \wedge B_{k^{\prime\cdots\prime}l^{\prime\cdots \prime}}  \big)  + \big( B_{lk} \wedge B_{ l^{\prime} k^{\prime}} \wedge \cdots \wedge B_{l^{\prime\cdots \prime} k^{\prime\cdots\prime} }  \big) \big| }    \bigg)  \\ \bigotimes  \bigg( \underset{1 \leq z \leq N-2}{\bigotimes}\textbf{I}_z \bigg)       \bigg)\ket{\psi_{N\mathrm{XOR} \wedge \cdots \wedge N\mathrm{XOR}}}              \bigg| \bigg| \text{, } \end{align*}

  \noindent where, before the operation of parallel repetition is performed, one distributes questions,

  \begin{align*}
   l, l^{\prime}, \cdots, l^{\prime\cdots\prime}, k, k^{\prime}, k^{\prime\cdots\prime}   \text{,}
  \end{align*}

\noindent for the players, drawn from the referee's probability distribution. In the expanded game, one can make use of identical constructions of the possible responses that each player can provide in respond to the referee's questions, under each repetition of the game.

\subsubsection{A set of four equivalent conditioons for exact, and approximate, optimality}

The distinction between exact, and approximate, optimality in several game-theoretic settings is always of interest to explore, and characterize. In previous circumstances, for the $\mathrm{CHSH} \big(n \big)$, and XOR, games, [42], approximate optimality was characterized with an, up to constants, $\epsilon$ upper bound dependent upon a polynomial of the number of participants. Besides the fact that there exists several useful analogies between error bounds for XOR, $\mathrm{XOR}^{*}$, and FFL, games and those of expanded games, similar analogies extend to deviations from optimality. We state a collection of such equivalent characterizations below. In the first characterization of approximately optimal strategies for expanded games presented below, recall that the condition $(0)$ states,

\begin{align*}
    \frac{1}{\sqrt{2}} \big( 1 - \epsilon \big) \leq    \underset{A_i , B_{jk} , \psi}{\mathrm{sup}}   \frac{1}{4 {n \choose 2}}   \underset{1\leq i \leq j \leq n}{\sum}  \bra{\psi} \bigg(       A_i B_{ij} +  A_j B_{ij}  +   A_i B_{ji}  -   A_j B_{ji}  \bigg)           \ket{\psi}  \leq \frac{1}{\sqrt{2}}          \text{. } \tag{$0$}
\end{align*}

\noindent An equivalent statement can be obtained to the equality above for the optimal value of the XOR game, for parallel repetition of expanded games. We denote this condition as $(0)-\mathrm{Expanded}$. Before stating the condition, denote,

\begin{align*}
  \omega \big( \mathrm{Expanded}^{\otimes n} \big) \equiv  \omega_{\mathrm{Expanded}^{\otimes n}}   \text{. }
\end{align*}

\noindent Explicitly, under parallel repetition, this condition stipulates,

\begin{align*}
  \omega_{\mathrm{Expanded}^{\otimes n}}  \big( 1 - \epsilon_{\mathrm{Expanded}} \big)  \lesssim   \underset{A_i , B_{jk} , \psi^{\prime}}{\mathrm{sup}}     \underset{1\leq i \leq j \leq n}{\sum}   \bigg(  \bra{\psi^{\prime}} \wedge \cdots  \wedge \bra{\psi^{\prime}}  \bigg) \\ \times \bigg( \underset{\mathrm{Players} \mathscr{P}}{\bigotimes}  \mathscr{P}_{\mathrm{Expanded}}   \bigg(   \big( -1 \big)^{\textbf{I}_{\mathrm{Win the Expanded game}}}  +  \big( -1 \big)^{\textbf{I}_{\mathrm{Lose the Expanded game}}}   \bigg)    \bigg)    \\ \times     \bigg(   \ket{\psi^{\prime}} \wedge \cdots \wedge \ket{\psi^{\prime}} \bigg)    \lesssim         \omega_{\mathrm{Expanded}^{\otimes n}}  \text{, } \tag{$(0)$-Expanded}
\end{align*}

\noindent for $\epsilon_{\mathrm{Expanded}}$ taken sufficiently small.

\bigskip

\noindent \textbf{Theorem} \textit{1} (\textit{equivalent characterizations of approximate, and exact, optimality in expanded games}). TFAE:

\begin{itemize}
 \item[$\bullet$] \underline{\textit{First characterization of approximate optimality}:} An $\epsilon$-approximate Expanded game strategy satisfies $(0)-\mathrm{Expanded}$.

 \item[$\bullet$] \underline{\textit{Second characterization of approximate optimality}:} For an $\epsilon$-approximate quantum strategy, 

  \begin{align*}
     \underset{1 \leq i < j \leq n}{\sum} \bigg[ \text{ }  \bigg| \bigg|   \bigg[         \big( \frac{A_i + A_j}{\sqrt{2}} \big) \otimes I \bigg] \ket{\psi}    -   \big[  I \otimes B_{ij} \big] \ket{\psi^{\prime}}    \bigg|\bigg|^2 + \bigg| \bigg|   \bigg[ \big( \frac{A_i - A_j}{\sqrt{2}} \big)  \otimes I  \bigg] \ket{\psi^{\prime}}     -   \big[  I \\ \otimes B_{ji} \big]  \ket{\psi^{\prime}}       \bigg|\bigg|^2 \text{ }  \bigg]  \leq 2n \big( n - 1 \big) \epsilon \text{. }
 \end{align*}
 \item[$\bullet$] \underline{\textit{Reversing the order of the tensor product for observables}:} Related to the inequality for $\epsilon$-approximate strategies above, another inequality,

\begin{align*}
    \underset{1 \leq i < j \leq n}{\sum} \bigg[  \text{ } \bigg| \bigg| \big[ A_i \otimes I \big]\ket{\psi^{\prime}}       - \bigg[ I \otimes \big( \frac{B_{ij} + B_{ji}}{\sqrt{2}} \big)  \bigg] \ket{\psi^{\prime}}    \bigg| \bigg|^2  + \bigg| \bigg|    \big[ A_j  \otimes I \big] \ket{\psi^{\prime}}    -   \bigg[   I  \\ \otimes \big( \frac{B_{ij} - B_{ji}}{\sqrt{2}} \big) \bigg]  \ket{\psi^{\prime}}     \big| \big|^2     \text{ }   \bigg] \leq 2n \big( n - 1 \big) \epsilon \text{, } 
\end{align*}

 \noindent also holds.

 \item[$\bullet$] \underline{\textit{Characterization of exact optimality}}: For $\epsilon \equiv 0$,

\begin{align*}
    \underset{1 \leq i < j \leq n}{\sum} \bigg[ \text{ }  \bigg| \bigg|  \bigg[  \big( \frac{A_i + A_j}{\sqrt{2}} \big) \otimes I \bigg] \ket{\psi^{\prime}}     -  \big[ I \otimes B_{ij} \big] \ket{\psi^{\prime}}     \bigg| \bigg|^2  \bigg]  =  -  \underset{1 \leq i < j \leq n}{\sum} \bigg[  \text{ }   \bigg| \bigg|     \bigg[  \big( \frac{A_i - A_j}{\sqrt{2}} \big) \\   \otimes I \bigg] \ket{\psi^{\prime}}    -   \big[ I \otimes B_{ji} \big] \ket{\psi^{\prime}}     \bigg| \bigg|^2  \bigg]      \text{, } 
\end{align*}

\noindent corresponding to the first inequality, and,

\begin{align*}
       \underset{1 \leq i < j \leq n}{\sum} \bigg[ \text{ }  \bigg| \bigg|       \big[  A_i \otimes I \big] \ket{\psi^{\prime}}     - \bigg[   I \otimes \big( \frac{B_{ij} + B_{ji}}{\sqrt{2}} \big)  \bigg] \ket{\psi^{\prime}}  \bigg|\bigg|^2 \bigg] = -  \underset{1 \leq i < j \leq n}{\sum}   \bigg[  \bigg| \bigg|  \big[   A_j  \otimes I \big] \ket{\psi^{\prime}}   \\  -  \bigg[    I  \otimes \big( \frac{B_{ij} - B_{ji}}{\sqrt{2}} \big) \bigg]  \ket{\psi^{\prime}}     \bigg|\bigg|^2  \bigg]             \text{, } 
\end{align*}

\noindent corresponding to the second inequality.

\end{itemize}

\bigskip

\noindent We provide the argument for the four equivalent characterizations of exact, and approximate, optimality for expanded games provided above, as it is a direct application of arguments for the 2-player setting provided in [42], which were adapted by the author to the multiplayer setting in [52]. To lighten the notation used in the following arguments, for convenience, denote,

\begin{align*}
\underline{\mathcal{Q}^{(1)}_1} \equiv \text{Set of     questions distributed to the first player of the expanded game} \text{, } \\ \\ \underline{\mathcal{Q}^{(1)}_2} \equiv \text{Set of     questions distributed to the second player of the expanded game}\text{, } \\ \\ \underline{\mathcal{Q}^{(1)}_1 \wedge \mathcal{Q}^{(1)}_2} \equiv \text{Parallel repetition between the question sets of the first and second} \\ \text{ players in the expanded game} \text{, } \\ \\ \underline{\mathcal{Q}^{(1)}_1 \wedge \cdots \wedge \mathcal{Q}^{(n)}_1} \equiv \text{Parallel repetition of the question sets from the first player} \\ \text{in the expanded game} \text{, } \\ \\    \underline{G_{\mathrm{Sym}, \mathrm{Expanded}}} \equiv \text{Symmetrized game tensor for the expanded game} \text{, } \end{align*}

\begin{align*}      \underline{G_{\mathrm{Sym}, \mathrm{Expanded}\wedge \cdots \wedge \mathrm{Expanded}}} \equiv \text{Symmetrized game tensor for parallel repeti-} \\ \text{tion of the expanded game}    \text{, } \\ \\      \underline{y_{\mathcal{Q}}}      \equiv      \text{Dual semidefinite program objective function} \text{. }
\end{align*}

\noindent Furthermore, denote,

\begin{align*}
 \bra{ \big( i_1 i_2 \big)  \wedge  \big(      i^{\prime}_1 i^{\prime}_2 \big) }       =   \bra{i_1 i_2} \wedge \bra{i^{\prime}_1 i^{\prime}_2}   \text{, } \\  \ket{ \big( i_1 i_2 \big)  \wedge  \big(      i^{\prime}_1 i^{\prime}_2 \big) }    = \ket{i_1 i_2 } \wedge \ket{ i^{\prime}_1 i^{\prime}_2 }           \text{, }
\end{align*}

\noindent corresponding to two applications of the parallel repetition operation to the responses from each player, and, similarly,

\begin{align*}
 \bra{ \big( i_1 i_2 \big)  \wedge  \big(      i^{\prime}_1 i^{\prime}_2 \big) \wedge \cdots \wedge \big( i^{\prime\cdots \prime}_1 i^{\prime\cdots \prime}_2  \big)  }       =   \bra{i_1 i_2} \wedge \bra{i^{\prime}_1 i^{\prime}_2} \wedge \cdots \wedge \bra{ i^{\prime\cdots \prime}_1 i^{\prime\cdots \prime}_2}  \text{, } \\  \ket{ \big( i_1 i_2 \big)  \wedge  \big(      i^{\prime}_1 i^{\prime}_2 \big) \wedge \cdots \wedge \big( i^{\prime\cdots \prime}_1 i^{\prime\cdots \prime}_2  \big)  }    =      \ket{i_1 i_2 } \wedge \ket{ i^{\prime}_1 i^{\prime}_2 }    \wedge \cdots \wedge \ket{i^{\prime\cdots \prime}_1 i^{\prime\cdots \prime}_2  }      \text{, }
\end{align*}

\noindent corresponding to an arbitrary number of parallel repetition operations.

\bigskip

\noindent \textit{Proof of Theorem 1}. The arguments for the result fundamentally rely upon the following condition, involving,

\begin{align*}
     y_i E_{ii} - G_{\mathrm{Sym}, \mathrm{Expanded}\wedge \cdots \wedge \mathrm{Expanded}} \text{, }
\end{align*}

\noindent where $y$ denotes the dual objective function associated to parallel repetition of semidefinite programs,

\begin{align*}
  \underset{1 \leq j \leq n}{\bigwedge} \bigg[  \underset{\forall Z_{\mathrm{Expanded}\wedge \cdots \wedge \mathrm{Expanded}} \succcurlyeq 0 , 1 \leq i \leq m, F^{(j)}_{\wedge i} \cdot Z_{\mathrm{Expanded} \wedge \cdots \wedge \mathrm{Expanded} } \equiv C^{(j)}_i}{\mathrm{sup}}    G_{\mathrm{Expanded}} Z_{\mathrm{XOR}}      \bigg]       \text{, }
\end{align*}

\noindent and the symmetrized game tensor for the expanded game,

\begin{align*}
\begin{bmatrix} 0 &  \underset{1 \leq j \leq n}{\wedge}  \big( G^{(j)}_{\mathrm{Expanded}} \big)^{\textbf{T}} \\ \underset{1 \leq j \leq n}{\wedge} G^{(j)}_{\mathrm{Expanded}} & 0   \end{bmatrix}  \text{, }
\end{align*}

\noindent which is taken with respect to the partial ordering, $\succcurlyeq$, of the positive semidefinite cone, is equivalent, by duality, to,

\begin{align*}
\underset{1 \leq j \leq n}{\bigwedge} \bigg[ \underset{ \underset{1 \leq j \leq n}{\wedge}   \underset{\mathcal{Q}^{(1)}_N \wedge \cdots \wedge \mathcal{Q}^{(i)}_N \wedge  \cdots \wedge \mathcal{Q}^{(n)}_N}{\underset{\vdots}{{\underset{\mathcal{Q}^{(1)}_1 \wedge \cdots \wedge \mathcal{Q}^{(i)}_N \wedge \cdots  \wedge \mathcal{Q}^{(n)}_1}{\sum}}}}    y_{\mathcal{Q}^{(j)}_i} E^{(j)}_{ii}     \succcurlyeq  G^{(j)}_{\mathrm{Sym}, \mathrm{Expanded} \wedge \cdots \wedge \mathrm{Expanded}}          }{\mathrm{inf}}  \bigg(    \underset{\mathcal{Q}^{(1)}_N \wedge \cdots \wedge \mathcal{Q}^{(i)}_N \wedge \cdots \wedge \mathcal{Q}^{(n)}_N}{\underset{\vdots}{\underset{\mathcal{Q}^{(1)}_1 \wedge \cdots \wedge \mathcal{Q}^{(i)}_N \wedge \cdots \wedge \mathcal{Q}^{(n)}_1}{\sum}}}       y_{\mathcal{Q}^{(j)}_i}   \bigg)         \bigg]  \text{, }
\end{align*}

\noindent The summation, over finitely many terms, of $y_i E_{ii} - G_{\mathrm{Sym}, \mathrm{Expanded}\wedge \cdots \wedge \mathrm{Expanded}}$, equals,

\begin{align*}
     \frac{1}{\bigg( C_{\mathrm{Expanded} \wedge \mathrm{Expanded}} \omega_{\mathrm{Expanded} \wedge \mathrm{Expanded}} \bigg) \bigg(   3  \underset{1 \leq j \leq 2}{\prod}  \big( n - j \big) \bigg)  }             \underset{i_3 \in \mathcal{Q}_3}{\underset{i_2 \in \mathcal{Q}_2 }{\underset{i_1 \in \mathcal{Q}_1}{\sum}}}   \bigg(  \big(    u^{\prime}_{i_1i_2 i_3}   -    v^{\prime}_{i_1i_2 i_3}  \big) \big(   u^{\prime}_{i_1 i_2 i_3}  \\  -    v^{\prime}_{i_1 i_2 i_3}    \big)^{\textbf{T}}      +  \big(    u^{\prime}_{i_2 i_1 i_3 }    -    v^{\prime}_{i_2 i_1 i_3}  \big)   \big( u^{\prime}_{i_2 i_1 i_3 }   -    v^{\prime}_{i_2 i_1 i_3  }  \big)^{\textbf{T}}  +     \big(    u^{\prime}_{i_1 i_3 i_2}   -    v^{\prime}_{i_1 i_3 i_2}   \big)   \big(   u^{\prime}_{i_1 i_3 i_2}     -    v^{\prime}_{i_1 i_3 i_2}   \big)^{\textbf{T}}     \bigg)      \text{,}
\end{align*}

\noindent where, in the above summation over the possible set of questions that can be distributed to each player,

\begin{align*}
        u^{\prime}_{i_1 i_2 i_3 } \equiv \frac{1}{\sqrt{3}} \bigg( \underset{1 \leq j \leq 3}{\sum} \ket{\text{Player } j \text{ state}} \bigg)      \text{, }  v^{\prime}_{i_1 i_2 i_3} \equiv \ket{i_1 i_2 i_3 } \\  u^{\prime}_{i_2 i_1 i_3 } \equiv \frac{1}{\sqrt{3}} \bigg( \ket{\text{Player 1 state}} - \ket{\text{Player 2 state}} + \ket{\text{Player } 3 \text{ state}} \bigg) \text{, } v^{\prime}_{i_2 i_1 i_3} \equiv \ket{i_2 i_1 i_3} \\   u^{\prime}_{i_1 i_3 i_2 } \equiv \frac{1}{\sqrt{3}} \bigg(  \underset{1 \leq j \leq 2}{\sum} \ket{\text{Player } j \text{ state}}  - \ket{\text{Player } 3 \text{ state}} \bigg)       \text{, } v^{\prime}_{i_1 i_3 i_2} \equiv \ket{i_1 i_3 i_2} \text{. }
 \end{align*}

\noindent We conclude the argument by observing that the equality,

\begin{align*}
     y_i E_{ii} - G_{\mathrm{Sym}, \mathrm{Expanded}\wedge \cdots \wedge \mathrm{Expanded}} =    \frac{1}{\bigg( C_{\mathrm{Expanded} \wedge \mathrm{Expanded}} \omega_{\mathrm{Expanded} \wedge \mathrm{Expanded}} \bigg) \bigg(   3  \underset{1 \leq j \leq 2}{\prod}  \big( n - j \big) \bigg)   }           \\ \times   \underset{i_3 \in \mathcal{Q}_3}{\underset{i_2 \in \mathcal{Q}_2 }{\underset{i_1 \in \mathcal{Q}_1}{\sum}}}   \bigg(  \big(    u^{\prime}_{i_1i_2 i_3}   -    v^{\prime}_{i_1i_2 i_3}  \big) \big(   u^{\prime}_{i_1 i_2 i_3}   -    v^{\prime}_{i_1 i_2 i_3}    \big)^{\textbf{T}}      +  \big(    u^{\prime}_{i_2 i_1 i_3 }    -    v^{\prime}_{i_2 i_1 i_3}  \big)   \big( u^{\prime}_{i_2 i_1 i_3 }   -    v^{\prime}_{i_2 i_1 i_3  }  \big)^{\textbf{T}}  \\ +     \big(    u^{\prime}_{i_1 i_3 i_2}   -    v^{\prime}_{i_1 i_3 i_2}   \big)   \big(   u^{\prime}_{i_1 i_3 i_2}     -    v^{\prime}_{i_1 i_3 i_2}   \big)^{\textbf{T}}     \bigg)          \text{, }
\end{align*}

\noindent holds by taking $C_{\mathrm{Expanded} \wedge\cdots \wedge \mathrm{Expanded}} \equiv \frac{1}{N! C_{\mathrm{Expanded}}}$. \boxed{}

\bigskip

\noindent Under two operations of parallel repetition in the expanded game, one expects that the exponential rate of decay for the optimal value would take the form, [16],

\begin{align*}
    \mathrm{val} \big( G_{\mathrm{Exp}}^{\otimes 2} \big) \leq \mathrm{exp} \bigg(  - \frac{2c \epsilon^5 }{ k^2\mathrm{log} \big| A \big|}       \bigg)  \text{. }
\end{align*}

\noindent Under the previously mentioned identification, under parallel repetition,

\begin{align*}
 \bra{ \big( i_1 i_2 \big)  \wedge  \big(      i^{\prime}_1 i^{\prime}_2 \big) }       =   \bra{i_1 i_2} \wedge \bra{i^{\prime}_1 i^{\prime}_2}   \text{, } \\  \ket{ \big( i_1 i_2 \big)  \wedge  \big(      i^{\prime}_1 i^{\prime}_2 \big) }    = \ket{i_1 i_2 } \wedge \ket{ i^{\prime}_1 i^{\prime}_2 }           \text{, }
\end{align*}

\noindent one can establish that the following result, as a simplifaction of the more general result under an arbitrary number of parallel repetition operations, also holds.

\bigskip

\noindent \textbf{Theorem} \textit{2} (\textit{set of equivalent characterizations under two operations of parallel repetition for the expanded game}). An analogous result, corresponding to that stated in \textbf{Theorem} \textit{1}, holds for two operations of parallel repetition, in place for an arbitrary number of parallel repetition operations.

\bigskip

\noindent \textit{Proof of Theorem 2}. Directly apply the argument for the previous result, incorporating the exponential rate of decay for $\mathrm{val} \big( G^{\otimes 2}_{\mathrm{Exp}} \big)$, from which we conclude the argument. \boxed{}

\bigskip

\noindent Less surprisingly, under an arbitrary number of parallel repetition operations previous arguments adapted by the author in [46] for parallel repetition also directly apply to expanded games. We state this result below.

\bigskip

\noindent \textbf{Lemma} \textit{5} (\textit{parallel repetition of }$\sqrt{\epsilon^{\wedge}}$- \textit{expanded game approximality}, \textbf{Lemma} \textit{8}, [50]). From the same quantities introduced in the previous result, one has, for constants $N^{\prime}$ and $\epsilon^{\wedge,\prime}$ taken sufficiently small,

\begin{align*}
   \bigg| \bigg|               \bigg( \big(  A_k  \wedge A_{k^{\prime}} \big) \otimes \textbf{I} \bigg) \bigg(  \ket{\psi^{\prime}}    \wedge  \ket{\psi^{\prime}}  \bigg)   -  \bigg( \textbf{I} \otimes \bigg(     \frac{\pm \big(  B_{kl} \wedge B_{k^{\prime} l^{\prime}} \big)  + \big( B_{lk} \wedge B_{k^{\prime} l^{\prime}} \big) }{\big| \pm \big(  B_{kl} \wedge B_{k^{\prime} l^{\prime}} \big)  + \big( B_{lk} \wedge B_{k^{\prime} l^{\prime}} \big) \big| }           \bigg) \bigg) \\ \times \bigg(  \ket{\psi^{\prime}}    \wedge  \ket{\psi^{\prime}}  \bigg)        \bigg| \bigg|   < \sqrt{N^{\prime} \epsilon^{\wedge,\prime}}     \text{. } 
\end{align*}

\noindent \textbf{Lemma} \textit{6} (\textit{generalized error bound for $\sqrt{\epsilon^{\wedge}}$ expanded game approximality under two parallel repetition operations}). Fix a parameter $\epsilon^{\wedge \wedge}_{\mathrm{Expanded}}$ sufficiently small. From the same quantities introduced in \textbf{Lemma} \textit{5}, for the expanded game one has that the quantities,

\begin{align*}
  \mathcal{I}^{\wedge \wedge}_1 \equiv  \bigg| \bigg|               \bigg( \big(  A_k  \wedge A_{k^{\prime}} \big) \bigotimes  \textbf{I}  \bigg) \bigg( \ket{\psi^{\prime}} \wedge  \ket{\psi^{\prime}} \bigg)     -  \bigg( \textbf{I}  \bigotimes \bigg(     \frac{\pm \big(  B_{kl} \wedge B_{k^{\prime} l^{\prime}} \big)  + \big( B_{lk} \wedge B_{ l^{\prime} k^{\prime}}  \big) }{\big| \pm \big(  B_{kl} \wedge B_{k^{\prime} l^{\prime}}  \big)  + \big( B_{lk} \wedge B_{ l^{\prime} k^{\prime}}\big) \big| }    \bigg)     \bigg) \end{align*}

      \begin{align*}  \times \bigg( \ket{\psi^{\prime}} \wedge \ket{\psi^{\prime}} \bigg) \bigg| \bigg| \\ \\ \mathcal{I}^{\wedge\wedge}_2 \equiv   \bigg| \bigg|    \bigg(    \frac{1}{\sqrt{\# \sigma^{\prime} }}  \bigg(    \underset{\text{Permutations } \sigma}{\sum}  \big(   B^{(2)}_{\sigma ( i_1, i_2)}         \wedge  B^{(2)}_{\sigma ( i^{\prime}_1, i^{\prime}_2)}        \big)      \bigg)  \bigotimes \textbf{I}  \bigg)      \bigg( \ket{\psi^{\prime}} \wedge  \ket{\psi^{\prime}} \bigg) \\ -  \bigg( \textbf{I} \bigotimes \frac{1}{\sqrt{\# \sigma }}  \bigg(    \underset{\text{Permutations } \sigma}{\sum}  \big(   B^{(2)}_{\sigma ( i_1, i_2)}         \wedge  B^{(2)}_{\sigma ( i^{\prime}_1, i^{\prime}_2)}   \big)     \bigg) \bigg)   \bigg( \ket{\psi^{\prime}}   \wedge \ket{\psi^{\prime}} \bigg)  \bigg| \bigg|      \text{, } 
\end{align*}

\noindent are strictly upper bounded by,

\begin{align*}
\underset{1 \leq j \leq 2}{\sum} \mathcal{I}^{\wedge\wedge}_j < 40 \sqrt{2 \epsilon^{\wedge \wedge}_{\mathrm{Expanded}}} \text{.}
 \end{align*}

\noindent \textbf{Lemma} \textit{7} (\textit{generalized error bound for $\sqrt{\epsilon^{\wedge}}$ expanded game approximality under an arbitrary number of parallel repetition operations}). Fix a parameter $\epsilon^{\wedge \cdots \wedge}_{\mathrm{Expanded}}$ sufficiently small. From the same quantities introduced in \textbf{Lemma} \textit{5}, for the expanded game one has that the quantities,

\begin{align*}
  \mathcal{I}^{\wedge\cdots\wedge}_1 \equiv  \bigg| \bigg|               \bigg( \big(  A_k  \wedge A_{k^{\prime}} \wedge \cdots \wedge A_{k^{\prime\cdots \prime}} \big) \bigotimes  \textbf{I}  \bigg) \bigg( \ket{\psi^{\prime}} \wedge \cdots \wedge \ket{\psi^{\prime}} \bigg)    \\  -  \bigg( \textbf{I}  \bigotimes \bigg(     \frac{\pm \big(  B_{kl} \wedge B_{k^{\prime} l^{\prime}} \wedge \cdots \wedge B_{k^{\prime\cdots\prime}l^{\prime\cdots \prime}}  \big)  + \big( B_{lk} \wedge B_{ l^{\prime} k^{\prime}} \wedge \cdots \wedge B_{l^{\prime\cdots \prime} k^{\prime\cdots\prime} }  \big) }{\big| \pm \big(  B_{kl} \wedge B_{k^{\prime} l^{\prime}} \wedge \cdots \wedge B_{k^{\prime\cdots\prime}l^{\prime\cdots \prime}}  \big)  + \big( B_{lk} \wedge B_{ l^{\prime} k^{\prime}} \wedge \cdots \wedge B_{l^{\prime\cdots \prime} k^{\prime\cdots\prime} }  \big) \big| }    \bigg)     \bigg) \\  \times \bigg( \ket{\psi^{\prime}} \wedge \cdots \wedge \ket{\psi^{\prime}} \bigg) \bigg| \bigg| \end{align*}

  \begin{align*} \mathcal{I}^{\wedge \cdots \wedge}_2 \equiv   \bigg| \bigg|    \bigg(    \frac{1}{\sqrt{\# \sigma^{\prime} }}  \bigg(    \underset{\text{Permutations } \sigma}{\sum}  \big(   B^{(2)}_{\sigma ( i_1, i_2)}         \wedge  B^{(2)}_{\sigma ( i^{\prime}_1, i^{\prime}_2)}      \wedge  \cdots \wedge   B^{(2)}_{\sigma ( i^{\prime\cdots\prime}_1, i^{\prime\cdots\prime}_2)}   \big)      \bigg)  \bigotimes \textbf{I}  \bigg)    \\ \times   \bigg( \ket{\psi^{\prime}} \wedge \cdots  \wedge \ket{\psi^{\prime}} \bigg) \\ -  \bigg( \textbf{I} \bigotimes \frac{1}{\sqrt{\# \sigma }}  \bigg(    \underset{\text{Permutations } \sigma}{\sum}  \big(   B^{(2)}_{\sigma ( i_1, i_2)}         \wedge  B^{(2)}_{\sigma ( i^{\prime}_1, i^{\prime}_2)}      \wedge  \cdots \wedge   B^{(2)}_{\sigma ( i^{\prime\cdots\prime}_1, i^{\prime\cdots\prime}_2)}   \big)      \bigg) \bigg)   \\ \times \bigg( \ket{\psi^{\prime}} \wedge \cdots   \wedge \ket{\psi^{\prime}} \bigg)  \bigg| \bigg|      \text{, } 
\end{align*}

\noindent are strictly upper bounded by,

\begin{align*}
\underset{1 \leq j \leq 2}{\sum} \mathcal{I}^{\wedge \cdots \wedge}_j < 40 \sqrt{2 \epsilon^{\wedge \cdots \wedge}_{\mathrm{Expanded}}} \text{.}
 \end{align*}

\noindent We conclude the subsection by providing an analog of the following approximality result of the bias under parallel repetition.

\bigskip

\noindent \textbf{Lemma} \textit{8} (\textit{$\epsilon$-approximality of the bias of 3-XOR, N-XOR, and strong parallel repetitions of XOR and FFL games}, \textbf{Lemma} \textit{2}, [42]). Fix $\epsilon_1$, $\epsilon_2$, $\epsilon_3$, and $\epsilon_4$ sufficiently small. One has,

\begin{align*}
     \overset{r}{\underset{k=1}{\sum} } \bigg|\bigg|    \bigg( u_k \cdot \vec{A} \bigotimes \bigg(   \textbf{I} \bigotimes \textbf{I} \bigg)  \bigg) \ket{\psi_{3\mathrm{XOR}}}       -              \bigg( \bigg(  \textbf{I} \bigotimes  v_k \bigotimes \textbf{I} \bigg) \cdot \vec{B} \bigg) \ket{\psi_{3\mathrm{XOR}}}   \bigg|\bigg|^2 \\ \leq \beta \big( G_{3\mathrm{XOR}} \big) \epsilon_1 \text{,} \\  \tag{1} \\   \overset{r}{\underset{k=1}{\sum} } \bigg|\bigg|    \bigg( u_k \cdot \vec{A} \bigotimes \bigg(  \underset{1 \leq z \leq N-1}{\bigotimes}\textbf{I}_z \bigg) \bigg) \ket{\psi_{N\mathrm{XOR}}}       -              \bigg( \bigg(  \underset{1 \leq z \leq N-1}{\bigotimes} \textbf{I}_z  \bigg) \bigotimes v_k \cdot \vec{B} \bigg)  \ket{\psi_{N\mathrm{XOR}}}    \bigg|\bigg|^2   \\ \leq \beta \big( G_{N\mathrm{XOR}} \big) \epsilon_3 \text{, } \\   \tag{2} \\  \overset{r}{\underset{k=1}{\sum} } \bigg|\bigg|    \bigg( u_k \cdot \bigg( \vec{A} \wedge \cdots \wedge \vec{A}^{\prime\cdots\prime} \bigg)  \bigotimes \bigg(  \underset{1 \leq z \leq N-1}{\bigotimes}\textbf{I}_z \bigg) \bigg) \ket{\psi_{N\mathrm{XOR}\wedge\cdots\wedge N\mathrm{XOR}}}       -              \bigg( \bigg(   \underset{1 \leq z \leq N-1} {\bigotimes} \textbf{I}_z  \bigg) \\ \bigotimes v_k \cdot  \bigg( \vec{B} \wedge \cdots  \wedge \vec{B}^{\prime\cdots\prime} \bigg)  \bigg)  \ket{\psi_{N\mathrm{XOR}\wedge\cdots\wedge N\mathrm{XOR}}}    \bigg|\bigg|^2    \\ \leq \beta \big( G_{N\mathrm{XOR}\wedge\cdots\wedge N\mathrm{XOR}} \big) \epsilon_2 \text{, }  \\  \tag{3} \\  \overset{r}{\underset{k=1}{\sum} } \bigg|\bigg|    \bigg( u_k \cdot \vec{A} \bigotimes \bigg( \textbf{I} \bigotimes \textbf{I} \bigg) \bigg) \ket{\psi_{\mathrm{FFL}\wedge\mathrm{FFL}}}       -              \bigg( \bigg(   \textbf{I} \bigotimes \textbf{I} \bigg)  v_k \cdot \vec{B} \bigg)  \ket{\psi_{\mathrm{FFL}\wedge\mathrm{FFL}}}    \bigg|\bigg|^2 \\    \leq \beta \big( G_{\mathrm{FFL}\wedge\mathrm{FFL}} \big) \epsilon_4 \text{,} \tag{4}
\end{align*}    

\noindent for the optimal strategies for each game.

\bigskip

\noindent We state the corresponding result for parallel repetition of the bias below.

\bigskip

\noindent \textbf{Lemma} \textit{9} (\textit{epsilon approximality of the expanded game bias under parallel repetition}). For $\epsilon \equiv \epsilon_{\mathrm{Expanded} \wedge \cdots \wedge \mathrm{Expanded}}$ taken sufficiently small, one has,

\begin{align*}
     \overset{r}{\underset{k=1}{\sum} } \bigg|\bigg|    \bigg( u_k \cdot \vec{A} \bigotimes    \textbf{I}    \bigg) \ket{\psi^{\prime}}      -              \bigg( \bigg(  \textbf{I} \bigotimes  v_k  \bigg) \cdot \vec{B} \bigg) \ket{\psi^{\prime}}  \bigg|\bigg|^2 \leq \beta \big( G_{\mathrm{Expanded}} \big) \epsilon\text{, }
\end{align*}

\noindent for the expanded game bias,

\begin{align*}
    \beta \big( G_{\mathrm{Expanded}} \big)  \equiv   \underset{\mathcal{Q}_{\mathrm{Expanded}}}{\sum}    \bra{\psi^{\prime}}             \bigg(   \underset{\text{Players}}{\bigotimes}     \text{Player Tensor Observables}    \bigg)                     \ket{\psi^{\prime}}         \text{, }
\end{align*}

\noindent given the question set,

\begin{align*}
 \mathcal{Q}_{\mathrm{Expanded} } \equiv \mathcal{Q}_A \cup \mathcal{Q}_B   \sim \pi_{\mathrm{Expanded}} \text{. }
\end{align*}

\section{Dependency-breaking arguments: symmetry-breaking of Quantum correlations in states of player strategies}

\noindent We provide general descriptions for each of the main three results corresponding to multiplayer parallel repetition under anchoring. After having provided a general description of each result, we provide the accompanying argument, specifically in obtaining the dominant term for each system of expectation values. First, recall that the system of expectation values which we desire to upper bound are,

\[  \widetilde{\mathscr{E}_{\Gamma_i}} \equiv  \left\{\!\begin{array}{ll@{}>{{}}l} 
   \underset{R_{-i} | W_C}{\textbf{E}} \underset{\mathcal{Q}_1}{\textbf{E}}       \big| \big|  \big[ \big( \mathcal{U}_1 \bigotimes \textbf{I}^{\otimes N-1} \big) - \Gamma_1  \big] \widetilde{\ket{\psi} }   \big| \big|\text{, } \\    \underset{R_{-i} | W_C}{\textbf{E}} \underset{\mathcal{Q}_2}{\textbf{E}}      \big| \big|  \big[ \big( \textbf{I} \bigotimes \mathcal{U}_2 \bigotimes \textbf{I}^{\otimes N-2} \big) - \Gamma_2 \big] \widetilde{\ket{\psi} }     \big| \big|\text{, } \\ \vdots \\    \underset{R_{-i} | W_C}{\textbf{E}} \underset{\mathcal{Q}_N}{\textbf{E}}     \big| \big|  \big[ \big( \textbf{I}^{\otimes N-1 } \bigotimes \mathcal{U}_N  \big) - \Gamma_N  \big] \widetilde{\ket{\psi} }    \big| \big|  \text{, } \\ \underset{R_{-i} | W_C}{\textbf{E}} \underset{\mathcal{Q}_1 \mathcal{Q}_2 }{\textbf{E}}     \big| \big|  \big[ \big( \textbf{I}^{\otimes N-1 } \bigotimes \mathcal{U}_N  \big) - \Gamma_1 \Gamma_2  \big] \widetilde{\ket{\psi} }   \big| \big| \text{, } \\ \vdots \\ \underset{R_{-i} | W_C}{\textbf{E}} \underset{\mathcal{Q}_1 \times \cdots \times \mathcal{Q}_N}{\textbf{E}}    \bigg| \bigg|  \bigg[ \big( \textbf{I}^{\otimes N-1 } \bigotimes \mathcal{U}_N  \big) - \bigg[ \underset{1 \leq i \leq N}{\prod} \Gamma_i \bigg] \bigg] \widetilde{\ket{\psi} }  \bigg| \bigg|  .
\end{array}\right. 
\]

\[  \mathscr{P}\mathscr{O}\mathscr{V} \mathscr{M} \mathscr{E} \equiv  \left\{\!\begin{array}{ll@{}>{{}}l} 
   \underset{I}{\textbf{E}} \bigg| \bigg|      \textbf{P}_{R_{-i}| W_C}   \textbf{P}_{\mathcal{Q}_1 \mathcal{Q}_2 \mathcal{Q}_3} \mathscr{P}\mathscr{O}\mathscr{V} \mathscr{M}_1 -   \textbf{P}_{\mathcal{Q}_1 \mathcal{Q}_2 \mathcal{Q}_3 R_{-i} \mathcal{A}_1 \mathcal{A}_2 \mathcal{A}_3 | W_C}           \bigg| \bigg| ,     \\     \underset{I}{\textbf{E}}  \bigg| \bigg|  \textbf{P}_{R_{-i}| W_C}   \textbf{P}_{\mathcal{Q}_1 \mathcal{Q}_2 \mathcal{Q}_3}  \mathscr{P}\mathscr{O}\mathscr{V} \mathscr{M}_2 - \textbf{P}_{\mathcal{Q}_1 \mathcal{Q}_2 \mathcal{Q}_3 \mathcal{Q}_4 R_{-i} \mathcal{A}_1 \mathcal{A}_2 \mathcal{A}_3 \mathcal{A}_4 | W_C}            \bigg| \bigg|  ,  \\ \vdots \\   \underset{I}{\textbf{E}}  \bigg| \bigg|  \textbf{P}_{R_{-i}| W_C}   \textbf{P}_{\mathcal{Q}_1 \mathcal{Q}_2 \mathcal{Q}_3}  \mathscr{P}\mathscr{O}\mathscr{V} \mathscr{M}_N - \textbf{P}_{\mathcal{Q}_1 \mathcal{Q}_2 \mathcal{Q}_3 \mathcal{Q}_4 \times \cdots \times \mathcal{Q}_N R_{-i} \mathcal{A}_1 \mathcal{A}_2 \mathcal{A}_3 \mathcal{A}_4 \times \cdots \times \mathcal{A}_N | W_C}            \bigg| \bigg|   . \end{array}\right. 
\]

\[ \mathcal{P} \mathcal{I} \equiv   \left\{\!\begin{array}{ll@{}>{{}}l} 
\underset{I}{\textbf{E}} \underset{R_{-i} | W_C}{\textbf{E}} \underset{\mathcal{Q}_1 \mathcal{Q}_2 \times \cdots \times \mathcal{Q}_N}{\textbf{E}}  \big| \big| \textbf{I}^{E_1}_{r_{-i}, q_1,q_2, \cdots, q_N}  -  \textbf{I}^{E_1}_{r_{-i}, q_1,\bot,q_3, \cdots, q_N} \big| \big|^2_1   , \\ \underset{I}{\textbf{E}} \underset{R_{-i} | W_C}{\textbf{E}} \underset{\mathcal{Q}_1 \mathcal{Q}_2 \times \cdots \times \mathcal{Q}_N}{\textbf{E}}  \big| \big| \textbf{2}^{E_2}_{r_{-i}, q_1,q_2, \cdots, q_N}  -  \textbf{2}^{E_2}_{r_{-i}, \bot, q_2, q_3, \cdots, q_N} \big| \big|^2_1  ,  \\ \vdots \\ \underset{I}{\textbf{E}} \underset{R_{-i} | W_C}{\textbf{E}} \underset{\mathcal{Q}_1 \mathcal{Q}_2 \times \cdots \times \mathcal{Q}_N}{\textbf{E}}  \big| \big| \textbf{N}^{E_N}_{r_{-i}, q_1,q_2, \cdots, q_N}  -  \textbf{N}^{E_N}_{r_{-i}, q_1,\cdots, q_{N-2}, \bot, q_N} \big| \big|^2_1  ,
\end{array}\right. 
\]

\noindent corresponding to the system of expected values corresponding to the action of the unitary operations for each player, $\mathcal{U}_1, \cdots, \mathcal{U}_N$, POVMs $\mathscr{P}\mathscr{O}\mathscr{V}\mathscr{M}_1, \cdots, \mathscr{P}\mathscr{O}\mathscr{V}\mathscr{M}_N$, and states $\textbf{I},\cdots, \textbf{N}$, respectively. To relate upper bounds for each system to one another, we make use of observations gathered from the previously mentioned thresholds, 

 \begin{align*}
         \frac{\xi^2 \epsilon^4}{14440000}   \text{. }
        \end{align*}

        \noindent and,

 \begin{align*}
              \frac{\xi^2 \epsilon^4}{14440000 \mathrm{log} \big[ \big| \mathcal{A} \big| \big| \mathcal{B} \big| \big] }    \text{, }
        \end{align*}

\noindent provided in [30] to generalize arguments provided for anchored parallel repetition in the two-player setting provided in [2, 3, 4]. For the second main result, provided in \textbf{Theorem} \textit{2}, one must incorporate additional structure into the collection of unitaries,

\[  \left\{\!\begin{array}{ll@{}>{{}}l} 
 \mathcal{U}_{q_1} \equiv \text{\textit{First player unitary}}   \text{, } \\  \mathcal{U}_{q_2} \equiv \text{\textit{Second player unitary}}  \text{, } \\ \vdots \\ \mathcal{U}_{q_N} \equiv \text{\textit{N th player unitary}}  \text{, } 
\end{array}\right. \equiv   \left\{\!\begin{array}{ll@{}>{{}}l} 
 \mathcal{U}_1 \equiv \text{\textit{First player unitary}}   \text{, } \\  \mathcal{U}_2 \equiv \text{\textit{Second player unitary}}  \text{, } \\ \vdots \\ \mathcal{U}_N \equiv \text{\textit{N th player unitary}}  \text{, } 
\end{array}\right. 
\]

\noindent introduced for each player, through the collection of unitaries,

\[   \left\{\!\begin{array}{ll@{}>{{}}l} 
 \mathcal{U}_{1,r_{-i}} \equiv \text{\textit{First player dependency-breaking unitary}}   \text{, } \\  \mathcal{U}_{2,r_{-i}} \equiv \text{\textit{Second player  dependency-breaking unitary}}  \text{, } \\ \vdots \\ \mathcal{U}_{N,r_{-i}} \equiv \text{\textit{N th player dependency-breaking unitary}}  \text{, } 
\end{array}\right. 
\] 

\noindent corresponding to \textit{dependency-breaking unitaries}. In comparison to the collection of unitaries $\mathcal{U}_j$ without \textit{dependency-breaking}, each unitary $\mathcal{U}_{j,r_{-i}}$, for $1 \leq j \leq N$, with \textit{dependency-breaking} allows for one to upper bound $\mathscr{P}\mathscr{O}\mathscr{V}\mathscr{M}\mathscr{E}$ from the system,

\[  \mathscr{P}\mathscr{O}\mathscr{V} \mathscr{M}  \equiv  \left\{\!\begin{array}{ll@{}>{{}}l} 
   \big| \big|      \textbf{P}_{R_{-i}| W_C}   \textbf{P}_{\mathcal{Q}_1 \mathcal{Q}_2 \mathcal{Q}_3} \mathscr{P}\mathscr{O}\mathscr{V} \mathscr{M}_1 -   \textbf{P}_{\mathcal{Q}_1 \mathcal{Q}_2 \mathcal{Q}_3 R_{-i} \mathcal{A}_1 \mathcal{A}_2 \mathcal{A}_3 | W_C}           \big| \big| ,     \\  \big| \big|  \textbf{P}_{R_{-i}| W_C}   \textbf{P}_{\mathcal{Q}_1 \mathcal{Q}_2 \mathcal{Q}_3}  \mathscr{P}\mathscr{O}\mathscr{V} \mathscr{M}_2 - \textbf{P}_{\mathcal{Q}_1 \mathcal{Q}_2 \mathcal{Q}_3 \mathcal{Q}_4 R_{-i} \mathcal{A}_1 \mathcal{A}_2 \mathcal{A}_3 \mathcal{A}_4 | W_C}            \big| \big|  ,  \\ \vdots \\     \big| \big|  \textbf{P}_{R_{-i}| W_C}   \textbf{P}_{\mathcal{Q}_1 \mathcal{Q}_2 \mathcal{Q}_3}  \mathscr{P}\mathscr{O}\mathscr{V} \mathscr{M}_N - \textbf{P}_{\mathcal{Q}_1 \mathcal{Q}_2 \mathcal{Q}_3 \mathcal{Q}_4 \times \cdots \times \mathcal{Q}_N R_{-i} \mathcal{A}_1 \mathcal{A}_2 \mathcal{A}_3 \mathcal{A}_4 \times \cdots \times \mathcal{A}_N | W_C}            \big| \big|   . \end{array}\right. 
\]

\noindent of expected values. The upper bound for the stochastic system above, and hence for the system of expected values in $\mathscr{P}\mathscr{O}\mathscr{V}\mathscr{M}\mathscr{E}$, can be upper bounded by straightforwardly adapting the argumented provided in [2, 3, 4].

\noindent Moreoever, while it is straightforward to obtain upper bounds for the system of expected values with Pinkser's inequality, which are provided in the last system described in \textbf{Theorem} \textit{3}, we formulate a system of expected values of the relative entropy,

\[ \mathcal{P} \mathcal{I}^{\prime} \equiv   \left\{\!\begin{array}{ll@{}>{{}}l} 
\frac{1}{N}  \underset{\textbf{R}|W_C}{\textbf{E}} S \bigg( \bigg(  r_{-i} \text{ } \textit{1 st player state}      \bigg) \bigg| \bigg| \bigg(   r_{-i} \text{ } \textit{2 nd player state}      \bigg) \bigg)   , \\ \frac{1}{N}  \underset{\textbf{R}|W_C}{\textbf{E}} S \bigg( \bigg(  r_{-i} \text{ } \textit{2 nd player state}      \bigg) \bigg| \bigg| \bigg(   r_{-i} \text{ } \textit{3 rd player state}      \bigg) \bigg)  ,  \\ \vdots \\ \frac{1}{N}  \underset{\textbf{R}|W_C}{\textbf{E}} S \bigg( \bigg(  r_{-i} \text{ } \textit{(N-1) th player state}      \bigg) \bigg| \bigg| \bigg(   r_{-i} \text{ } \textit{N th player state}      \bigg) \bigg)  ,
\end{array}\right. 
\]

\noindent which can be upper bounded, in expectation with relative-min entropies which are then related to $\delta_{\mathrm{Multiplayer}}$.

\subsection{Preliminary Lemma}

\noindent We make use of the following result:

\bigskip

\noindent \textbf{Lemma} (\textbf{Lemma} \textit{4.6}, [3], \textit{determining the value of the multiplayer $\delta$ threshold for differences of probabilities}). One has that:

\begin{itemize}
\item[$\bullet$] \textit{The average of the probabilities of sampling dependency-breaking variables is strictly upper bounded by a fraction of $\delta_{\mathrm{Multiplayer}}$}:

\begin{align*}
\frac{1}{N} \underset{1 \leq i \leq N}{\sum}  \big| \big| \textbf{P} _{\Omega_i \mathcal{Q}_1 \times \cdots \times \mathcal{Q}_N | W_C} - \textbf{P}_{\Omega_i \mathcal{Q}_1 \times \cdots \times \mathcal{Q}_N}  \big| \big|       \leq \sqrt{\delta^{\frac{N}{300}}_{\mathrm{Multiplayer}}} .
\end{align*}

\item[$\bullet$] \textit{The average of the probabilities of sampling $R$ is strictly upper bounded by a fraction of $\delta_{\mathrm{Multiplayer}}$}:

\begin{align*}
   \frac{1}{N} \underset{1 \leq i \leq N}{\sum}  \big| \big| \textbf{P}_{R \mathcal{Q}_1 \times \cdots \times \mathcal{Q}_N | W_C} - \textbf{P}_{R \mathcal{Q}_1 \times \cdots \times \mathcal{Q}_N}  \big| \big|       \leq \sqrt{\delta^{\frac{N}{300}}_{\mathrm{Multiplayer}}} .
\end{align*}

\item[$\bullet$] \textit{The average of the probabilities of sampling $R$, conditioned upon all player responding with anchored responses $\bot$, is strictly upper bounded by a fraction of $\frac{\delta_{\mathrm{Multiplayer}}}{\alpha_{\mathrm{Multiplayer}}}$}:

\begin{align*}
   \frac{1}{N} \underset{1 \leq i \leq N}{\sum}  \big| \big| \textbf{P}_{\Omega_i | W_C} \textbf{P}_{R_{-i} | \mathcal{Q}_1 \equiv \bot, \cdots, \mathcal{Q}_N \equiv \bot , W_C} -  \textbf{P}_{\Omega_i | W_C} \textbf{P}_{R_{-i} | \Omega_{-i} W_C }  \big| \big|       \\ \leq  \mathrm{O} \bigg(  \frac{\sqrt{\delta^{\frac{N}{300}}_{\mathrm{Multiplayer}}}}{\alpha^{N+1}_{\mathrm{Multiplayer}}} \bigg)   .
\end{align*}

\item[$\bullet$] \textit{The average of the probabilities of sampling $\mathcal{Q}_1, \cdots, \mathcal{Q}_N$, conditioned upon all player responding with anchored responses $\bot$, is strictly upper bounded by a fraction of $\frac{\delta_{\mathrm{Multiplayer}}}{\alpha_{\mathrm{Multiplayer}}}$}:

\begin{align*}
   \frac{1}{N} \underset{1 \leq i \leq N}{\sum}  \big| \big| \textbf{P}_{\Omega_i | W_C} \textbf{P}_{R_{-i} | \mathcal{Q}_1 \equiv \bot, \cdots, \mathcal{Q}_N \equiv \bot , W_C} -  \textbf{P}_{\mathcal{Q}_1 \times \cdots \times \mathcal{Q}_N | W_C} \textbf{P}_{R_{-i} | \mathcal{Q}_1 \times \cdots \times \mathcal{Q}_N W_C }  \big| \big|     \\   \leq                    \mathrm{O} \bigg(  \frac{\sqrt{\delta^{\frac{N}{300}}_{\mathrm{Multiplayer}}}}{\alpha^{N+1}_{\mathrm{Multiplayer}}} \bigg)   .
\end{align*}

\end{itemize}

\noindent \textit{Proof of Lemma}. Directly apply the arguments presented for \textbf{Lemma} \textit{4.6} in [3], from which we conclude the argument. \boxed{}

\bigskip

\noindent \textbf{Remark}. The following inequalities provided in the previous item above are related to the following probabilistic statements:

\begin{itemize}
    \item[$\bullet$] \textit{Conditionally upon $W_C$, the probability of sampling $R$ given any $\mathcal{Q}_i$ is upper bounded by a fraction of $\frac{\delta_{\mathrm{Multiplayer}}}{\alpha_{\mathrm{Multiplayer}}}$}. In expectation,

\[   \left\{\!\begin{array}{ll@{}>{{}}l} 
   \underset{I}{\textbf{E}} \big| \big|   \textbf{P}_{R \mathcal{Q}_1 | W_C}  - \textbf{P}_{\Omega_i | W_C } \textbf{P}_{R_{-i} | W_C} \textbf{P}_{\mathcal{Q}_1 | \Omega_{-i}}    \big| \big|   \leq      \mathrm{O} \bigg(  \frac{\sqrt{\delta^{\frac{N}{300}}_{\mathrm{Multiplayer}}}}{\alpha^{N+1}_{\mathrm{Multiplayer}}} \bigg)        ,  \\ \vdots \\    \underset{I}{\textbf{E}} \big| \big|   \textbf{P}_{R \mathcal{Q}_N | W_C}  - \textbf{P}_{\Omega_i | W_C } \textbf{P}_{R_{-i} | W_C} \textbf{P}_{\mathcal{Q}_N | \Omega_{-i}}    \big| \big|   \leq      \mathrm{O} \bigg(  \frac{\sqrt{\delta^{\frac{N}{300}}_{\mathrm{Multiplayer}}}}{\alpha^{N+1}_{\mathrm{Multiplayer}}} \bigg)   \text{, }
\end{array}\right. 
\]

      \item[$\bullet$] \textit{Conditionally upon $W_C$, the difference of the probabilities of sampling $R$, and $\Omega$, is upper bounded by a fraction of $\frac{\delta_{\mathrm{Multiplayer}}}{\alpha_{\mathrm{Multiplayer}}}$}. In expectation,

\[   \left\{\!\begin{array}{ll@{}>{{}}l} 
 \underset{I}{\textbf{E}} \big| \big|   \textbf{P}_{R \mathcal{Q}_1 | W_C}    -  \textbf{P}_{\Omega_i | W_C} \textbf{P}_{R_{-i} | W_C}  \textbf{P}_{\mathcal{Q}_1 | \Omega_i}      \big| \big|   \leq  \sqrt{\delta^{\frac{N}{300}}_{\mathrm{Multiplayer}}}  ,  \\ \vdots \\    \underset{I}{\textbf{E}} \big| \big|   \textbf{P}_{R \mathcal{Q}_N | W_C}    -  \textbf{P}_{\Omega_i | W_C} \textbf{P}_{R_{-i} | W_C}  \textbf{P}_{\mathcal{Q}_N | \Omega_i}    \big| \big|  \leq \sqrt{\delta^{\frac{N}{300}}_{\mathrm{Multiplayer}}}   \text{, }
\end{array}\right. 
\]

  \item[$\bullet$] \textit{Conditionally upon $W_C$, and $\Omega_{-i}$, the difference of conditional probabilities dependent upon sampling $\Omega_{-i}, R_{-i},$ and each $\mathcal{Q}_i$ is upper bounded by a fraction of $\frac{\delta_{\mathrm{Multiplayer}}}{\alpha_{\mathrm{Multiplayer}}}$}. In expectation, 

\[   \left\{\!\begin{array}{ll@{}>{{}}l} 
 \underset{I}{\textbf{E}} \big| \big|  \textbf{P}_{\Omega_{-i} | W_C} \textbf{P}_{R_{-i} | \Omega_{-i} W_C} \textbf{P}_{\mathcal{Q}_1 | \Omega_i}      - \textbf{P}_{\Omega_{-i} | W_C} \textbf{P}_{R_{-i} | W_C} \textbf{P}_{\mathcal{Q}_1 | \Omega_i}     \big| \big|  \leq    \mathrm{O} \bigg(  \frac{\sqrt{\delta^{\frac{N}{300}}_{\mathrm{Multiplayer}}}}{\alpha^{N+1}_{\mathrm{Multiplayer}}} \bigg)     ,  \\ \vdots \\ \underset{I}{\textbf{E}} \big| \big|  \textbf{P}_{\Omega_{-i} | W_C} \textbf{P}_{R_{-i} | \Omega_{-i} W_C} \textbf{P}_{\mathcal{Q}_N | \Omega_i}      - \textbf{P}_{\Omega_{-i} | W_C} \textbf{P}_{R_{-i} | W_C} \textbf{P}_{\mathcal{Q}_N | \Omega_i}        \big| \big|  \leq     \mathrm{O} \bigg(  \frac{\sqrt{\delta^{\frac{N}{300}}_{\mathrm{Multiplayer}}}}{\alpha^{N+1}_{\mathrm{Multiplayer}}} \bigg)      \text{. }
\end{array}\right. 
\]

\end{itemize}

\subsection{General description of arguments for Theorem $1$}

\noindent To argue that \textbf{Theorem} \textit{1} holds, recall that the set of all dependency-breaking variables,

\begin{align*}
   \Omega_{\mathrm{Multiplayer} } \equiv  \underset{\omega \in \Omega_{\mathrm{Multiplayer}}}{\bigcup} \big\{ \omega  : \omega       \text{ is a dependency breaking variable for} \\ \text{position j}       \big\} \text{, }
\end{align*}

\noindent can be utilized to remove quantum correlations in strategies used by each player, which are related to the probabilities,

\[   \left\{\!\begin{array}{ll@{}>{{}}l} 
     \textbf{P} \big[ \Omega_{\mathrm{Multiplayer}} \big| \textbf{X}_{1,2,\cdots,N-1} \equiv x \backslash \big\{ N \big\} , \textbf{X}_N \equiv \bot ,  W_C  \big]  \approx       \textbf{P} \big[ \Omega_{\mathrm{Multiplayer}} \\ \big| \textbf{X}_{1,2,\cdots,N-2} \equiv x \backslash \big\{ N -1 , N \big\} , \textbf{X}_{N-1} \equiv \bot  , \textbf{X}_N \equiv \bot ,  W_C  \big] \\ \vdots \\  \approx  \textbf{P} \big[ \Omega_{\mathrm{Multiplayer}}  \big| \textbf{X}_{1 } \equiv x \backslash \big\{ 2, \cdots , N -1 , N \big\} , \textbf{X}_2 \dots \equiv \textbf{X}_N \equiv \bot ,  W_C  \big]         \text{, }
\end{array}\right. 
\]

\noindent of sampling a multiplayer dependency-breaking variable. In comparison to previous results provided in [2, 3, 4] which determined whether upper bounds, of the order,

\begin{align*}
    \mathrm{O} \bigg( \frac{\delta^{\frac{1}{16}}}{\alpha^{\frac{5}{4}}} \bigg) , \\   \mathrm{O} \bigg( \frac{\delta^{\frac{1}{16}}}{\alpha^2} \bigg), \\ \mathrm{O} \bigg( \frac{\delta^{\frac{1}{8}}}{\alpha^2} \bigg) ,
\end{align*}

\noindent existed for anchored parallel repetition of two-player game-theoretic settings, we suitably normalized dependecy-breaking states, with,

\[    \left\{\!\begin{array}{ll@{}>{{}}l} 
\widetilde{\ket{\psi_{r_{-i},\bot , \cdots , \bot }}}  \equiv \frac{\ket{\psi_{r_{-i},\bot , \cdots , \bot }}}{\big| \big|\ket{\psi_{r_{-i},\bot , \cdots , \bot }}  \big| \big|  } , \\ \widetilde{\ket{\psi_{r_{-i},q_1 , \bot \cdots , \bot}}} \equiv \frac{\ket{\psi_{r_{-i},q_1 , \bot \cdots , \bot}}}{\big| \big| \ket{\psi_{r_{-i},q_1 , \bot \cdots , \bot}} \big| \big|} , \\ \widetilde{\ket{\psi_{r_{-i},\bot , \cdots , \bot }}}  \equiv \frac{\ket{\psi_{r_{-i},\bot , \cdots , \bot }} }{\big| \big| \ket{\psi_{r_{-i},\bot , \cdots , \bot }}  \big| \big| } , \\  \widetilde{\ket{\psi_{r_{-i},\bot , \cdots , \bot,  q_N }}}  \equiv \frac{\ket{\psi_{r_{-i},\bot , \cdots , \bot,  q_N }}}{\big| \big| \ket{\psi_{r_{-i},\bot , \cdots , \bot,  q_N }}  \big| \big| } ,  \\  \widetilde{\ket{\psi_{r_{-i},\bot \backslash q_1 , q_2, \cdots, q_N}}}  \equiv \frac{\ket{\psi_{r_{-i},\bot \backslash q_1 , q_2, \cdots, q_N}} }{\big| \big| \ket{\psi_{r_{-i},\bot \backslash q_1 , q_2, \cdots, q_N}}  \big| \big| }       , \\  \widetilde{\ket{\psi_{r_{-i},\bot \backslash q_1 ,\bot, q_3, \cdots, q_N}}}  \equiv \frac{\ket{\psi_{r_{-i},\bot \backslash q_1 ,\bot, q_3, \cdots, q_N}}}{\big| \big|\ket{\psi_{r_{-i},\bot \backslash q_1 ,\bot, q_3, \cdots, q_N}}          \big| \big| } , \\ \vdots  
\end{array}\right. 
\]

\noindent as given in the system,

\[  \widetilde{\mathscr{E}_{\bot}} \equiv  \left\{\!\begin{array}{ll@{}>{{}}l} 
   \underset{R_{-i} | W_C}{\textbf{E}} \underset{\mathcal{Q}_1}{\textbf{E}}       \big| \big| \big( \mathcal{U}_{1,r_{-i}} \bigotimes \textbf{I}^{\otimes N-1} \big)  \widetilde{\ket{\psi_{r_{-i},\bot , \cdots , \bot }}}   - \widetilde{\ket{\psi_{r_{-i},q_1 , \bot \cdots , \bot}}}    \big| \big| \text{, } \\    \underset{R_{-i} | W_C}{\textbf{E}} \underset{\mathcal{Q}_2}{\textbf{E}}      \big| \big|\big( \textbf{I} \bigotimes \mathcal{U}_{2,r_{-i}} \bigotimes \textbf{I}^{\otimes N-2} \big)\widetilde{\ket{\psi_{r_{-i},\bot , \cdots , \bot }}}    -  \widetilde{\ket{\psi_{r_{-i}, \bot , q_2, \bot , \cdots , \bot}}}   \big| \big|\text{, } \\ \vdots \\    \underset{R_{-i} | W_C}{\textbf{E}} \underset{\mathcal{Q}_N}{\textbf{E}}     \big| \big| \big( \textbf{I}^{\otimes N-1 } \bigotimes \mathcal{U}_{N,r_{-i}}  \big) \widetilde{\ket{\psi_{r_{-i},\bot , \cdots , \bot }}}  - \widetilde{\ket{\psi_{r_{-i},\bot , \cdots , \bot,  q_N }}}    \big| \big| \text{, } \\ \underset{R_{-i} | W_C}{\textbf{E}} \underset{\mathcal{Q}_1 \mathcal{Q}_2 }{\textbf{E}}    \big| \big| \big( \textbf{I}^{\otimes N-1 } \bigotimes \mathcal{U}_{N,r_{-i}}  \big) \widetilde{\ket{\psi_{r_{-i},\bot \backslash q_1 , q_2, \cdots, q_N}}}  - \widetilde{\ket{\psi_{r_{-i},\bot \backslash q_1 ,\bot, q_3, \cdots, q_N}}}   \big| \big|  \text{, } \\ \vdots \\ \underset{R_{-i} | W_C}{\textbf{E}} \underset{\mathcal{Q}_1 \times \cdots \times \mathcal{Q}_N}{\textbf{E}}    \big| \big|\big( \textbf{I}^{\otimes N-1 } \bigotimes \mathcal{U}_{N,r_{-i}}  \big) \widetilde{\ket{\psi_{r_{-i}, q_1 \backslash \bot, q_2 , \cdots ,  q_N} }} \\  - \widetilde{\ket{\psi_{r_{-i}, q_1 \backslash \bot, q_2 \backslash \bot, \cdots , q_{N-1} \backslash \bot, q_N} }}  \big| \big| ,
\end{array}\right. 
\]

\noindent of expected values.

\subsection{Argument}

\noindent \textit{Proof of Theorem 1}. To upper bound the system of expectation values provided in $\widetilde{\mathscr{E}_{\bot}}$, recall that it is related to the systems,

\[  \widetilde{\mathscr{E}} \equiv  \left\{\!\begin{array}{ll@{}>{{}}l} 
   \underset{R_{-i} | W_C}{\textbf{E}} \underset{\mathcal{Q}_1}{\textbf{E}}       \big| \big|  \big[ \big( \mathcal{U}_1 \bigotimes \textbf{I}^{\otimes N-1} \big) - \textbf{I} \big] \widetilde{\ket{\psi} }   \big| \big|\text{, } \\    \underset{R_{-i} | W_C}{\textbf{E}} \underset{\mathcal{Q}_2}{\textbf{E}}      \big| \big|  \big[ \big( \textbf{I} \bigotimes \mathcal{U}_2 \bigotimes \textbf{I}^{\otimes N-2} \big) - \textbf{I} \big] \widetilde{\ket{\psi} }     \big| \big|\text{, } \\ \vdots \\    \underset{R_{-i} | W_C}{\textbf{E}} \underset{\mathcal{Q}_N}{\textbf{E}}     \big| \big|  \big[ \big( \textbf{I}^{\otimes N-1 } \bigotimes \mathcal{U}_N  \big) - \textbf{I} \big] \widetilde{\ket{\psi} }    \big| \big|  \text{, } \\ \underset{R_{-i} | W_C}{\textbf{E}} \underset{\mathcal{Q}_1 \mathcal{Q}_2 }{\textbf{E}}     \big| \big|  \big[  \big( \textbf{I}^{\otimes N-1 } \bigotimes \mathcal{U}_N  \big) - \textbf{I} \big] \widetilde{\ket{\psi} }   \big| \big| \text{, } \\ \vdots \\ \underset{R_{-i} | W_C}{\textbf{E}} \underset{\mathcal{Q}_1 \times \cdots \times \mathcal{Q}_N}{\textbf{E}}    \big| \big|  \big[  \big( \textbf{I}^{\otimes N-1 } \bigotimes \mathcal{U}_N  \big) - \textbf{I} \big] \widetilde{\ket{\psi} }  \big| \big|  ,
\end{array}\right. 
\]

\[   \left\{\!\begin{array}{ll@{}>{{}}l} 
   \underset{R_{-i} | W_C}{\textbf{E}} \underset{\mathcal{Q}_1}{\textbf{E}}       \big| \big| \big( \mathcal{U}_{1,r_{-i}} \bigotimes \textbf{I}^{\otimes N-1} \big) \ket{\psi_{r_{-i},\bot , \cdots , \bot }}   - \ket{\psi_{r_{-i},q_1 , \bot \cdots , \bot}}    \big| \big| \text{, } \\    \underset{R_{-i} | W_C}{\textbf{E}} \underset{\mathcal{Q}_2}{\textbf{E}}      \big| \big|\big( \textbf{I} \bigotimes \mathcal{U}_{2,r_{-i}} \bigotimes \textbf{I}^{\otimes N-2} \big)\ket{\psi_{r_{-i},\bot , \cdots , \bot }}    -  \ket{\psi_{r_{-i}, \bot , q_2, \bot , \cdots , \bot}}    \big| \big|\text{, } \\ \vdots \\    \underset{R_{-i} | W_C}{\textbf{E}} \underset{\mathcal{Q}_N}{\textbf{E}}     \big| \big| \big( \textbf{I}^{\otimes N-1 } \bigotimes \mathcal{U}_{N,r_{-i}}  \big) \ket{\psi_{r_{-i},\bot , \cdots , \bot }}  - \ket{\psi_{r_{-i},\bot , \cdots , \bot,  q_N }}    \big| \big| \text{, } 
\\ 
\underset{R_{-i} | W_C}{\textbf{E}} \underset{\mathcal{Q}_1 \mathcal{Q}_2 }{\textbf{E}}    \big| \big| \big( \textbf{I}^{\otimes N-1 } \bigotimes \mathcal{U}_{N,r_{-i}}  \big) \ket{\psi_{r_{-i},\bot \backslash q_1 , q_2, \cdots, q_N}}  - \ket{\psi_{r_{-i},\bot \backslash q_1 ,\bot, q_3, \cdots, q_N}}   \big| \big|  \text{, } \end{array}\right. 
\]

\[   \left\{\!\begin{array}{ll@{}>{{}}l}  \vdots \\ \underset{R_{-i} | W_C}{\textbf{E}} \underset{\mathcal{Q}_1 \times \cdots \times \mathcal{Q}_N}{\textbf{E}}    \big| \big|\big( \textbf{I}^{\otimes N-1 } \bigotimes \mathcal{U}_{N,r_{-i}}  \big) \ket{\psi_{r_{-i}, q_1 \backslash \bot, q_2 , \cdots ,  q_N} } \\  - \ket{\psi_{r_{-i}, q_1 \backslash \bot, q_2 \backslash \bot, \cdots , q_{N-1} \backslash \bot, q_N} }  \big| \big| ,
\end{array}\right. 
\]

\noindent of expectation values. It suffices to demonstrate that a stochastic domination of the form,

\begin{align*}
    \mathrm{O} \bigg( \frac{\delta^{M_1}}{\alpha^{N+1}_{\mathrm{Multiplayer}}} \bigg) \equiv           \underset{R_{-i} | \bot, \bot , \cdots, \bot, W_C}{\textbf{E}} \underset{\mathcal{Q}_1}{\textbf{E}} \big| \big|        \ket{\widetilde{\Psi}_{q_1, \bot, \cdots, \bot}} - \big(    \mathcal{U}_1 \otimes \textbf{I} \big)   \ket{\widetilde{\Psi}_{\bot,\cdots, \bot}}             \big| \big|       \\   <           \underset{R_{-i}, \bot,\cdots,\bot}{\textbf{E}} \underset{\mathcal{Q}_1}{\textbf{E}}  \big| \big|     \big( \mathcal{U}_{r_{-i},1} \otimes \textbf{I} \big) \ket{\widetilde{\Psi}_{r_{-i}, \bot, \cdots, \bot}}   -      \ket{\widetilde{\Psi}_{r_{-i}, q_1, \bot, \cdots, \bot}}       \big| \big|               \\    \equiv \mathrm{O} \bigg( \frac{\delta^{M_2}}{\alpha^{N+1}_{\mathrm{Multiplayer}}} \bigg)   \Longleftrightarrow M_1 < M_2 ,   
 \end{align*}

\noindent holds, for question sets $\mathcal{Q}_i$ corresponding to player $i$. Over joint probability measures for the questions of each player, the above stochastic domination would instead take the form,

\begin{align*}
    \mathrm{O} \bigg( \frac{\delta^{M_1}}{\alpha^{N+1}_{\mathrm{Multiplayer}}} \bigg) \equiv           \underset{R_{-i} | \bot, \bot , \cdots, \bot, W_C}{\textbf{E}} \underset{\mathcal{Q}_1\times \cdots \times \mathcal{Q}_N}{\textbf{E}} \big| \big|        \ket{\widetilde{\Psi}_{q_1, \bot, \cdots, \bot}} - \big(    \mathcal{U}_1 \otimes \textbf{I} \big)   \ket{\widetilde{\Psi}_{\bot,\cdots, \bot}}             \big| \big|       \\   <           \underset{R_{-i}, \bot,\cdots,\bot}{\textbf{E}} \underset{\mathcal{Q}_1\times \cdots \times \mathcal{Q}_N}{\textbf{E}}  \big| \big|     \big( \mathcal{U}_{r_{-i},1} \otimes \textbf{I} \big) \ket{\widetilde{\Psi}_{r_{-i}, \bot, \cdots, \bot}}   -      \ket{\widetilde{\Psi}_{r_{-i}, q_1, \bot, \cdots, \bot}}       \big| \big|               \\    \equiv \mathrm{O} \bigg( \frac{\delta^{M_2}}{\alpha^{N+1}_{\mathrm{Multiplayer}}} \bigg)   \Longleftrightarrow M_1 < M_2 .   
 \end{align*}

\noindent Obtaining analytical expressions for $M_1$ and $M_2$, which are reliant upon $N$ and,

\begin{align*}
   \bigg|  \underset{1 \leq i \leq N}{\prod} \big| \mathcal{Q}_i \big|  \bigg|    , 
\end{align*}

\noindent is dependent upon computations with expectations of the form,

\[  \widetilde{\widetilde{\mathscr{E}}} \equiv  \left\{\!\begin{array}{ll@{}>{{}}l} 
   \underset{R_{-i}| q_1, \bot, \cdots, \bot, W_C}{\textbf{E}} \underset{\mathcal{Q}_1}{\textbf{E}}  \big| \big| \big[ \mathcal{U}_1 \otimes \textbf{I}^{\otimes N-1} \big]  \ket{\widetilde{\Psi}_{\bot, q_2, \cdots, q_N}} -  \ket{\widetilde{\Psi}_{q_1, \cdots, q_N}}   \big| \big|   ,  \\ \vdots \\  \underset{R_{-i}| q_1, \bot, \cdots, \bot, W_C}{\textbf{E}} \underset{\mathcal{Q}_1\times \cdots \times \mathcal{Q}_N}{\textbf{E}}  \big| \big| \big[ \mathcal{U}_1 \otimes \cdots \otimes \mathcal{U}_N \big]  \ket{\widetilde{\Psi}_{\bot, \cdots, \bot}} -  \ket{\widetilde{\Psi}_{q_1, \cdots, q_N}}   \big| \big|              . 
\end{array}\right. 
\]

\noindent To demonstrate,

\begin{align*}
   \widetilde{\widetilde{\mathscr{E}}} = \widetilde{\mathscr{E}_{\bot}}= \mathrm{O} \bigg(  \frac{\delta^{\frac{N}{300}}_{\mathrm{Multiplayer}}}{\alpha^{N+1}_{\mathrm{Multiplayer}}}     \bigg) ,
\end{align*}

\noindent write,

\begin{align*}
    \underset{R_{-i}| q_1, \bot, \cdots, \bot, W_C}{\textbf{E}} \underset{\mathcal{Q}_1}{\textbf{E}}  \big| \big| \ket{\widetilde{\Psi_{r_{-i}, q_1, \bot,\cdots, \bot}}}  - \mathcal{U}_{r_{-i},1}\ket{\widetilde{\Psi_{r_{-i}, \bot, \bot, \cdots, \bot}}}  \big| \big| \\ \leq   \sqrt{    \underset{R_{-i}| q_1, \bot, \cdots, \bot, W_C}{\textbf{E}} \underset{\mathcal{Q}_1}{\textbf{E}}  \big| \big| \ket{\widetilde{\Psi_{r_{-i}, q_1, \bot,\cdots, \bot}}}  - \mathcal{U}_{r_{-i},1}\ket{\widetilde{\Psi_{r_{-i}, \bot, \bot, \cdots, \bot}}}  \big| \big|^2 }   \\ =  \mathrm{O} \bigg(  \frac{\delta^{\frac{N}{300}}_{\mathrm{Multiplayer}}}{\alpha^{N+1}_{\mathrm{Multiplayer}}}     \bigg)  ,
\end{align*}

\noindent Computations of the above form are used to upper bound the system,

\[   \left\{\!\begin{array}{ll@{}>{{}}l}      \underset{R_{-i}|  \bot,q_2,  \cdots, \bot, W_C}{\textbf{E}} \underset{\mathcal{Q}_1}{\textbf{E}}  \big| \big| \ket{\widetilde{\Psi_{r_{-i},  \bot, q_2, \bot , \cdots, \bot}}}  - \mathcal{U}_{r_{-i},1}\ket{\widetilde{\Psi_{r_{-i}, \bot, \bot, \cdots, \bot}}}  \big| \big|          ,              \\ \vdots \\    \underset{R_{-i}| \bot , \bot, \cdots, \bot,q_N,  W_C}{\textbf{E}} \underset{\mathcal{Q}_1}{\textbf{E}}  \big| \big| \ket{\widetilde{\Psi_{r_{-i}, \bot,\cdots, \bot, q_N}}}  - \mathcal{U}_{r_{-i},1}\ket{\widetilde{\Psi_{r_{-i}, \bot, \bot, \cdots, \bot}}}  \big| \big|                , \end{array}\right. 
\]

\noindent of expectation values, with,

\[   \left\{\!\begin{array}{ll@{}>{{}}l}                  \sqrt{    \underset{R_{-i}| q_1, \bot, \cdots, \bot, W_C}{\textbf{E}} \underset{\mathcal{Q}_1}{\textbf{E}}  \big| \big| \ket{\widetilde{\Psi_{r_{-i}, \bot, q_2 , \bot,\cdots, \bot}}}  - \mathcal{U}_{r_{-i},2}\ket{\widetilde{\Psi_{r_{-i}, \bot, \bot, \cdots, \bot}}}  \big| \big|^2 } \\ \vdots \\              \sqrt{    \underset{R_{-i}| , \bot, \cdots, \bot, q_N , W_C}{\textbf{E}} \underset{\mathcal{Q}_1}{\textbf{E}}  \big| \big| \ket{\widetilde{\Psi_{r_{-i}, \bot, q_2 , \bot,\cdots, \bot}}}  - \mathcal{U}_{r_{-i},N}\ket{\widetilde{\Psi_{r_{-i}, \bot, \bot, \cdots, \bot}}}  \big| \big|^2 }    , \end{array}\right. 
\]

\noindent In particular, the upper bound to the above system not only determines how the fractional power to which $\delta_{\mathrm{Multiplayer}}$ is incorporated into the $\mathrm{O} \bigg( \cdot \bigg)$ upper bound, but also reflects upon the normalization in the denomination from the multiplayer anchoring probability, $\alpha_{\mathrm{Multiplayer}}$, as,

\[  \widetilde{\mathscr{E}_{\Gamma_i}} \equiv  \left\{\!\begin{array}{ll@{}>{{}}l} 
   \underset{R_{-i} | W_C}{\textbf{E}} \underset{\mathcal{Q}_1}{\textbf{E}}       \big| \big|  \big[ \big( \mathcal{U}_1 \bigotimes \textbf{I}^{\otimes N-1} \big) - \Gamma_1  \big] \widetilde{\ket{\psi} }   \big| \big|\text{, } \\    \underset{R_{-i} | W_C}{\textbf{E}} \underset{\mathcal{Q}_2}{\textbf{E}}      \big| \big|  \big[ \big( \textbf{I} \bigotimes \mathcal{U}_2 \bigotimes \textbf{I}^{\otimes N-2} \big) - \Gamma_2 \big] \widetilde{\ket{\psi} }     \big| \big|\text{, } \\ \vdots \\    \underset{R_{-i} | W_C}{\textbf{E}} \underset{\mathcal{Q}_N}{\textbf{E}}     \big| \big|  \big[ \big( \textbf{I}^{\otimes N-1 } \bigotimes \mathcal{U}_N  \big) - \Gamma_N  \big] \widetilde{\ket{\psi} }    \big| \big|  \text{, } \\ \underset{R_{-i} | W_C}{\textbf{E}} \underset{\mathcal{Q}_1 \mathcal{Q}_2 }{\textbf{E}}     \big| \big|  \big[ \big( \textbf{I}^{\otimes N-1 } \bigotimes \mathcal{U}_N  \big) - \Gamma_1 \Gamma_2  \big] \widetilde{\ket{\psi} }   \big| \big| \text{, } \\ \vdots \\ \underset{R_{-i} | W_C}{\textbf{E}} \underset{\mathcal{Q}_1 \times \cdots \times \mathcal{Q}_N}{\textbf{E}}    \bigg| \bigg|  \bigg[ \big( \textbf{I}^{\otimes N-1 } \bigotimes \mathcal{U}_N  \big) - \bigg[ \underset{1 \leq i \leq N}{\prod} \Gamma_i \bigg] \bigg] \widetilde{\ket{\psi} }  \bigg| \bigg|  .
\end{array}\right. 
\]

\[  \leq  \left\{\!\begin{array}{ll@{}>{{}}l}      \underset{R_{-i}|  q_1, \bot,  \cdots, \bot, W_C}{\textbf{E}} \underset{\mathcal{Q}_1}{\textbf{E}}  \big| \big| \ket{\widetilde{\Psi_{r_{-i}, q_1, \bot,\cdots, \bot}}}  - \mathcal{U}_{r_{-i},1}\ket{\widetilde{\Psi_{r_{-i}, \bot, \bot, \cdots, \bot}}}  \big| \big|          ,              \\ \vdots \\    \underset{R_{-i}| \bot , \bot, \cdots, \bot,q_N,  W_C}{\textbf{E}} \underset{\mathcal{Q}_1}{\textbf{E}}  \big| \big| \ket{\widetilde{\Psi_{r_{-i}, \bot,\cdots, \bot, q_N}}}  - \mathcal{U}_{r_{-i},1}\ket{\widetilde{\Psi_{r_{-i}, \bot, \bot, \cdots, \bot}}}  \big| \big|                , \end{array}\right. 
\]

\[  \leq  \left\{\!\begin{array}{ll@{}>{{}}l}                  \sqrt{    \underset{R_{-i}| q_1, \bot, \cdots, \bot, W_C}{\textbf{E}} \underset{\mathcal{Q}_1}{\textbf{E}}  \big| \big| \ket{\widetilde{\Psi_{r_{-i}, q_1,  \bot,\cdots, \bot}}}  - \mathcal{U}_{r_{-i},2}\ket{\widetilde{\Psi_{r_{-i}, \bot, \bot, \cdots, \bot}}}  \big| \big|^2 } \\ \vdots \\              \sqrt{    \underset{R_{-i}| , \bot, \cdots, \bot, q_N , W_C}{\textbf{E}} \underset{\mathcal{Q}_1}{\textbf{E}}  \big| \big| \ket{\widetilde{\Psi_{r_{-i}, \bot, q_2 , \bot,\cdots, \bot}}}  - \mathcal{U}_{r_{-i},N}\ket{\widetilde{\Psi_{r_{-i}, \bot, \bot, \cdots, \bot}}}  \big| \big|^2 }    , \end{array}\right. 
\] 

\[
\equiv \mathrm{O} \bigg(  \frac{\delta^{\frac{N}{300}}_{\mathrm{Multiplayer}}}{\alpha^{N+1}_{\mathrm{Multiplayer}}}     \bigg)
\]

\noindent from which we conclude the argument. \boxed{}

\bigskip

\noindent \textbf{Corollary} (\textit{purification from Uhlmann's Theorem}). The state $\ket{\Psi_{r_{-i}, q_1 , \bot , \cdots , \bot}}^{E_1 E_2 \times \cdots \times E_N}$ is a purification of $\textbf{I}^{E_2 \times \cdots \times  E_N}_{r_{-i}, q_1, \bot, \cdots, \bot}$.

\bigskip

\noindent \textit{Proof of Corollary}. Apply Uhlmann's Theorem, as in \textbf{Claim} \textit{5.16} from [3], from which we conclude the argument. \boxed{}

\subsection{General description of arguments for Theorem $2$}

\noindent To argue that \textbf{Theorem} \textit{2} holds, recall that to upper bound the POVM system,

\[  \mathscr{P}\mathscr{O}\mathscr{V} \mathscr{M} \mathscr{E} \equiv  \left\{\!\begin{array}{ll@{}>{{}}l} 
   \underset{I}{\textbf{E}} \bigg| \bigg|      \textbf{P}_{R_{-i}| W_C}   \textbf{P}_{\mathcal{Q}_1 \mathcal{Q}_2 \mathcal{Q}_3} \mathscr{P}\mathscr{O}\mathscr{V} \mathscr{M}_1 -   \textbf{P}_{\mathcal{Q}_1 \mathcal{Q}_2 \mathcal{Q}_3 R_{-i} \mathcal{A}_1 \mathcal{A}_2 \mathcal{A}_3 | W_C}           \bigg| \bigg| ,     \\     \underset{I}{\textbf{E}}  \bigg| \bigg|  \textbf{P}_{R_{-i}| W_C}   \textbf{P}_{\mathcal{Q}_1 \mathcal{Q}_2 \mathcal{Q}_3}  \mathscr{P}\mathscr{O}\mathscr{V} \mathscr{M}_2 - \textbf{P}_{\mathcal{Q}_1 \mathcal{Q}_2 \mathcal{Q}_3 \mathcal{Q}_4 R_{-i} \mathcal{A}_1 \mathcal{A}_2 \mathcal{A}_3 \mathcal{A}_4 | W_C}            \bigg| \bigg|  ,  \\ \vdots \\   \underset{I}{\textbf{E}}  \bigg| \bigg|  \textbf{P}_{R_{-i}| W_C}   \textbf{P}_{\mathcal{Q}_1 \mathcal{Q}_2 \mathcal{Q}_3}  \mathscr{P}\mathscr{O}\mathscr{V} \mathscr{M}_N - \textbf{P}_{\mathcal{Q}_1 \mathcal{Q}_2 \mathcal{Q}_3 \mathcal{Q}_4 \times \cdots \times \mathcal{Q}_N R_{-i} \mathcal{A}_1 \mathcal{A}_2 \mathcal{A}_3 \mathcal{A}_4 \times \cdots \times \mathcal{A}_N | W_C}            \bigg| \bigg|   , \end{array}\right. 
\]

\noindent it suffices to obtain upper bounds of the form,

\[   \left\{\!\begin{array}{ll@{}>{{}}l}  2    \underset{I}{\textbf{E}} \underset{R_{-i} | W_C}{\textbf{E}} \underset{\mathcal{Q}_1 \mathcal{Q}_2 \mathcal{Q}_3 \mathcal{Q}_4}{\textbf{E}} \big| \big|   \big( \mathcal{U}_{1,r_{-i}}  \otimes  \mathcal{U}_{2,r_{-i}}   \otimes  \mathcal{U}_{3,r_{-i}}  \otimes  \mathcal{U}_{4,r_{-i}}  \big)   \ket{\widetilde{\Psi_{r_{-i}, \bot,\bot,\bot, \bot}}}   - \ket{\widetilde{\Psi_{r_{-i},  q_1,q_2,q_3, q_4}}}      \big| \big| ,    \\ \vdots \\   \sqrt{N}    \underset{I}{\textbf{E}} \underset{R_{-i} | W_C}{\textbf{E}} \underset{\mathcal{Q}_1 \mathcal{Q}_2 \mathcal{Q}_3 \mathcal{Q}_4}{\textbf{E}} \big| \big|   \big( \mathcal{U}_{1,r_{-i}}  \otimes  \mathcal{U}_{2,r_{-i}}   \otimes  \mathcal{U}_{3,r_{-i}}  \otimes  \cdots \otimes \mathcal{U}_{N,r_{-i}}  \big)   \ket{\widetilde{\Psi_{r_{-i}, \bot,\bot,\bot, \cdots, \bot}}} \\   - \ket{\widetilde{\Psi_{r_{-i},  q_1,q_2,q_3, \cdots, q_N}}}      \big| \big|      . \end{array}\right. 
\]

\noindent The above system of expected values is related to the following two POVM systems, the first of which is,

\[  \mathscr{P}\mathscr{O}\mathscr{V}\mathscr{M}\mathscr{E}^{\prime} \equiv  \left\{\!\begin{array}{ll@{}>{{}}l} \big| \big| \mathscr{P}\mathscr{O}\mathscr{V} \mathscr{M}_1 -    \textbf{P}_{\mathcal{A}_1\mathcal{A}_2 \mathcal{A}_3 | r_{-i}, q_1, q_2, q_3}  \big| \big|  , \\ \big| \big| 
\mathscr{P}\mathscr{O}\mathscr{V} \mathscr{M}_2 -  \textbf{P}_{\mathcal{A}_1\mathcal{A}_2 \mathcal{A}_3 \mathcal{A}_4 | r_{-i}, q_1, q_2, q_3, q_4}  \big| \big|  , \\ \vdots \\  
\big| \big| 
\mathscr{P}\mathscr{O}\mathscr{V} \mathscr{M}_N -  \textbf{P}_{\mathcal{A}_1\mathcal{A}_2 \mathcal{A}_3 \cdots \mathcal{A}_N | r_{-i}, q_1, q_2, q_3, \cdots, q_N}  \big| \big| 
 , \end{array}\right. 
\] 

\noindent and the second of which is,

\[  \mathscr{P}\mathscr{O}\mathscr{V}\mathscr{M}\mathscr{E}^{\prime\prime} \equiv  \left\{\!\begin{array}{ll@{}>{{}}l} \underset{I}{\textbf{E}} \underset{R_{-i}| W_C}{\textbf{E}} \underset{\mathcal{Q}_1 \mathcal{Q}_2 \mathcal{Q}_3}{\textbf{E}} \big| \big| \mathscr{P}\mathscr{O}\mathscr{V} \mathscr{M}_1 -    \textbf{P}_{\mathcal{A}_1\mathcal{A}_2 \mathcal{A}_3 | r_{-i}, q_1, q_2, q_3}  \big| \big|  , \\ \underset{I}{\textbf{E}} \underset{R_{-i}| W_C}{\textbf{E}} \underset{\mathcal{Q}_1 \mathcal{Q}_2 \mathcal{Q}_3 \mathcal{Q}_4}{\textbf{E}} \big| \big| 
\mathscr{P}\mathscr{O}\mathscr{V} \mathscr{M}_2 -  \textbf{P}_{\mathcal{A}_1\mathcal{A}_2 \mathcal{A}_3 \mathcal{A}_4 | r_{-i}, q_1, q_2, q_3, q_4}  \big| \big|  , \\ \vdots \\  \underset{I}{\textbf{E}} \underset{R_{-i}| W_C}{\textbf{E}} \underset{\mathcal{Q}_1 \mathcal{Q}_2 \mathcal{Q}_3\times \cdots \times \mathcal{Q}_N}{\textbf{E}}
\big| \big| 
\mathscr{P}\mathscr{O}\mathscr{V} \mathscr{M}_N -  \textbf{P}_{\mathcal{A}_1\mathcal{A}_2 \mathcal{A}_3 \cdots \mathcal{A}_N | r_{-i}, q_1, q_2, q_3, \cdots, q_N}  \big| \big| 
 ,  \end{array}\right. 
\]

\noindent that are together used to conclude that the desired upper bound for the POVM system holds.

\subsection{Argument}

\noindent \textit{Proof of Theorem 2}. To upper bound the system of expectation values provided in $\mathscr{P}\mathscr{O}\mathscr{V} \mathscr{M} \mathscr{E}$, observe,

\begin{align*}
 \big| \big| \mathscr{P}\mathscr{O}\mathscr{V} \mathscr{M}_1 - \textbf{P}_{\mathcal{A}_1 \mathcal{A}_2 \mathcal{A}_3 | r_{-i} , q_1, q_2, q_3} \big| \big|  \leq   \big| \big|       \big( \mathcal{U}_{1,r_{-i}}  \otimes  \mathcal{U}_{2,r_{-i}}   \otimes  \mathcal{U}_{3,r_{-i}}   \big) \widetilde{\Psi_{r_{-i}, \bot,\bot,\bot}}  \big( \mathcal{U}_{1,r_{-i}}  \otimes  \mathcal{U}_{2,r_{-i}}   \otimes  \mathcal{U}_{3,r_{-i}}   \big)^{\dagger} \\  - \widetilde{\Psi_{r_{-i},  q_1,q_2,q_3}}  \big| \big|_1  \\ \leq \sqrt{3} \big| \big| \big( \mathcal{U}_{1,r_{-i}}  \otimes  \mathcal{U}_{2,r_{-i}}   \otimes  \mathcal{U}_{3,r_{-i}}   \big)   \ket{\widetilde{\Psi_{r_{-i}, \bot,\bot,\bot}}}   - \ket{\widetilde{\Psi_{r_{-i},  q_1,q_2,q_3}}} \big| \big|    .
\end{align*}

\noindent Expectedly, for the remaining expectation values within $\mathscr{P}\mathscr{O}\mathscr{V}\mathscr{M}$,

\[   \left\{\!\begin{array}{ll@{}>{{}}l}   \big| \big|  \textbf{P}_{R_{-i}| W_C}   \textbf{P}_{\mathcal{Q}_1 \mathcal{Q}_2 \mathcal{Q}_3}  \mathscr{P}\mathscr{O}\mathscr{V} \mathscr{M}_2 - \textbf{P}_{\mathcal{Q}_1 \mathcal{Q}_2 \mathcal{Q}_3 \mathcal{Q}_4 R_{-i} \mathcal{A}_1 \mathcal{A}_2 \mathcal{A}_3 \mathcal{A}_4 | W_C}            \big| \big|  ,  \\ \vdots \\     \big| \big|  \textbf{P}_{R_{-i}| W_C}   \textbf{P}_{\mathcal{Q}_1 \mathcal{Q}_2 \mathcal{Q}_3}  \mathscr{P}\mathscr{O}\mathscr{V} \mathscr{M}_N - \textbf{P}_{\mathcal{Q}_1 \mathcal{Q}_2 \mathcal{Q}_3 \mathcal{Q}_4 \times \cdots \times \mathcal{Q}_N R_{-i} \mathcal{A}_1 \mathcal{A}_2 \mathcal{A}_3 \mathcal{A}_4 \times \cdots \times \mathcal{A}_N | W_C}            \big| \big|   . \end{array}\right. 
\]

\noindent one has that,

\begin{align*}
 \big| \big| \mathscr{P}\mathscr{O}\mathscr{V} \mathscr{M}_2 - \textbf{P}_{\mathcal{A}_1 \mathcal{A}_2 \mathcal{A}_3 \mathcal{A}_4 | r_{-i} , q_1, q_2, q_3, q_4} \big| \big|  \leq   \big| \big|       \big( \mathcal{U}_{1,r_{-i}}  \otimes  \mathcal{U}_{2,r_{-i}}   \otimes  \mathcal{U}_{3,r_{-i}}   \big) \widetilde{\Psi_{r_{-i}, \bot,\bot,\bot}}  \big( \mathcal{U}_{1,r_{-i}}  \otimes  \mathcal{U}_{2,r_{-i}}   \\ \otimes  \mathcal{U}_{3,r_{-i}} \otimes \mathcal{U}_{4,r_{-i}}   \big)^{\dagger}   - \widetilde{\Psi_{r_{-i},  q_1,q_2,q_3}}  \big| \big|_1  \\ \leq 2 \big| \big| \big( \mathcal{U}_{1,r_{-i}}  \otimes  \mathcal{U}_{2,r_{-i}}   \otimes  \mathcal{U}_{3,r_{-i}}   \otimes \mathcal{U}_{4,r_{-i}}  \big)   \ket{\widetilde{\Psi_{r_{-i}, \bot,\bot,\bot}}}   - \ket{\widetilde{\Psi_{r_{-i},  q_1,q_2,q_3}}} \big| \big| \\ \vdots \end{align*}
 
 \begin{align*} \big| \big| \mathscr{P}\mathscr{O}\mathscr{V} \mathscr{M}_N - \textbf{P}_{\mathcal{A}_1 \mathcal{A}_2 \mathcal{A}_3 \cdots \mathcal{A}_N| r_{-i} , q_1, q_2, q_3, \cdots, q_N} \big| \big|  \leq   \bigg| \bigg|       \bigg( \mathcal{U}_{1,r_{-i}}  \otimes  \mathcal{U}_{2,r_{-i}}   \otimes  \mathcal{U}_{3,r_{-i}}  \otimes \cdots \otimes \mathcal{U}_{N,r_{-i}}     \bigg) \end{align*}

 \begin{align*} \times \widetilde{\Psi_{r_{-i}, \bot,\bot,\bot}}   \big( \mathcal{U}_{1,r_{-i}}  \otimes  \mathcal{U}_{2,r_{-i}}   \otimes  \mathcal{U}_{3,r_{-i}} \otimes \mathcal{U}_{4,r_{-i}} \otimes \cdots \otimes \mathcal{U}_{N,r_{-i}}    \big)^{\dagger}   - \widetilde{\Psi_{r_{-i},  q_1,q_2,q_3,\cdots, q_N}}  \bigg| \bigg|_1  \\ \leq \sqrt{N} \bigg| \bigg| \bigg( \mathcal{U}_{1,r_{-i}}  \otimes  \mathcal{U}_{2,r_{-i}}   \otimes  \mathcal{U}_{3,r_{-i}}   \otimes \mathcal{U}_{4,r_{-i}} \otimes \cdots \otimes \mathcal{U}_{N,r_{-i}}  \bigg)   \ket{\widetilde{\Psi_{r_{-i}, \bot,\bot,\bot}}}   - \ket{\widetilde{\Psi_{r_{-i},  q_1,q_2,q_3}}} \bigg| \bigg|   .
\end{align*}

\noindent Altogether, to make use of the upper bound, 

\begin{align*}
    \sqrt{3} \big| \big| \big( \mathcal{U}_{1,r_{-i}}  \otimes  \mathcal{U}_{2,r_{-i}}   \otimes  \mathcal{U}_{3,r_{-i}}   \big)   \ket{\widetilde{\Psi_{r_{-i}, \bot,\bot,\bot}}}   - \ket{\widetilde{\Psi_{r_{-i},  q_1,q_2,q_3}}} \big| \big|   , 
\end{align*}

\noindent for,

\begin{align*}
 \big| \big| \mathscr{P}\mathscr{O}\mathscr{V} \mathscr{M}_1 - \textbf{P}_{\mathcal{A}_1 \mathcal{A}_2 \mathcal{A}_3 | r_{-i} , q_1, q_2, q_3} \big| \big| ,
\end{align*}

\noindent to upper bound,

\begin{align*}
     \underset{I}{\textbf{E}} \bigg| \bigg|      \textbf{P}_{R_{-i}| W_C}   \textbf{P}_{\mathcal{Q}_1 \mathcal{Q}_2 \mathcal{Q}_3} \mathscr{P}\mathscr{O}\mathscr{V} \mathscr{M}_1 -   \textbf{P}_{\mathcal{Q}_1 \mathcal{Q}_2 \mathcal{Q}_3 R_{-i} \mathcal{A}_1 \mathcal{A}_2 \mathcal{A}_3 | W_C}           \bigg| \bigg| , 
\end{align*}

\noindent write,

\begin{align*}
 \underset{I}{\textbf{E}} \bigg| \bigg|      \textbf{P}_{R_{-i}| W_C}   \textbf{P}_{\mathcal{Q}_1 \mathcal{Q}_2 \mathcal{Q}_3} \mathscr{P}\mathscr{O}\mathscr{V} \mathscr{M}_1 -   \textbf{P}_{\mathcal{Q}_1 \mathcal{Q}_2 \mathcal{Q}_3 R_{-i} \mathcal{A}_1 \mathcal{A}_2 \mathcal{A}_3 | W_C}           \bigg| \bigg| \equiv     \underset{I}{\textbf{E}} \bigg| \bigg|      \textbf{P}_{R_{-i}| W_C}   \textbf{P}_{\mathcal{Q}_1 \mathcal{Q}_2 \mathcal{Q}_3} \mathscr{P}\mathscr{O}\mathscr{V} \mathscr{M}_1 \\ -   \textbf{P}_{\mathcal{Q}_1 \mathcal{Q}_2 \mathcal{Q}_3 R_{-i}} \textbf{P}_{ \mathcal{A}_1 \mathcal{A}_2 \mathcal{A}_3 | W_C}           \bigg| \bigg| \\ \equiv \underset{I}{\textbf{E}} \bigg| \bigg|      \textbf{P}_{R_{-i}| W_C}   \textbf{P}_{\mathcal{Q}_1 \mathcal{Q}_2 \mathcal{Q}_3} \mathscr{P}\mathscr{O}\mathscr{V} \mathscr{M}_1  -   \textbf{P}_{\mathcal{Q}_1 \mathcal{Q}_2 \mathcal{Q}_3 } \textbf{P}_{ \mathcal{A}_1 \mathcal{A}_2 \mathcal{A}_3 | r_{-i}, q_1, q_2, q_3}           \bigg| \bigg| \\ \equiv      \underset{I}{\textbf{E}} \bigg| \bigg|      \textbf{P}_{R_{-i}| W_C}   \textbf{P}_{\mathcal{Q}_1 \mathcal{Q}_2 \mathcal{Q}_3} \bigg[ \mathscr{P}\mathscr{O}\mathscr{V} \mathscr{M}_1  -    \textbf{P}_{ \mathcal{A}_1 \mathcal{A}_2 \mathcal{A}_3 | r_{-i}, q_1, q_2, q_3}       \bigg]     \bigg| \bigg| \\ \leq     \underset{I}{\textbf{E}} \underset{R_{-i} | W_C}{\textbf{E}} \underset{\mathcal{Q}_1 \mathcal{Q}_2 \mathcal{Q}_3}{\textbf{E}}  \big| \big|     \mathscr{P}\mathscr{O}\mathscr{V} \mathscr{M}_1  -    \textbf{P}_{ \mathcal{A}_1 \mathcal{A}_2 \mathcal{A}_3 | r_{-i}, q_1, q_2, q_3}       \big| \big| \\  \leq \sqrt{3}     \underset{I}{\textbf{E}} \underset{R_{-i} | W_C}{\textbf{E}} \underset{\mathcal{Q}_1 \mathcal{Q}_2 \mathcal{Q}_3}{\textbf{E}} \big| \big|   \big( \mathcal{U}_{1,r_{-i}}  \otimes  \mathcal{U}_{2,r_{-i}}   \otimes  \mathcal{U}_{3,r_{-i}}   \big)   \ket{\widetilde{\Psi_{r_{-i}, \bot,\bot,\bot}}}   - \ket{\widetilde{\Psi_{r_{-i},  q_1,q_2,q_3}}}      \big| \big|  . 
\end{align*}

\noindent One can obtain closely related upper bounds, namely,

\[   \left\{\!\begin{array}{ll@{}>{{}}l}  2    \underset{I}{\textbf{E}} \underset{R_{-i} | W_C}{\textbf{E}} \underset{\mathcal{Q}_1 \mathcal{Q}_2 \mathcal{Q}_3 \mathcal{Q}_4}{\textbf{E}} \big| \big|   \big( \mathcal{U}_{1,r_{-i}}  \otimes  \mathcal{U}_{2,r_{-i}}   \otimes  \mathcal{U}_{3,r_{-i}}  \otimes  \mathcal{U}_{4,r_{-i}}  \big)   \ket{\widetilde{\Psi_{r_{-i}, \bot,\bot,\bot, \bot}}}   - \ket{\widetilde{\Psi_{r_{-i},  q_1,q_2,q_3, q_4}}}      \big| \big| ,    \\ \vdots \\   \sqrt{N}    \underset{I}{\textbf{E}} \underset{R_{-i} | W_C}{\textbf{E}} \underset{\mathcal{Q}_1 \mathcal{Q}_2 \mathcal{Q}_3 \mathcal{Q}_4}{\textbf{E}} \big| \big|   \big( \mathcal{U}_{1,r_{-i}}  \otimes  \mathcal{U}_{2,r_{-i}}   \otimes  \mathcal{U}_{3,r_{-i}}  \otimes  \cdots \otimes \mathcal{U}_{N,r_{-i}}  \big)   \ket{\widetilde{\Psi_{r_{-i}, \bot,\bot,\bot, \cdots, \bot}}} \\   - \ket{\widetilde{\Psi_{r_{-i},  q_1,q_2,q_3, \cdots, q_N}}}      \big| \big|      , \end{array}\right. 
\] 

\noindent for upper bounding $\mathscr{P}\mathscr{O}\mathscr{V}\mathscr{M}\mathscr{E}$, through the intermediate systems,

\[  \mathscr{P}\mathscr{O}\mathscr{V}\mathscr{M}\mathscr{E}^{\prime} \equiv  \left\{\!\begin{array}{ll@{}>{{}}l} \big| \big| \mathscr{P}\mathscr{O}\mathscr{V} \mathscr{M}_1 -    \textbf{P}_{\mathcal{A}_1\mathcal{A}_2 \mathcal{A}_3 | r_{-i}, q_1, q_2, q_3}  \big| \big|  , \\ \big| \big| 
\mathscr{P}\mathscr{O}\mathscr{V} \mathscr{M}_2 -  \textbf{P}_{\mathcal{A}_1\mathcal{A}_2 \mathcal{A}_3 \mathcal{A}_4 | r_{-i}, q_1, q_2, q_3, q_4}  \big| \big|  , \\ \vdots \\  
\big| \big| 
\mathscr{P}\mathscr{O}\mathscr{V} \mathscr{M}_N -  \textbf{P}_{\mathcal{A}_1\mathcal{A}_2 \mathcal{A}_3 \cdots \mathcal{A}_N | r_{-i}, q_1, q_2, q_3, \cdots, q_N}  \big| \big| 
 , \end{array}\right. 
\] 

\noindent and,

\[  \mathscr{P}\mathscr{O}\mathscr{V}\mathscr{M}\mathscr{E}^{\prime\prime} \equiv  \left\{\!\begin{array}{ll@{}>{{}}l} \underset{I}{\textbf{E}} \underset{R_{-i}| W_C}{\textbf{E}} \underset{\mathcal{Q}_1 \mathcal{Q}_2 \mathcal{Q}_3}{\textbf{E}} \big| \big| \mathscr{P}\mathscr{O}\mathscr{V} \mathscr{M}_1 -    \textbf{P}_{\mathcal{A}_1\mathcal{A}_2 \mathcal{A}_3 | r_{-i}, q_1, q_2, q_3}  \big| \big|  , \\ \underset{I}{\textbf{E}} \underset{R_{-i}| W_C}{\textbf{E}} \underset{\mathcal{Q}_1 \mathcal{Q}_2 \mathcal{Q}_3 \mathcal{Q}_4}{\textbf{E}} \big| \big| 
\mathscr{P}\mathscr{O}\mathscr{V} \mathscr{M}_2 -  \textbf{P}_{\mathcal{A}_1\mathcal{A}_2 \mathcal{A}_3 \mathcal{A}_4 | r_{-i}, q_1, q_2, q_3, q_4}  \big| \big|  , \\ \vdots \\  \underset{I}{\textbf{E}} \underset{R_{-i}| W_C}{\textbf{E}} \underset{\mathcal{Q}_1 \mathcal{Q}_2 \mathcal{Q}_3\times \cdots \times \mathcal{Q}_N}{\textbf{E}}
\big| \big| 
\mathscr{P}\mathscr{O}\mathscr{V} \mathscr{M}_N -  \textbf{P}_{\mathcal{A}_1\mathcal{A}_2 \mathcal{A}_3 \cdots \mathcal{A}_N | r_{-i}, q_1, q_2, q_3, \cdots, q_N}  \big| \big| 
 ,  \end{array}\right. 
\]

\noindent of expectation values.

\bigskip

\noindent The final expectation value,

\begin{align*}
    \sqrt{3}     \underset{I}{\textbf{E}} \underset{R_{-i} | W_C}{\textbf{E}} \underset{\mathcal{Q}_1 \mathcal{Q}_2 \mathcal{Q}_3}{\textbf{E}} \big| \big|   \big( \mathcal{U}_{1,r_{-i}}  \otimes  \mathcal{U}_{2,r_{-i}}   \otimes  \mathcal{U}_{3,r_{-i}}   \big)   \ket{\widetilde{\Psi_{r_{-i}, \bot,\bot,\bot}}}   - \ket{\widetilde{\Psi_{r_{-i},  q_1,q_2,q_3}}}      \big| \big|  ,
\end{align*}

\noindent provided in the upper bound for,

\begin{align*}
 \underset{I}{\textbf{E}} \bigg| \bigg|      \textbf{P}_{R_{-i}| W_C}   \textbf{P}_{\mathcal{Q}_1 \mathcal{Q}_2 \mathcal{Q}_3} \mathscr{P}\mathscr{O}\mathscr{V} \mathscr{M}_1 -   \textbf{P}_{\mathcal{Q}_1 \mathcal{Q}_2 \mathcal{Q}_3 R_{-i} \mathcal{A}_1 \mathcal{A}_2 \mathcal{A}_3 | W_C}           \bigg| \bigg| , 
\end{align*}

\noindent of the three-player multiplayer game can be achieved with the following result:

\bigskip

\noindent \textbf{Proposition} (\textit{upper bounding the expected value from dependency-breaking unitaries for three players}). Given \textit{dependency-breaking unitaries} acting on $V^1, V^2, V^3$, one has that,

\begin{align*}
      \underset{I}{\textbf{E}} \underset{R_{-i} | W_C}{\textbf{E}} \underset{\mathcal{Q}_1 \mathcal{Q}_2 \mathcal{Q}_3 \mathcal{Q}_4 }{\textbf{E}} \big| \big|   \big( \mathcal{U}_{1,r_{-i}}  \otimes  \mathcal{U}_{2,r_{-i}}   \otimes  \mathcal{U}_{3,r_{-i}}   \big)   \ket{\widetilde{\Psi_{r_{-i}, \bot,\bot,\bot}}}   - \ket{\widetilde{\Psi_{r_{-i},  q_1,q_2,q_3}}}      \big| \big|  \leq \mathrm{O} \bigg(    \frac{\delta^{\frac{2}{150}}_{\mathrm{Multiplayer}}}{\alpha^{4}_{\mathrm{Multiplayer}}}   \bigg)  .
\end{align*}

\noindent \textit{Proof of Proposition}. The result follows from directly applying the computations provided in \textbf{Proposition} \textit{5.1} in [3].  The remaning computations from which it suffices to obtain the desired result, which claims that the system of expected values can be upper bounded with,

\begin{align*}
    \mathrm{O} \bigg(    \frac{\delta^{\frac{2}{150}}_{\mathrm{Multiplayer}}}{\alpha^{4}_{\mathrm{Multiplayer}}}   \bigg) , 
\end{align*}

\noindent is related to computations in the system,

\[   \left\{\!\begin{array}{ll@{}>{{}}l}   \underset{\mathcal{Q}_1 \times \cdots \times \mathcal{Q}_N}{\textbf{E}} \underset{R_{-i} | \bot , \cdots , \bot, W_C}{\textbf{E}}   \bigg| 1 -   \frac{\big| \big|             \ket{\psi_{r_{-i}, q_1, q_2, \cdots, q_N}}        \big| \big| }{\big| \big|    \ket{\psi_{r_{-i}, \bot, \cdots, \bot}}     \big|  \big| }    \bigg|   , \\  \underset{\mathcal{Q}_1 \times \cdots \times \mathcal{Q}_N}{\textbf{E}} \underset{R_{-i} | \bot , \cdots , \bot, W_C}{\textbf{E}}   \bigg| 1 -   \frac{\big| \big|             \ket{\psi_{r_{-i}, q_1 \backslash \bot, q_2, \cdots, q_N}}        \big| \big| }{\big| \big|    \ket{\psi_{r_{-i}, \bot, \cdots, \bot}}     \big|  \big| }    \bigg|          , \\ \underset{\mathcal{Q}_1 \times \cdots \times \mathcal{Q}_N}{\textbf{E}} \underset{R_{-i} | \bot , \cdots , \bot, W_C}{\textbf{E}}   \bigg| 1 -   \frac{\big| \big|             \ket{\psi_{r_{-i}, q_1 \backslash \bot, q_2 \backslash \bot , \cdots, q_N}}        \big| \big| }{\big| \big|    \ket{\psi_{r_{-i}, \bot, \cdots, \bot}}     \big|  \big| }    \bigg|   ,   \\ \vdots \\  \underset{\mathcal{Q}_1 \times \cdots \times \mathcal{Q}_N}{\textbf{E}} \underset{R_{-i} | \bot , \cdots , \bot, W_C}{\textbf{E}}   \bigg| 1 -   \frac{\big| \big|             \ket{\psi_{r_{-i}, q_1 \backslash \bot, q_2 \backslash \bot , \cdots, q_N \backslash \bot}}        \big| \big| }{\big| \big|    \ket{\psi_{r_{-i}, \bot, \cdots, \bot}}     \big|  \big| }    \bigg|    , \\  \underset{\mathcal{Q}_1 \times \cdots \times \mathcal{Q}_N}{\textbf{E}} \underset{R_{-i} | \bot , \cdots , \bot, W_C}{\textbf{E}}   \bigg| 1 -   \frac{\big| \big|             \ket{\psi_{r_{-i}, q_1, q_2 \backslash \bot , \cdots, q_N}}        \big| \big| }{\big| \big|    \ket{\psi_{r_{-i}, \bot, \cdots, \bot}}     \big|  \big| }    \bigg|   , \\ \vdots \\ \underset{\mathcal{Q}_1 \times \cdots \times \mathcal{Q}_N}{\textbf{E}} \underset{R_{-i} | \bot , \cdots , \bot, W_C}{\textbf{E}}   \bigg| 1 -   \frac{\big| \big|             \ket{\psi_{r_{-i}, q_1, q_2 \backslash \bot , \cdots, q_N \backslash \bot }}        \big| \big| }{\big| \big|    \ket{\psi_{r_{-i}, \bot, \cdots, \bot}}     \big|  \big| }    \bigg|     ,   \\ \vdots \\     \underset{\mathcal{Q}_1 \times \cdots \times \mathcal{Q}_N}{\textbf{E}} \underset{R_{-i} | \bot , \cdots , \bot, W_C}{\textbf{E}}      \bigg| 1 -    \frac{\big| \big| \ket{\psi_{r_{-i}, q_1 , q_2 , \cdots , q_{n-1} \backslash \bot , q_N }}                \big| \big| }{\big| \big|  \ket{\psi_{r_{-i}, \bot, \cdots, \bot}}          \big| \big| }     \bigg|   , \\ \underset{\mathcal{Q}_1 \times \cdots \times \mathcal{Q}_N}{\textbf{E}} \underset{R_{-i} | \bot , \cdots , \bot, W_C}{\textbf{E}}      \bigg| 1 -    \frac{\big| \big| \ket{\psi_{r_{-i}, q_1 , q_2 , \cdots , q_{n-1} \backslash \bot , q_N \backslash \bot }}                \big| \big| }{\big| \big|  \ket{\psi_{r_{-i}, \bot, \cdots, \bot}}          \big| \big| }     \bigg|                  . \end{array}\right. 
\]

\noindent To upper bound the above system, it suffices to argue that a result of the following form holds,

\begin{align*}
 \underset{\mathcal{Q}_1 \times \cdots \times \mathcal{Q}_N}{\textbf{E}} \bigg[ \underset{r_{-i} \in W_C}{\sum}   \bigg|   \textbf{P}_{r_{-i}, \bot, \cdots, \bot} \big( r_{-i} \big) -   \textbf{P}_{r_{-i}, q_1, \cdots,  q_N} \big( r_{-i} \big) \bigg|   \bigg] =  \mathrm{O} \bigg( \frac{\delta^{\frac{N}{300}}_{\mathrm{Multiplayer}}}{\alpha^{N+5}_{\mathrm{Multiplayer}}} \bigg) \textbf{P} \big[ W_C \big| \mathcal{Q} \equiv \bot \big] \\ \equiv   \mathrm{O} \bigg( \frac{\delta^{\frac{N}{300}}_{\mathrm{Multiplayer}}}{\alpha^{N+5}_{\mathrm{Multiplayer}}} \bigg) \textbf{P} \big[ W_C \big| \mathcal{Q}_1 \equiv \bot , \cdots , \mathcal{Q}_N \equiv \bot \big]  , 
\end{align*}

\noindent namely the fact that the above expectation occurs with probability at least $\mathrm{O}$ (the above claim is a generalization of \textbf{Claim} \textit{5.18}, [3]). For game-theoretic settings with less than $N$ players, the above statement takes the form,

\begin{align*}
 \underset{\mathcal{Q}_1 \mathcal{Q}_2 \mathcal{Q}_3}{\textbf{E}} \bigg[ \underset{r_{-i} \in W_C}{\sum}   \bigg|   \textbf{P}_{r_{-i}, \bot, \bot, \bot} \big( r_{-i} \big) -   \textbf{P}_{r_{-i}, q_1, q_2,  q_3} \big( r_{-i} \big) \bigg|   \bigg] =  \mathrm{O} \bigg( \frac{\delta^{\frac{2}{150}}_{\mathrm{Multiplayer}}}{\alpha^{4}_{\mathrm{Multiplayer}}} \bigg) \textbf{P} \big[ W_C \big| \mathcal{Q}  \equiv \bot \big] \\ \equiv   \mathrm{O} \bigg( \frac{\delta^{\frac{2}{150}}_{\mathrm{Multiplayer}}}{\alpha^{4}_{\mathrm{Multiplayer}}} \bigg) \textbf{P} \big[ W_C \big| \mathcal{Q}_1 \equiv \bot , \mathcal{Q}_2 \equiv \bot , \mathcal{Q}_3 \equiv \bot \big]  . 
\end{align*}

\noindent To demonstrate that the above statement holds, from which the generalized version with $N$ participants can also be shown to hold, observe,

\begin{align*}
 \underset{I}{\textbf{E}} \bigg[    \textbf{P}_{\mathcal{Q}_1 \mathcal{Q}_2 \mathcal{Q}_3}  \big[ q_1 , q_2, q_3 \big] \big|    \textbf{P} \big[ W_C \big| \mathcal{Q}_1 \equiv q_1, \mathcal{Q}_2 \equiv q_2, \mathcal{Q}_3 \equiv q_3  \big]  -   \textbf{P} \big[ W_C \big]   \big|   \bigg] \leq \mathrm{O} \bigg( \delta^{\frac{1}{200}}_{\mathrm{Multiplayer}} \bigg)     , \\ \\ \underset{I}{\textbf{E}}  \underset{\mathcal{Q}_1 \mathcal{Q}_2 \mathcal{Q}_3}{\textbf{E}}        \big|    \textbf{P} \big[ W_C \big| \mathcal{Q}_1 \equiv q_1, \mathcal{Q}_2 \equiv q_2, \mathcal{Q}_3 \equiv q_3  \big]  -   \textbf{P} \big[ W_C \big]   \big|      \leq \mathrm{O} \bigg( \frac{\delta^{\frac{1}{100}}_{\mathrm{Multiplayer}}}{\alpha^{4}_{\mathrm{Multiplayer}}} \bigg)    , \\  \\   \underset{I}{\textbf{E}}        \underset{\mathcal{Q}_1 \mathcal{Q}_2 \mathcal{Q}_3}{\textbf{E}}         \bigg[       \underset{r_{-i} \in W_C}{\sum} \bigg|   \textbf{P} \big[ W_C \big]  \textbf{P}_{R_{-i}, q_1, q_2,q_3,  W_C} -  \textbf{P}_{R_{-i}, q_1, q_2,q_3,  W_C}      \bigg|  \bigg]       \leq  \mathrm{O} \bigg( \delta^{\frac{1}{100}}_{\mathrm{Multiplayer}} \bigg) \textbf{P} \big[ W_C \big] , \\ \\ \underset{I}{\textbf{E}}        \underset{\mathcal{Q}_1 \mathcal{Q}_2 \mathcal{Q}_3}{\textbf{E}}         \bigg[     \bigg|   \textbf{P} \big[ W_C \big]  \textbf{P}_{R_{-i} | q_1, q_2,q_3,  W_C} -  \textbf{P}_{R_{-i} | q_1, q_2,q_3,  W_C}     \bigg|       + \bigg|   \textbf{P}_{R_{-i}, W_C | q_1, q_2,q_3,  } -  \textbf{P}_{R_{-i}, W_C | q_1, q_2,q_3,  W_C}           \bigg|      \bigg]     \\   \leq  \mathrm{O} \bigg( \delta^{\frac{1}{100}}_{\mathrm{Multiplayer}} \bigg) \textbf{P} \big[ W_C \big]  , 
\end{align*}

\noindent Hence,

\begin{align*}
  \underset{R_{-i}, \bot, \bot, \bot}{\textbf{E}} \underset{\mathcal{Q}_1 \mathcal{Q}_2 \mathcal{Q}_3}{\textbf{E}} \frac{1}{\big| \big| \ket{\psi_{r_{-i}, \bot, \bot , \bot }} \big| \big| } \bigg| \bigg|  \mathcal{U}_{1,2,3} \ket{\Psi_{q_1 \backslash \bot , q_2}}    -  \ket{\Psi_{q_1 \backslash \bot , \bot , q_3 }}  \bigg| \bigg| \\   \leq      \underset{R_{-i}, \bot, \bot, \bot}{\textbf{E}} \underset{\mathcal{Q}_1 \mathcal{Q}_2 \mathcal{Q}_3}{\textbf{E}} \frac{1}{\big| \big| \ket{\psi_{r_{-i}, \bot, \bot , \bot }} \big| \big| } \bigg| \bigg|   \frac{\big| \big| \ket{\Psi_{\bot, \bot , \bot, \bot}}\big| \big| }{\big| \big|  \ket{\Psi_{ q_1 \backslash \bot , q_2, q_3, q_4}}     \big| \big| } \mathcal{U}_{1,2,3}        \ket{\widetilde{\Psi_{q_1 \backslash \bot , q_2}}}  \\   -  \frac{\big| \big| \ket{\Psi_{\bot, \bot , \bot, \bot}}\big| \big| }{\big| \big|  \ket{\Psi_{ q_1 \backslash \bot , \bot , \bot , \bot }}     \big| \big| } \ket{\widetilde{\Psi_{q_1 \backslash \bot , \bot , q_3 }}}        \bigg| \bigg|  \\ \leq      \underset{R_{-i}, \bot, \bot, \bot}{\textbf{E}} \underset{\mathcal{Q}_1 \mathcal{Q}_2 \mathcal{Q}_3}{\textbf{E}} \bigg[ \frac{1}{\big| \big| \ket{\psi_{r_{-i}, \bot, \bot , \bot }} \big| \big| } \bigg| \bigg|   \frac{\big| \big| \ket{\Psi_{\bot, \bot , \bot, \bot}}\big| \big| }{\big| \big|  \ket{\Psi_{ q_1 \backslash \bot , q_2, q_3, q_4}}     \big| \big| } \mathcal{U}_{1,2,3}        \ket{\widetilde{\Psi_{q_1 \backslash \bot , q_2}}}  \\   +  \ket{\widetilde{\Psi_{q_1 \backslash \bot , \bot , \bot }}}  - \ket{\widetilde{\Psi_{q_1 \backslash \bot , \bot , \bot }}}   -  \frac{\big| \big| \ket{\Psi_{\bot, \bot , \bot, \bot}}\big| \big| }{\big| \big|  \ket{\Psi_{ q_1 \backslash \bot , \bot , \bot , \bot }}     \big| \big| } \ket{\widetilde{\Psi_{q_1 \backslash \bot , \bot , q_3 }}}        \bigg| \bigg|  \bigg] \\ \leq    \underset{R_{-i}, \bot, \bot, \bot}{\textbf{E}} \underset{\mathcal{Q}_1 \mathcal{Q}_2 \mathcal{Q}_3}{\textbf{E}} \bigg[ \frac{1}{\big| \big| \ket{\psi_{r_{-i}, \bot, \bot , \bot }} \big| \big| } \bigg| \bigg|   \frac{\big| \big| \ket{\Psi_{\bot, \bot , \bot, \bot}}\big| \big| }{\big| \big|  \ket{\Psi_{ q_1 \backslash \bot , q_2, q_3, q_4}}     \big| \big| } \mathcal{U}_{1,2,3}        \ket{\widetilde{\Psi_{q_1 \backslash \bot , q_2}}}  \\   +  \ket{\widetilde{\Psi_{q_1 \backslash \bot , \bot , \bot }}}  \bigg| \bigg| -  \bigg| \bigg| \ket{\widetilde{\Psi_{q_1 \backslash \bot , \bot , \bot }}}   -  \frac{\big| \big| \ket{\Psi_{\bot, \bot , \bot, \bot}}\big| \big| }{\big| \big|  \ket{\Psi_{ q_1 \backslash \bot , \bot , \bot , \bot }}     \big| \big| } \ket{\widetilde{\Psi_{q_1 \backslash \bot , \bot , q_3 }}}        \bigg| \bigg|  \bigg] \\  \equiv \underset{R_{-i}, \bot, \bot, \bot}{\textbf{E}} \underset{\mathcal{Q}_1 \mathcal{Q}_2 \mathcal{Q}_3}{\textbf{E}} \bigg[ \frac{1}{\big| \big| \ket{\psi_{r_{-i}, \bot, \bot , \bot }} \big| \big| } \bigg| \bigg|   \frac{\big| \big| \ket{\Psi_{\bot, \bot , \bot, \bot}}\big| \big| }{\big| \big|  \ket{\Psi_{ q_1 \backslash \bot , q_2, q_3, q_4}}     \big| \big| } \mathcal{U}_{1,2,3}        \ket{\widetilde{\Psi_{q_1 \backslash \bot , q_2}}}  \\   +  \ket{\widetilde{\Psi_{q_1 \backslash \bot , \bot , \bot }}}  \bigg| \bigg| \bigg] - \underset{R_{-i}, \bot, \bot, \bot}{\textbf{E}} \underset{\mathcal{Q}_1 \mathcal{Q}_2 \mathcal{Q}_3}{\textbf{E}} \bigg[  \bigg| \bigg| \ket{\widetilde{\Psi_{q_1 \backslash \bot , \bot , \bot }}}  \\  -  \frac{\big| \big| \ket{\Psi_{\bot, \bot , \bot, \bot}}\big| \big| }{\big| \big|  \ket{\Psi_{ q_1 \backslash \bot , \bot , \bot , \bot }}     \big| \big| } \ket{\widetilde{\Psi_{q_1 \backslash \bot , \bot , q_3 }}}        \bigg| \bigg|  \bigg]   
\end{align*}

\begin{align*}
  \equiv   \underset{R_{-i}, \bot, \bot, \bot}{\textbf{E}} \underset{\mathcal{Q}_1 \mathcal{Q}_2 \mathcal{Q}_3}{\textbf{E}} \bigg[ \frac{1}{\big| \big| \ket{\psi_{r_{-i}, \bot, \bot , \bot }} \big| \big| } \bigg| \bigg|   \frac{\big| \big| \ket{\Psi_{\bot, \bot , \bot, \bot}}\big| \big| }{\big| \big|  \ket{\Psi_{ q_1 \backslash \bot , q_2, q_3, q_4}}     \big| \big| } \mathcal{U}_{1,2,3}        \ket{\widetilde{\Psi_{q_1 \backslash \bot , q_2}}}  \\   +  \ket{\widetilde{\Psi_{q_1 \backslash \bot , \bot , \bot }}}        +  \ket{\widetilde{\Psi_{q_1 \backslash \bot, q_2 \backslash \bot, \bot }}} -  \ket{\widetilde{\Psi_{q_1 \backslash \bot, q_2 \backslash \bot, \bot }}}        \bigg| \bigg| \bigg] - \underset{R_{-i}, \bot, \bot, \bot}{\textbf{E}} \underset{\mathcal{Q}_1 \mathcal{Q}_2 \mathcal{Q}_3}{\textbf{E}} \bigg[  \bigg| \bigg| \ket{\widetilde{\Psi_{q_1 \backslash \bot , \bot , \bot }}}  \\  -  \frac{\big| \big| \ket{\Psi_{\bot, \bot , \bot, \bot}}\big| \big| }{\big| \big|  \ket{\Psi_{ q_1 \backslash \bot , \bot , \bot , \bot }}     \big| \big| } \ket{\widetilde{\Psi_{q_1 \backslash \bot , \bot , q_3 }}}        \bigg| \bigg|  \bigg] \\   \equiv   \underset{R_{-i}, \bot, \bot, \bot}{\textbf{E}} \underset{\mathcal{Q}_1 \mathcal{Q}_2 \mathcal{Q}_3}{\textbf{E}} \bigg[ \frac{1}{\big| \big| \ket{\psi_{r_{-i}, \bot, \bot , \bot }} \big| \big| } \bigg| \bigg|   \frac{\big| \big| \ket{\Psi_{\bot, \bot , \bot, \bot}}\big| \big| }{\big| \big|  \ket{\Psi_{ q_1 \backslash \bot , q_2, q_3, q_4}}     \big| \big| } \mathcal{U}_{1,2,3}        \ket{\widetilde{\Psi_{q_1 \backslash \bot , q_2}}}  \\   +  \ket{\widetilde{\Psi_{q_1 \backslash \bot , \bot , \bot }}}        +  \ket{\widetilde{\Psi_{q_1 \backslash \bot, q_2 \backslash \bot, \bot }}} -  \ket{\widetilde{\Psi_{q_1 \backslash \bot, q_2 \backslash \bot, \bot }}}        \bigg| \bigg| \bigg] - \underset{R_{-i}, \bot, \bot, \bot}{\textbf{E}} \underset{\mathcal{Q}_1 \mathcal{Q}_2 \mathcal{Q}_3}{\textbf{E}} \bigg[  \bigg| \bigg| \ket{\widetilde{\Psi_{q_1 \backslash \bot , \bot , \bot }}}  \\  -  \frac{\big| \big| \ket{\Psi_{\bot, \bot , \bot, \bot}}\big| \big| }{\big| \big|  \ket{\Psi_{ q_1 \backslash \bot , \bot , \bot , \bot }}     \big| \big| } \ket{\widetilde{\Psi_{q_1 \backslash \bot , \bot , q_3 }}}       +  \ket{\widetilde{\Psi_{q_1 \backslash \bot, q_2 \backslash \bot, q_3 \backslash \bot }}} -  \ket{\widetilde{\Psi_{q_1 \backslash \bot, q_2 \backslash \bot, q_3 \backslash \bot }}}     \bigg| \bigg|  \bigg]  \\ \leq  \underset{1 \leq k^{\prime} \leq 4}{\sum} \mathrm{O} \bigg(     \frac{\delta^{\frac{k^{\prime}}{300}}_{\mathrm{Multiplayer}}}{\alpha^{k^{\prime}}_{\mathrm{Multiplayer}}} \bigg) \equiv     \mathrm{O} \bigg(     \frac{\delta^{\frac{2}{150}}_{\mathrm{Multiplayer}}}{\alpha^{4}_{\mathrm{Multiplayer}}} \bigg)        ,
\end{align*}

\noindent from which one also has,

\begin{align*}
   \frac{2}{\sqrt{\eta}} \underset{R_{-i}, \bot, \bot, \bot}{\textbf{E}} \underset{\mathcal{Q}_1 \mathcal{Q}_2 \mathcal{Q}_3}{\textbf{E}} \frac{1}{\big| \big| \ket{\psi_{r_{-i}, \bot, \bot , \bot }} \big| \big| } \bigg| \bigg|  \mathcal{U}_{1,2,3} \ket{\Psi_{q_1 \backslash \bot , q_2}}    -  \ket{\Psi_{q_1 \backslash \bot , \bot , q_3 }}  \bigg| \bigg| \\   \leq     \frac{2}{\sqrt{\eta}}     \underset{R_{-i}, \bot, \bot, \bot}{\textbf{E}} \underset{\mathcal{Q}_1 \mathcal{Q}_2 \mathcal{Q}_3}{\textbf{E}} \frac{1}{\big| \big| \ket{\psi_{r_{-i}, \bot, \bot , \bot }} \big| \big| } \bigg| \bigg|   \frac{\big| \big| \ket{\Psi_{\bot, \bot , \bot, \bot}}\big| \big| }{\big| \big|  \ket{\Psi_{ q_1 \backslash \bot , q_2, q_3, q_4}}     \big| \big| } \mathcal{U}_{1,2,3}        \ket{\widetilde{\Psi_{q_1 \backslash \bot , q_2}}}  \\   -  \frac{\big| \big| \ket{\Psi_{\bot, \bot , \bot, \bot}}\big| \big| }{\big| \big|  \ket{\Psi_{ q_1 \backslash \bot , \bot , \bot , \bot }}     \big| \big| } \ket{\widetilde{\Psi_{q_1 \backslash \bot , \bot , q_3 }}}        \bigg| \bigg|  \\ \leq       \frac{2}{\sqrt{\eta}}   \underset{R_{-i}, \bot, \bot, \bot}{\textbf{E}} \underset{\mathcal{Q}_1 \mathcal{Q}_2 \mathcal{Q}_3}{\textbf{E}} \bigg[ \frac{1}{\big| \big| \ket{\psi_{r_{-i}, \bot, \bot , \bot }} \big| \big| } \bigg| \bigg|   \frac{\big| \big| \ket{\Psi_{\bot, \bot , \bot, \bot}}\big| \big| }{\big| \big|  \ket{\Psi_{ q_1 \backslash \bot , q_2, q_3, q_4}}     \big| \big| } \mathcal{U}_{1,2,3}        \ket{\widetilde{\Psi_{q_1 \backslash \bot , q_2}}}  \\   +  \ket{\widetilde{\Psi_{q_1 \backslash \bot , \bot , \bot }}}  - \ket{\widetilde{\Psi_{q_1 \backslash \bot , \bot , \bot }}}   -  \frac{\big| \big| \ket{\Psi_{\bot, \bot , \bot, \bot}}\big| \big| }{\big| \big|  \ket{\Psi_{ q_1 \backslash \bot , \bot , \bot , \bot }}     \big| \big| } \ket{\widetilde{\Psi_{q_1 \backslash \bot , \bot , q_3 }}}        \bigg| \bigg|  \bigg] \\ \leq     \frac{2}{\sqrt{\eta}}  \underset{R_{-i}, \bot, \bot, \bot}{\textbf{E}} \underset{\mathcal{Q}_1 \mathcal{Q}_2 \mathcal{Q}_3}{\textbf{E}} \bigg[ \frac{1}{\big| \big| \ket{\psi_{r_{-i}, \bot, \bot , \bot }} \big| \big| } \bigg| \bigg|   \frac{\big| \big| \ket{\Psi_{\bot, \bot , \bot, \bot}}\big| \big| }{\big| \big|  \ket{\Psi_{ q_1 \backslash \bot , q_2, q_3, q_4}}     \big| \big| } \mathcal{U}_{1,2,3}        \ket{\widetilde{\Psi_{q_1 \backslash \bot , q_2}}}  \\   +  \ket{\widetilde{\Psi_{q_1 \backslash \bot , \bot , \bot }}}  \bigg| \bigg| -  \bigg| \bigg| \ket{\widetilde{\Psi_{q_1 \backslash \bot , \bot , \bot }}}   -  \frac{\big| \big| \ket{\Psi_{\bot, \bot , \bot, \bot}}\big| \big| }{\big| \big|  \ket{\Psi_{ q_1 \backslash \bot , \bot , \bot , \bot }}     \big| \big| } \ket{\widetilde{\Psi_{q_1 \backslash \bot , \bot , q_3 }}}        \bigg| \bigg|  \bigg] \\  \equiv   \frac{2}{\sqrt{\eta}}  \underset{R_{-i}, \bot, \bot, \bot}{\textbf{E}} \underset{\mathcal{Q}_1 \mathcal{Q}_2 \mathcal{Q}_3}{\textbf{E}} \bigg[ \frac{1}{\big| \big| \ket{\psi_{r_{-i}, \bot, \bot , \bot }} \big| \big| } \bigg| \bigg|   \frac{\big| \big| \ket{\Psi_{\bot, \bot , \bot, \bot}}\big| \big| }{\big| \big|  \ket{\Psi_{ q_1 \backslash \bot , q_2, q_3, q_4}}     \big| \big| } \mathcal{U}_{1,2,3}        \ket{\widetilde{\Psi_{q_1 \backslash \bot , q_2}}}  \end{align*}

  \begin{align*}     +  \ket{\widetilde{\Psi_{q_1 \backslash \bot , \bot , \bot }}}  \bigg| \bigg| \bigg] - \underset{R_{-i}, \bot, \bot, \bot}{\textbf{E}} \underset{\mathcal{Q}_1 \mathcal{Q}_2 \mathcal{Q}_3}{\textbf{E}} \bigg[  \bigg| \bigg| \ket{\widetilde{\Psi_{q_1 \backslash \bot , \bot , \bot }}}  \\  -  \frac{\big| \big| \ket{\Psi_{\bot, \bot , \bot, \bot}}\big| \big| }{\big| \big|  \ket{\Psi_{ q_1 \backslash \bot , \bot , \bot , \bot }}     \big| \big| } \ket{\widetilde{\Psi_{q_1 \backslash \bot , \bot , q_3 }}}        \bigg| \bigg|  \bigg]   
\\ 
  \equiv  \frac{2}{\sqrt{\eta}}   \underset{R_{-i}, \bot, \bot, \bot}{\textbf{E}} \underset{\mathcal{Q}_1 \mathcal{Q}_2 \mathcal{Q}_3}{\textbf{E}} \bigg[ \frac{1}{\big| \big| \ket{\psi_{r_{-i}, \bot, \bot , \bot }} \big| \big| } \bigg| \bigg|   \frac{\big| \big| \ket{\Psi_{\bot, \bot , \bot, \bot}}\big| \big| }{\big| \big|  \ket{\Psi_{ q_1 \backslash \bot , q_2, q_3, q_4}}     \big| \big| } \mathcal{U}_{1,2,3}        \ket{\widetilde{\Psi_{q_1 \backslash \bot , q_2}}}  \\ +  \ket{\widetilde{\Psi_{q_1 \backslash \bot , \bot , \bot }}}        +  \ket{\widetilde{\Psi_{q_1 \backslash \bot, q_2 \backslash \bot, \bot }}} -  \ket{\widetilde{\Psi_{q_1 \backslash \bot, q_2 \backslash \bot, \bot }}}        \bigg| \bigg| \bigg] - \underset{R_{-i}, \bot, \bot, \bot}{\textbf{E}} \underset{\mathcal{Q}_1 \mathcal{Q}_2 \mathcal{Q}_3}{\textbf{E}} \bigg[  \bigg| \bigg| \ket{\widetilde{\Psi_{q_1 \backslash \bot , \bot , \bot }}}  \\  -  \frac{\big| \big| \ket{\Psi_{\bot, \bot , \bot, \bot}}\big| \big| }{\big| \big|  \ket{\Psi_{ q_1 \backslash \bot , \bot , \bot , \bot }}     \big| \big| } \ket{\widetilde{\Psi_{q_1 \backslash \bot , \bot , q_3 }}}        \bigg| \bigg|  \bigg] \\   \equiv  \frac{2}{\sqrt{\eta}}   \underset{R_{-i}, \bot, \bot, \bot}{\textbf{E}} \underset{\mathcal{Q}_1 \mathcal{Q}_2 \mathcal{Q}_3}{\textbf{E}} \bigg[ \frac{1}{\big| \big| \ket{\psi_{r_{-i}, \bot, \bot , \bot }} \big| \big| } \bigg| \bigg|   \frac{\big| \big| \ket{\Psi_{\bot, \bot , \bot, \bot}}\big| \big| }{\big| \big|  \ket{\Psi_{ q_1 \backslash \bot , q_2, q_3, q_4}}     \big| \big| } \mathcal{U}_{1,2,3}        \ket{\widetilde{\Psi_{q_1 \backslash \bot , q_2}}}  \\   +  \ket{\widetilde{\Psi_{q_1 \backslash \bot , \bot , \bot }}}        +  \ket{\widetilde{\Psi_{q_1 \backslash \bot, q_2 \backslash \bot, \bot }}} -  \ket{\widetilde{\Psi_{q_1 \backslash \bot, q_2 \backslash \bot, \bot }}}        \bigg| \bigg| \bigg] - \underset{R_{-i}, \bot, \bot, \bot}{\textbf{E}} \underset{\mathcal{Q}_1 \mathcal{Q}_2 \mathcal{Q}_3}{\textbf{E}} \bigg[  \bigg| \bigg| \ket{\widetilde{\Psi_{q_1 \backslash \bot , \bot , \bot }}}  \\  -  \frac{\big| \big| \ket{\Psi_{\bot, \bot , \bot, \bot}}\big| \big| }{\big| \big|  \ket{\Psi_{ q_1 \backslash \bot , \bot , \bot , \bot }}     \big| \big| } \ket{\widetilde{\Psi_{q_1 \backslash \bot , \bot , q_3 }}}       +  \ket{\widetilde{\Psi_{q_1 \backslash \bot, q_2 \backslash \bot, q_3 \backslash \bot }}} -  \ket{\widetilde{\Psi_{q_1 \backslash \bot, q_2 \backslash \bot, q_3 \backslash \bot }}}     \bigg| \bigg|  \bigg]  \\ \leq  \frac{2}{\sqrt{\eta}}   \underset{1 \leq k^{\prime} \leq 4}{\sum} \mathrm{O} \bigg(     \frac{\delta^{\frac{k^{\prime}}{300}}_{\mathrm{Multiplayer}}}{\alpha^{k^{\prime}}_{\mathrm{Multiplayer}}} \bigg) \equiv     \mathrm{O} \bigg(     \frac{\delta^{\frac{2}{150}}_{\mathrm{Multiplayer}}}{\alpha^{4}_{\mathrm{Multiplayer}}} \bigg)      ,
\end{align*}

\noindent which is related to the series of computations,

\begin{align*}
  \left\{\!\begin{array}{ll@{}>{{}}l}  
\underset{R_{-i}, \bot, \bot, \bot}{\textbf{E}} \underset{\mathcal{Q}_1 \mathcal{Q}_2 \mathcal{Q}_3}{\textbf{E}} \frac{1}{\big| \big| \ket{\psi_{r_{-i}, \bot, \bot , \bot }} \big| \big| } \bigg| \bigg|  \mathcal{U}_{1,2,3} \ket{\Psi_{q_1 \backslash \bot , q_2}}    -  \ket{\Psi_{q_1 \backslash \bot , \bot , q_3 }}  \bigg| \bigg|  ,   \\ \underset{R_{-i}, \bot, \bot, \bot}{\textbf{E}} \underset{\mathcal{Q}_1 \mathcal{Q}_2 \mathcal{Q}_3}{\textbf{E}} \frac{1}{\big| \big| \ket{\psi_{r_{-i}, \bot, \bot , \bot }} \big| \big| } \bigg| \bigg|  \mathcal{U}_{1,2,3} \ket{\Psi_{q_1 \backslash \bot , q_2}}    -  \ket{\Psi_{q_1 \backslash \bot , \bot , \bot }}  \bigg| \bigg|  , \\ 
   \underset{R_{-i}, \bot, \bot, \bot, \bot }{\textbf{E}} \underset{\mathcal{Q}_1 \mathcal{Q}_2 \mathcal{Q}_3 \mathcal{Q}_4 }{\textbf{E}} \frac{1}{\big| \big| \ket{\psi_{r_{-i}, \bot, \bot , \bot, \bot  }} \big| \big| } \bigg| \bigg|   \mathcal{U}_{1,2,3,4}    \ket{\Psi_{q_1 \backslash \bot, q_2 \backslash \bot, q_3 \backslash \bot, q_4 }}         -     \ket{\Psi_{q_1 \backslash \bot, q_2 \backslash \bot, q_3 \backslash \bot, \bot }}         \bigg| \bigg|  , 
 \\ \vdots  \\  \underset{R_{-i}, \bot, \bot, \bot, \cdots, \bot }{\textbf{E}} \underset{\mathcal{Q}_1 \mathcal{Q}_2 \mathcal{Q}_3 \times \cdots \times \mathcal{Q}_N}{\textbf{E}} \frac{1}{\big| \big| \ket{\psi_{r_{-i}, \bot, \bot , \bot , \cdots, \bot}} \big| \big| }  \bigg| \bigg|  \mathcal{U}_{1,2, \cdots, N}  \ket{\Psi_{q_1 \backslash \bot, q_2 \backslash \bot, \cdots, q_N \backslash \bot} }  \\ - \ket{\Psi_{q_1 \backslash \bot , \cdots , q_{N-1} \backslash \bot , \bot }} \bigg| \bigg|  , 
  \end{array}\right. . 
\end{align*}

\noindent Namely, one must insert additional factors of the form,

\[ \left\{\!\begin{array}{ll@{}>{{}}l}  \ket{\widetilde{\Psi}_{q_1 \backslash \bot, \bot , \cdots , \bot}} -  \ket{\widetilde{\Psi}_{q_1 \backslash \bot, \bot , \cdots , \bot}}  , \\ \vdots \\ \ket{\widetilde{\Psi}_{q_1 \backslash \bot, q_2 \backslash \bot , \cdots ,  q_N  \backslash \bot}} -  \ket{\widetilde{\Psi}_{q_1 \backslash \bot, q_2 \backslash \bot , cdots ,  q_N  \backslash \bot}}  , 
 \end{array}\right. 
\] 

\noindent from which we conclude the argument, as, the desired upper bound takes the form,

\begin{align*}
    \mathrm{O} \bigg(    \frac{\delta^{\frac{2}{150}}_{\mathrm{Multiplayer}}}{\alpha^{4}_{\mathrm{Multiplayer}}}   \bigg) . \boxed{} 
\end{align*}

\bigskip

\noindent The above argument can be straightforwardly adapted to conclude that a similar result holds. Namely, in place of,

\begin{align*}
    \mathrm{O} \bigg(    \frac{\delta^{\frac{1}{100}}_{\mathrm{Multiplayer}}}{\alpha^{4}_{\mathrm{Multiplayer}}}   \bigg) , 
\end{align*}

\noindent one would have that the upper bound takes the form,

\begin{align*}
  \mathrm{O} \bigg( \frac{\delta^{\frac{N}{300}}_{\mathrm{Multiplayer}}}{\alpha^{N+5}_{\mathrm{Multiplayer}}} \bigg)   ,
\end{align*}

\noindent from the observations,

\begin{align*}
 \underset{I}{\textbf{E}} \bigg[    \textbf{P}_{\mathcal{Q}_1 \mathcal{Q}_2 \times \mathcal{Q}_N}  \big[ q_1 , q_2, \cdots , q_N \big] \big|    \textbf{P} \big[ W_C \big| \mathcal{Q}_1 \equiv q_1, \cdots , \mathcal{Q}_N \equiv q_N \big]  -   \textbf{P} \big[ W_C \big]   \big|   \bigg] \leq \mathrm{O} \bigg( \delta^{\frac{N}{200}}_{\mathrm{Multiplayer}} \bigg)     , \\ \\ \underset{I}{\textbf{E}}  \underset{\mathcal{Q}_1 \mathcal{Q}_2 \mathcal{Q}_3}{\textbf{E}}        \big|    \textbf{P} \big[ W_C \big| \mathcal{Q}_1 \equiv q_1, \cdots, \mathcal{Q}_N \equiv q_N \big]  -   \textbf{P} \big[ W_C \big]   \big|      \leq   \mathrm{O} \bigg( \frac{\delta^{\frac{N}{300}}_{\mathrm{Multiplayer}}}{\alpha^{N+5}_{\mathrm{Multiplayer}}} \bigg)     , \end{align*}

 \begin{align*} \underset{I}{\textbf{E}}        \underset{\mathcal{Q}_1 \mathcal{Q}_2 \times \cdots \times \mathcal{Q}_N}{\textbf{E}}         \bigg[       \underset{r_{-i} \in W_C}{\sum} \bigg|   \textbf{P} \big[ W_C \big]  \textbf{P}_{R_{-i}, q_1, \cdots , q_N ,  W_C} -  \textbf{P}_{R_{-i}, q_1, \cdots, q_N ,  W_C}      \bigg|  \bigg]       \leq   \mathrm{O} \bigg( \frac{\delta^{\frac{N}{300}}_{\mathrm{Multiplayer}}}{\alpha^{N+5}_{\mathrm{Multiplayer}}} \bigg)   \textbf{P} \big[ W_C \big] , \\ \\ \underset{I}{\textbf{E}}        \underset{\mathcal{Q}_1 \mathcal{Q}_2 \times \cdots \times \mathcal{Q}_N}{\textbf{E}}             \bigg[     \bigg|   \textbf{P} \big[ W_C \big]  \textbf{P}_{R_{-i} | q_1, \cdots, q_N ,  W_C} -  \textbf{P}_{R_{-i} | q_1, \cdots , q_N ,  W_C}     \bigg|       + \bigg|   \textbf{P}_{R_{-i}, W_C | q_1, \cdots , q_N ,  }  \\ -  \textbf{P}_{R_{-i}, W_C | q_1, \cdots , q_N ,  W_C}           \bigg|      \bigg]       \leq  \mathrm{O} \bigg( \delta^{\frac{N}{200}}_{\mathrm{Multiplayer}} \bigg)  \textbf{P} \big[ W_C \big]  , 
\end{align*}

\bigskip

\noindent The \textbf{Proposition} above can be extended for multiplayer game-theoretic settings, which allows for the second main result to be obtained. In particular,

\[   \left\{\!\begin{array}{ll@{}>{{}}l}  2    \underset{I}{\textbf{E}} \underset{R_{-i} | W_C}{\textbf{E}} \underset{\mathcal{Q}_1 \mathcal{Q}_2 \mathcal{Q}_3 \mathcal{Q}_4}{\textbf{E}} \big| \big|   \big( \mathcal{U}_{1,r_{-i}}  \otimes  \mathcal{U}_{2,r_{-i}}   \otimes  \mathcal{U}_{3,r_{-i}}  \otimes  \mathcal{U}_{4,r_{-i}}  \big)   \ket{\widetilde{\Psi_{r_{-i}, \bot,\bot,\bot, \bot}}}   - \ket{\widetilde{\Psi_{r_{-i},  q_1,q_2,q_3, q_4}}}      \big| \big| \\  \leq  \mathrm{O} \bigg(  \frac{\delta^{\frac{1}{60}}_{\mathrm{Multiplayer}}}{\alpha^{5}_{\mathrm{Multiplayer}}} \bigg)    ,    \\ \vdots \\   \sqrt{N}    \underset{I}{\textbf{E}} \underset{R_{-i} | W_C}{\textbf{E}} \underset{\mathcal{Q}_1 \mathcal{Q}_2 \mathcal{Q}_3 \mathcal{Q}_4}{\textbf{E}} \big| \big|   \big( \mathcal{U}_{1,r_{-i}}  \otimes  \mathcal{U}_{2,r_{-i}}   \otimes  \mathcal{U}_{3,r_{-i}}  \otimes  \cdots \otimes \mathcal{U}_{N,r_{-i}}  \big)   \ket{\widetilde{\Psi_{r_{-i}, \bot,\bot,\bot, \cdots, \bot}}} \\   - \ket{\widetilde{\Psi_{r_{-i},  q_1,q_2,q_3, \cdots, q_N}}}      \big| \big|    \leq  \mathrm{O} \bigg(  \frac{\delta^{\frac{N}{300}}_{\mathrm{Multiplayer}}}{\alpha^{N+1}_{\mathrm{Multiplayer}}} \bigg)     , \end{array}\right. 
\]

\noindent can be obtained by following the same argument for the \textbf{Proposition} above, which is itself an adaptation of \textbf{Proposition} \textit{5.1} in [4].

\bigskip

\noindent To conclude the argument for $\textbf{Theorem}$ \textit{2}, we make use of the following result, specifically a Quantum Raz's Lemma (\textbf{Proposition} \textit{5.10}, [3]), which states:

\bigskip

\noindent \textbf{Proposition}. Let $\rho$ and $\sigma$ be two Quantum states for which $\rho^{X_1 X_2 \cdots X_N A} = \underset{1 \leq i \leq n}{\bigotimes} \rho^{X_i} \otimes \rho^A$, and $\sigma^{XA} \equiv \underset{1 \leq i \leq N}{\bigotimes} \sigma^{X_i} \otimes \sigma^A$, for classical $X_i$. One has that,

\begin{align*}
    \underset{1 \leq i \leq N}{\sum} I \big( X_i : A \big)_{\rho} \leq D \big( \rho^{XA } \big| \big| \sigma^{XA} \big)  .
\end{align*}

\noindent With the Quantum Raz's Lemma, the above computation, in combination with Fusch-van de Graff, and Markov's, inequalities, imply the desired result,

\begin{align*}
  \mathscr{P}\mathscr{O}\mathscr{V} \mathscr{M} \mathscr{E} \equiv \mathrm{O} \bigg( \frac{\sqrt{\sqrt{\delta^{\frac{N}{300}}_{\mathrm{Multiplayer}}}}}{\alpha^{N+5}_{\mathrm{Multiplayer}}} \bigg)  ,
\end{align*}

\noindent from which we conclude the argument. \boxed{}

\bigskip

\noindent We provide the following \textbf{Corollary} below.

\bigskip

\noindent \textbf{Corollary} (\textit{relating the fidelity between pairs of Quantum states to the existence of suitable unitaries}). Denote $\ket{\Psi_1}^{12\cdots N}, \ket{\Psi_2}^{12\cdots N}, \cdots, \ket{\Psi_N}^{12 \cdots N}$ as bipartite states, with respective density matrices $\big( \rho_1 \big)^{12\cdots N} , \big( \rho_2 \big)^{12\cdots N}, \cdots, \big( \rho_N\big)^{12\cdots N}$. Then there exists unitaries,

\[  \left\{\!\begin{array}{ll@{}>{{}}l} 
 \mathcal{U}_{q_1} \equiv \text{\textit{First player unitary}}   \text{, } \\  \mathcal{U}_{q_2} \equiv \text{\textit{Second player unitary}}  \text{, } \\ \vdots \\ \mathcal{U}_{q_N} \equiv \text{\textit{N th player unitary}}  \text{, } 
\end{array}\right. \equiv   \left\{\!\begin{array}{ll@{}>{{}}l} 
 \mathcal{U}_1 \equiv \text{\textit{First player unitary}}   \text{, } \\  \mathcal{U}_2 \equiv \text{\textit{Second player unitary}}  \text{, } \\ \vdots \\ \mathcal{U}_N \equiv \text{\textit{N th player unitary}}  \text{, } 
\end{array}\right. 
\]

\noindent for which,

\[  \left\{\!\begin{array}{ll@{}>{{}}l} 
 \bra{ \Psi_1}^{12 \cdots N} \bigg[ \textbf{I} \otimes \mathcal{U}_2 \otimes \textbf{I}^{\otimes N-2} \bigg] \ket{ \Psi_2}^{12 \cdots N} =         F \big( \big( \rho_1 \big)^{12 \cdots N},  \big( \rho_2 \big)^{12 \cdots N} \big) \equiv  F \big( \rho_1 , \rho_2 \big)    \text{, } \\   \bra{ \Psi_1}^{12 \cdots N} \bigg[ \textbf{I} \otimes \textbf{I} \otimes \mathcal{U}_3 \otimes \textbf{I}^{\otimes N-3} \bigg] \ket{ \Psi_2}^{12 \cdots N} =         F \big( \big( \rho_1 \big)^{12 \cdots N},  \big( \rho_3 \big)^{12 \cdots N} \big) \equiv  F \big( \rho_1 , \rho_3 \big)  \text{, } \\ \vdots \\     \bra{ \Psi_1}^{12 \cdots N} \bigg[ \textbf{I}^{\otimes N-1} \otimes \mathcal{U}_N \bigg] \ket{ \Psi_N}^{12 \cdots N} =         F \big( \big( \rho_1 \big)^{12 \cdots N},  \big( \rho_N \big)^{12 \cdots N} \big) \equiv  F \big( \rho_1 , \rho_N \big)      \text{. } 
\end{array}\right.\]

\noindent \textit{Proof of Corollary.} The above statement is a multi-dimensional analog of Uhlmann's Theorem, see \textbf{Theorem} \textit{5.11} in [2]. \boxed{}

\subsection{General description of arguments for Theorem $3$}

\noindent To argue that \textbf{Theorem} \textit{3} holds, namely to obtain upper bounds for,

\[ \mathcal{P} \mathcal{I} \equiv   \left\{\!\begin{array}{ll@{}>{{}}l} 
\underset{I}{\textbf{E}} \underset{R_{-i} | W_C}{\textbf{E}} \underset{\mathcal{Q}_1 \mathcal{Q}_2 \times \cdots \times \mathcal{Q}_N}{\textbf{E}}  \big| \big| \textbf{I}^{E_1}_{r_{-i}, q_1,q_2, \cdots, q_N}  -  \textbf{I}^{E_1}_{r_{-i}, q_1,\bot,q_3, \cdots, q_N} \big| \big|^2_1   , \\ \underset{I}{\textbf{E}} \underset{R_{-i} | W_C}{\textbf{E}} \underset{\mathcal{Q}_1 \mathcal{Q}_2 \times \cdots \times \mathcal{Q}_N}{\textbf{E}}  \big| \big| \textbf{2}^{E_2}_{r_{-i}, q_1,q_2, \cdots, q_N}  -  \textbf{2}^{E_2}_{r_{-i}, \bot, q_2, q_3, \cdots, q_N} \big| \big|^2_1  ,  \\ \vdots \\ \underset{I}{\textbf{E}} \underset{R_{-i} | W_C}{\textbf{E}} \underset{\mathcal{Q}_1 \mathcal{Q}_2 \times \cdots \times \mathcal{Q}_N}{\textbf{E}}  \big| \big| \textbf{N}^{E_N}_{r_{-i}, q_1,q_2, \cdots, q_N}  -  \textbf{N}^{E_N}_{r_{-i}, q_1,\cdots, q_{N-2}, \bot, q_N} \big| \big|^2_1  ,
\end{array}\right. 
\]

\noindent we manipulate expectation values from the system,

\[ \mathcal{P} \mathcal{I}^{\prime} \equiv   \left\{\!\begin{array}{ll@{}>{{}}l} 
\frac{1}{N}  \underset{\textbf{R}|W_C}{\textbf{E}} S \bigg( \bigg(  r_{-i} \text{ } \textit{1 st player state}      \bigg) \bigg| \bigg| \bigg(   r_{-i} \text{ } \textit{2 nd player state}      \bigg) \bigg)   , \\ \frac{1}{N}  \underset{\textbf{R}|W_C}{\textbf{E}} S \bigg( \bigg(  r_{-i} \text{ } \textit{2 nd player state}      \bigg) \bigg| \bigg| \bigg(   r_{-i} \text{ } \textit{3 rd player state}      \bigg) \bigg)  ,  \\ \vdots \\ \frac{1}{N}  \underset{\textbf{R}|W_C}{\textbf{E}} S \bigg( \bigg(  r_{-i} \text{ } \textit{(N-1) th player state}      \bigg) \bigg| \bigg| \bigg(   r_{-i} \text{ } \textit{N th player state}      \bigg) \bigg)  ,
\end{array}\right. 
\]

\noindent namely, the system of expected values of relative entropies. By adding additional on additional terms with relative-min entropies, as described previously through,

\begin{align*}
\underset{1 \leq i \leq N}{\sum}  \underset{\textbf{P}_{1_C 2_C 3_C \times \cdots \times N_C D G | \mathcal{E}}}{\textbf{E}}     S_{\infty} \bigg(  \bigg(  \textit{(i-1) th player state} \bigg)  \bigg | \bigg| \bigg(  \textit{i th player state}  \bigg)  \bigg), 
\end{align*}

\noindent we obtain the desired bound for $\mathcal{P}\mathcal{I}$ by upper bounding $\mathcal{P}\mathcal{I}^{\prime}$ with,

\[ \mathcal{P} \mathcal{I}^{\prime\prime} \equiv   \left\{\!\begin{array}{ll@{}>{{}}l} 
\frac{1}{N}  \underset{\textbf{R}|W_C}{\textbf{E}} S \bigg( \bigg(  r_{-i} \text{ } \textit{1 st player state}      \bigg) \bigg| \bigg| \bigg(   r_{-i} \text{ } \textit{2 nd player state}      \bigg) \bigg) \\ + \frac{1}{N}  \underset{\textbf{P}_{(\mathcal{Q}_1)_C (\mathcal{Q}_2)_C (\mathcal{Q}_3)_C D G | \mathcal{E}}}{\textbf{E}}    S_{\infty} \bigg(  \bigg(  \textit{1 st player state} \bigg)  \bigg | \bigg| \bigg(  \textit{2 nd player state}  \bigg)  \bigg)    , \\ \frac{1}{N}  \underset{\textbf{R}|W_C}{\textbf{E}} S \bigg( \bigg(  r_{-i} \text{ } \textit{2 nd player state}      \bigg) \bigg| \bigg| \bigg(   r_{-i} \text{ } \textit{3 rd player state}      \bigg) \bigg) \\ + \frac{1}{N}  \underset{1 \leq i \leq 3}{\sum} \underset{\textbf{P}_{X_C Y_C Z_C D G | \mathcal{E}}}{\textbf{E}}    S_{\infty} \bigg(  \bigg(  \textit{(i-1) th player state} \bigg)  \bigg | \bigg| \bigg(  \textit{i th player state}  \bigg)  \bigg) ,  \\ \vdots \\ \frac{1}{N}  \underset{\textbf{R}|W_C}{\textbf{E}} S \bigg( \bigg(  r_{-i} \text{ } \textit{(N-1) th player state}      \bigg) \bigg| \bigg| \bigg(   r_{-i} \text{ } \textit{N th player state}      \bigg) \bigg)  \\ + \frac{1}{N}  \underset{1 \leq i \leq N}{\sum} \underset{\textbf{P}_{X_C Y_C Z_C D G | \mathcal{E}}}{\textbf{E}}    S_{\infty} \bigg(  \bigg(  \textit{(i-1) th player state} \bigg)  \bigg | \bigg| \bigg(  \textit{i th player state}  \bigg)  \bigg) .
\end{array}\right. 
\]

\subsection{Argument}

\noindent \textit{Proof of Theorem 3}. To upper bound the system of expectation values provided in $\mathcal{P}\mathcal{I}$, observe that it suffices to obtain an upper bound of the form,

\begin{align*}
  \underset{I}{\textbf{E}} \underset{R_{-i}| W_C}{\textbf{E}} \underset{\mathcal{Q}_1 \mathcal{Q}_2 \mathcal{Q}_3 \times \cdots \times \mathcal{Q}_N}{\textbf{E}} \big| \big| \textbf{I}^{E^1}_{r_{-i}, q_1, q_2, \cdots, q_N} - \textbf{2}^{E^1}_{r_{-i},q_1, \bot, q_3, \cdots , q_N} \big| \big|^2_1= \mathrm{O} \bigg(     \frac{\sqrt{\delta^{\frac{N}{300}}_{\mathrm{Multiplayer}}}}{\alpha^{N+1}_{\mathrm{Multiplayer}}}    \bigg)   ,
\end{align*}

\noindent corresponding to the expected value of the square of the $l$-1 norm. Expectedly, the remaining collection of squares of $l$-1 norms take the form,

\[   \left\{\!\begin{array}{ll@{}>{{}}l}   \underset{I}{\textbf{E}} \underset{R_{-i}| W_C}{\textbf{E}} \underset{\mathcal{Q}_1 \mathcal{Q}_2 \mathcal{Q}_3 \times \cdots \times \mathcal{Q}_N}{\textbf{E}} \big| \big| \textbf{I}^{E^2}_{r_{-i}, q_1, q_2, \cdots, q_N} - \textbf{2}^{E^2}_{r_{-i},q_1, \bot, q_3, \cdots , q_N} \big| \big|^2_1= \mathrm{O} \bigg(     \frac{\sqrt{\delta^{\frac{N}{300}}_{\mathrm{Multiplayer}}}}{\alpha^{N+1}_{\mathrm{Multiplayer}}}    \bigg)    ,    \\ \vdots \\    \underset{I}{\textbf{E}} \underset{R_{-i}| W_C}{\textbf{E}} \underset{\mathcal{Q}_1 \mathcal{Q}_2 \mathcal{Q}_3 \times \cdots \times \mathcal{Q}_N}{\textbf{E}} \big| \big| \textbf{I}^{E^N}_{r_{-i}, q_1, q_2, \cdots, q_N} - \textbf{2}^{E^N}_{r_{-i},q_1, \bot, q_3, \cdots , q_N} \big| \big|^2_1= \mathrm{O} \bigg(     \frac{\sqrt{\delta^{\frac{N}{300}}_{\mathrm{Multiplayer}}}}{\alpha^{N+1}_{\mathrm{Multiplayer}}}    \bigg)   , \end{array}\right. 
\]

\noindent which in itself is related to the broader system,

\[   \left\{\!\begin{array}{ll@{}>{{}}l}   \underset{I}{\textbf{E}} \underset{R_{-i}| W_C}{\textbf{E}} \underset{\mathcal{Q}_1 \mathcal{Q}_2 \mathcal{Q}_3\times \cdots \times \mathcal{Q}_N}{\textbf{E}} \big| \big| \textbf{I}^{E^2}_{r_{-i}, q_1, q_2, \cdots, q_N} - \textbf{2}^{E^2}_{r_{-i},q_1, \bot, \bot, q_4,  \cdots , q_N} \big| \big|^2_1= \mathrm{O} \bigg(     \frac{\sqrt{\delta^{\frac{N}{300}}_{\mathrm{Multiplayer}}}}{\alpha^{N+1}_{\mathrm{Multiplayer}}}    \bigg)    , \\ \vdots \\      \underset{I}{\textbf{E}} \underset{R_{-i}| W_C}{\textbf{E}} \underset{\mathcal{Q}_1 \mathcal{Q}_2 \mathcal{Q}_3 \times \cdots \times \mathcal{Q}_N}{\textbf{E}} \big| \big| \textbf{I}^{E^2}_{r_{-i}, q_1, q_2, \cdots, q_N} - \textbf{2}^{E^2}_{r_{-i},q_1, \bot, \bot, \bot,  \cdots, \bot , q_N} \big| \big|^2_1= \mathrm{O} \bigg(     \frac{\sqrt{\delta^{\frac{N}{300}}_{\mathrm{Multiplayer}}}}{\alpha^{N+1}_{\mathrm{Multiplayer}}}    \bigg)     ,  \\  \underset{I}{\textbf{E}} \underset{R_{-i}| W_C}{\textbf{E}} \underset{\mathcal{Q}_1 \mathcal{Q}_2 \mathcal{Q}_3 \times \cdots \times \mathcal{Q}_N}{\textbf{E}} \big| \big| \textbf{I}^{E^3}_{r_{-i}, q_1, q_2, \cdots, q_N} - \textbf{2}^{E^3}_{r_{-i},q_1, q_2, \bot, q_4, \cdots , q_N} \big| \big|^2_1= \mathrm{O} \bigg(     \frac{\sqrt{\delta^{\frac{N}{300}}_{\mathrm{Multiplayer}}}}{\alpha^{N+1}_{\mathrm{Multiplayer}}}    \bigg)      , \\  \underset{I}{\textbf{E}} \underset{R_{-i}| W_C}{\textbf{E}} \underset{\mathcal{Q}_1 \mathcal{Q}_2 \mathcal{Q}_3 \times \cdots \times \mathcal{Q}_N}{\textbf{E}} \big| \big| \textbf{I}^{E^3}_{r_{-i}, q_1, q_2, \cdots, q_N} - \textbf{2}^{E^3}_{r_{-i},q_1, \bot, \bot, \bot,  \cdots, \bot,  , q_N} \big| \big|^2_1= \mathrm{O} \bigg(     \frac{\sqrt{\delta^{\frac{N}{300}}_{\mathrm{Multiplayer}}}}{\alpha^{N+1}_{\mathrm{Multiplayer}}}    \bigg)      ,   ,  \\ \vdots \\    \underset{I}{\textbf{E}} \underset{R_{-i}| W_C}{\textbf{E}} \underset{\mathcal{Q}_1 \mathcal{Q}_2 \mathcal{Q}_3\times \cdots \times \mathcal{Q}_N }{\textbf{E}} \big| \big| \textbf{I}^{E^N}_{r_{-i}, q_1, q_2, \cdots, q_N} - \textbf{2}^{E^N}_{r_{-i},q_1, \bot, q_3, \cdots , q_N} \big| \big|^2_1= \mathrm{O} \bigg(     \frac{\sqrt{\delta^{\frac{N}{300}}_{\mathrm{Multiplayer}}}}{\alpha^{N+1}_{\mathrm{Multiplayer}}}    \bigg)   , \\ \vdots \\         \underset{I}{\textbf{E}} \underset{R_{-i}| W_C}{\textbf{E}} \underset{\mathcal{Q}_1 \mathcal{Q}_2 \mathcal{Q}_3\times \cdots \times \mathcal{Q}_N }{\textbf{E}} \big| \big| \textbf{I}^{E^N}_{r_{-i}, q_1, q_2, \cdots, q_N} - \textbf{2}^{E^N}_{r_{-i},q_1, \bot, \cdots , \bot,  q_N} \big| \big|^2_1= \mathrm{O} \bigg(     \frac{\sqrt{\delta^{\frac{N}{300}}_{\mathrm{Multiplayer}}}}{\alpha^{N+1}_{\mathrm{Multiplayer}}}    \bigg)   ,        \end{array}\right. 
\]

\noindent of expected values with the desired upper bounds dependent upon $\mathrm{O} \bigg( \cdot \bigg) $. Similar to arguments provided in the previous main result, $\textbf{Theorem}$ \textit{2}, we seek to upper bound the desired expected value within $\mathcal{P}\mathcal{I}$ from the observation,

\begin{align*}
  \underset{I}{\textbf{E}} \underset{R_{-i}| W_C}{\textbf{E}} \underset{\mathcal{Q}_1 \mathcal{Q}_2 \mathcal{Q}_3 \times \cdots \times \mathcal{Q}_N}{\textbf{E}} \big| \big| \textbf{I}^{E^1}_{r_{-i}, q_1, q_2, \cdots, q_N} - \textbf{2}^{E^1}_{r_{-i},q_1, \bot, q_3, \cdots , q_N} \big| \big|^2_1 \\ \leq 2 \mathrm{ln} 2 \text{ } \underset{I}{\textbf{E}}    \underset{R_{-i} \mathcal{Q}_2 | W_C}{\textbf{E}}  D \big( \textbf{I}^{E^1}_{r_{-i}, q_2} \big| \big|   \textbf{I}^{E^1}_{r_{-i}, \bot}  \big) \equiv \mathcal{E} \end{align*}

\noindent obtained from an application of Pinkser's inequality. Proceeding, we make use of a previous remark, provided in \textit{1.5.3}, namely that it suffices to upper bound the summation of relative-min entropies,

\begin{align*}
\underset{1 \leq i \leq N}{\sum} \underset{\textbf{P}_{X_C Y_C Z_C D G | \mathcal{E}}}{\textbf{E}}    S_{\infty} \bigg(  \bigg(  \textit{(i-1) th player state} \bigg)  \bigg | \bigg| \bigg(  \textit{i th player state}  \bigg)  \bigg), 
\end{align*}

\noindent one can obtain the desired upper bound for,

\begin{align*}
  \underset{I}{\textbf{E}} \underset{R_{-i}| W_C}{\textbf{E}} \underset{\mathcal{Q}_1 \mathcal{Q}_2 \mathcal{Q}_3 \times \cdots \times \mathcal{Q}_N}{\textbf{E}} \big| \big| \textbf{I}^{E^1}_{r_{-i}, q_1, q_2, \cdots, q_N} - \textbf{2}^{E^1}_{r_{-i},q_1, \bot, q_3, \cdots , q_N} \big| \big|^2_1  ,
\end{align*}

\noindent and hence for,

\[ \mathcal{P} \mathcal{I} \equiv   \left\{\!\begin{array}{ll@{}>{{}}l} 
\underset{I}{\textbf{E}} \underset{R_{-i} | W_C}{\textbf{E}} \underset{\mathcal{Q}_1 \mathcal{Q}_2 \times \cdots \times \mathcal{Q}_N}{\textbf{E}}  \big| \big| \textbf{I}^{E_1}_{r_{-i}, q_1,q_2, \cdots, q_N}  -  \textbf{I}^{E_1}_{r_{-i}, q_1,\bot,q_3, \cdots, q_N} \big| \big|^2_1   , \\ \underset{I}{\textbf{E}} \underset{R_{-i} | W_C}{\textbf{E}} \underset{\mathcal{Q}_1 \mathcal{Q}_2 \times \cdots \times \mathcal{Q}_N}{\textbf{E}}  \big| \big| \textbf{2}^{E_2}_{r_{-i}, q_1,q_2, \cdots, q_N}  -  \textbf{2}^{E_2}_{r_{-i}, \bot, q_2, q_3, \cdots, q_N} \big| \big|^2_1  ,  \\ \vdots \\ \underset{I}{\textbf{E}} \underset{R_{-i} | W_C}{\textbf{E}} \underset{\mathcal{Q}_1 \mathcal{Q}_2 \times \cdots \times \mathcal{Q}_N}{\textbf{E}}  \big| \big| \textbf{N}^{E_N}_{r_{-i}, q_1,q_2, \cdots, q_N}  -  \textbf{N}^{E_N}_{r_{-i}, q_1,\cdots, q_{N-2}, \bot, q_N} \big| \big|^2_1  ,
\end{array}\right. 
\]

\noindent from the computation,

  \begin{align*} \frac{1}{N} \mathcal{E} \leq \frac{1}{N}  \underset{\textbf{R}|W_C}{\textbf{E}} S \bigg( \bigg(  r_{-i} \text{ } \textit{1 st player state}      \bigg) \bigg| \bigg| \bigg(   r_{-i} \text{ } \textit{2 nd player state}      \bigg) \bigg) \\ \\  \leq  \frac{1}{N}  \underset{\textbf{R}|W_C}{\textbf{E}} S \bigg( \bigg( \omega  \text{ } \textit{1 st player state}      \bigg) \bigg| \bigg| \bigg(   \omega  \text{ } \textit{2 nd player state}      \bigg) \bigg)       \\   \\ \leq  \frac{1}{N}  \underset{\textbf{R}|W_C}{\textbf{E}} S \bigg( \bigg( \omega  \text{ } \textit{1 st player state}      \bigg) \bigg| \bigg| \bigg(   \omega  \text{ } \textit{2 nd player state}      \bigg) \bigg)    \\   + \frac{1}{N}  \underset{\textbf{P}_{(\mathcal{Q}_1)_C (\mathcal{Q}_2)_C (\mathcal{Q}_3)_C D G | \mathcal{E}}}{\textbf{E}}    S_{\infty} \bigg(  \bigg(  \textit{1 st player state} \bigg)  \bigg | \bigg| \bigg(  \textit{2 nd player state}  \bigg)  \bigg) \\   \vdots \\ \leq  \frac{1}{N}  \underset{\textbf{R}|W_C}{\textbf{E}} S \bigg( \bigg( \omega  \text{ } \textit{1 st player state}      \bigg) \bigg| \bigg| \bigg(   \omega  \text{ } \textit{2 nd player state}      \bigg) \bigg) \\   + \frac{1}{N} \underset{\textbf{P}_{(\mathcal{Q}_1)_C (\mathcal{Q}_2)_C (\mathcal{Q}_3)_C D G | \mathcal{E}}}{\textbf{E}}    S_{\infty} \bigg(  \bigg(  \textit{1 st player state} \bigg)  \bigg | \bigg| \bigg(  \textit{2 nd player state}  \bigg)  \bigg) \\   + \cdots + \frac{1}{N}  \underset{\textbf{P}_{(\mathcal{Q}_1)_C (\mathcal{Q}_2)_C (\mathcal{Q}_3)_C D G | \mathcal{E}}}{\textbf{E}}    S_{\infty} \bigg(  \bigg(  \textit{(N-1) th player state} \bigg)  \bigg | \bigg| \bigg(  \textit{N th player state}  \bigg)  \bigg) \\     \equiv    \frac{1}{N}  \underset{\textbf{R}|W_C}{\textbf{E}} S \bigg( \bigg( \omega  \text{ } \textit{1 st player state}      \bigg) \bigg| \bigg| \bigg(   \omega  \text{ } \textit{2 nd player state}      \bigg) \bigg)  \\   +     \frac{1}{N}      \underset{1 \leq i \leq N}{\sum} \underset{\textbf{P}_{(\mathcal{Q}_1)_C (\mathcal{Q}_2)_C (\mathcal{Q}_3)_C D G | \mathcal{E}}}{\textbf{E}}    S_{\infty} \bigg(  \bigg(  \textit{(i-1) th player state} \bigg)   \\   \bigg | \bigg| \bigg(  \textit{i th player state}  \bigg)  \bigg)  \\ \\    \equiv \frac{1}{N}  \underset{\textbf{R}|W_C}{\textbf{E}} S \bigg( \bigg( \omega  \text{ } \textit{1 st player state}      \bigg) \bigg| \bigg| \bigg(   \omega  \text{ } \textit{2 nd player state}      \bigg) \bigg)    \\  +\frac{1}{N}    \underset{\textbf{P}_{(\mathcal{Q}_1)_C (\mathcal{Q}_2)_C (\mathcal{Q}_3)_C D G | \mathcal{E}}}{\textbf{E}}  \underset{1 \leq i \leq N}{\sum}     S_{\infty} \bigg(  \bigg(  \textit{(i-1) th player state} \bigg)   \end{align*}

\begin{align*}         \bigg | \bigg| \bigg(  \textit{i th player state}  \bigg)  \bigg)   \\  \\  \equiv   \frac{1}{N} \bigg[   \underset{\textbf{R}|W_C}{\textbf{E}} S \bigg( \bigg( \omega  \text{ } \textit{1 st player state}      \bigg) \bigg| \bigg| \bigg(   \omega  \text{ } \textit{2 nd player state}      \bigg) \bigg) \\  +    \underset{\textbf{P}_{(\mathcal{Q}_1)_C (\mathcal{Q}_2)_C (\mathcal{Q}_3)_C D G | \mathcal{E}}}{\textbf{E}}  \underset{1 \leq i \leq N}{\sum}     S_{\infty} \bigg(  \bigg(  \textit{(i-1) th player state} \bigg) \\   \bigg | \bigg| \bigg(  \textit{i th player state}  \bigg)  \bigg)    \bigg] \\ \\ \leq \frac{1}{N} \bigg[      \big| C \big| \mathrm{log} \bigg[ \bigg| \underset{1 \leq i \leq N}{\prod} \big| \mathcal{Q}_i \big| \bigg|  \bigg]         +    \mathrm{log} \bigg[  \bigg| \frac{1}{\textbf{P}\big[ W_C \big] } \bigg|    \bigg]    \bigg]  \\ \\  \equiv \delta_{\mathrm{Multiplayer}}      ,
\end{align*}

\noindent which can be used to readily conclude that,

\begin{align*}
  \mathcal{P}\mathcal{I} \equiv  \mathrm{O} \bigg(  \frac{\sqrt{\delta^{\frac{N}{300}}_{\mathrm{Multiplayer}}}}{\alpha^{2(N+1)}_{\mathrm{Multiplayer}}} \bigg)  ,
\end{align*}

\noindent for,

\[  \left\{\!\begin{array}{ll@{}>{{}}l} 
r_{-i} \text{ } \textit{1 st player state} \equiv      \textit{Dependency-breaking variable} \text{ } \cup \big( \mathcal{Q}_1 \big)_C     , \\ \vdots \\ r_{-i} \text{ } \textit{N th player state} \equiv     \textit{Dependency-breaking variable} \text{ } \cup    \bigg( \underset{1 \leq i \leq N}{\bigcup} \mathcal{Q}_i \bigg)_C     ,
\end{array}\right. 
\] 

\noindent and,

\[  \left\{\!\begin{array}{ll@{}>{{}}l} 
\omega \text{ } \textit{1 st player state} \equiv \textit{Dependency-breaking states for the first player}, \\ \vdots \\ \omega \text{ } \textit{N th player state} \equiv  \textit{Dependency-breaking states for the N th player},
\end{array}\right. 
\]

\noindent from the fact that, as previously stated,

\[   \left\{\!\begin{array}{ll@{}>{{}}l} 
   \underset{I}{\textbf{E}} \big| \big|   \textbf{P}_{R \mathcal{Q}_1 | W_C}  - \textbf{P}_{\Omega_i | W_C } \textbf{P}_{R_{-i} | W_C} \textbf{P}_{\mathcal{Q}_1 | \Omega_{-i}}    \big| \big|   \leq      \mathrm{O} \bigg(  \frac{\sqrt{\delta^{\frac{N}{300}}_{\mathrm{Multiplayer}}}}{\alpha^{N+1}_{\mathrm{Multiplayer}}} \bigg)        ,  \\ \vdots \\    \underset{I}{\textbf{E}} \big| \big|   \textbf{P}_{R \mathcal{Q}_N | W_C}  - \textbf{P}_{\Omega_i | W_C } \textbf{P}_{R_{-i} | W_C} \textbf{P}_{\mathcal{Q}_N | \Omega_{-i}}    \big| \big|   \leq      \mathrm{O} \bigg(  \frac{\sqrt{\delta^{\frac{N}{300}}_{\mathrm{Multiplayer}}}}{\alpha^{N+1}_{\mathrm{Multiplayer}}} \bigg)   \text{, }
\end{array}\right. 
\]

\[   \left\{\!\begin{array}{ll@{}>{{}}l} 
 \underset{I}{\textbf{E}} \big| \big|   \textbf{P}_{R \mathcal{Q}_1 | W_C}    -  \textbf{P}_{\Omega_i | W_C} \textbf{P}_{R_{-i} | W_C}  \textbf{P}_{\mathcal{Q}_1 | \Omega_i}      \big| \big|   \leq  \sqrt{\delta^{\frac{N}{300}}_{\mathrm{Multiplayer}}}  ,  \\ \vdots \\    \underset{I}{\textbf{E}} \big| \big|   \textbf{P}_{R \mathcal{Q}_N | W_C}    -  \textbf{P}_{\Omega_i | W_C} \textbf{P}_{R_{-i} | W_C}  \textbf{P}_{\mathcal{Q}_N | \Omega_i}    \big| \big|  \leq \sqrt{\delta^{\frac{N}{300}}_{\mathrm{Multiplayer}}}   \text{, }
\end{array}\right. 
\]

\[   \left\{\!\begin{array}{ll@{}>{{}}l} 
 \underset{I}{\textbf{E}} \big| \big|  \textbf{P}_{\Omega_{-i} | W_C} \textbf{P}_{R_{-i} | \Omega_{-i} W_C} \textbf{P}_{\mathcal{Q}_1 | \Omega_i}      - \textbf{P}_{\Omega_{-i} | W_C} \textbf{P}_{R_{-i} | W_C} \textbf{P}_{\mathcal{Q}_1 | \Omega_i}     \big| \big|  \leq    \mathrm{O} \bigg(  \frac{\sqrt{\delta^{\frac{N}{300}}_{\mathrm{Multiplayer}}}}{\alpha^{N+1}_{\mathrm{Multiplayer}}} \bigg)    ,  \\ \vdots \\ \underset{I}{\textbf{E}} \big| \big|  \textbf{P}_{\Omega_{-i} | W_C} \textbf{P}_{R_{-i} | \Omega_{-i} W_C} \textbf{P}_{\mathcal{Q}_N | \Omega_i}      - \textbf{P}_{\Omega_{-i} | W_C} \textbf{P}_{R_{-i} | W_C} \textbf{P}_{\mathcal{Q}_N | \Omega_i}        \big| \big|  \leq     \mathrm{O} \bigg(  \frac{\sqrt{\delta^{\frac{N}{300}}_{\mathrm{Multiplayer}}}}{\alpha^{N+1}_{\mathrm{Multiplayer}}} \bigg)   \text{. }
\end{array}\right. 
\]

\noindent We conclude the argument, as,

\[ \mathcal{P} \mathcal{I} \equiv   \left\{\!\begin{array}{ll@{}>{{}}l} 
\underset{I}{\textbf{E}} \underset{R_{-i} | W_C}{\textbf{E}} \underset{\mathcal{Q}_1 \mathcal{Q}_2 \times \cdots \times \mathcal{Q}_N}{\textbf{E}}  \big| \big| \textbf{I}^{E_1}_{r_{-i}, q_1,q_2, \cdots, q_N}  -  \textbf{I}^{E_1}_{r_{-i}, q_1,\bot,q_3, \cdots, q_N} \big| \big|^2_1   \equiv \mathrm{O} \bigg(  \frac{\sqrt{\delta^{\frac{N}{300}}_{\mathrm{Multiplayer}}}}{\alpha^{2(N+1)}_{\mathrm{Multiplayer}}} \bigg)  . \\ \underset{I}{\textbf{E}} \underset{R_{-i} | W_C}{\textbf{E}} \underset{\mathcal{Q}_1 \mathcal{Q}_2 \times \cdots \times \mathcal{Q}_N}{\textbf{E}}  \big| \big| \textbf{2}^{E_2}_{r_{-i}, q_1,q_2, \cdots, q_N}  -  \textbf{2}^{E_2}_{r_{-i}, \bot, q_2, q_3, \cdots, q_N} \big| \big|^2_1 \equiv  \mathrm{O} \bigg(  \frac{\sqrt{\delta^{\frac{N}{300}}_{\mathrm{Multiplayer}}}}{\alpha^{2(N+1)}_{\mathrm{Multiplayer}}} \bigg)   .  \\ \vdots \\ \underset{I}{\textbf{E}} \underset{R_{-i} | W_C}{\textbf{E}} \underset{\mathcal{Q}_1 \mathcal{Q}_2 \times \cdots \times \mathcal{Q}_N}{\textbf{E}}  \big| \big| \textbf{N}^{E_N}_{r_{-i}, q_1,q_2, \cdots, q_N}  -  \textbf{N}^{E_N}_{r_{-i}, q_1,\cdots, q_{N-2}, \bot, q_N} \big| \big|^2_1  \equiv  \mathrm{O} \bigg(  \frac{\sqrt{\delta^{\frac{N}{300}}_{\mathrm{Multiplayer}}}}{\alpha^{2(N+1)}_{\mathrm{Multiplayer}}} \bigg) .
\end{array}\right.  \boxed{}
\]

\subsection{General description of arguments for Theorem 4}

\noindent One can conclude that the desired exponential decay for parallel repetition of the anchored optimal value,

\begin{align*}
   \omega_{\mathrm{Multiplayer}} \big( \big( G_{\bot} \big)    \big)^{\otimes n} \leq  \frac{10}{\epsilon_{\mathrm{Multiplayer}}} \mathrm{exp} \bigg[  -  \frac{c_{\mathrm{Multiplayer}} \alpha^{20N + 1}_{\mathrm{Multiplayer}}        \epsilon^{20N}_{\mathrm{Multiplayer}}  N }{s_{\mathrm{Multiplayer}}  }\bigg]          \text{, }
\end{align*}

\noindent holds from the following observations. Recall:

\begin{itemize}
\item[$\bullet$] \textit{Square root computation from the Proof of Theorem 1},

\begin{align*}
 \sqrt{    \underset{R_{-i}| q_1, \bot, \cdots, \bot, W_C}{\textbf{E}} \underset{\mathcal{Q}_1}{\textbf{E}}  \big| \big| \ket{\widetilde{\Psi_{r_{-i}, q_1, \bot,\cdots, \bot}}}  - \mathcal{U}_{r_{-i},1}\ket{\widetilde{\Psi_{r_{-i}, \bot, \bot, \cdots, \bot}}}  \big| \big|^2 }   \\ =  \mathrm{O} \bigg(  \frac{\delta^{\frac{N}{300}}_{\mathrm{Multiplayer}}}{\alpha^{N+1}_{\mathrm{Multiplayer}}}     \bigg)  ,
\end{align*}

\item[$\bullet$] \textit{POVM expectations from the Proof of Theorem 2},

\begin{align*}
 \underset{I}{\textbf{E}} \bigg| \bigg|      \textbf{P}_{R_{-i}| W_C}   \textbf{P}_{\mathcal{Q}_1 \mathcal{Q}_2 \mathcal{Q}_3} \mathscr{P}\mathscr{O}\mathscr{V} \mathscr{M}_1 -   \textbf{P}_{\mathcal{Q}_1 \mathcal{Q}_2 \mathcal{Q}_3 R_{-i} \mathcal{A}_1 \mathcal{A}_2 \mathcal{A}_3 | W_C}           \bigg| \bigg| , 
\end{align*}

\item[$\bullet$]

\begin{align*}
 \sqrt{N}    \underset{I}{\textbf{E}} \underset{R_{-i} | W_C}{\textbf{E}} \underset{\mathcal{Q}_1 \mathcal{Q}_2 \mathcal{Q}_3 \mathcal{Q}_4}{\textbf{E}} \big| \big|   \big( \mathcal{U}_{1,r_{-i}}  \otimes  \mathcal{U}_{2,r_{-i}}   \otimes  \mathcal{U}_{3,r_{-i}}  \otimes  \cdots \otimes \mathcal{U}_{N,r_{-i}}  \big)   \ket{\widetilde{\Psi_{r_{-i}, \bot,\bot,\bot, \cdots, \bot}}} \\   - \ket{\widetilde{\Psi_{r_{-i},  q_1,q_2,q_3, \cdots, q_N}}}      \big| \big|    \leq  \mathrm{O} \bigg(  \frac{\delta^{\frac{N}{300}}_{\mathrm{Multiplayer}}}{\alpha^{N+1}_{\mathrm{Multiplayer}}} \bigg)     ,    , 
\end{align*}

\end{itemize}

\subsection{Argument}

\noindent \textit{Proof of Lemma 4}. One can argue that the desired expression from the optimal value holds, either from computations involving the tensor product representation,

\begin{align*}
    \theta_{1_C \bar{1_C } 2_C \bar{2_C} \times \cdots \times N_C \bar{N_C} A M E^B}   =   \underset{q_1 q_2 \cdots q_N}{\sum} P_{Q_1\cdots Q_N} \big(q_1 q_2 \cdots q_N \big) \ket{q_1 q_2 \cdots q_N} \bra{q_1 q_2 \cdots q_N} \\  \bigotimes \ket{\theta} \bra{\theta}  \text{, }
\end{align*}

\noindent provided in [30], or by directly applying the arguments for anchored parallel repetition in [3]. For convenience, we discuss, in the arguments below, how computations initially provided in [3] yield the desired exponential decay for the parallel repetition of the anchored multiplayer optimal value.

\bigskip

\noindent \textit{Proposition} (\textit{conditional probability of winning on one coordinate, depending upon winning on all coordinates}, \textbf{Proposition} \textit{6.5}, [3]). Denote $W$ as the indicator function of winning across all $n$ coordinates, and $W_i$ the indicator function of winning at the $i$ th coordinate, and,

\begin{align*} n \geq \frac{6N}{\epsilon_{\mathrm{Multiplayer}}} \mathrm{log} \bigg[ \bigg| \frac{\sqrt{6N}}{\epsilon_{\mathrm{Multiplayer}} \textbf{P} \big[ W \big]}      \bigg| \bigg] . \end{align*}

\noindent Then there exists a set of coordinates $C$, $C \subseteq [n]$, with cardinality at most,

\begin{align*} t \equiv \frac{3 N}{\epsilon_{\mathrm{Multiplayer}}} \mathrm{log} \bigg[ \bigg| \frac{\frac{3}{2}N }{\epsilon_{\mathrm{Multiplayer}} \textbf{P} \big[ W \big]} \bigg| \bigg] , \end{align*}

\noindent such that,

\begin{align*}
    \underset{I}{\textbf{E}} \text{ } \textbf{P} \bigg[ W_i \big| W_C  \bigg] \equiv   \underset{I}{\textbf{E}} \text{ } \textbf{P} \bigg[ W_i \bigg|  \underset{1 \leq i^{\prime} \leq N}{\prod} W_{C_{i^{\prime}}}   \bigg]  \geq 1 - \frac{\epsilon_{\mathrm{Multiplater}}}{2} .
\end{align*}

\noindent \textit{Proof of Proposition}. Directly apply the argument provided in \textbf{Proposition} of [2], from which we conclude the argument. \boxed{}

\bigskip

\noindent \textit{Proposition} (\textit{computation of the power of decay in the exponential upper bound for parallel repetition of the multiplayer anchored optimal value}, \textbf{Theorem} \textit{6.1}, [3]). With the choice of parameters provided in the previous result above,  

\begin{align*}
  \epsilon_{\mathrm{Multiplayer}} \geq   \bigg[ \frac{\delta_{\mathrm{Multiplayer}}}{c_{\mathrm{Multiplayer}}}    \cdot \frac{1}{40N } \cdot \frac{1}{\mathrm{log} \big[ e \big]} \cdot \frac{1}{\alpha^{20N+1}_{\mathrm{Multiplayer}}} \bigg]^{\frac{1}{6N}} ,
\end{align*}

\noindent hence implying the desired exponential rate of decay for parallel repetition of the anchored multiplayer optimal value.

\bigskip

\noindent \textit{Proof of Proposition}. By direct computation, observe,

\begin{align*}
  n \geq \frac{n}{s_{\mathrm{Multiplayer}}} \geq \frac{1}{c_{\mathrm{Multiplayer}} \alpha^{20N+1}_{\mathrm{Multiplayer}} \epsilon^{6N}_{\mathrm{Multiplayer}}}  \mathrm{log} \bigg[ \bigg|   \frac{N n}{\epsilon_{\mathrm{Multiplayer}} \textbf{P} \big[ W \big]}    \bigg| \bigg] \\ \geq  \frac{2}{\epsilon_{\mathrm{Multiplayer}}} \mathrm{log} \bigg[ \bigg|   \frac{N n}{\epsilon_{\mathrm{Multiplayer}} \textbf{P} \big[ W \big]}    \bigg| \bigg] ,
\end{align*}

\noindent Relatedly, observe,

\begin{align*}
  \delta_{\mathrm{Multiplayer}} \equiv          \frac{1}{n-t} \bigg[ \mathrm{log} \bigg[ \bigg| \frac{ N n}{\textbf{P} \big[ W_C \big]} \bigg| \bigg] + t \cdot s_{\mathrm{Multiplayer}} \bigg] \\ \leq  \frac{2}{n} \bigg[ \frac{4  N n \cdot s_{\mathrm{Multiplayer}}}{\epsilon_{\mathrm{Multiplayer}}}  \mathrm{log} \bigg[ \bigg| \frac{N}{\textbf{P} \big[ W_C \big]} \bigg| \bigg]  \bigg]  \\ \leq   \frac{2}{n}  \bigg[    \frac{6N s_{\mathrm{Multiplayer}}}{\epsilon_{\mathrm{Multiplayer}}}   \cdot  \frac{c_{\mathrm{Multiplayer}} \mathrm{log} \big[ e \big]\alpha^{20N+1}_{\mathrm{Multiplayer}} \epsilon^{6N}_{\mathrm{Multiplayer}} n   }{s_{\mathrm{Multiplayer}}}      \bigg]   \\  \equiv 40 N    c_{\mathrm{Multiplayer}}     \mathrm{log} \big[ e \big]   \alpha^{20N + 1}_{\mathrm{Multiplayer}} \epsilon^{6N}_{\mathrm{Multiplayer}}      ,
\end{align*}

\noindent which implies that the multiplayer winning probability equals,

\begin{align*}
  \textbf{P} \big[ N \text{ } \textit{players win the multiplayer game} \big] \equiv \omega_{\mathrm{Multiplayer}} \equiv 1 - \frac{\epsilon_{\mathrm{Multiplayer}}}{2} - \beta^{\prime} \frac{\delta^{\frac{N}{300}}}{\alpha^{N+1}_{\mathrm{Multiplayer}}} ,   
\end{align*}

\noindent for some universal constant $\beta^{\prime}$, from which we conclude the argument. \boxed{}

\bigskip

\noindent With the two above results, we readily obtain the desired rate of decay of the exponential upper bound for parallel repetition of the anchored optimal value is obtained, from which we conclude the argument. \boxed{}

\section{Appendix}

\begin{tabular}{|l|l|}
\hline\parbox[t]{0.25\textwidth}{
\begin{itemize}
\item $\textbf{Theorem}$ \textit{1}
\item $\textbf{Theorem}$ \textit{2}
\item $\textbf{Theorem}$ \textit{3}
\item $\textbf{Theorem}$ \textit{4}
\item $\textbf{Theorem}$ \textit{5}
\item $\textbf{Theorem}$ \textit{6}
\item $\textbf{Theorem}$ $\textit{1}^{*}$
\item $\textbf{Theorem}$ $\textit{2}^{*}$
\item $\textbf{Theorem}$ $\textit{3}^{*}$
\item $\textbf{Theorem}$ $\textit{4}^{*}$
\item $\textbf{Theorem}$ $\textit{5}^{*}$
\item $\textbf{Theorem}$ $\textit{6}^{*}$
\end{itemize}}& 
\parbox[t]{0.73\textwidth}{
\begin{itemize}
\item $3$-XOR primal feasible solutions and duality gaps
\item $4$-XOR primal feasible solutions and duality gaps
\item $5$-XOR primal feasible solutions and duality gaps
\item $N$-XOR primal feasible solutions and duality gaps
\item $N$-XOR strong parallel repetition primal feasible solutions and duality gaps
\item FFL strong parallel repetition primal feasible solutions and duality gaps
\item Strong parallel repetition of $\textbf{Theorem}$ $\textit{1}$
\item Strong parallel repetition of $\textbf{Theorem}$ $\textit{2}$
\item Strong parallel repetition of $\textbf{Theorem}$ $\textit{3}$
\item Strong parallel repetition of $\textbf{Theorem}$ $\textit{4}$
\item Strong parallel repetition of $\textbf{Theorem}$ $\textit{5}$
\item Strong parallel repetition of $\textbf{Theorem}$ $\textit{6}$
\end{itemize}}\\
\hline
\end{tabular}
\noindent \textit{Table *}. An overview of the main results provided in previous work of the author, [46], which provide statements of error bounds, and related objects, which will also be characterized in forthcoming arguments in the Appendix for Expanded games.

\subsection{Arguments for error bounds in expanded games}

\subsubsection{Exponentially dependent Frobenius norm upper bounds}

\noindent \textit{Proof of Lemma 1}. Obtaining the desired collection of upper bounds for the Frobenius norm amounts to upper bounding the each Frobenius norm in the statement of the result above, implying,

\begin{align*}
 \bigg| \bigg|    \bigg[             \bigg( \bigg( \underset{1 \leq i \leq n}{\prod}  A^{j_i}_i \bigg) \bigotimes \bigg( \underset{1 \leq k \leq n-1}{\bigotimes} \textbf{I}_k \bigg) \bigg)   -  \bigg(  \omega_{\mathrm{Expanded}} \bigg(         \pm \mathrm{sign} \big( i_1, \cdots, i_n \big)            \\ \times     \bigg( \underset{1 \leq i \leq n}{\prod}  A^{j_i}_i \bigg)           \bigg)    \bigotimes \bigg( \underset{1 \leq k \leq n-1}{\bigotimes} \textbf{I}_k \bigg)\bigg)                       \bigg]   \ket{\psi_{\mathrm{Expanded}}}        \bigg| \bigg|_F  \\  <    \bigg( n_1 +       \big(    n_1 + 2    \big) \omega_{\mathrm{Expanded}}^{-1}                \bigg) n^N \sqrt{\epsilon}        \text{,}         \\ \\  \bigg| \bigg|    \bigg[ \bigg(  \textbf{I} \bigotimes            \bigg( \underset{1 \leq i \leq n}{\prod}  A^{1,j_i}_i \bigg) \bigotimes \bigg( \underset{1 \leq k \leq n-2}{\bigotimes} \textbf{I}_k \bigg) \bigg)   -  \bigg(  \textbf{I} \bigotimes  \bigg( \omega_{\mathrm{Expanded}} \\ \times \bigg(         \pm \mathrm{sign} \big( i_1, j_1, \cdots, i_n, \cdots, j_n \big)        \bigg( \underset{1 \leq i_2 \leq m}{\underset{1 \leq i_1 \leq n}{\prod}}  A^{1,j_{i_1,i_2}}_{i_1,i_2} \bigg)           \bigg)  \bigg) \\    \bigotimes \bigg( \underset{1 \leq k \leq n-2}{\bigotimes} \textbf{I}_k \bigg)\bigg)                       \bigg]  \ket{\psi_{\mathrm{Expanded}}}       \bigg| \bigg|_F   <   \bigg( n_2 +  \big( n_2 + 2 \big) \omega_{\mathrm{Expanded}}^{-1}            \bigg) n^N \sqrt{\epsilon}  \\  \vdots \\ \bigg| \bigg|    \bigg[             \bigg(\bigg( \underset{1 \leq k \leq n-1}{\bigotimes} \textbf{I}_k \bigg) \bigotimes   \bigg( \underset{1 \leq i \leq n}{\prod}  A^{(n-1),j_{i_1,\cdots,i_n}}_{i_1,\cdots, i_n} \bigg)  \bigg)   -  \bigg( \bigg( \underset{1 \leq k \leq n-1}{\bigotimes} \textbf{I}_k \bigg) \bigotimes \bigg(    \omega_{\mathrm{Expanded}} \\ \times  \bigg(         \pm \mathrm{sign} \big( i_1, \cdots, i_n, j_1  , \cdots, j_n \big)                         \bigg)   \bigg( \underset{1 \leq i \leq n}{\prod}  A^{(n-1),j_{i_1,\cdots,i_n}}_{i_1,\cdots, i_n} \bigg)  \bigg)                 \bigg)   \bigg] \ket{\psi_{\mathrm{Expanded}}}    \bigg| \bigg|_F   \\  <   \bigg( n_N        +   \big(  n_N + 2\big)     \omega_{\mathrm{Expanded}}^{-1}            \bigg) n^N \sqrt{\epsilon}  \text{, }
\end{align*}

\noindent for the quantum optimal strategy,

\begin{align*}
       \ket{\psi_{\mathrm{Expanded}}} \equiv 
 \ket{\psi_{\mathrm{Expanded}} \big( \mathcal{S} \big) } \equiv \underset{\mathcal{S}}{\mathrm{sup}}  \big\{ \text{payoff for all players with some quantum strategy } \mathcal{S}             \big\}                  \text{,}
\end{align*}

\noindent and suitable $n_i$. Hence the desired upper bound,

\begin{align*}
  \mathscr{C}  \text{, }
\end{align*}

\noindent for the collection of Frobenius norms over all players of the Expanded game takes the form,

\begin{align*}
          \underset{1 \leq i \leq N}{\sum}      \bigg(                  n_i + \big( n_i + 2 \big) \omega^{-1}_{\mathrm{Expanded}}    \bigg) n^N \sqrt{\epsilon}  \equiv \bigg( \big( n_1 + \cdots + n_N \big) +  \big( \big( n_1 + \cdots + n_N \big) +   2N  \big) \\ \times \omega^{-1}_{\mathrm{Expanded}}            \bigg) n^N \sqrt{\epsilon }   <  \bigg(    \big( n_1 + \cdots + n_N \big) +   \big(  \big( n_1 + \cdots + n_N \big) + 2N \big)   \bigg) \omega^{-1}_{\mathrm{Expanded}} n^N \sqrt{\epsilon}  \\ <      \bigg(   \big( n_1 + \cdots + n_N \big)   \bigg( 2 + 2 N \bigg)       \bigg) \omega^{-1}_{\mathrm{Expanded}} n^N \sqrt{\epsilon}           <       \bigg(   \big( n_1 + \cdots + n_N \big)   \bigg( 2 + 2 N \bigg)       \bigg) \\ \times  \big( \omega^{-1}_{\mathrm{Expanded}} n^N \big)^2 \sqrt{\epsilon}  
      \equiv  \bigg(  2 \big( n_1 + \cdots + n_N \big) \big( \omega^{-1}_{\mathrm{Expanded}} n^N \big)^2 + 2N \big( n_1 + \cdots + n_N \big) \\ \times  \big( \omega^{-1}_{\mathrm{Expanded}} n^N \big)^2 \bigg)        \sqrt{\epsilon}        <   2 \big( n_1 + \cdots n_N \big)   N n^N \sqrt{\epsilon}    + 2N \big( n_1 + \cdots + n_N \big)    N n^N \sqrt{\epsilon}       \\   <     5 N^2 n_1 n^N \sqrt{\epsilon}    <        5 \big( N n^N \big)^2 \sqrt{\epsilon}    \equiv \mathscr{C}^{\prime}       \text{, }
\end{align*}

\noindent Setting the upper bound to,

\begin{align*}
    \mathscr{C} \equiv   \mathscr{C}^{\prime}   \text{, }
\end{align*}

\noindent yields the desired constant, from which we conclude the argument. \boxed{}

\subsubsection{Signed product expansion for odd $n$}

\noindent \textit{Proof of Lemma 3}. By direct computation, along the lines of a previous adaptation from the two-player setting for $\mathrm{XOR}^{*}$ and $\mathrm{FFL}$ games in [50],

\begin{align*}
            2 \underset{1 \leq i < j \leq n }{\sum} \bigg| \bigg| \bigg(   \bigg(   \frac{A_i A_j + A_j A_i}{2}                          \bigg) \bigotimes \bigg( \underset{1\leq k \leq N-1}{\bigotimes} \textbf{I}_k\bigg) \bigg) \ket{\psi_{\mathrm{Expanded}}}              \bigg| \bigg|^2   \leq       \big( 1 + \omega^{-1}_{\mathrm{Expanded}} \big)^2    \\ \times \underset{1 \leq i < j  \leq n}{\sum}  \bigg[      \bigg| \bigg|        \bigg(     \bigg(  \frac{A_i + A_j}{\sqrt{2}} \bigg)   \bigg( \underset{1 \leq k \leq N-1}{\bigotimes} \textbf{I}_k \bigg) \bigg) \ket{\psi_{\mathrm{Expanded}}}   
 - \bigg(  \textbf{I}    \bigotimes B_{ij}  \bigotimes \bigg( \underset{1 \leq k \leq N-2}{\bigotimes} \textbf{I}_k \bigg) \bigg)  \\ \times \ket{\psi_{\mathrm{Expanded}}}  \bigg| \bigg|^2  +      \bigg| \bigg|      \bigg(          \bigg( \frac{A_i - A_j}{\sqrt{2}} \bigg)   \bigotimes  \bigg( \underset{1 \leq k \leq N-1}{\bigotimes}           \textbf{I}_k \bigg) \bigg) \ket{\psi_{\mathrm{Expanded}}}  - \bigg( \textbf{I}   \bigotimes B_{ji}   \\ \bigotimes \bigg( \underset{1 \leq k \leq N-2}{\bigotimes}           \textbf{I}_k \bigg)          \bigg)  \ket{\psi_{\mathrm{Expanded}}}           \bigg| \bigg|^2          \bigg]                   \text{, }
\end{align*}

\noindent implies the desired upper bound is obtained from the observations,

\begin{align*}
        \underset{1 \leq i < j  \leq n}{\sum}      \bigg| \bigg|        \bigg(     \bigg(  \frac{A_i + A_j}{\sqrt{2}} \bigg)  \bigg( \underset{1 \leq k \leq N-1}{\bigotimes} \textbf{I}_k \bigg) \bigg) \ket{\psi_{\mathrm{Expanded}}}   
 - \bigg(  \textbf{I}   \bigotimes B_{ij}  \bigotimes \bigg( \underset{1 \leq k \leq N-2}{\bigotimes} \textbf{I}_k \bigg) \bigg)  \\ \times \ket{\psi_{\mathrm{Expanded}}}  \bigg| \bigg|^2   <                   \frac{1}{100}       n    \bigg( \underset{1 \leq j \leq N-1}{\prod}     \big( n - j \big) \bigg)    \epsilon_{\mathrm{Expanded}} 
          \text{, }
\end{align*}

\noindent and also that,

\begin{align*}
       \underset{1 \leq i < j  \leq n}{\sum}     \bigg| \bigg|      \bigg(          \bigg( \frac{A_i - A_j}{\sqrt{2}} \bigg)   \bigotimes  \bigg( \underset{1 \leq k \leq N-1}{\bigotimes}           \textbf{I}_k \bigg) \bigg) \ket{\psi_{\mathrm{Expanded}}}  - \bigg( \textbf{I}  \bigotimes B_{ji}   \bigotimes \bigg( \underset{1 \leq k \leq N-2}{\bigotimes}           \textbf{I}_k \bigg)          \bigg)  \\ \times  \ket{\psi_{\mathrm{Expanded}}}           \bigg| \bigg|^2  <                \frac{1}{100}   n  \bigg( \underset{1 \leq j \leq N-1}{\prod}     \big( n - j \big) \bigg)  \epsilon_{\mathrm{Expanded}}                   \text{, }
\end{align*}

\noindent which together yield the desired upper bound, upon taking,

\begin{align*}
  C_1 > \frac{1}{50}    \text{, }
\end{align*}

\noindent from the observation that the desired upper bound holds iff,

\begin{align*}
     \underset{i,j}{\sum}     \bigg| \bigg|     \bigg[            \bigg( \bigg(    \frac{A_i A_j + A_j A_i}{2}        \bigg) \bigotimes  \bigg( \underset{1\leq k \leq N-1}{\bigotimes} \textbf{I}_k \bigg)  \bigg)      \bigg] \ket{\psi_{\mathrm{Expanded}}}                           \bigg| \bigg|^2  <    \frac{1}{50} n  \bigg( \underset{1 \leq j \leq N-1}{\prod}     \big( n - j \big) \bigg) \\ \times \epsilon_{\mathrm{Expanded}}  \Longleftrightarrow C_1 > \frac{1}{50}            \text{, }
\end{align*}

\noindent from which we conclude the argument. \boxed{}

\subsubsection{Upper bound in error bounds of the first type}

\noindent \textit{Proof of Lemma $5^{**}$}. Set $\epsilon \equiv \epsilon_{\mathrm{Expanded}\wedge \cdots \wedge  \mathrm{Expanded}}$. The desired upper bound is of the form,

\begin{align*}
        n^N_{\wedge} \sqrt{\epsilon} \bigg(     \bigg(               \omega_{\mathrm{Expanded}\wedge \cdots \wedge \mathrm{Expanded}}^{-2}         +        \omega_{\mathrm{Expanded}\wedge \cdots \wedge \mathrm{Expanded}}^{-1}        \bigg)        +     \frac{1}{\sqrt{n^{N-1}}}   \bigg(  \omega_{\mathrm{Expanded}\wedge \cdots \wedge \mathrm{Expanded}}^{-2}    \\      +        \omega_{\mathrm{Expanded}\wedge \cdots \wedge \mathrm{Expanded}}^{-1}          \bigg)         \bigg)     \\ \equiv    n^N_{\wedge} \sqrt{\epsilon} \bigg(     \bigg(                \bigg( \frac{1}{\sqrt{2}} \bigg)^{-2N} + \bigg( \frac{1}{\sqrt{2}} \bigg)^{-N}         \bigg)        +     \frac{1}{\sqrt{n^{N-1}}}   \bigg(    \bigg( \frac{1}{\sqrt{2}} \bigg)^{-2N} + \bigg( \frac{1}{\sqrt{2}} \bigg)^{-N}            \bigg)         \bigg) \\ \equiv             \bigg( \frac{1}{\sqrt{2}} \bigg)^{-2N} \bigg[    n^N_{\wedge} \sqrt{\epsilon} \bigg(     \bigg(               1 + \bigg( \frac{1}{\sqrt{2}} \bigg)^{N}         \bigg)        +     \frac{1}{\sqrt{n^{N-1}}}   \bigg(    1 + \bigg( \frac{1}{\sqrt{2}} \bigg)^{N}             \bigg)         \bigg)         \bigg]        \\   <      \bigg( \frac{1}{\sqrt{2}} \bigg)^{-2N} \bigg[    n^{N+\epsilon}_{\wedge} \bigg(     \bigg(               1  + \bigg( \frac{1}{\sqrt{2}} \bigg)^{N}         \bigg)        +     \frac{1}{\sqrt{n^{N-1}}}   \bigg(    1 + \bigg( \frac{1}{\sqrt{2}} \bigg)^{N}             \bigg)         \bigg)         \bigg]    \\  <   \bigg( \frac{1}{\sqrt{2}} \bigg)^{-N} \bigg[    n^{N+\epsilon}_{\wedge} \bigg(     \bigg(               1 + \bigg( \frac{1}{\sqrt{2}} \bigg)^{N}         \bigg)        +     \frac{1}{\sqrt{n^{N-1}}}   \bigg(    1 + \bigg( \frac{1}{\sqrt{2}} \bigg)^{N}             \bigg)         \bigg)         \bigg]    \\ <      \bigg( \frac{1}{\sqrt{2}} \bigg)^{-N} \bigg[    n^{N+\epsilon}_{\wedge} \bigg(   \bigg( \frac{2}{\sqrt{2}} \bigg)^{N}          +     \frac{1}{\sqrt{n^{N-1}}}  \bigg( \frac{2}{\sqrt{2}} \bigg)^{N}                    \bigg)         \bigg]      \\  \equiv    \bigg( \frac{1}{\sqrt{2}} \bigg)^{-N}  \bigg( \frac{2}{\sqrt{2}} \bigg)^{N}     \bigg[    n^{N+\epsilon}_{\wedge} \bigg(   1      +     \frac{1}{\sqrt{n^{N-1}}}                   \bigg)         \bigg]  \\  <               \bigg( \frac{2}{\sqrt{2}} \bigg)^{-N}  \bigg( \frac{2}{\sqrt{2}} \bigg)^{N}     \bigg[    n^{N+\epsilon}_{\wedge} \bigg(   1      +     \frac{1}{\sqrt{n^{N-1}}}                   \bigg)         \bigg]  \\  \equiv    \bigg[    n^{N+\epsilon}_{\wedge} \bigg(   1      +     \frac{1}{\sqrt{n^{N-1}}}                   \bigg)         \bigg]  \\  <    n^{N+\epsilon}_{\wedge } + \bigg(  \frac{50 n^{N+\epsilon}_{\wedge}}{\sqrt{n^{N-1}}}  \bigg)    \omega_{\mathrm{Expanded} \wedge \cdots \wedge \mathrm{Expanded}}    \text{, }
\end{align*}

\noindent from which we conclude the argument. \boxed{}

\subsection{Adaptation of Lemma $6$}

\subsubsection{$\sqrt{\epsilon}$ upper bound after performing an interchanging the order of tensor observables from players}

\noindent \textit{Proof of Lemma 5B}. From previous arguments used in \textbf{Lemma} \textit{4}, observe from the two-player setting the proof, with upper bound $17 \sqrt{n \epsilon}$, crucially,

\begin{align*}
          \frac{\bigg| \textbf{I} \otimes \bigg( \frac{\pm B_{kl} + B_{lk}}{\sqrt{2}} \bigg)     \bigg| }{\bigg| \textbf{I} \otimes \bigg( \frac{\pm B_{kl} + B_{lk}}{\big| \pm B_{kl } + B_{lk} \big| }   \bigg)    \bigg| }    \equiv \frac{\bigg|   \bigg[ \textbf{I} \otimes \bigg( \frac{\pm B_{kl} + B_{lk}}{\sqrt{2}} \bigg)    \bigg]  \bigg|   \frac{\big| \pm B_{kl} + B_{lk} \big|}{\big| \pm B_{kl} + B_{lk} \big|}   }{\bigg|   \textbf{I} \otimes \bigg( \frac{\pm B_{kl} + B_{lk}}{\big| \pm B_{kl } + B_{lk} \big| }  \bigg)          \bigg| }  \approx          \frac{\big| \pm B_{kl} + B_{lk} \big| }{\sqrt{2}}      \approx 1    \text{, } 
\end{align*}

\noindent from which, from the computations in the $N$-player setting from the expression,

\begin{align*}
    \underset{\text{Questions }k,l}{\sum}    \frac{\bigg| \textbf{I} \bigotimes \bigg( \frac{\pm B_{kl} + B_{lk}}{\sqrt{2}} \bigg)  \bigotimes \bigg( \underset{1 \leq k^{\prime} \leq N-2}{\bigotimes} \textbf{I}_{k^{\prime}} \bigg)    \bigg| }{ \bigg| \textbf{I} \bigotimes \bigg( \frac{\pm B_{kl} + B_{lk}}{\big| \pm B_{kl} + B_{lk} \big| } \bigg)  \bigotimes \bigg( \underset{1 \leq k^{\prime} \leq N-2}{\bigotimes} \textbf{I}_{k^{\prime}} \bigg)    \bigg| } \\  \equiv    \underset{\text{Questions }k,l}{\sum}    \frac{\bigg| \textbf{I} \bigotimes \bigg( \frac{\pm B_{kl} + B_{lk}}{\sqrt{2}} \bigg)  \bigotimes \bigg( \underset{1 \leq k^{\prime} \leq N-2}{\bigotimes} \textbf{I}_{k^{\prime}} \bigg)    \bigg| }{ \bigg| \textbf{I} \bigotimes \bigg( \frac{\pm B_{kl} + B_{lk}}{\big| \pm B_{kl} + B_{lk} \big| } \bigg)  \bigotimes \bigg( \underset{1 \leq k^{\prime} \leq N-2}{\bigotimes} \textbf{I}_{k^{\prime}} \bigg)    \bigg| }     \bigg[   \frac{\big| \pm B_{kl} + B_{lk} \big|}{\big| \pm B_{kl} + B_{lk} \big|}   \bigg]  \\ \approx    \underset{\text{Questions }k,l}{\sum}  \bigg[   \frac{\big| \pm B_{kl} + B_{lk} \big|}{\sqrt{2}} \bigg]    \text{, }
\end{align*}

\noindent and from strong parallel repetition in the $\mathrm{Expanded} \wedge \mathrm{Expanded}$ setting,

\begin{align*}
    \underset{\text{Questions }k,l,k^{\prime},l^{\prime}}{\sum}    \frac{\bigg| \textbf{I} \bigotimes \bigg( \frac{\pm\big( B_{kl} \wedge B_{k^{\prime} l^{\prime}} \big)  + \big( B_{lk} \wedge B_{l^{\prime} k^{\prime}} \big) }{\sqrt{2}} \bigg) \bigg|  }{ \bigg| \textbf{I} \bigotimes \bigg( \frac{\pm\big( B_{kl} \wedge B_{k^{\prime} l^{\prime}} \big)  + \big( B_{lk} \wedge B_{l^{\prime} k^{\prime}} \big)}{\big|  \pm\big( B_{kl} \wedge B_{k^{\prime} l^{\prime}} \big)  + \big( B_{lk} \wedge B_{l^{\prime} k^{\prime}} \big) \big| } \bigg)    \bigg| } \\   \equiv    \underset{\text{Questions }k,l}{\sum}    \frac{\bigg| \textbf{I} \bigotimes \bigg( \frac{\pm\big( B_{kl} \wedge B_{k^{\prime} l^{\prime}} \big)  + \big( B_{lk} \wedge B_{l^{\prime} k^{\prime}} \big)}{\sqrt{2}} \bigg)    \bigg| }{ \bigg| \textbf{I} \bigotimes \bigg( \frac{\pm\big( B_{kl} \wedge B_{k^{\prime} l^{\prime}} \big)  + \big( B_{lk} \wedge B_{l^{\prime} k^{\prime}} \big)}{\big| \pm\big( B_{kl} \wedge B_{k^{\prime} l^{\prime}} \big)  + \big( B_{lk} \wedge B_{l^{\prime} k^{\prime}} \big) \big| } \bigg)  \bigg| }      \bigg[   \frac{\big( B_{kl} \wedge B_{k^{\prime} l^{\prime}} \big)  + \big( B_{lk} \wedge B_{l^{\prime} k^{\prime}} \big) }{\big( B_{kl} \wedge B_{k^{\prime} l^{\prime}} \big)  + \big( B_{lk} \wedge B_{l^{\prime} k^{\prime}} \big) }   \bigg]  \\ \approx    \underset{\text{Questions }k,l,k^{\prime},l^{\prime}}{\sum}  \bigg[   \frac{\big| \pm \big( B_{kl} \wedge B_{k^{\prime} l^{\prime}} \big)  + \big( B_{lk} \wedge B_{l^{\prime} k^{\prime}} \big)  \big|}{\sqrt{2}} \bigg]    \text{, }
\end{align*}

\noindent Hence, the numerator of the previous expression,

\begin{align*}
 \big| \pm \big( B_{kl} \wedge B_{k^{\prime} l^{\prime}} \big)  + \big( B_{lk} \wedge B_{l^{\prime} k^{\prime}} \big)  \big|       \text{, }
\end{align*}

\noindent can be used to obtain the desired upper bound, from upper bounding the product norm,

\begin{align*}
           \bigg| \bigg|  \bigg[        \bigg( \bigg( \big( A_k \wedge A_{k^{\prime}} \big)  \otimes \textbf{I} \bigg) + \bigg( \textbf{I} \otimes   \bigg(   \frac{\pm \big( B_{kl} \wedge B_{k^{\prime}l^{\prime}} \big) + \big( B_{lk} \wedge B_{l^{\prime}k^{\prime}}\big) }{\sqrt{2}}    \bigg) \bigg)  \\ \times \bigg(   \bigg( \big( A_k \wedge A_{k^{\prime}} \big)   \otimes \textbf{I} \bigg) - \bigg( \textbf{I}  \otimes  \bigg( \frac{\pm \big( B_{kl} \wedge B_{k^{\prime}l^{\prime}} \big) + \big( B_{lk} \wedge B_{l^{\prime}k^{\prime}}\big) }{\sqrt{2}} \bigg)        \bigg)     \bigg]  \ket{\psi_{\mathrm{Expanded}\wedge\mathrm{Expanded}}}                    \bigg| \bigg| \\    <    \bigg| \bigg|  \bigg[        \bigg( \bigg( \big(  A_k \wedge A_{k^{\prime}} \big)  \otimes \textbf{I} \bigg) + \bigg( \textbf{I}  \otimes   \bigg(   \frac{\pm \big( B_{kl} \wedge B_{k^{\prime}l^{\prime}} \big) + \big( B_{lk} \wedge B_{l^{\prime}k^{\prime}}\big)}{\sqrt{2}}    \bigg) \bigg) \bigg|  \bigg| \\ \times  
        \bigg| \bigg|  \bigg(   \bigg( \big(  A_k \wedge A_{k^{\prime}} \big)  \otimes \textbf{I} \bigg)  \\ -  \bigg( \textbf{I}  \otimes \bigg( \frac{\pm \big( B_{kl} \wedge B_{k^{\prime}l^{\prime}} \big) + \big( B_{lk} \wedge B_{l^{\prime}k^{\prime}}\big)}{\sqrt{2}} \bigg)        \bigg)     \bigg] \ket{\psi_{\mathrm{Expanded}\wedge\mathrm{Expanded}}}                    \bigg| \bigg|     \text{, }
\end{align*}

\noindent from which we conclude the argument, as the upper bound to the above expression takes the form,

\begin{align*}
         3 \big( \omega_{\mathrm{Expanded} \wedge \mathrm{Expanded}} \big)^{-2} \big( 1 + \big( \omega_{\mathrm{Expanded}\wedge \mathrm{Expanded}} \big)^{-2} \big) \sqrt{n\epsilon^{\wedge}}     \equiv 3 \big( \frac{3}{2} \big) \big( 1 + \frac{3}{2} \big) \sqrt{n \epsilon^{\wedge}} \\ < 20 \sqrt{N \epsilon^{\wedge}}             \text{. } \boxed{}
\end{align*}

\subsection{$\sqrt{\epsilon}$ upper bound for an arbitrary number of parallel repetition operations}

\noindent \textit{Proof of Lemma $\textit{7}$}. Given observations in the computations for the previous result, in the $N$-player setting,

\begin{align*}
    \underset{\text{Questions }k,l}{\sum}    \frac{\bigg| \textbf{I} \bigotimes \bigg( \frac{\pm B_{kl} + B_{lk}}{\sqrt{2}} \bigg)  \bigotimes \bigg( \underset{1 \leq k^{\prime} \leq N-2}{\bigotimes} \textbf{I}_{k^{\prime}} \bigg)    \bigg| }{ \bigg| \textbf{I} \bigotimes \bigg( \frac{\pm B_{kl} + B_{lk}}{\big| \pm B_{kl} + B_{lk} \big| } \bigg)  \bigotimes \bigg( \underset{1 \leq k^{\prime} \leq N-2}{\bigotimes} \textbf{I}_{k^{\prime}} \bigg)    \bigg| } \\ \equiv    \underset{\text{Questions }k,l}{\sum}    \frac{\bigg| \textbf{I} \bigotimes \bigg( \frac{\pm B_{kl} + B_{lk}}{\sqrt{2}} \bigg)  \bigotimes \bigg( \underset{1 \leq k^{\prime} \leq N-2}{\bigotimes} \textbf{I}_{k^{\prime}} \bigg)    \bigg| }{ \bigg| \textbf{I} \bigotimes \bigg( \frac{\pm B_{kl} + B_{lk}}{\big| \pm B_{kl} + B_{lk} \big| } \bigg)  \bigotimes \bigg( \underset{1 \leq k^{\prime} \leq N-2}{\bigotimes} \textbf{I}_{k^{\prime}} \bigg)    \bigg| }    \bigg[   \frac{\big| \pm B_{kl} + B_{lk} \big|}{\big| \pm B_{kl} + B_{lk} \big|}   \bigg]  \\ \approx    \underset{\text{Questions }k,l}{\sum}  \bigg[   \frac{\big| \pm B_{kl} + B_{lk} \big|}{\sqrt{2}} \bigg]    \text{, }
\end{align*}

\noindent the desired upper bound,

\begin{align*}
  20 N \sqrt{N \epsilon^{\wedge}_{{\mathrm{Expanded}\wedge \cdots \wedge \mathrm{Expanded}}}}  \text{, }
\end{align*}

\noindent follows from the previous arguments, in which $\mathcal{I}_1, \cdots, \mathcal{I}_N$ can each be individually upper bounded by,

\begin{align*}
    20 \sqrt{N \epsilon^{\wedge}_{{\mathrm{Expanded} \wedge \cdots \wedge \mathrm{Expanded}}}} \text{, }
\end{align*}

\noindent corresponding to the maximum contribution that \textit{each player} can contribute to the upper bound with a factor of $20N$, after multiplying the following collection of suitable factors,

\begin{align*}
 \mathcal{I}^{\prime}_1 \equiv \bigg|   \frac{\big( B_{kl} \wedge B_{k^{\prime} l^{\prime}} \wedge \cdots \wedge B_{k^{\prime\cdots\prime}l^{\prime\cdots\prime}} \big)  + \big( B_{lk} \wedge B_{l^{\prime} k^{\prime}} \wedge \cdots \wedge B_{l^{\prime\cdots\prime}k^{\prime\cdots\prime}} \big) }{ \pm \big( B_{kl} \wedge B_{k^{\prime} l^{\prime}} \wedge \cdots \wedge B_{k^{\prime\cdots\prime}l^{\prime\cdots\prime}}\big)  + \big( B_{lk} \wedge B_{l^{\prime} k^{\prime}} \wedge \cdots \wedge B_{l^{\prime\cdots\prime}k^{\prime\cdots\prime}}   \big)  } \bigg|    \\    \equiv      \frac{\bigg| \big( B_{kl} \wedge B_{k^{\prime} l^{\prime}} \wedge \cdots \wedge B_{k^{\prime\cdots\prime}l^{\prime\cdots\prime}} \big)  + \big( B_{lk} \wedge B_{l^{\prime} k^{\prime}} \wedge \cdots \wedge B_{l^{\prime\cdots\prime}k^{\prime\cdots\prime}} \big)\bigg| }{ \bigg|  \pm \big( B_{kl} \wedge B_{k^{\prime} l^{\prime}} \wedge \cdots \wedge B_{k^{\prime\cdots\prime}l^{\prime\cdots\prime}}\big)  + \big( B_{lk} \wedge B_{l^{\prime} k^{\prime}} \wedge \cdots \wedge B_{l^{\prime\cdots\prime}k^{\prime\cdots\prime}}   \big) \bigg| }      \end{align*}

 \begin{align*} \Longleftrightarrow    \bigg|  \pm \big( B_{kl} \wedge B_{k^{\prime} l^{\prime}} \wedge \cdots \wedge B_{k^{\prime\cdots\prime}l^{\prime\cdots\prime}}\big)  + \big( B_{lk} \wedge B_{l^{\prime} k^{\prime}} \wedge \cdots \\ \wedge B_{l^{\prime\cdots\prime}k^{\prime\cdots\prime}}   \big) \bigg|  \neq 0   \text{, } \\ \vdots \\  \mathcal{I}^{\prime}_N \equiv               \bigg|     \frac{ \mathcal{I}^1_N  + \mathcal{I}^2_N}{ \pm \mathcal{I}^1_N + \mathcal{I}^2_N   }  \bigg| \equiv               \frac{  \big|  \mathcal{I}^1_N  + \mathcal{I}^2_N  \big|  }{  \big|  \pm \mathcal{I}^1_N + \mathcal{I}^2_N   \big|   }      \Longleftrightarrow   \big|  \pm \mathcal{I}^1_N + \mathcal{I}^2_N   \big|   \neq 0       \text{, }
\end{align*}

\noindent where, in the last expression,

\begin{align*}
   \mathcal{I}^1_N \equiv  \frac{1}{\sqrt{\# \sigma }}  \bigg(    \underset{\text{Permutations } \sigma}{\sum}  \big(   B^{(N-1)}_{\sigma ( i_1, \cdots, i_{N-1})}         \wedge   B^{(N-1)}_{\sigma ( i^{\prime}_1, \cdots, i^{\prime}_{N-1})}  \wedge  \cdots \wedge   B^{(N-1)}_{\sigma ( i^{\prime\cdots\prime}_1, \cdots, i^{\prime\cdots\prime}_{N-1})} \big)      \bigg) \text{, } \\    \mathcal{I}^2_N \equiv \frac{1}{\sqrt{\# \sigma^{\prime} }}  \bigg(    \underset{\text{Permutations } \sigma^{\prime}}{\sum}  \big(   B^{(N-1)}_{\sigma^{\prime} ( i_1, \cdots, i_{N-1})}         \wedge   B^{(N-1)}_{\sigma^{\prime} ( i^{\prime}_1, \cdots, i^{\prime}_{N-1})}  \wedge  \cdots \wedge   B^{(N-1)}_{\sigma^{\prime} ( i^{\prime\cdots\prime}_1, \cdots, i^{\prime\cdots\prime}_{N-1})} \big)      \bigg)
\end{align*}

\noindent for upper bounding each $\mathcal{I}_j$, for $1 \leq j \leq N$, provided in the statement of the result,

\begin{align*}
  \bigg| \bigg|               \bigg( \big(  A_k  \wedge A_{k^{\prime}} \wedge \cdots \wedge A_{k^{\prime\cdots \prime}} \big) \bigotimes  \bigg( \underset{1 \leq z \leq N-1}{\bigotimes}\textbf{I}_z \bigg)  \bigg) \ket{\psi_{\mathrm{Expanded} \wedge \cdots \wedge \mathrm{Expanded}}}    \\  -  \bigg( \textbf{I}  \bigotimes \bigg(     \frac{\pm \big(  B_{kl} \wedge B_{k^{\prime} l^{\prime}} \wedge \cdots \wedge B_{k^{\prime\cdots\prime}l^{\prime\cdots \prime}}  \big)  + \big( B_{lk} \wedge B_{ l^{\prime} k^{\prime}} \wedge \cdots \wedge B_{l^{\prime\cdots \prime} k^{\prime\cdots\prime} }  \big) }{\big| \pm \big(  B_{kl} \wedge B_{k^{\prime} l^{\prime}} \wedge \cdots \wedge B_{k^{\prime\cdots\prime}l^{\prime\cdots \prime}}  \big)  + \big( B_{lk} \wedge B_{ l^{\prime} k^{\prime}} \wedge \cdots \wedge B_{l^{\prime\cdots \prime} k^{\prime\cdots\prime} }  \big) \big| }    \bigg) \\  \bigotimes  \bigg( \underset{1 \leq z \leq N-2}{\bigotimes}\textbf{I}_z \bigg)       \bigg) \ket{\psi_{\mathrm{Expanded} \wedge \cdots \wedge \mathrm{Expanded}}}             \bigg| \bigg| \\  \vdots  \end{align*}

  \begin{align*}  \bigg| \bigg|    \bigg(   \bigg( \underset{1 \leq z \leq N-2}{\bigotimes} \textbf{I}_z \bigg) \bigotimes   \frac{1}{\sqrt{\# \sigma^{\prime} }}  \bigg(    \underset{\text{Permutations } \sigma^{\prime}}{\sum}  \big(   B^{(N-1)}_{\sigma^{\prime} ( i_1, \cdots, i_{N-1})}         \wedge   B^{(N-1)}_{\sigma^{\prime} ( i^{\prime}_1, \cdots, i^{\prime}_{N-1})}  \wedge  \cdots \\ \wedge   B^{(N-1)}_{\sigma^{\prime} ( i^{\prime\cdots\prime}_1, \cdots, i^{\prime\cdots\prime}_{N-1})} \big)      \bigg)     \bigotimes \textbf{I}  \bigg)     \ket{\psi_{\mathrm{Expanded} \wedge \cdots \wedge \mathrm{Expanded}}}  \\ -  \bigg( \bigg( \underset{1 \leq z \leq N-1}{\bigotimes} \textbf{I}_z \bigg) \bigotimes \frac{1}{\sqrt{\# \sigma }}  \bigg(    \underset{\text{Permutations } \sigma}{\sum}  \big(   B^{(N-1)}_{\sigma ( i_1, \cdots, i_{N-1})}         \wedge   B^{(N-1)}_{\sigma ( i^{\prime}_1, \cdots, i^{\prime}_{N-1})}  \wedge  \cdots \\ \wedge   B^{(N-1)}_{\sigma ( i^{\prime\cdots\prime}_1, \cdots, i^{\prime\cdots\prime}_{N-1})} \big)      \bigg) \bigg) \ket{\psi_{\mathrm{Expanded} \wedge \cdots \wedge \mathrm{Expanded}}}           \bigg| \bigg|     \text{. }
\end{align*}

\noindent Therefore, to finish the computation for obtaining the desired upper bound, expressing the summation,

\begin{align*}
     \underset{\text{Questions }k,l,k^{\prime},l^{\prime}, \cdots, k^{\prime\cdots\prime}, l^{\prime\cdots \prime}}{\sum}   \bigg[   \frac{\big| \pm \big(  B_{kl} \wedge B_{k^{\prime} l^{\prime}} \wedge \cdots \wedge B_{k^{\prime\cdots\prime}l^{\prime\cdots \prime}}  \big)  + \big( B_{lk} \wedge B_{ l^{\prime} k^{\prime}} \wedge \cdots \wedge B_{l^{\prime\cdots \prime} k^{\prime\cdots\prime} }  \big)  \big|}{\sqrt{2}} \bigg]    \text{, }
\end{align*}

\noindent from the previously obtained bound for two operations of strong parallel repetition for the $\mathrm{FFL}$ game, in the setting of an arbitrary number of operations for strong parallel repetition in $\mathrm{Expanded} \wedge \cdots \wedge \mathrm{Expanded}$ games, takes the desired form from the observation,

\begin{align*}
  \underset{\# \text{ Player Observables}}{\sum}  \underset{\text{Players}}{\mathrm{sup}} \big( \mathcal{I}_j \big) <   \underset{\text{Players}}{\mathrm{sup}} \bigg\{    \underset{1 \leq j \leq \big( \# \text{ Player Observables} \big) }{\sum}  \mathcal{I}_j \bigg\}   <    N \underset{1 \leq j \leq N}{\mathrm{sup}} \big\{ \mathcal{I}_j \big\}    \\     <   20 N  \sqrt{N \epsilon^{\wedge}_{{\mathrm{Expanded}\wedge \cdots \wedge \mathrm{Expanded}}}} \text{, }
\end{align*}

\noindent  from the summation over player observables,

\begin{align*}
\underset{\# \text{ Player Observables}}{\sum} \mathcal{I}_j  \equiv \underset{1 \leq j \leq N}{\sum}   \mathcal{I}_j     \text{, }
\end{align*}

\noindent from which we conclude the argument. \boxed{}

\subsection{$\sqrt{\epsilon}$ upper bound for two operations of parallel repetition}

\noindent \textit{Proof of Lemma 6}. Directly apply the previous argument, for two operations of parallel repetition instead of an arbitrary number of operations of parallel repetition, from which we conclude the argument. \boxed{}

\subsection{$\epsilon$ approximality of the bias}

\noindent \textit{Proof of Lemma 7}. The results for each of the games above follows from the fact that the corresponding primal feasible solutions for each setting,

\begin{align*}
  G_{\mathrm{Expanded}} \cdot Z_{\mathrm{Expanded}} \equiv \bra{\psi_{\mathrm{Expanded}}}  \bigg( \underset{1 \leq i \leq 3}{\bigotimes } i \text{ th player tensor observable} \bigg)                \ket{\psi_{\mathrm{Expanded}}}  \text{, } \end{align*}

\noindent satisfy the equalities,

\begin{align*}
      (\textit{1}) \equiv  \beta \big( G_{\mathrm{Expanded}} \big) \epsilon_{\mathrm{Expanded}}  - G_{\mathrm{Expanded}} Z_{\mathrm{Expanded}}   \text{, } 
\end{align*}

\noindent respectively, from which a straightforward adaptation of arguments from \textbf{Lemma} \textit{5}, [50], yield the desired result, from which we conclude the argument. \boxed{}

\subsection{Bell states}

\noindent As a generalization of the Bell states, or equivalently, the EPR pairs, arising through entanglement in tensor products of two operators in the $n\equiv 2$ Bell states for the CHSH$\big( n\big)$ game,

\begin{align*}
\bigg(  \textbf{I} \otimes \textbf{I}  \bigg) \bigg(  \frac{\ket{00} + \ket{11}}{\sqrt{2}}  \bigg) = \frac{\ket{00} + \ket{11}}{\sqrt{2}}   \text{ } \text{ , }     \bigg(     \sigma_x \otimes \textbf{I}  \bigg) \bigg( \frac{\ket{00} + \ket{11}}{\sqrt{2}} \bigg)   = \frac{\ket{10} + \ket{01}}{\sqrt{2}}   \text{, }   \\   \bigg(   \sigma_z    \otimes \textbf{I}  \bigg)  \bigg( \frac{\ket{00} + \ket{11}}{\sqrt{2}} \bigg)  = \frac{\ket{00} - \ket{11}}{\sqrt{2}}       \text{ } \text{ , }    \bigg( \sigma_x \sigma_z     \otimes   \textbf{I}   \bigg) \bigg( \frac{\ket{00}+\ket{11}}{\sqrt{2}} \bigg)  = \frac{\ket{10}- \ket{01}}{\sqrt{2}}     \text{, } 
\end{align*}

\noindent the Bell states for parallel repetition of multiplayer games can be generalized from the following collection of relations for Bell states in the 3 multiplayer game,

 \begin{align*}
 \big( \textbf{I} \otimes \textbf{I} \otimes \textbf{I} \big) \\ \times  \bigg( \frac{1}{\sqrt{3}} \bigg( \big( \ket{\text{$Player^{(1)}$ 1 state}} \wedge \cdots \wedge  \ket{\text{$Player^{(n)}$ 1 state}}  \big) + \big( \ket{\text{$Player^{(1)}$ 2 state}} \wedge \cdots \\ \wedge  \ket{\text{$Player^{(n)}$2  state}}  \big)    + \big( \ket{\text{$Player^{(1)}$ 3 state}} \wedge \cdots \wedge  \ket{\text{$Player^{(n)}$3 state}}  \big)  \bigg) \bigg)    \\  =      \frac{1}{\sqrt{3}} \bigg( \big( \ket{\text{$Player^{(1)}$ 1 state}} \wedge \cdots \wedge  \ket{\text{$Player^{(n)}$ 1 state}}  \big)   + \big( \ket{\text{$Player^{(1)}$ 2 state}} \wedge \cdots \\ \wedge  \ket{\text{$Player^{(n)}$2  state}}  \big)  + \big( \ket{\text{$Player^{(1)}$ 3 state}} \wedge \cdots   \wedge  \ket{\text{$Player^{(n)}$3 state}}  \big)  \bigg)                 \text{, } \\ \\  \big( \sigma_x  \otimes \textbf{I} \otimes \textbf{I} \big) \\     \times  \bigg( \frac{1}{\sqrt{3}} \bigg( \big( \ket{\text{$Player^{(1)}$ 1 state}} \wedge \cdots \wedge  \ket{\text{$Player^{(n)}$ 1 state}}  \big) + \big( \ket{\text{$Player^{(1)}$ 2 state}} \wedge \cdots \\ \wedge  \ket{\text{$Player^{(n)}$2  state}}  \big)  + \big( \ket{\text{$Player^{(1)}$ 3 state}} \wedge \cdots \wedge  \ket{\text{$Player^{(n)}$3 state}}  \big)  \bigg) \bigg)  \\  =              \frac{1}{\sqrt{3}} \bigg( \big( \widetilde{\ket{\text{$Player^{(1)}$ 1 state}}} \wedge \cdots \wedge  \widetilde{\ket{\text{$Player^{(n)}$ 1 state}}}  \big)      + \big( \ket{\text{$Player^{(1)}$ 2 state}} \wedge \cdots \\ \wedge  \ket{\text{$Player^{(n)}$2  state}}  \big)  + \big( \widetilde{\ket{\text{$Player^{(1)}$ 3 state}}} \wedge \cdots   \wedge  \widetilde{\ket{\text{$Player^{(n)}$3 state}}}  \big)  \bigg)     \text{, } \\ \\  \end{align*}

     \begin{align*}    \big( \textbf{I} \otimes \sigma_x \otimes \textbf{I} \big) \\  \times  \bigg( \frac{1}{\sqrt{3}} \bigg( \big( \ket{\text{$Player^{(1)}$ 1 state}} \wedge \cdots   \wedge  \ket{\text{$Player^{(n)}$ 1 state}}  \big)   + \big( \ket{\text{$Player^{(1)}$ 2 state}} \wedge \cdots  \\  \wedge  \ket{\text{$Player^{(n)}$2  state}}  \big)  + \big( \ket{\text{$Player^{(1)}$ 3 state}} \wedge \cdots   \wedge  \ket{\text{$Player^{(n)}$3 state}}  \big)  \bigg) \bigg)  \\    =           \frac{1}{\sqrt{3}} \bigg( \big( \widetilde{\ket{\text{$Player^{(1)}$ 1 state}}} \wedge \cdots \wedge  \widetilde{\ket{\text{$Player^{(n)}$ 1 state}}} \big)   + \big( \widetilde{\ket{\text{$Player^{(1)}$ 2 state}}} \wedge \cdots \\ \wedge  \widetilde{\ket{\text{$Player^{(n)}$2  state}}}  \big)  + \big( {\ket{\text{$Player^{(1)}$ 3 state}}} \wedge \cdots \wedge  {\ket{\text{$Player^{(n)}$3 state}}}  \big)  \bigg)                     \text{, } \\ \\   \big( \textbf{I} \otimes \textbf{I}  \otimes \sigma_x \big) \\   \times  \bigg( \frac{1}{\sqrt{3}} \bigg( \big( \ket{\text{$Player^{(1)}$ 1 state}} \wedge \cdots \wedge  \ket{\text{$Player^{(n)}$ 1 state}}  \big) + \big( \ket{\text{$Player^{(1)}$ 2 state}} \wedge \cdots  \\ \wedge  \ket{\text{$Player^{(n)}$2  state}}  \big)    + \big( \ket{\text{$Player^{(1)}$ 3 state}} \wedge \cdots \wedge  \ket{\text{$Player^{(n)}$3 state}}  \big)  \bigg) \bigg) \\  =    \frac{1}{\sqrt{3}} \bigg( \big( \widetilde{\ket{\text{$Player^{(1)}$ 1 state}}} \wedge \cdots \wedge  \widetilde{\ket{\text{$Player^{(n)}$ 1 state}}}  \big) + \big( \widetilde{\ket{\text{$Player^{(1)}$ 2 state}}} \wedge \cdots \\  \wedge  \widetilde{\ket{\text{$Player^{(n)}$2  state}} } \big)   + \big( \ket{\text{$Player^{(1)}$ 3 state}} \wedge \cdots \wedge  \ket{\text{$Player^{(n)}$3 state}}  \big)  \bigg)  \\ \\       \big( \sigma_z \otimes \textbf{I}  \otimes \textbf{I} \big) \\  \times  \bigg( \frac{1}{\sqrt{3}} \bigg( \big( \ket{\text{$Player^{(1)}$ 1 state}} \wedge \cdots \wedge  \ket{\text{$Player^{(n)}$ 1 state}}  \big) + \big( \ket{\text{$Player^{(1)}$ 2 state}} \wedge \cdots \\ \wedge  \ket{\text{$Player^{(n)}$2  state}}  \big)   + \big( \ket{\text{$Player^{(1)}$ 3 state}} \wedge \cdots \wedge  \ket{\text{$Player^{(n)}$3 state}}  \big)  \bigg) \bigg) \\  =    \frac{1}{\sqrt{3}} \bigg( \big( \ket{\text{$Player^{(1)}$ 1 state}} \wedge \cdots \wedge  \ket{\text{$Player^{(n)}$ 1 state}}  \big) - \big( \ket{\text{$Player^{(1)}$ 2 state}} \wedge \cdots \\ \wedge  \ket{\text{$Player^{(n)}$2  state}}  \big)    - \big( \ket{\text{$Player^{(1)}$ 3 state}} \wedge \cdots \wedge  \ket{\text{$Player^{(n)}$3 state}}  \big)  \bigg)        \text{, } \\ \\ 
    \big( \textbf{I} \otimes \sigma_z   \otimes \textbf{I} \big)  \\  \times  \bigg( \frac{1}{\sqrt{3}} \bigg( \big( \ket{\text{$Player^{(1)}$ 1 state}} \wedge \cdots \wedge  \ket{\text{$Player^{(n)}$ 1 state}}  \big) + \big( \ket{\text{$Player^{(1)}$ 2 state}} \wedge \cdots \\ \wedge  \ket{\text{$Player^{(n)}$2  state}}  \big)     + \big( \ket{\text{$Player^{(1)}$ 3 state}} \wedge \cdots \wedge  \ket{\text{$Player^{(n)}$3 state}}  \big)  \bigg) \bigg) \\  =                        \frac{1}{\sqrt{3}} \bigg( - \big( \ket{\text{$Player^{(1)}$ 1 state}} \wedge \cdots \wedge  \ket{\text{$Player^{(n)}$ 1 state}}  \big)     + \big( \ket{\text{$Player^{(1)}$ 2 state}} \wedge \cdots \\ \wedge  \ket{\text{$Player^{(n)}$2  state}}  \big)   - \big( \ket{\text{$Player^{(1)}$ 3 state}} \wedge \cdots \wedge  \ket{\text{$Player^{(n)}$3 state}}  \big)  \bigg)                                                                     \text{, }  \\ \\           \big( \textbf{I} \otimes \textbf{I}  \otimes \sigma_z  \big) \\ \times  \bigg( \frac{1}{\sqrt{3}} \bigg( \big( \ket{\text{$Player^{(1)}$ 1 state}} \wedge \cdots \wedge  \ket{\text{$Player^{(n)}$ 1 state}}  \big) + \big( \ket{\text{$Player^{(1)}$ 2 state}} \wedge \cdots \\  \wedge  \ket{\text{$Player^{(n)}$2  state}}  \big)   + \big( \ket{\text{$Player^{(1)}$ 3 state}} \wedge \cdots \wedge  \ket{\text{$Player^{(n)}$3 state}}  \big)  \bigg) \bigg)  \\  =  \frac{1}{\sqrt{3}} \bigg( - \big( \ket{\text{$Player^{(1)}$ 1 state}} \wedge \cdots \wedge  \ket{\text{$Player^{(n)}$ 1 state}}  \big)    - \big( \ket{\text{$Player^{(1)}$ 2 state}} \wedge \cdots \\ \wedge  \ket{\text{$Player^{(n)}$2  state}}  \big)   + \big( \ket{\text{$Player^{(1)}$ 3 state}} \wedge \cdots  \wedge  \ket{\text{$Player^{(n)}$3 state}}  \big)  \bigg)                   \text{, }    \\ \\        \big( \sigma_x  \otimes \textbf{I}  \otimes \sigma_z  \big)  \\  \times  \bigg( \frac{1}{\sqrt{3}} \bigg( \big( \ket{\text{$Player^{(1)}$ 1 state}} \wedge \cdots \wedge  \ket{\text{$Player^{(n)}$ 1 state}}  \big) + \big( \ket{\text{$Player^{(1)}$ 2 state}} \wedge \cdots \\ \wedge  \ket{\text{$Player^{(n)}$2  state}}  \big)     + \big( \ket{\text{$Player^{(1)}$ 3 state}} \wedge \cdots \wedge  \ket{\text{$Player^{(n)}$3 state}}  \big)  \bigg) \bigg) \\  =    \frac{1}{\sqrt{3}} \bigg(  \big( \widetilde{\ket{\text{$Player^{(1)}$ 1 state}}} \wedge \cdots \wedge  \widetilde{\ket{\text{$Player^{(n)}$ 1 state}}}  \big) + \big( \ket{\text{$Player^{(1)}$ 2 state}} \\ \wedge \cdots \wedge  \ket{\text{$Player^{(n)}$2  state}}  \big)   - \big( \widetilde{\ket{\text{$Player^{(1)}$ 3 state}}} \wedge \cdots \wedge  \widetilde{\ket{\text{$Player^{(n)}$3 state}}}  \big)  \bigg)                    \text{, }  \\ \\    \big(  \textbf{I} \otimes \sigma_x  \otimes \sigma_z  \big) \\ \times \bigg( \frac{1}{\sqrt{3}} \bigg( \big( \ket{\text{$Player^{(1)}$ 1 state}} \wedge \cdots \wedge  \ket{\text{$Player^{(n)}$ 1 state}}  \big) + \big( \ket{\text{$Player^{(1)}$ 2 state}} \wedge \cdots \\ \wedge  \ket{\text{$Player^{(n)}$2  state}}  \big)   + \big( \ket{\text{$Player^{(1)}$ 3 state}} \wedge \cdots \wedge  \ket{\text{$Player^{(n)}$3 state}}  \big)  \bigg) \bigg)  \\  =  \frac{1}{\sqrt{3}} \bigg(  \big( {\ket{\text{$Player^{(1)}$ 1 state}}} \wedge \cdots \wedge  {\ket{\text{$Player^{(n)}$ 1 state}}}  \big)    + \big( \widetilde{\ket{\text{$Player^{(1)}$ 2 state}}} \wedge \cdots \\ \wedge  \widetilde{\ket{\text{$Player^{(n)}$2  state}}}  \big)    - \big( \widetilde{\ket{\text{$Player^{(1)}$ 3 state}}} \wedge \cdots \wedge  \widetilde{\ket{\text{$Player^{(n)}$3 state}}}  \big)  \bigg) \text{, }  \\  \\ 
              \big(  \sigma_z \otimes \textbf{I} \otimes \sigma_x  \big) \\ \times \bigg( \frac{1}{\sqrt{3}} \bigg( \big( \ket{\text{$Player^{(1)}$ 1 state}} \wedge \cdots \wedge  \ket{\text{$Player^{(n)}$ 1 state}}  \big) + \big( \ket{\text{$Player^{(1)}$ 2 state}} \wedge \cdots \\ \wedge  \ket{\text{$Player^{(n)}$2  state}}  \big)   + \big( \ket{\text{$Player^{(1)}$ 3 state}} \wedge \cdots \wedge  \ket{\text{$Player^{(n)}$3 state}}  \big)  \bigg) \bigg)      \\  =  \frac{1}{\sqrt{3}} \bigg(   -   \big( \widetilde{\ket{\text{$Player^{(1)}$ 1 state}}} \wedge \cdots \wedge  \widetilde{\ket{\text{$Player^{(n)}$ 1 state}}}  \big)    + \big( \ket{\text{$Player^{(1)}$ 2 state}} \wedge \cdots \\  \wedge  \ket{\text{$Player^{(n)}$2  state}}  \big)  + \big( \widetilde{\ket{\text{$Player^{(1)}$ 3 state}}} \wedge \cdots \wedge  \widetilde{\ket{\text{$Player^{(n)}$3 state}} } \big) \bigg) \text{, }  \\ \\    \big(  \textbf{I} \otimes \sigma_z  \otimes \sigma_x  \big) \\ \times \bigg( \frac{1}{\sqrt{3}} \bigg( \big( \ket{\text{$Player^{(1)}$ 1 state}} \wedge \cdots \wedge  \ket{\text{$Player^{(n)}$ 1 state}}  \big) + \big( \ket{\text{$Player^{(1)}$ 2 state}} \wedge \cdots \\ \wedge  \ket{\text{$Player^{(n)}$2  state}}  \big)  + \big( \ket{\text{$Player^{(1)}$ 3 state}} \wedge \cdots \wedge  \ket{\text{$Player^{(n)}$3 state}}  \big)  \bigg) \bigg)     \\   =        \frac{1}{\sqrt{3}} \bigg( \big( \widetilde{\ket{\text{$Player^{(1)}$ 1 state}}} \wedge \cdots   \wedge  \widetilde{\ket{\text{$Player^{(n)}$ 1 state}}}  \big)  - \big( \ket{\text{$Player^{(1)}$ 2 state}} \wedge \cdots \\ \wedge  \ket{\text{$Player^{(n)}$2  state}}  \big)   + \big( \widetilde{\ket{\text{$Player^{(1)}$ 3 state}}} \wedge \cdots    \wedge  \widetilde{\ket{\text{$Player^{(n)}$3 state}}}  \big)  \bigg)     \text{. }                     
\end{align*}

\section{References}

\noindent [1] Amr, A., Villanueva, I. Quantum one way vs. classical two way communication in XOR games. \textit{Quantum Information Processing} \textbf{20}, 79 (2021).

\bigskip

\noindent [2] Bavarian, M. Parallel repetition of Multi-party and Quantum Games via Anchoring and Fortification. PhD Thesis, Massachussetts Institute of Technology (2017).

\bigskip

\noindent [3] Bavarian, M., Vidick, T., Yuen, H. Anchored Parallel Repetition for Nonlocal Games. \textit{SIAM Journal on Computing} \textbf{51}:2 (2022). https://doi.org/10.1137/21M1405927.

\bigskip

\noindent [4] Bavarian, M., Vidick, T., Yuen, H. Hardness amplification for entangled games via anchoring. \textit{STOC 2017: Proceedings of the 49th Annual ACM SIGACT Symposium on Theory of Computing}, 303-316. https://doi.org/10.1145/3055399.3055433.

\bigskip

\noindent [5] Briet, J. Buhrman, H., Toner, B. A generalized Grothendieck inequality and entanglement in XOR games. \textit{Comm. Math. Phys.} \textbf{305}: 827-843 (2011). $\mathrm{https://doi.org/ 10.1007/s00220-011-1280-3}$.

\bigskip

\noindent [6] Broadbent, A., Methot, A.A. On the power of non-local boxes. \textit{Theoretical Computer Science} \textbf{358}: 3-14 (2006). $\mathrm{
https://doi.org/10.1016/j.tcs.2005.08.035}$.

\bigskip

\noindent [7] Brassard, G., Broadbent, A., Tapp, A. Quantum Pseudo-Telepathy. \textit{Found. Phys.} \textbf{35}: 1877-1907 (2005). $\mathrm{https://philpapers.org/rec/BRAQP}$.

\bigskip

\noindent [8] Benedetti, M. and Coyle, B. and Fiorentini, M. and Lubasch, M. and Rosenkranz, M. Variational Inference with a Quantum Computer. \textit{Phys Rev Applied} 16: 044057 (2021) https://doi.org/10.1103/PhysRevApplied

\noindent .16.044057. 

\bigskip



\noindent [9] Bittel, L. and Kliesch, M. Training Variational Quantum Algorithms is NP-Hard. \textit{Physical Review Letters} 127: 120502 (2021). https://doi.org/10.1103/PhysRevLett.127.120502



\bigskip

\noindent [10] Catani, L. and Faleiro, R. and Emeriau,P.E. and Mansfield,S. and Pappa, A. Connecting XOR and XOR* games. \textit{Phys. Rev. A.} 109: 012427 (2024). https://doi.org/10.1103/PhysRevA.109.012427. 



\bigskip

\noindent [11] Chen, H. and Vives, M. and Metcalf, M. Parametric amplification of an optomechanical quantum interconnect. \textit{Physical Review Research} 4: 043119 (2022). https://doi.org/10.1103/PhysRevResearch.4.043119



\bigskip

\noindent [12] Cong, I. and Duan, L. Quantum discriminant analysis for dimensionality reduction and classification. \textit{New Journal of Physics} 18: 073011 (2016). https://doi.org/10.1088/1367-2630/18/7/073011. 


\bigskip

\noindent [13] Cleve, R., Hoyer, P., Toner, B., Watrous, J. Consequences and Limits of Nonlocal Strategies. \textit{19th IEEE Annual Conference on Computational Complexity Proceedings}: 236-249 (2004).

$\mathrm{https://doi.org/10.1109}$$\mathrm{/CCC.2004.1313847}$.

\bigskip

\noindent [14] Culf, E., Mousavi, H., and Spirig, T. "Approximation Algorithms for Noncommutative CSPs," 2024 \textit{IEEE 65th Annual Symposium on Foundations of Computer Science (FOCS)}, 920-929 (2024). $\mathrm{https://doi.org}$ $\mathrm{/ 10.1109/FOCS61266.2024.00061}$.

\bigskip

\noindent [15] Cui, D., Malavolta, G., Mehta, A., Natarajan, A., Paddock, C., Schmidt, S., Walter, M., Zhang, T. A Computational Tsireslson's Theorem for the Value of Compiled XOR games. \textit{arXiv: 2402.17301} (2024).

\bigskip

\noindent [16] Dinur, I., Harsha, P., Venkat, R., Yuen, H. Multiplayer Parallel Repetition for Expanding Games. 8th Innovations in Theoretical Computer Science Conference (ITCS 2017). Leibniz International Proceedings in Informatics (LIPIcs), Volume 67, pp. 37:1-37:16, Schloss Dagstuhl – Leibniz-Zentrum für Informatik (2017). https://doi.org/10.4230/LIPIcs.ITCS.2017.37.

\bigskip

\noindent [17] Doherty, A.C., Liang, Y.C., Toner, B., Wehner, S. The Quantum Moment Problem and Bounds on Entangled Multi-Prover Games. \textit{23rd Annual IEEE Conference on Computational Complexity} \textbf{8}: 1093-0159/08 (2018).

\bigskip

\noindent [18] Drmota, P., Main, D., Ainley, E.M., Agrawal, A., Araneda, G., Nadlinger, Srinivas, R., Cabello, A.,  et al. Experimental Quantum Advantage in the Odd-Cycle Game. \textit{Phys. Rev. Lett.} \textbf{134}: 070201 (2025).
$\mathrm{https://doi.org/10.1103/PhysRevLett.134.070201}$.


\bigskip

\noindent [19] Ewe, W-B. and Koh, D. E. and Goh, S. T. and Chu, H-S and Png, C. E. Variational Quantum-Based Simulation of Waveguide Modes. \textit{IEEE Transactions on Microwave Theory and Techniques} 70 (5): 2517-2525 (2022). https://doi.org/10.1109/TMTT.2022.3151510.

\bigskip

\noindent [20] Pierre-Emmanuel Emeriau, P-E., Howard, M. and Mansfield, S. Quantum Advantage in Information Retrieval. \textit{PRX Quantum} \textbf{3}, 020307 (2022). $\mathrm{ https://doi.org/10.1103/PRXQuantum.3.020307}$.



\bigskip

\noindent [21] Faleiro, R. Quantum strategies for simple 2-player XOR games. \textit{Quantum Inf Process} \textbf{19}, 229 (2020). $\mathrm{doi:10.1007/s11128-020-02717-2}$

\bigskip

\noindent [22] Garg, D. and Ikbal, S. and Srivastava, S.K. and Vishwakarma, H. and Karanam, H. and Subramaniam, L.V. Quantum Embedding of Knowledge for Reasoning. \textit{Advance in Neural Information Processing Systems} 32 (2019). 

\noindent $\mathrm{https://papers.nips.cc/paper_files/paper/2019/hash/cb12d7f933e7d102c52231bf62b8a678-Abstract.html}$.


\bigskip

\noindent [23] Genoni, M.G. and Tufarelli, T. Non-orthogonal bases for quantum metrology. \textit{Journal of Physics A: Mathematical and Theoretical} 52: 43 (2019). https://doi.org/10.1088/1751-8121/ab3fe0.


\bigskip


\noindent [24] Gidi, J.A. and Candia, B. and Munoz-Moller, A.D. and Rojas, A. and Pereira, L. and Munoz, M. and Zambrano, L. and Delgado, A. Stochastic optimization algorithms for quantum applications. \textit{Phys.Rev.A} 108: 032409 (2023). https://doi.org/10.1103/PhysRevA.108.032409. 



\bigskip

\noindent [25] Givi, P. and Daley, A.J. and Mavriplis, D. and Malik, M. Quantum Speedup for Aeroscience and Engineering. \textit{AIAA} 58:8 (2020). 

https://ntrs.nasa.gov/api/citations/20200003505/downloads/20200003505.pdf.


\bigskip

\noindent [26] Helton, J.W., Mousavi, H., Nezhadi, S.S. et al. Synchronous Values of Games. \textit{Ann. Henri Poincaré} \textbf{25}, 4357–4397 (2024). https://doi.org/10.1007/s00023-024-01426-1

\bigskip

\noindent [27] Hadiashar, S.B. and Nayak, A. and Sinha, P. Optimal lower bounds for Quantum Learning via Information Theory. \textit{IEEE Transactions on Information Theory} 70(3): 1876--1896 (2024). https://doi.org/10.1109/

\noindent TIT.2023.3324527. 



\bigskip

\noindent [28] Hur, T. and Kim, L. and Park, D.K. Quantum convolutional neural network for classical data classification. \textit{Quantum Machine Intelligence} 4: 3 (2022). https://doi.org/10.1007/s42484-021-00061-x. 



\bigskip

\noindent [29] Holmes, Z. and Coble, N.J. and Sornborger, A.T. and Subasi, Y. On nonlinear transformations in quantum computation. \textit{Phys. Rev. Research} 5: 013105 (2023). https://doi.org/10.1103/PhysRevResearch.5.013105. 



\bigskip

\noindent [30] Jain, R., Kundu, S. A Direct Product Theorem for One-Way Quantum Communication. \textit{36 th Computational Complexity Conference (CCC 2021)} \textbf{27}, 1-28 (2021). https://doi.org/10.48550/arXiv.2008.08963.

\bigskip

\noindent [31] Jing, H. and Wang, Y. and Li, Y. Data-Driven Quantum Approximate Optimization Algorithm for Cyber-Physical Power Systems. \textit{arXiv}: 2204.00738 (2022). https://doi.org/10.48550/arXiv.2204.00738.


\bigskip

\noindent [32] Junge, M., Palazuelos, C. On the power of quantum entanglement in multipartite quantum XOR games. \textit{Journal of the London Mathematical Society} \textbf{110}(5) (2024).

\bigskip

\noindent [33] Kubo, K. and Nakagawa, Y.O. and Endo, S. and Nagayama, S. Variational quantum simulations of stochastic differential equations. \textit{Physical Review A} 103: 052425 (2021). https://doi.org/10.1103/PhysRevA.

\noindent 103.052425.


\bigskip


\noindent [34] Kribs, D.W. A quantum computing primer for operator theorists. \textit{Linear Algebra and its Applications} 400: 147-167 (2005). https://doi.org/10.48550/arXiv.math/0404553.


\bigskip

\noindent [35] Li, R. Y. and Di Felice, R. and Rohs, R. and Lidar, D.A. Quantum annealing versus classical machine learning applied to a simplied computational biology problem. \textit{NPJ Quantum Information} 4: 14 (2008). https://doi.org/10.1038/s41534-018-0060-8. 


%

\bigskip

\noindent [36] Mahdian, M. and Yeganeh, H.D. Toward a quantum computing algorithm to quantify classical and quantum correlation of system states. \textit{Quantum Information Processing} 20: 393 (2021). https://doi.org/10.1007/

\noindent s11128-021-03331-6.


\bigskip

\noindent [37] Maldonado, T.J. and Flick, J. and Krastanov, S. and Galda, A. Error rate reduction of single-qubit gates via noise-aware decomposition into native gates. \textit{Scientific Reports} 12: 6379 (2022). https://doi.org/10.1038

\noindent /s41598-022-10339-0.



\bigskip

\noindent [38] Manby, F.R. and Stella, M. and Goodpaster, J.D. and Miller, T.F. A Simple, Exact Density-Functional-Theory Embedding Scheme. \textit{Journal of Chemical Theory and Computation} 8 (8): 2564-2568 (2012). https://doi.org/10.1021/ct300544e.


\bigskip


\noindent [39] Marwah, A., Dupuis, F. Security proof for parallel DIQKD. \textit{arXiv: 2507.03991} (2025). https://doi.org/10.48550/arXiv.2507.03991.

\bigskip

\noindent [40] Mensa, S. and Sahin, E. and Tacchino, F. and Barkoutsos, P.K. and Tavernelli, I. Quantum Machine Learning Framework for Virtual Screening in Drug Discovery: a Prospective Quantum Advantage. \textit{Mach. Learn.: Sci. Technol.} 4: 015023 (2023) https://doi.org/10.1088/2632-2153/acb900.


\bigskip


\noindent [41] Nan Sheng, H.M. and Govono, M. and Galli, G. Quantum Embedding Theory for Strongly-Correlated States in Materials. \textit{J. Chem. Theory Comput.} 17 (4): 2116-2125 (2021). https://doi.org/10.1021/acs.jctc.

\noindent 0c01258. 


\bigskip

\noindent [42] Ostrev, D. The structure of nearly-optimal quantum strategies for the $\mathrm{CHSH(n)}$ XOR games. \textit{Quantum Information and Computation} 16 (13-14): 1191-1211 (2016). https://doi.org/10.26421/QIC16.13-14-6.


\bigskip

\noindent [43] Ostrev, D. Composable, Unconditionally Secure Message Authentication without any Secret Key. \textit{IEEE International Symposium on Information Theory} \textbf{10}: 1109, 622-626 (2019).https://doi.org/10.1109/ISIT.2019.8849510"

\bigskip


\noindent [44] Paine, A.E. and Elfving, V.E. and Kyriienko, O. Quantum Kernel Methods for Solving Differential Equations. \textit{Physical Review A} 107: 032428 (2023). https://doi.org/10.1103/PhysRevA.107.032428. 


\bigskip


\noindent [45] Paudel, H.P., Syamlal, M., Crawford, S.E., Lee, Y-L, Shugayev, R.A., Lu, P., Ohodnicki, P.R., Mollot, D., Duan, Y. \textit{Quantum Computing and Simulations for Energy Applications: Review and Perspective. ACS Eng. Au}: 3 151-196 (2022). $\mathrm{https://doi.org/10.1021/ac}$$\mathrm{sengineeringau.1c00033}$.


\bigskip


\noindent [46] Przhiyalkovskiy, Y.V. Quantum process in probability representation of quantum mechanics. \textit{Journal of Physics A: Mathematical and Theoretical} 55: 085301 (2022). https://doi.org/10.1088/1751-8121/ac4b15.


\bigskip

\bigskip

\noindent [47] Perc, M. Statistical physics of human cooperation. \textit{Physics Reports} \textbf{687}: 1-51 (2017). $\mathrm{https://papers.ssrn}$ $\mathrm{.com/sol3/papers.cfm?abstract_id=2972841}$.

\bigskip

\noindent [48] 
Oded, R. and Vidick, T. 2015. Quantum XOR Games. \textit{ACM Transactions on Computation Theory} ]\textbf{7}(4): Art. No. 15. https://doi.org/10.1145/2799560.

\bigskip

\noindent [49] Ravishankar Ramanathan, R., Augusiak, R., and Murta, G. Generalized XOR games with $d$ outcomes and the task of nonlocal computation. \textit{Phys. Rev. A} \textbf{93}, 022333 (2016). $\mathrm{https://doi.org/10.1103/PhysRevA}$ $\mathrm{.93.022333}$.

\bigskip

\noindent [50] Rigas, P. Optimal, and approximately optimal, quantum strategies for $\mathrm{XOR^{*}}$ and $\mathrm{FFL}$ games. \textit{arXiv: 2311.12887} (2023), submitted.

\bigskip


\noindent [51] Rigas, P. Variational quantum algorithm for measurement extraction from the Navier-Stokes, Einstein, Maxwell, B-type, Lin-Tsien, Camassa-Holm, DSW, H-S, KdV-B, non-homogeneous KdV, generalized KdV, KdV, translational KdV, sKdV, B-L and Airy equations, shortened version including simulation results from only 3 PDEs submitted.


\bigskip

\noindent [52] Rigas, P. Quantum error bounds, optimality, and duality gaps for multiplayer XOR, XOR*, compiled XOR, XOR*, and strong parallel repetition of XOR, XOR*, and FFL games. Submitted (2025).

\bigskip

\bigskip

\noindent [53] Rigas, P. Error correction, authentication, and false acceptance, probabilities for communication over noisy quantum channels: converse upper bounds on the bit transmission rate. \textit{arXiv: 	arXiv:2507.03035}. Submitted (2025).

\bigskip


\noindent [54] van Dam, W. and Sasaki, Y. Quantum Algorithms for Problems in Number Theory, Algebraic Geometry, and Group Theory. \textit{Diversities in Quantum Computation and Quantum Information}: 79-105 (2012). https://doi.org/10.1142/9789814425988$\mathrm{-}$0003.


\bigskip


\noindent [55] Wang, Y. and Krstic, P.S. Multistate Transition Dynamics by Strong Time-Dependent Perturbation in NISQ era. \textit{J. Phys.} Commun.7: 075004 (2023). https://doi.org/10.1088/2399-6528/ace67a.

\bigskip

\noindent [56] Yuen, H. A Parallel Repetition Theorem for All Entangled Games. \textit{43 rd International Colloqium on Automata, Languages, and Programming (ICALP 2016)} \textbf{77}, 1-13 (2016). https://doi.org/10.48550/arXiv.1604.04340.


\bigskip


\noindent [57] Zhao, L. and Zhao, Z. and Rebentrost, P. and Fitzsimons, J. Compiling basic linear algebra subroutines for quantum computers,
 \textit{Quantum Machine Intelligence} 3: 21 (2021). https://doi.org/10.1007/s42484-021-00048-8"



\end{document}